\documentclass[journal]{IEEEtran}
\usepackage{multirow}
\usepackage{color}

\def\L{{\mathcal L}}

\usepackage{ifpdf}
\usepackage{cite}
\ifCLASSINFOpdf
   \usepackage[pdftex]{graphicx}
\else
\fi
\usepackage{amsmath}
\usepackage{amssymb}
\usepackage{array}

\usepackage[justification=centering]{caption}

\ifCLASSOPTIONcompsoc
  \usepackage[caption=false,font=normalsize,labelfont=sf,textfont=sf]{subfig}
\else
  \usepackage[caption=false,font=footnotesize]{subfig}
\fi
\usepackage{fixltx2e}

\usepackage{stfloats}

\ifCLASSOPTIONcaptionsoff
  \usepackage[nomarkers]{endfloat}
 \let\MYoriglatexcaption\caption
 \renewcommand{\caption}[2][\relax]{\MYoriglatexcaption[#2]{#2}}
\fi
\usepackage{url}

\begin{document}
\title{Extended (Conventional) Co-Prime Arrays and Difference Set Analysis: Low Latency Approach}
\author{Usham~V.~Dias\\ Department of Electrical Engineering, Indian Institute of Technology Delhi, India.
}%

\maketitle
\begin{abstract}
The co-prime array is a sub-Nyquist acquisition scheme for the estimation of second order statistics. It cannot generate all the difference values in the co-prime range and hence, one of the sub-array is extended to enable the estimation of the second order statistics at each difference value in the co-prime range. Recently, the difference set for the co-prime array was studied and low latency temporal spectrum estimation was demonstrated. 
In this paper, the fundamentals of the difference set of the extended co-prime array, also known as the conventional co-prime array, is developed. The closed-form equations for the weight function, correlogram bias window, and the variance are described. This is provided for the entire difference set, continuous difference set and for the prototype co-prime period. It is shown that the choice of $M\approx \frac{N}{2}$ generates a bias function with a large relative amplitude between the main-lobe and side-lobe peaks, where $M$ and $N$ are co-prime pairs. The expressions for the number of multiplications and additions required to compute the autocorrelation for the entire, continuous, and prototype range is derived. Simulation results demonstrate low latency spectrum estimation using the correlogram method.
\end{abstract}


\IEEEpeerreviewmaketitle

\section{Introduction}
\IEEEPARstart{C}{o-prime} sensing scheme provides a framework for sub-Nyquist acquisition and estimation of the second order statistics of a signal~\cite{4.7}. It has several applications which include direction of arrival estimation~\cite{4.45, F2, F3,F5}, power spectrum estimation~\cite{4.28,4.32}, system identification~\cite{4.40}, beamforming~\cite{F1,F4, 16.1}, cross-correlation estimation, time-delay, range, velocity, and acceleration estimation~\cite{U_S_NCC2018}.

Low latency co-prime estimation was reported for modal analysis~\cite{I1} and frequency estimation~\cite{S1,U_S_1}. A detailed analysis of correlogram spectral estimation method with low latency is considered in~\cite{UVD_PHD}. It will form the basis for the discussions on autocorrelation and correlogram estimation in this paper.

However, the co-prime array cannot generate a difference set containing all the elements in the co-prime range $[-MN+1, MN-1]$ and hence, an extended co-prime array was proposed in~\cite{4.62}. It has one of the sub-arrays extended by an additional co-prime period. The extended co-prime array generates all the values in the co-prime range but this range does not represent the largest continuous range. The weight function of the extended co-prime array is given in Table IV of~\cite{4.68} for $M=4$ and $N=5$, but does not provide a general expression for arbitrary values of $M$ and $N$. This paper develops the fundamentals of the difference set for the extended (also known as conventional) co-prime array and demonstrates low latency estimation. The closed-form expressions for the weight function and the bias window for the correlogram method are also provided. It may be noted that both the samplers (instead of only one) can be extended for two or more periods. This concept is referred to as the prototype co-prime sampling with multiple periods and is well studied for low latency estimation~\cite{UVD_PHD}.

\section{Extended Co-Prime Concept}
\begin{figure}[!t]
	\centering
	\includegraphics[width=0.45\textwidth]{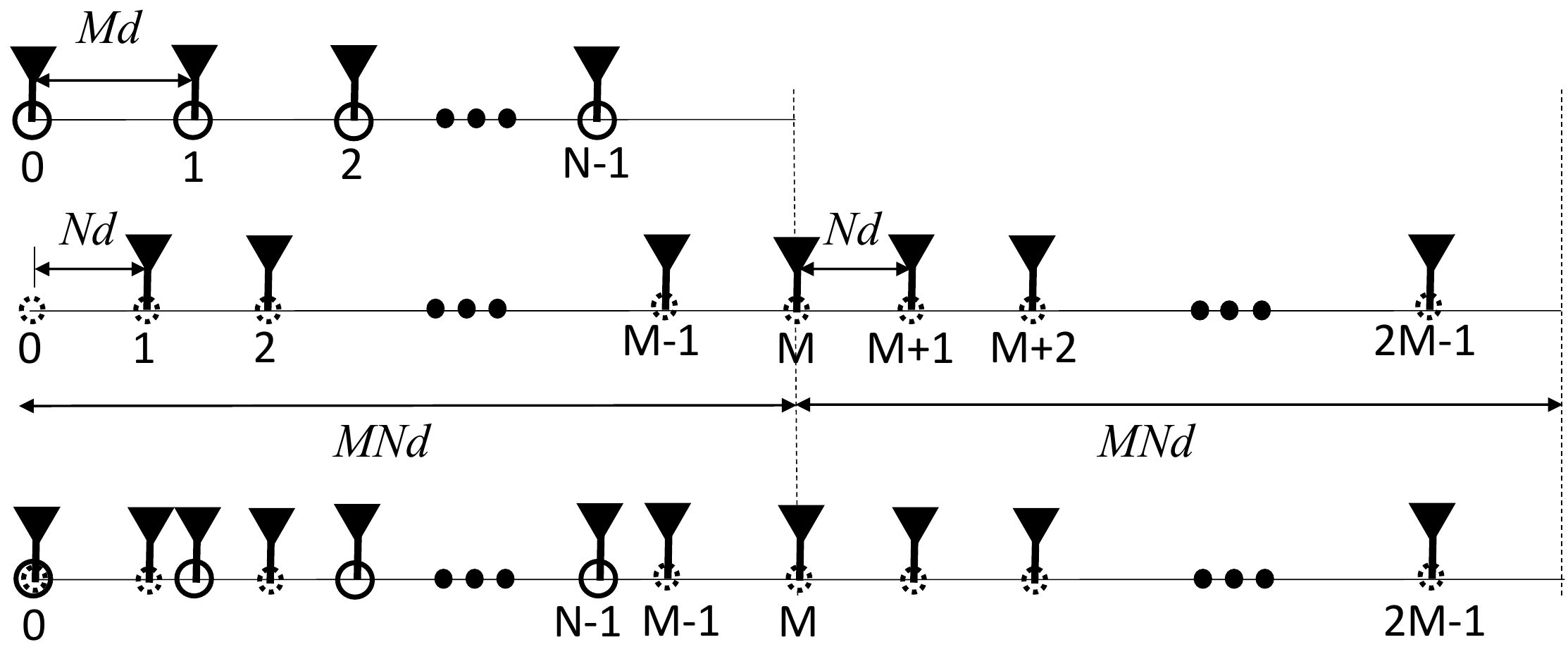}
	\caption{Extended co-prime structure}
	\label{fig_extended_concept}
\end{figure}
\begin{figure*}[!t]
	\centering
	\includegraphics[width=0.96\textwidth]{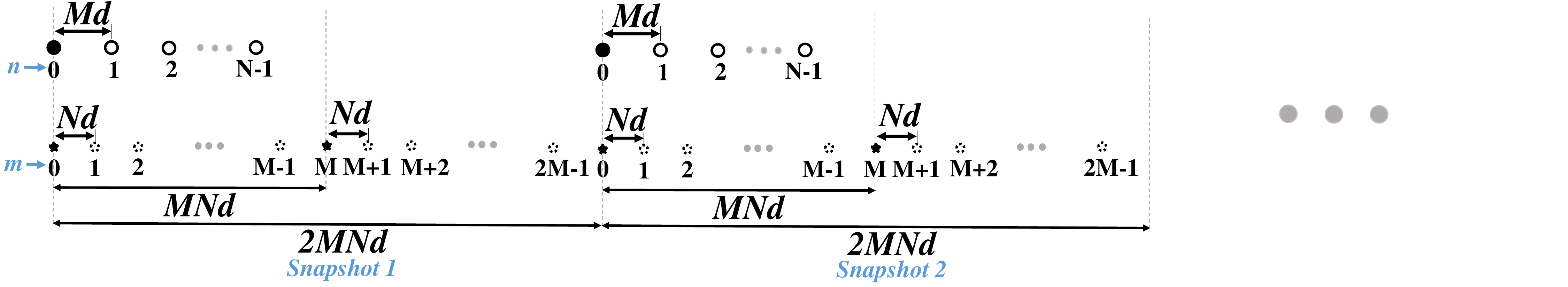}
	\caption{Extended co-prime temporal sampling.}
	\label{fig_extended_temporal}
\end{figure*}

An extended co-prime array is shown in Fig.~\ref{fig_extended_concept} where $M$ and $N$ are co-prime. The  Nyquist distance between the antennas is denoted by $d$. It has two sub-arrays and the antenna locations are given by:
\begin{eqnarray*}\label{eq:positions}
	\nonumber
	P1 &=& \{Mnd,\; 0\leq n \leq N-1\} \\
	P2 &=& \{Nmd,\; 0\leq m \leq 2M-1\}
\end{eqnarray*}
Note that the antenna at the zeroth location is common to both the sub-arrays.
In this paper, we assume that the second array with inter-element spacing of $Nd$ is extended. However, the first array could have been extended, without affecting the concept that governs this scheme. In the temporal or spacial domain, one prototype co-prime period is $[0, MN-1]$ while the extended co-prime period is $[0, 2MN-1]$. It may be noted that the co-prime and extended co-prime range that corresponds to one period for autocorrelation (or other second order statistics) is $[-(MN-1), MN-1]$ and $[-(2MN-1), 2MN-1]$ respectively. Fig.~\ref{fig_extended_temporal} describes this concept from a sampling perspective. Here $d$ represents the Nyquist sampling period. The structure implies that one of the sampler is turned off every alternate co-prime period, while the other sampler uniformly samples the incoming signal. Every extended co-prime period is referred to as a snapshot. The second order statistics (e.g. autocorrelation, power spectrum, etc.) is estimated for each snapshot. Averaging across several snapshots improves the estimate. In this paper, we demonstrate the results based on the sampling framework.
\section{Difference Set Analysis}
The difference sets defined for the prototype co-prime array~\cite{U_S_1}\cite{UVD_PHD}, is also applicable to the extended co-prime array. However, the extension of one of the sub-array needs to be taken into account. In this section, we briefly describe the difference sets before analyzing the degrees of freedom and the range of continuous difference values.

\begin{figure}
	\centering
	\subfloat[${\L}^{+}_{SM} \cup {\L}^{-}_{SM}$]{\includegraphics[width=0.24\textwidth]{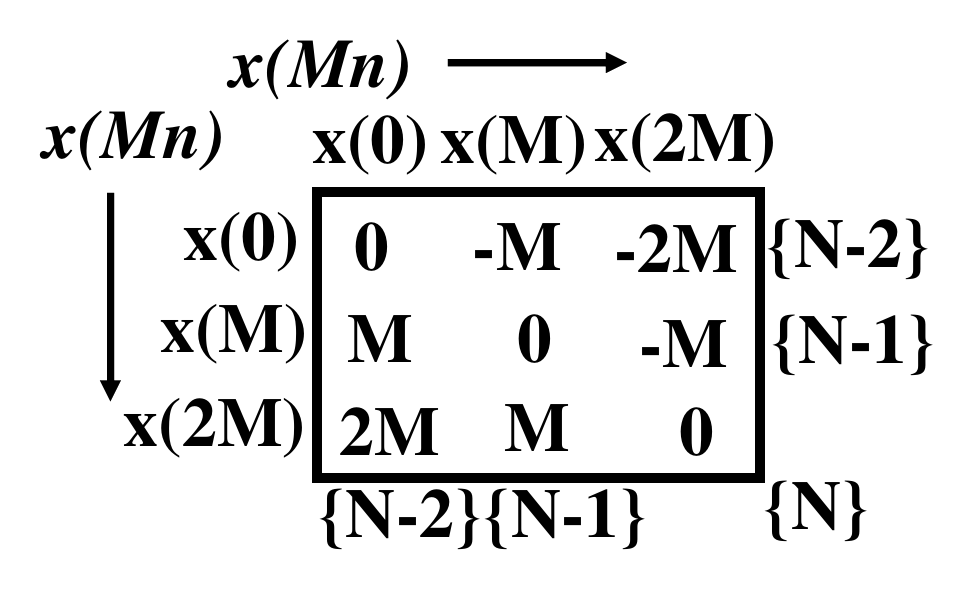}%
		\label{fig_self_M}}
	\hfill
	\subfloat[${\L}^{+}_{SN} \cup {\L}^{-}_{SN}$]{\includegraphics[width=0.24\textwidth]{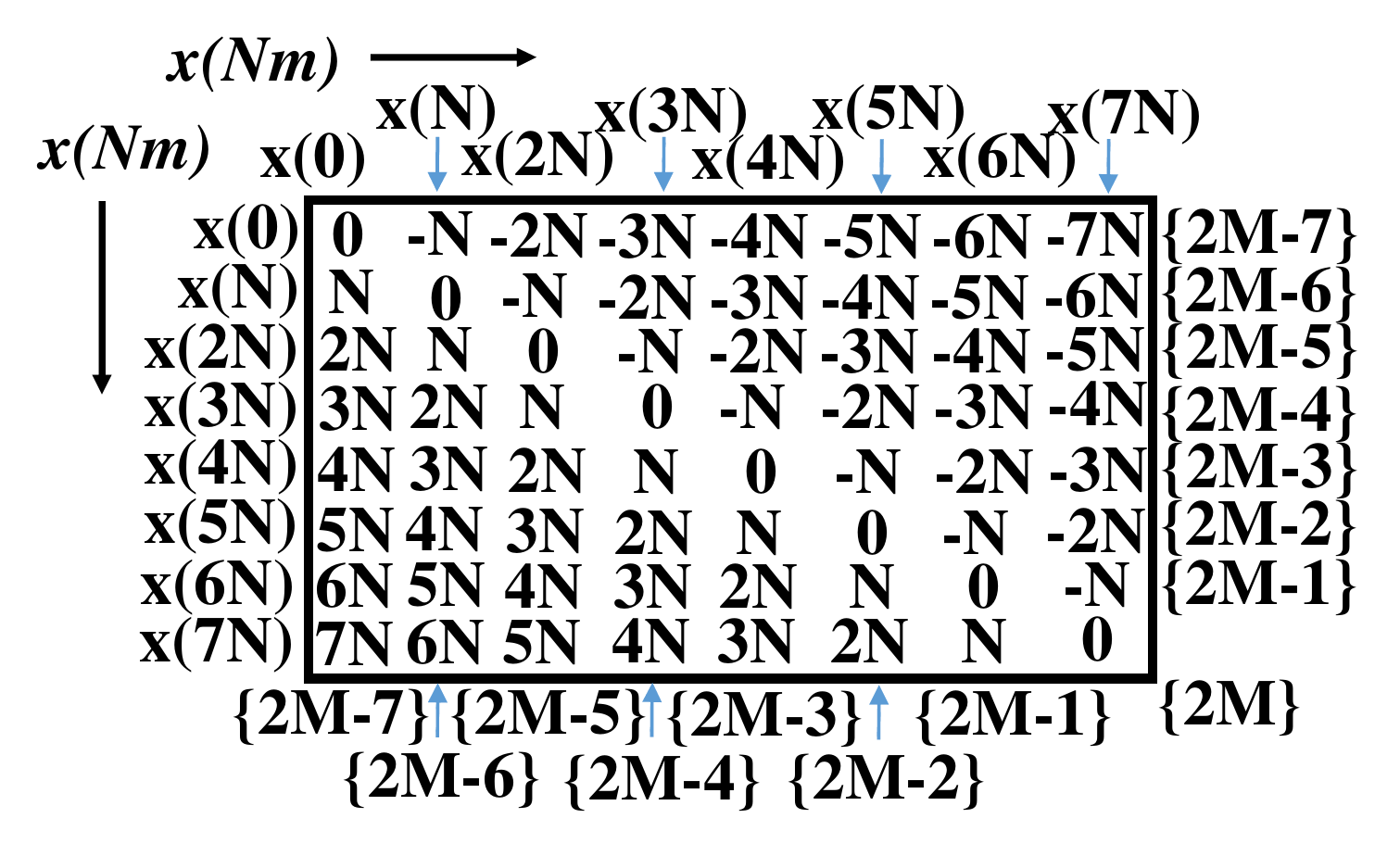}%
		\label{fig_self_N}}
	\caption{Self differences along with number of contributors per difference value indicated in $\{\cdot\}$ for $M=4$, $N=3$.}
	\label{fig_self}
\end{figure}

Let $x(Mn)$ and $x(Nm)$ represent the outputs obtained from the two sub-arrays with inter-element spacings $Md$ and $Nd$ respectively. Their self difference set is given by: ${\L}^+_{SM} = \{l_s|l_s=Mn\} \;\text{and}\; {\L}^+_{SN} = \{l_s|l_s=Nm\}$, while the cross difference set is given by: ${\L}^+_{C} = \{l_{c}|l_{c}=Mn-Nm\}$,
where $0\leq m\leq 2M-1$ and $0\leq n\leq N-1$. The sets containing the mirrored locations of the elements in the above defined sets are denoted by $\L^-_{SM}$, $\L^-_{SN}$ and $\L^-_{C}$ respectively. In addition, we define two union sets as: ${\L}_{S} = {\L}^+_{S} \cup {\L}^-_{S} \;\text{and}\; {\L}_{C} = {\L}^+_{C} \cup {\L}^-_{C}$,
where ${\L}^+_{S}= {\L}^+_{SM} \cup {\L}^+_{SN}$ and ${\L}^-_{S}= {\L}^-_{SM} \cup {\L}^-_{SN}$. These sets are shown in Fig.~\ref{fig_self}. It is obvious that ${\L}^+_{SM}$ and ${\L}^+_{SN}$ have $N$ and $2M$ unique differences respectively, which also holds true for the corresponding mirrored sets. The sets  ${\L}^+_{S}$ and  ${\L}^-_{S}$ have $2M+N-1$ unique differences, hence the union set $\L_{S}$ has $2(2M+N-1)-1$ unique differences. The difference value of zero is common to each set and needs to be considered only once in the union set. This justifies the negation of one in the equation for unique differences. The set $\L^+_{C}$ has $2MN$ unique differences and the same holds true for its mirrored counterpart. As in the case of the prototype co-prime array, the self differences of the extended array are a sub-set of the cross differences. However, the number of unique differences in the union set $\L_{C}$ is not trivial. Hence, to provide a better understanding we define two new sets $\L^+_{A}$ and $\L^+_{B}$ with $\L^-_{A}$ and $\L^-_{B}$ representing their mirrored counterpart:
\begin{eqnarray}\label{eq:pos_cross_new_sets}
\nonumber {\L}^+_{A} = \{l_{c}|l_{c}=Mn-Nm,n\in[0,N-1],m\in[0,M-1]\}\\
\nonumber {\L}^+_{B} = \{l_{c}|l_{c}=Mn-Nm,n\in[0,N-1],m\in[M,2M-1]\}
\end{eqnarray}
It is obvious that $\L^+_{A}$ is same as the cross-difference set of the prototype co-prime array. Hence, the properties of the cross difference set of a prototype co-prime array~\cite{U_S_1}\cite{UVD_PHD}, hold true even for the set $\L^+_{A}$.

For the extended co-prime array, $\L^+_{C}=\L^+_{A} \cup \L^+_{B}$ and $\L^-_{C}=\L^-_{A} \cup \L^-_{B}$. Since the set $\L^+_{A}$ is well understood, the focus here would be to understand the set $\L^+_{B}$. Some of the  properties of the set $\L^+_{B}$ are given below:\\

\textbf{Property I}:
\begin{enumerate}
    \item
    $\{l_{c}<0, \forall l_{c} \in \L^+_{B} \} \;\text{and}\; \{l_{c}>0, \forall l_{c} \in \L^-_{B} \}$
    \item
    $\L^-_{SM}-\{0\} \subseteq \L^+_{B} \;\text{and}\; \L^+_{SM}-\{0\} \subseteq \L^-_{B}$
    \item
    $\{l_{s}|l_{s} \in \L^-_{SN} \forall m \in [M, 2M-1] \} \subseteq \L^+_{B}$\\
    $\{l_{s}|l_{s} \in \L^+_{SN} \forall m \in [M, 2M-1] \} \subseteq \L^-_{B}$
\end{enumerate}
\begin{figure}
	\centering
	\subfloat[Set ${\L}^+_{C}$]{\includegraphics[width=0.28\textwidth]{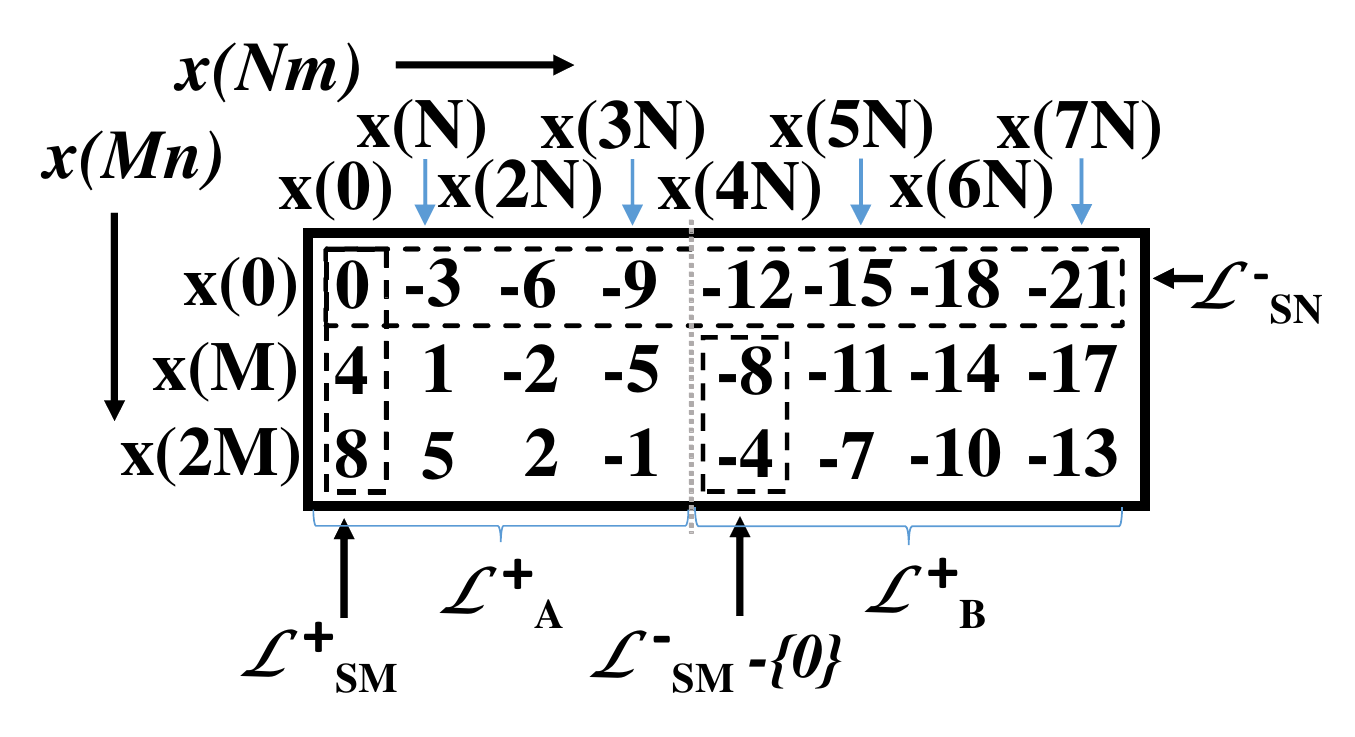}
		\label{fig_cross_MN}}
	\hfill
	\subfloat[Set ${\L}^-_{C}$]{\includegraphics[width=0.2\textwidth]{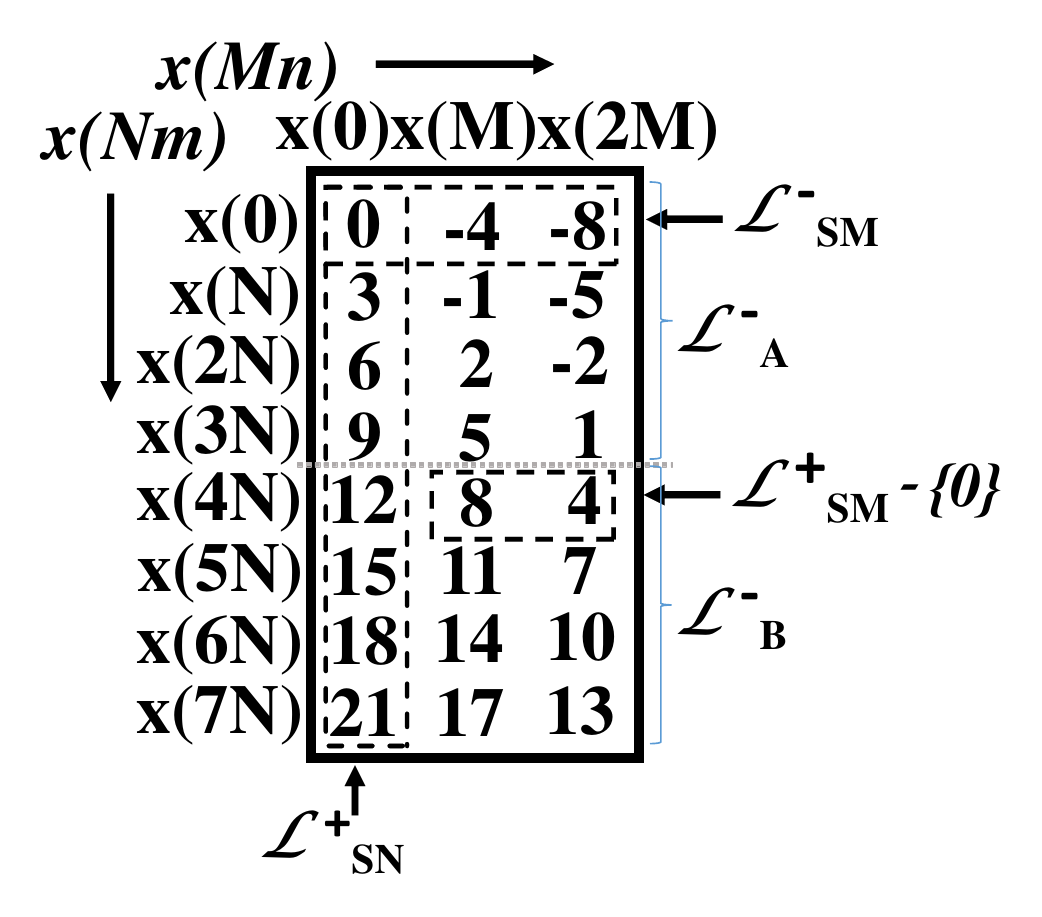}%
		\label{fig_cross_NM}}
	\caption{Cross differences having $2MN$ distinct values, each appearing once for $M=4$, $N=3$.}
	\label{fig_cross}
\end{figure}
According to Property I-1, the sets $\L^+_{B}$ and $\L^-_{B}$ do not have any common values. But they do have overlapping values with the set $\L^-_{A}$ and $\L^+_{A}$ respectively, which is captured in Property I-2. The Property I-2 implies that $(N-1)$ values in $\L^+_{B}$ overlap with $(N-1)$ values in $\L^-_{A}$ and the same holds true for the sets $\L^-_{B}$ and $\L^+_{A}$. The self differences of the array with inter-element spacing of $Nd$ are partly present in $\L^+_{A}$ and $\L^-_{A}$ ($m \in [0, M-1]$) and partly in $\L^+_{B}$ and $\L^-_{B}$ ($m \in [M, 2M-1]$). This is also evident from Fig.~\ref{fig_cross}. It implies that $2(MN-(N-1))$ unique values are present in $\L^+_{B} \cup \L^-_{B}$ which is not available in set $\L^+_{A} \cup \L^-_{A}$. Therefore, the set $\L_{C}$ has $MN+M+N-2+2(MN-(N-1))=3MN+M-N$ unique values. A summary of the unique differences or degrees of freedom (\textit{dof}) in each difference set for the extended array is given in Table~\ref{table_unique_diffs_extended}. The unique differences for the prototype co-prime array is also provided for comparison.
\begin{table*}[!t]
\caption{Summary of unique differences (or \textit{dof}) per difference set}
\label{table_unique_diffs_extended}
\centering
\resizebox{\textwidth}{!}{
\begin{tabular}{|c|c|c|c|c|c|c|c|c|c|c|}
\hline
Set & ${\L}^+_{SM}$ & ${\L}^-_{SM}$ & ${\L}^+_{SN}$ & ${\L}^-_{SN}$ & ${\L}^+_{S}$ & ${\L}^-_{S}$ & ${\L}_{S}$ & ${\L}^+_{C}$ & ${\L}^-_{C}$ & ${\L}_{C}$\\
\hline
\# Unique diffs. (Extended) & $N$ & $N$ & $2M$ & $2M$ & $2M+N-1$ & $2M+N-1$ & $2(2M+N-1)-1$ & $2MN$ & $2MN$ & $3MN+M-N$\\
\hline
\# Unique diffs. (Prototype) & $N$ & $N$ & $M$ & $M$ & $M+N-1$ & $M+N-1$ & $2(M+N-1)-1$ & $MN$ & $MN$ & $MN+M+N-2$\\
\hline
\end{tabular}}
\end{table*}
The prototype co-prime array had holes in the co-prime range, while the extended co-prime array generates all the difference values in the co-prime range $[-MN+1, MN-1]$. However, it does not represent the maximum continuous range of the extended co-prime array. We present Proposition I which provides the expressions for the range of integers in the cross difference sets of an extended co-prime array and their continuity (i.e. the range without holes). The proposition holds for one extended co-prime period $[0, 2MN-1]$ with $0\leq n \leq N-1$ and $0\leq m \leq 2M-1$.\\
\textbf{Proposition I}:
 \begin{enumerate}
   \item ${\L}^+_{C}$ has $2MN$ distinct integers in the range $-N(2M-1)\leq l_c \leq M(N-1)$
   \item ${\L}^-_{C}$ has $2MN$ distinct integers in the range $-M(N-1)\leq l_c \leq N(2M-1)$
   \item ${\L}^+_{C}$ has consecutive integers in the range $-(MN+M-1)\leq l_c \leq N-1$
   \item ${\L}^-_{C}$ has consecutive integers in the range $-(N-1)\leq l_c \leq (MN+M-1)$
   \item ${\L_C}$ has consecutive integers in the range $-(MN+M-1) \leq l_c \leq (MN+M-1)$ which implies that this set has its first hole at $\mid l_c\mid$ =$(MN+M)$.
 \end{enumerate}
Proposition I-(1) and I-(2) can be easily proved by substituting the values of $n$ and $m$ in $(Mn-Nm)$ and $(Nm-Mn)$ respectively, to generate the maximum and minimum values in the set.\\
\textit{Proof of Proposition I-(3):}
Let $l_c=Mn-Nm$ be an element in set ${\L}^+_{C}$ satisfying $-(MN+M-1)\leq l_c \leq N-1$.
  Since $n \in [0,N-1]$, the range of $Mn$ is $0\leq Mn \leq M(N-1)$.
  Using the range of $Mn$ and $l_c$, we need to prove that $m \in [0,2M-1]$, from the equation $Nm=Mn-l_c$. Therefore:
  \begin{eqnarray}\label{eq:proof_propI_3}
\nonumber   & -(N-1)\leq Nm\leq M(N-1)+(MN+M-1) \\
\nonumber    & -1 < m < 2M \\
\nonumber     & 0 \leq m \leq 2M-1
  \end{eqnarray}
  which satisfies the required range. Proposition I-(4) can be proved along similar lines as Proposition I-(3).\\
\textit{Proof of Proposition I-(5):}
Let $l_c=\pm (Mn-Nm)$ be an element in the set ${\L_{C}}$ satisfying: $-(MN+M-1) \leq l_c \leq (MN+M-1)$,
  where $m \in [0,2M-1]$ and $n \in [0,N-1]$. First we need to prove that $\pm (MN+M)$ is indeed a hole and then show that the range $ -(MN+M-1) \leq l_c \leq (MN+M-1)$ is continuous.
  To prove that $\pm (MN+M)$ is a hole in the set ${\L_{C}}$, it is sufficient to prove that it is a hole in ${\L}^+_C$. This holds true because ${\L}^-_C$ is a flipped version of ${\L}^+_C$ and ${\L_{C}} ={\L}^+_C \cup {\L}^-_C$.
  Let $l_c=Mn-Nm$ be an element in set ${\L}^+_C$, and let us assume that $\pm (MN+M)$ is not a hole and exists in set ${\L}^+_C$. This implies that: $ Mn-Nm=\pm (MN+M)$ or $M(n\mp 1)=N(m\pm M)$, hence: $\frac{M}{N}=\frac{(m\pm M)}{(n\mp 1)}$.
  Since $n-1<N$ and $m-M<M$, the ratios $\frac{(m+M)}{(n-1)}$ and $\frac{(m-M)}{(n+1)}$ cannot  be satisfied.
  The proof of continuity in the range $-(MN+M-1) \leq l_c \leq (MN+M-1)$ follows directly from Proposition I-(3) and I-(4). 
\section{Weight Function}

Proposition III in~\cite{U_S_1} gives the expression for the weight function or  the number of sample pairs that contribute to estimate the autocorrelation at each value in the difference set. This parameter affects the convergence, accuracy, latency and bias of the estimate. In this section we present the number of sample pairs that contribute to estimate the autocorrelation for the extended co-prime array and is given as Proposition II.

\textbf{Proposition II}: Let $z(l)$ denote the number of elements contributing to the estimation at difference value $l$.
 \begin{enumerate}
   \item For $l \in {\L}^+_{SM} \cup {\L}^-_{SM}-\{0\}$:
    \begin{equation}\label{eq:contributors_self_Mn}
 \nonumber       z(l)=(N-i)+1, \{1\leq i \leq N-1, l=\pm Mi\}
    \end{equation}
   \item For $l \in {\L}^+_{SN} \cup {\L}^-_{SN}-\{0\}$:
    \begin{equation}\label{eq:contributors_self_Nm}
\nonumber        z(l)=(2M-i), \{1\leq i \leq 2M-1, l=\pm Ni\}
    \end{equation}
   \item For $l \in \{l=0\}$: $z(l)=2M+N-1$
   \item For $l \in {\L}_{C} -{\L}_{S}$:
   \begin{equation}\label{eq:contributors_A}
\nonumber        z(l)=2,  \{l \in \{{\L}^+_{A} \cup {\L}^-_{A}\} -{\L}_{S}\}
    \end{equation}
       \begin{equation}\label{eq:contributors_B}
\nonumber        z(l)=1,  \{l \in \{{\L}^+_{B} \cup {\L}^-_{B}\} -{\L}_{S}\}
    \end{equation}
 \end{enumerate}
 It is easy to conclude that the number of sample pairs that map to each difference value in the self difference set of signal $x(Mn)$ is given by $N-i$ as shown in Fig.~\ref{fig_self_M}. In addition, the cross difference set also has pairs of samples that contribute to the estimate at these self differences except at `0'. Thus justifying Proposition II-(1). Proposition II-(2) can be easily inferred from Fig.\ref{fig_self_N}. The number of contributors at difference value `0', Proposition II-(3), is the sum of the contributors of the self difference set of the two arrays at zero minus one pair which is common to both the self difference sets.
 ${\L^+_{A}}$ and ${\L^-_{A}}$ represent the cross difference set of the prototype co-prime array and Proposition III-(4) in~\cite{U_S_1} shows that $z(l)=2$ for$\{l \in \{{\L}^+_{A} \cup {\L}^-_{A}\} -{\L}_{S}\}$. On the other hand, we can establish that $z(l)=1$ for $\{l \in \{{\L}^+_{B} \cup {\L}^-_{B}\} -{\L}_{S}\}$ from Fig.~\ref{fig_cross} and Property I. In Fig.~\ref{fig:wts_M>N} and~\ref{fig:wts_M<N}, we provide some examples of the weight function of the extended co-prime array for $M>N$ and $N>M$ respectively. These cases of $M$ and $N$ will also be used as examples for the bias analysis. The weight function forms the basis for the bias and variance analysis of the correlogram spectral estimate of the extended co-prime array.

\section{Bias Analysis}
\label{Bias Analysis}
The fundamentals of the correlogram method for co-prime based spectral estimation was developed in~\cite{UVD_PHD}. This section derives the bias of the correlogram spectral estimate for the extended co-prime array for the entire difference set, the continuous difference set and for the prototype co-prime period. The reason for investigating it over the prototype co-prime period is because the initial motivation for the extension was to fill the holes in the region $[-MN+1,MN-1]$.
The closed-form expression for the weight function of the extended co-prime array denoted by $z_{e_f}(l)$,  $z_{e_c}(l)$ and $z_{e_p}(l)$, for the three cases mentioned above, is given by equations \eqref{eq:extend_entire}-\eqref{eq:extend_proto_period} respectively. It may be noted that in this paper $f$, $c$ and $p$ will be used to represent the entire/full, continuous and the prototype range respectively.
%
\begin{figure*}[!t]
	\centering
	\subfloat[$M=4$, $N=3$]{\includegraphics[width=0.45\textwidth]{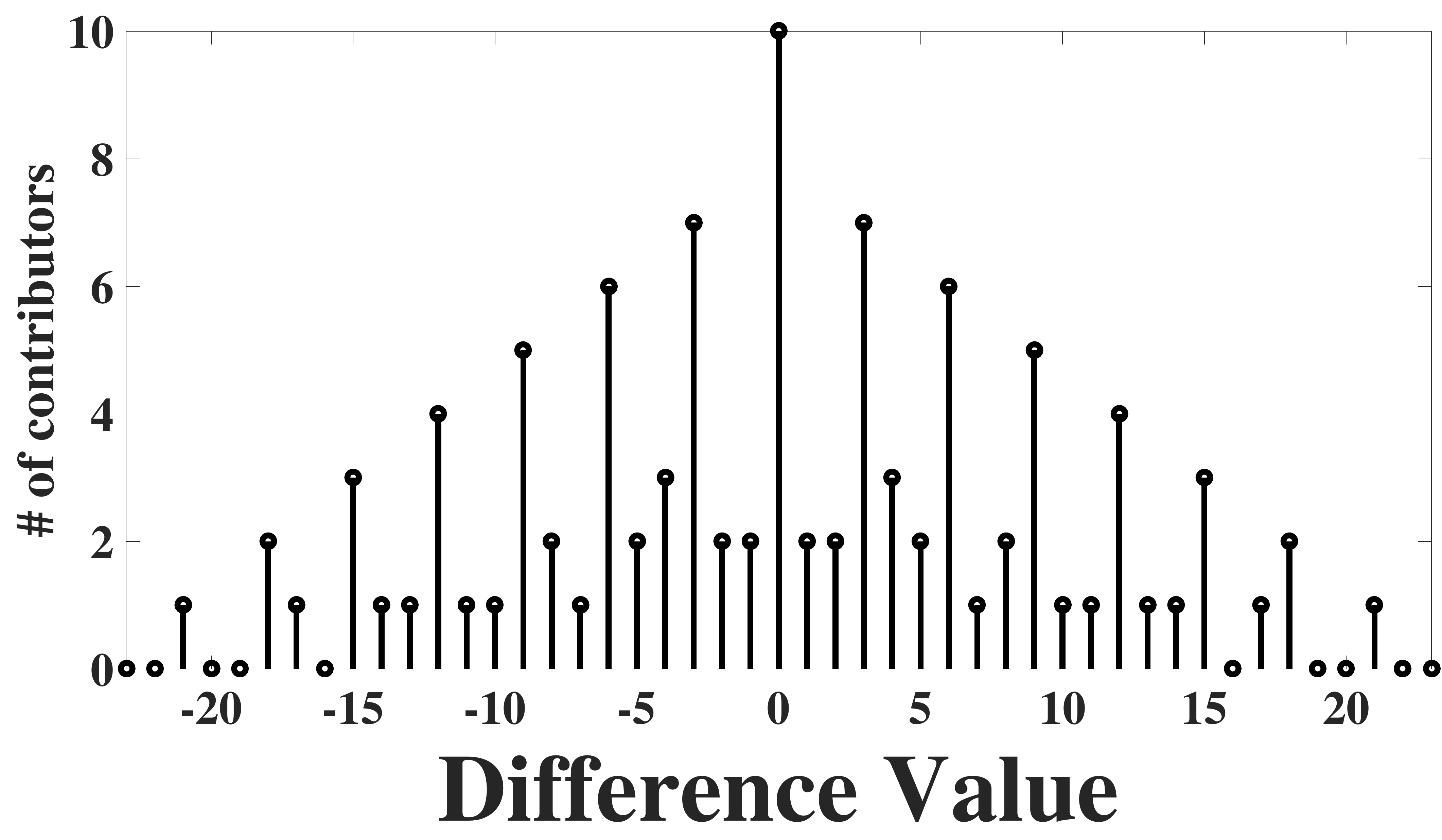}\label{ext_M4N3_wts}}
	\hfil
	\subfloat[$M=5$, $N=3$]{\includegraphics[width=0.45\textwidth]{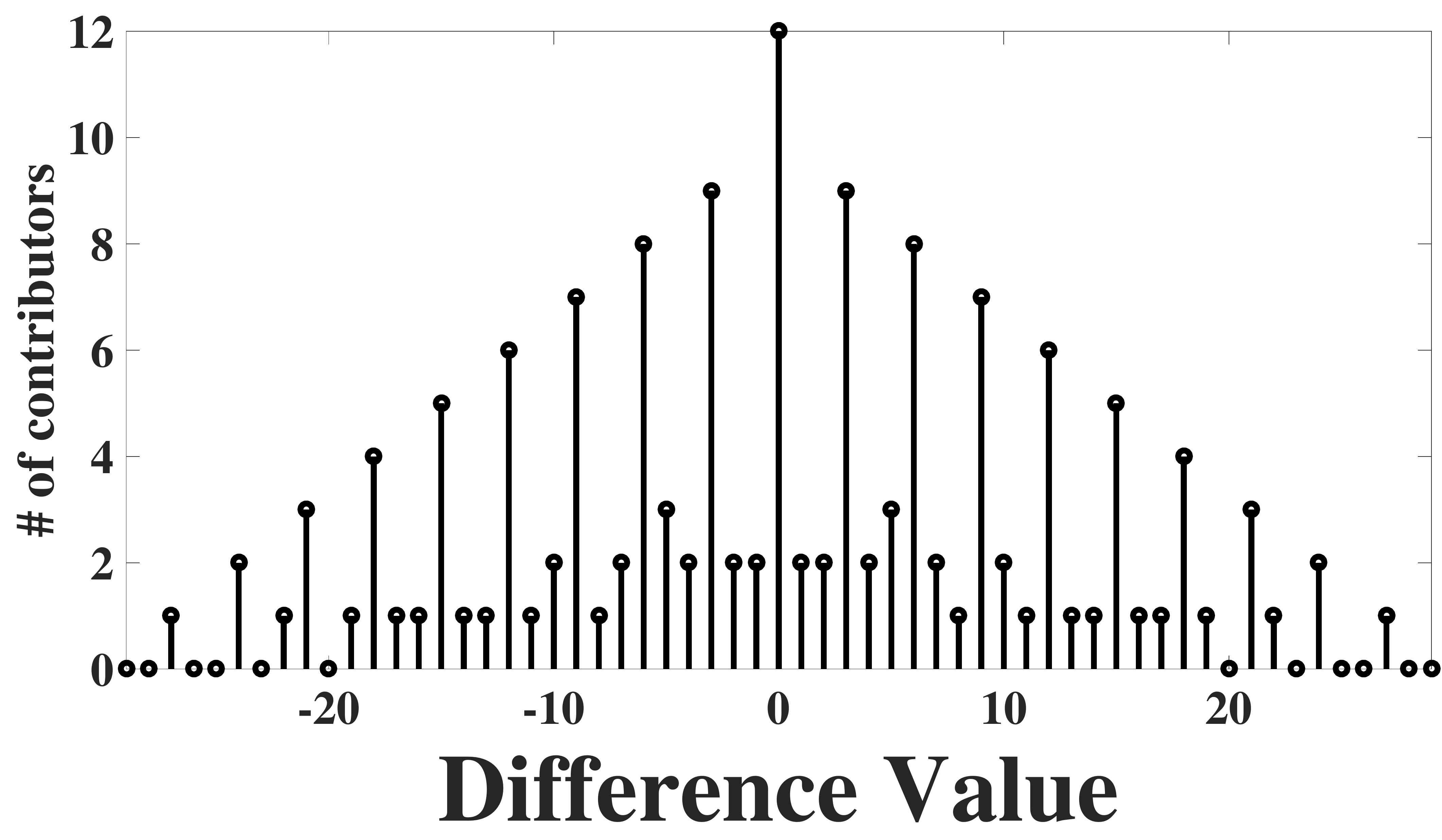}\label{ext_M5N3_wts}}
	\hfil
	\subfloat[$M=7$, $N=3$]{\includegraphics[width=0.45\textwidth]{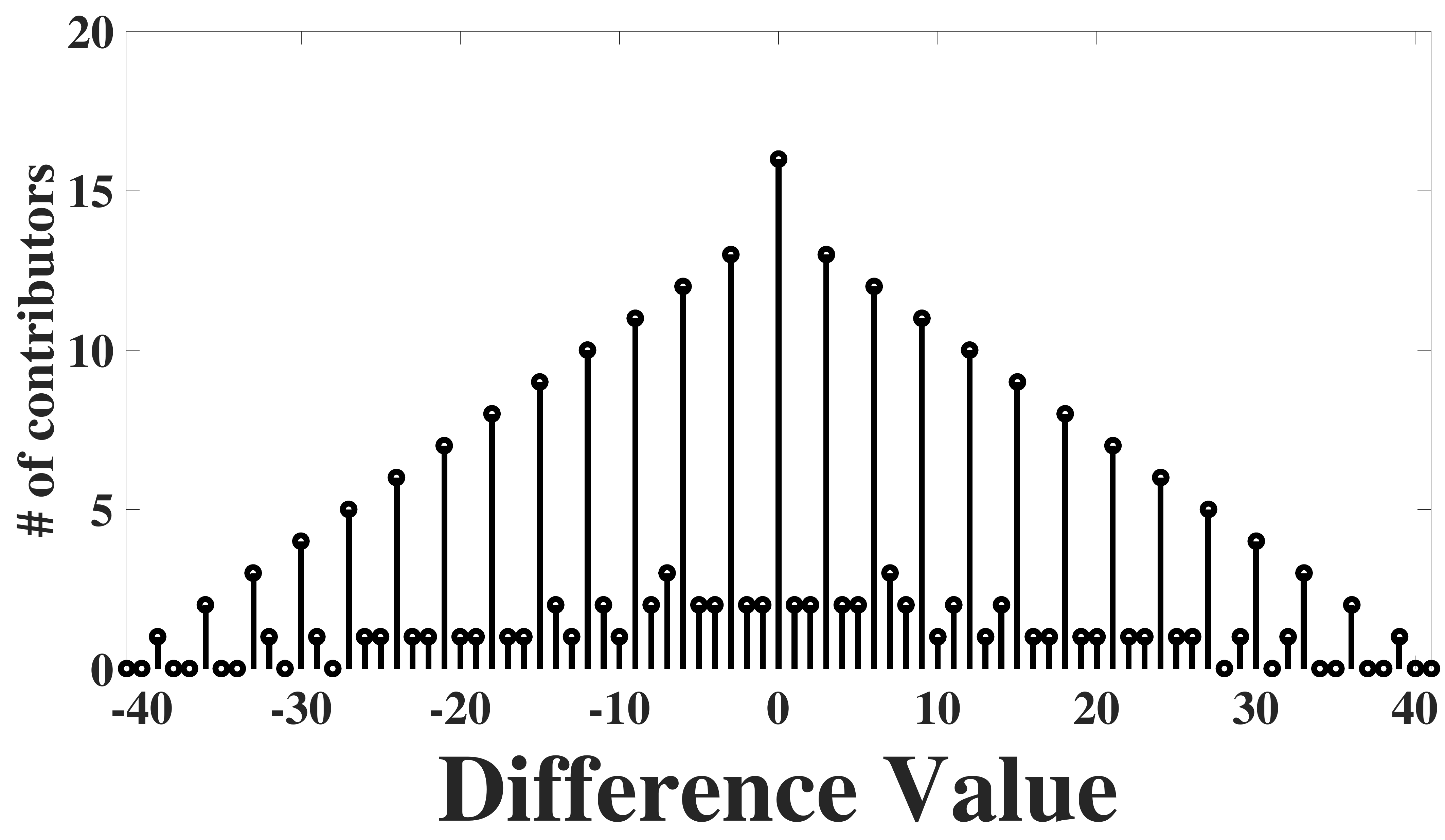}\label{ext_M7N3_wts}}
	\hfil
	\subfloat[$M=8$, $N=3$]{\includegraphics[width=0.45\textwidth]{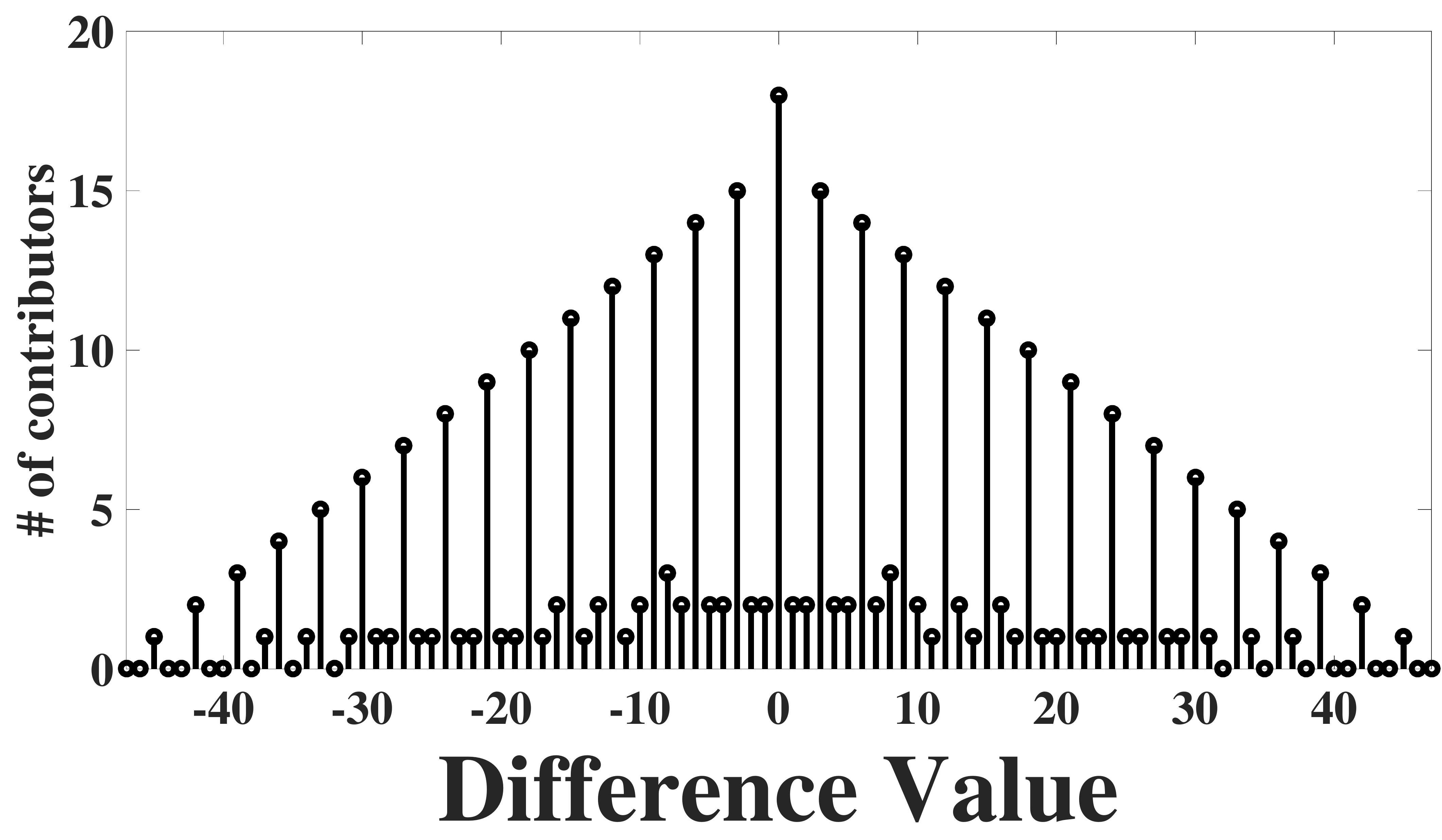}\label{ext_M8N3_wts}}
	\hfil
	\subfloat[$M=5$, $N=4$]{\includegraphics[width=0.45\textwidth]{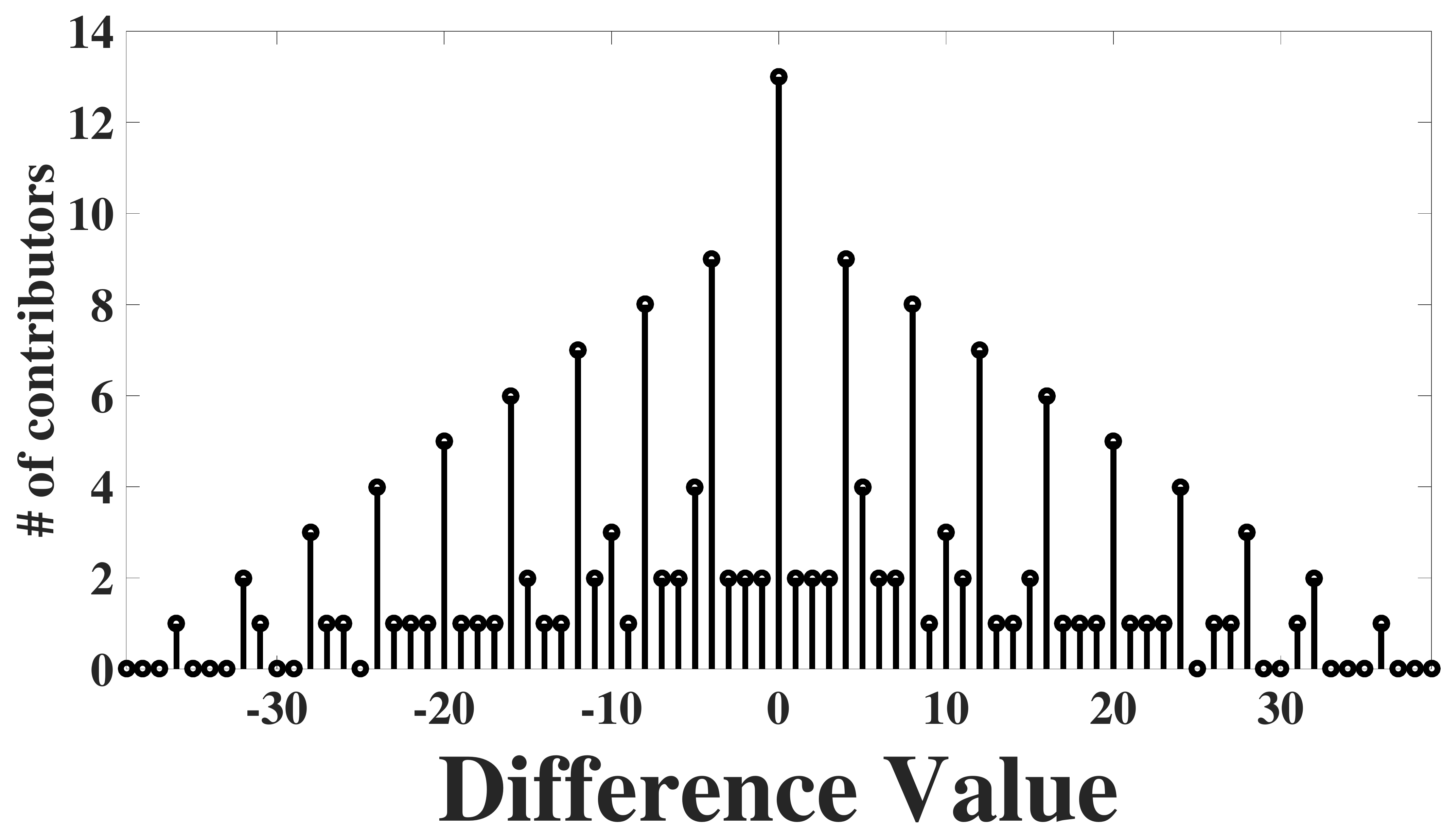}\label{ext_M5N4_wts}}
	\hfil
	\subfloat[$M=7$, $N=4$]{\includegraphics[width=0.45\textwidth]{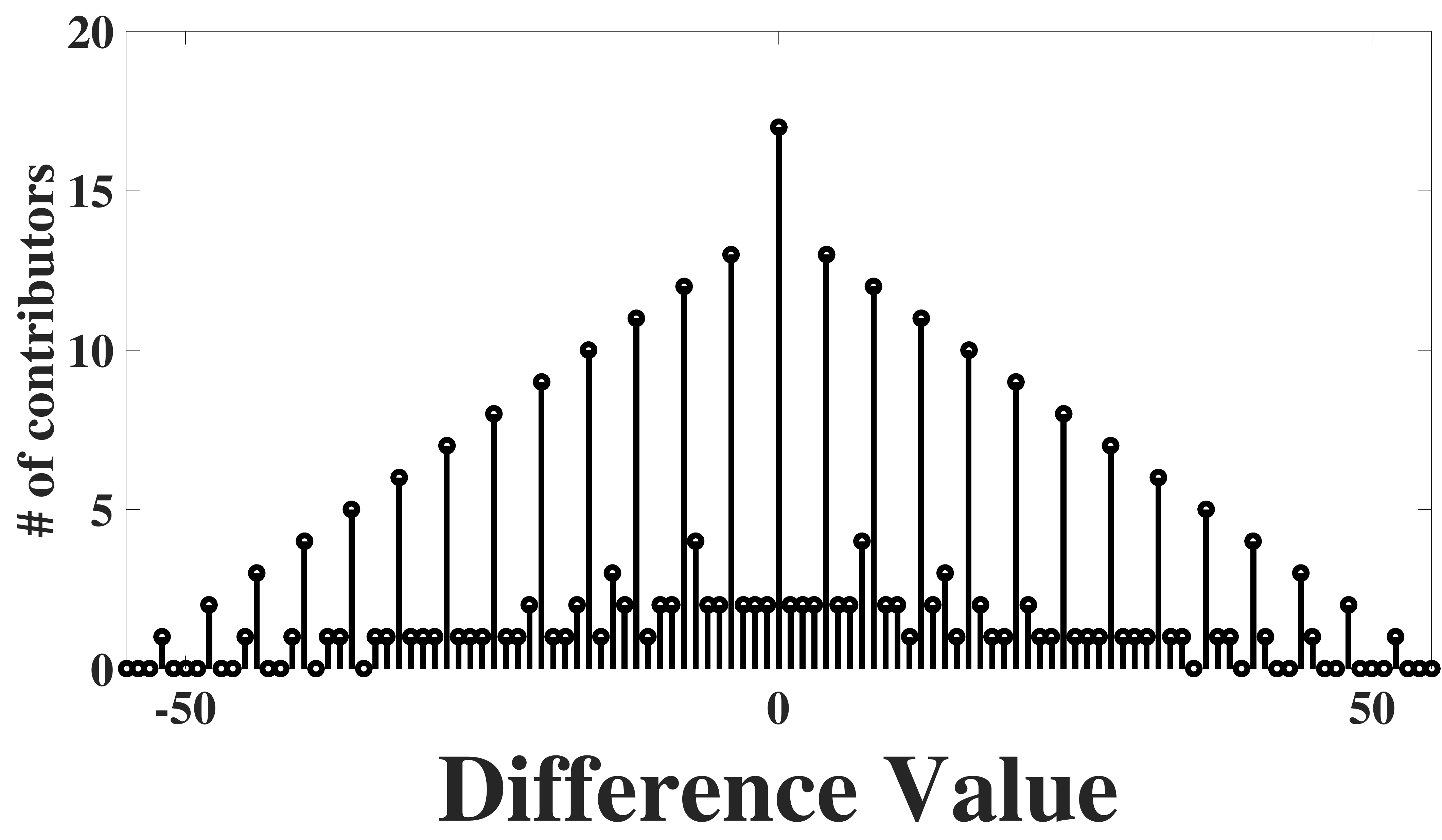}\label{ext_M7N4_wts}}
	\hfil
	\caption{Weight function: $M>N$.}
	\label{fig:wts_M>N}
\end{figure*}
\begin{figure*}[!t]
	\centering
	\subfloat[$M=3$, $N=4$]{\includegraphics[width=0.45\textwidth]{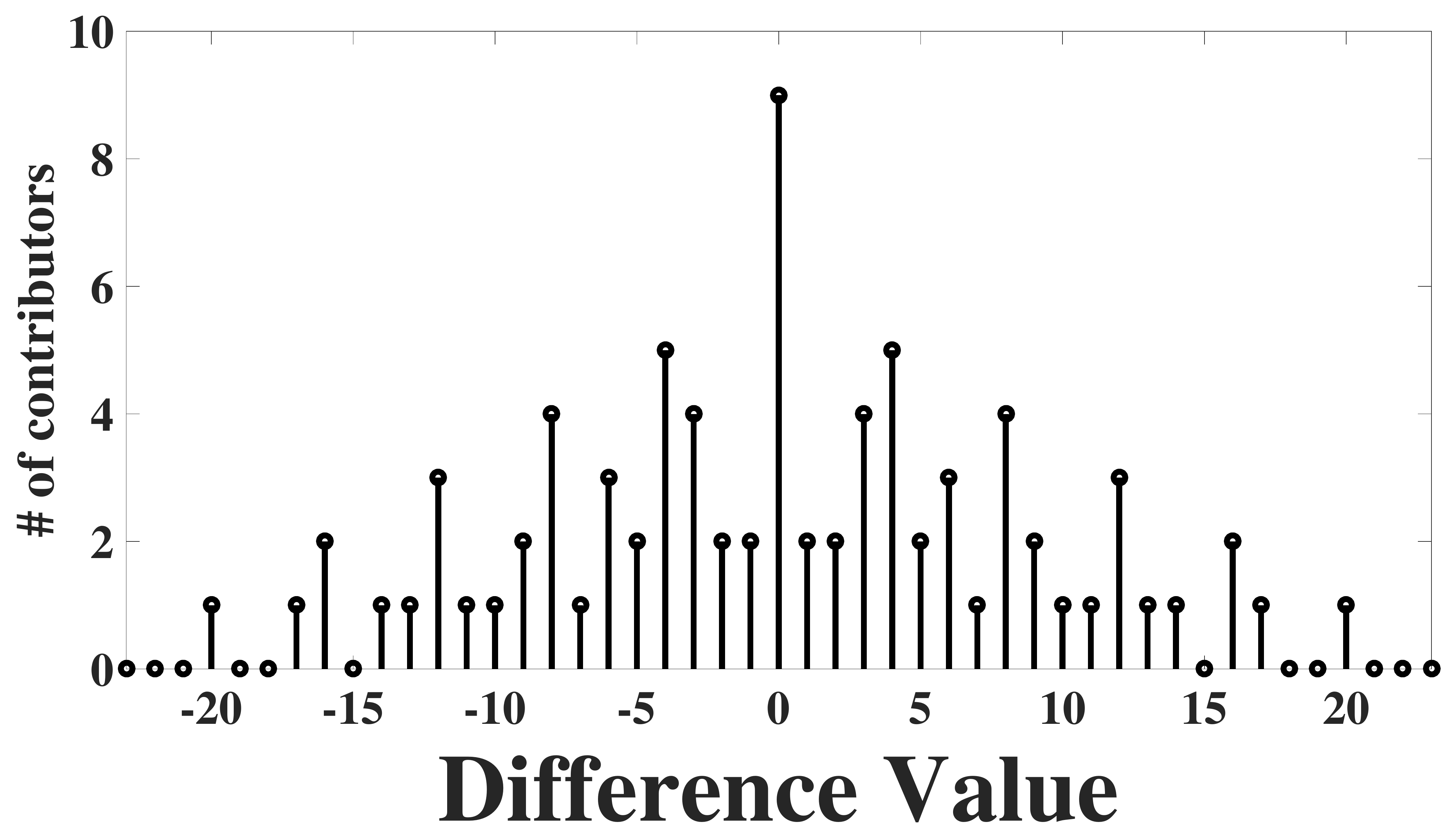}\label{ext_M3N4_wts}}
	\hfil
	\subfloat[$M=3$, $N=5$]{\includegraphics[width=0.45\textwidth]{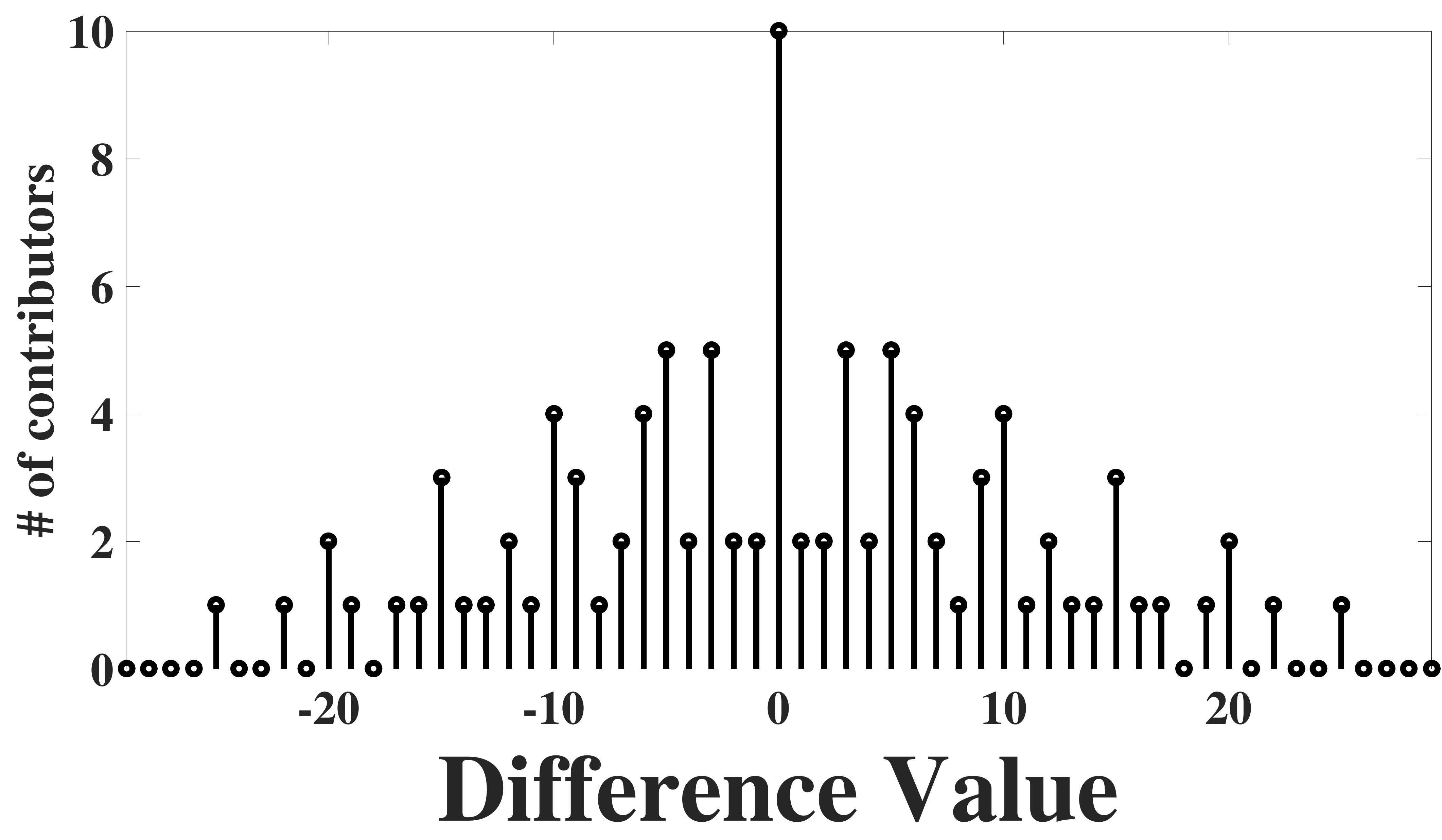}\label{ext_M3N5_wts}}
	\hfil
	\subfloat[$M=3$, $N=7$]{\includegraphics[width=0.45\textwidth]{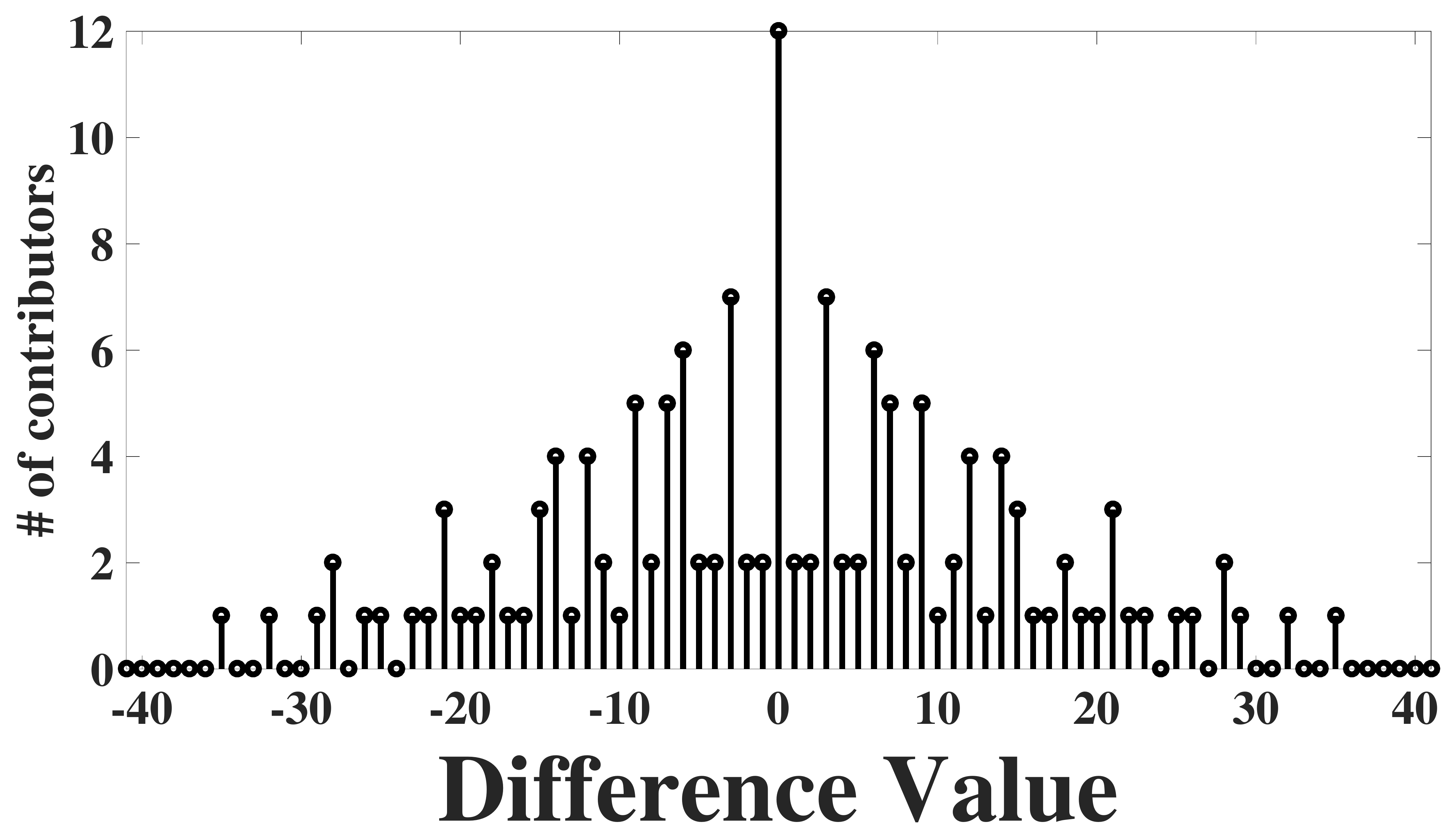}\label{ext_M3N7_wts}}
	\hfil
	\subfloat[$M=3$, $N=8$]{\includegraphics[width=0.45\textwidth]{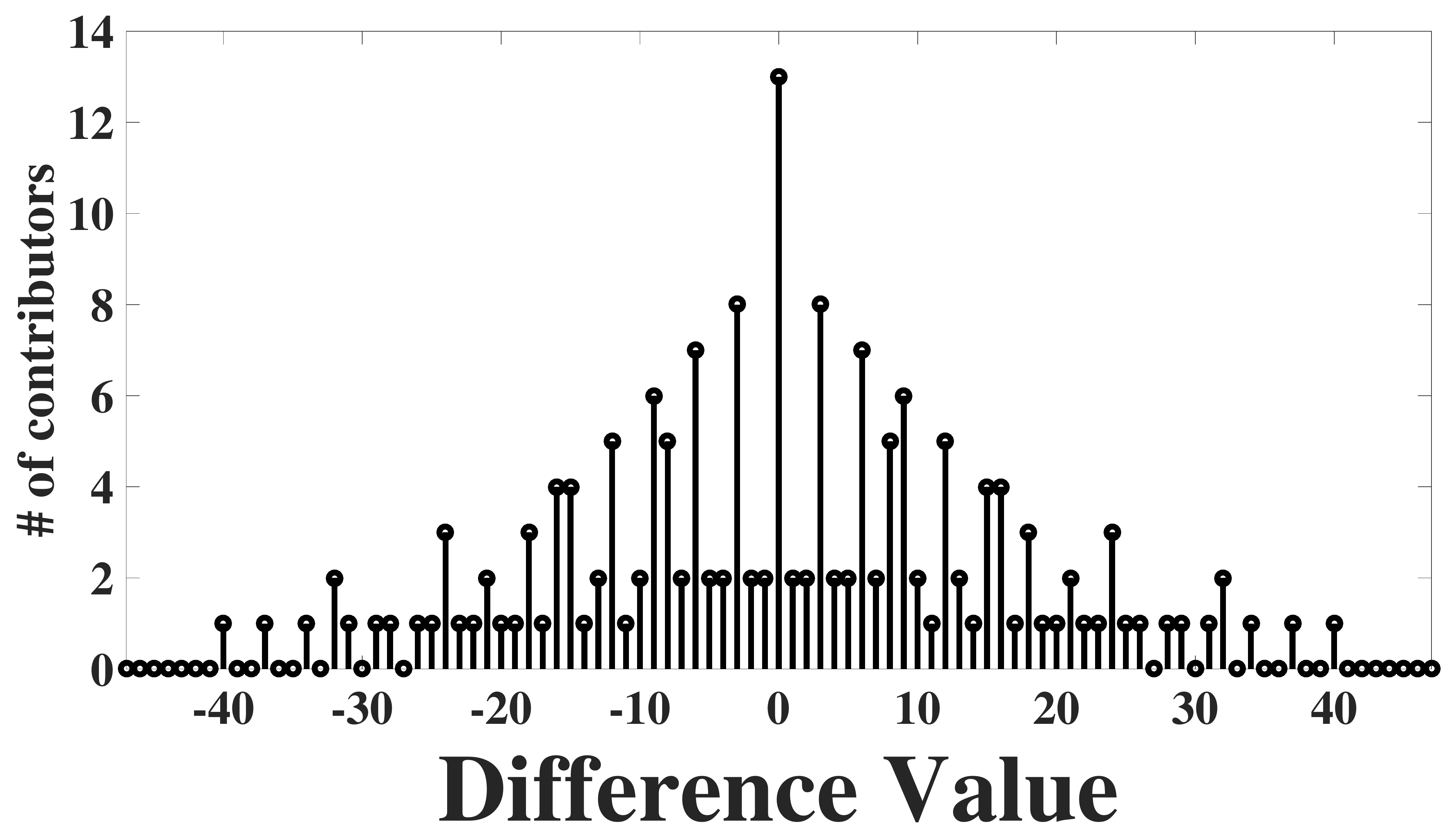}\label{ext_M3N8_wts}}
	\hfil
	\subfloat[$M=4$, $N=5$]{\includegraphics[width=0.45\textwidth]{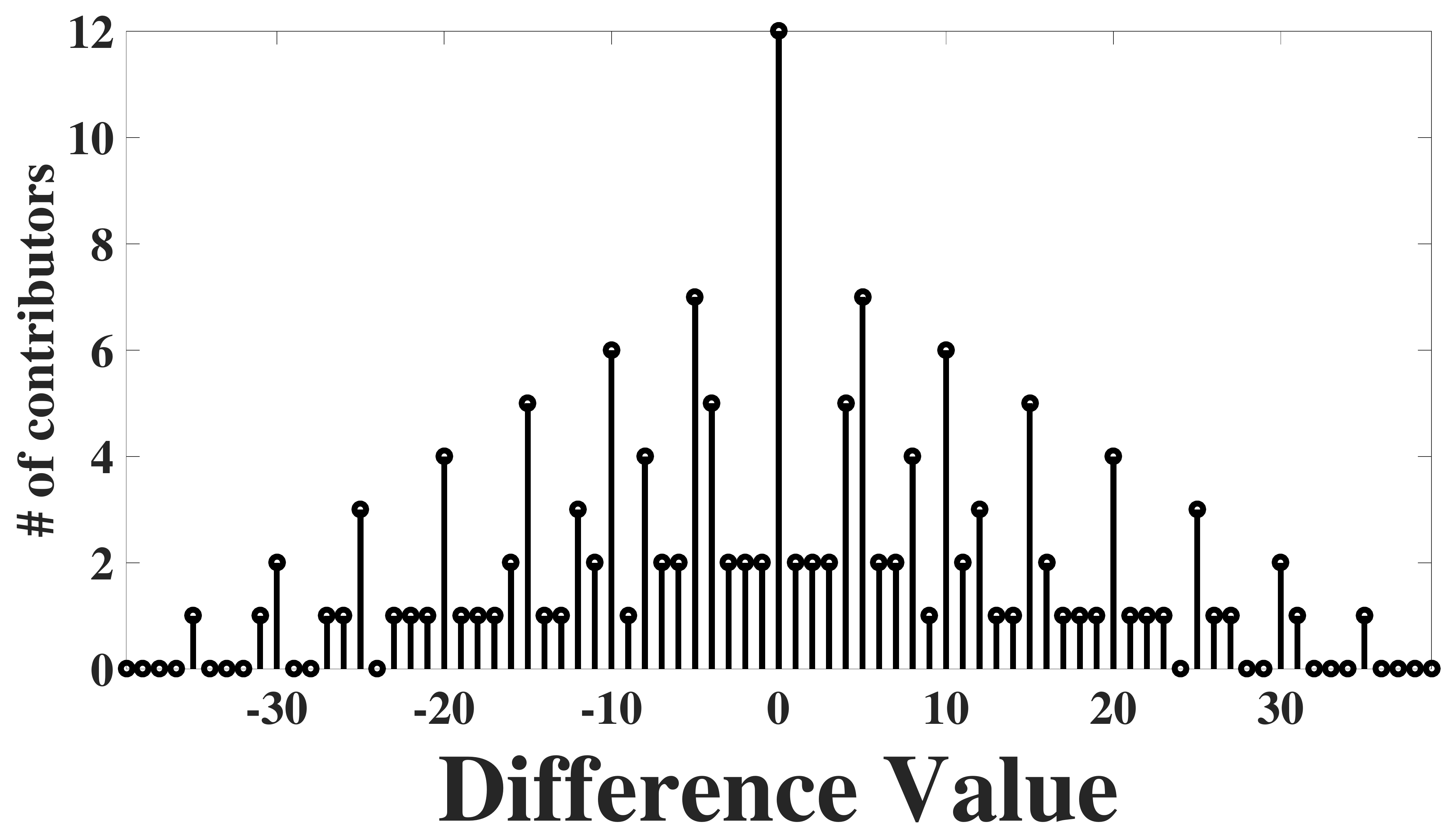}\label{ext_M4N5_wts}}
	\hfil
	\subfloat[$M=4$, $N=7$]{\includegraphics[width=0.45\textwidth]{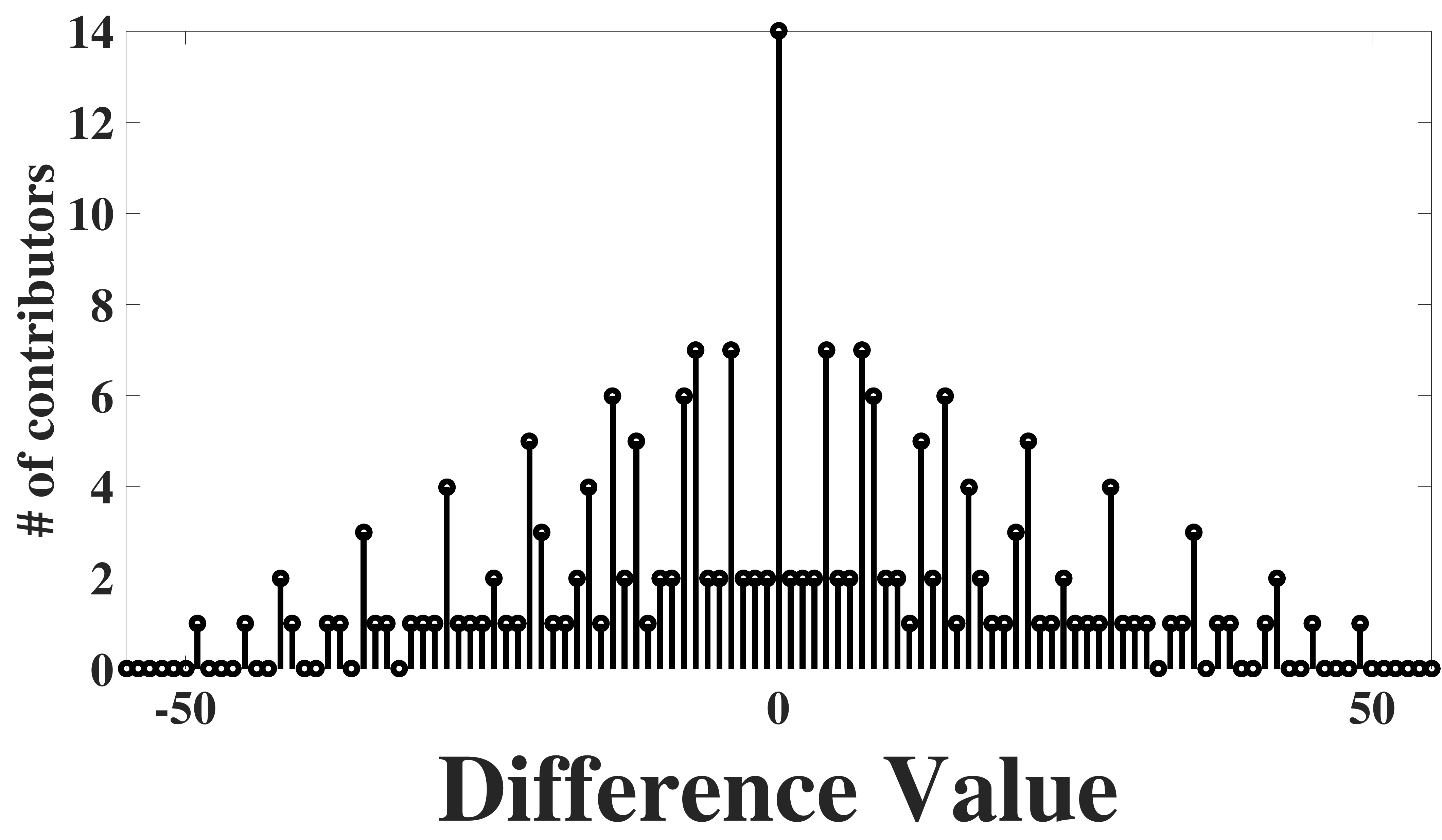}\label{ext_M4N7_wts}}
	\hfil
	\caption{Weight function: $M<N$.}
	\label{fig:wts_M<N}
\end{figure*}
\begin{equation}\label{eq:extend_entire}
\begin{split}
  z_{e_f}(l)=\underbrace{\sum\limits_{n=-(N-1)}^{N-1} (N-\mid n\mid +1)\delta(l-Mn)}_\text{A\textsubscript{f}}\\
        +\underbrace{\sum\limits_{m=-(2M-1)}^{2M-1} (2M-\mid m\mid)\delta(l-Nm)}_\text{B\textsubscript{f}}\\
        +\underbrace{\sum\limits_{n=1}^{N-1}\sum\limits_{m=1}^{M-1} 2 \delta(l-(Mn-Nm))-\delta(l)}_\text{C\textsubscript{f}}\\
        +\underbrace{\sum\limits_{n=1}^{N-1}\sum\limits_{m=M+1}^{2M-1} \delta(\mid l\mid-|Mn-Nm|)-\delta(l)}_\text{D\textsubscript{f}}
\end{split}
\end{equation}
\begin{equation}\label{eq:extend_continuous}
\begin{split}
  z_{e_c}(l)=\underbrace{\sum\limits_{n=-(N-1)}^{N-1} (N-\mid n\mid +1)\delta(l-Mn)}_\text{A\textsubscript{c}}\\
        +\underbrace{\sum\limits_{m=-\lfloor\frac{MN+M-1}{N}\rfloor}^{\lfloor\frac{MN+M-1}{N}\rfloor} (2M-\mid m\mid)\delta(l-Nm)}_\text{B\textsubscript{c}}\\
        +\underbrace{\sum\limits_{n=1}^{N-1}\sum\limits_{m=1}^{M-1} 2 \delta(l-(Mn-Nm))-\delta(l)}_\text{C\textsubscript{c}}\\
        +\underbrace{\sum\limits_{n=1}^{N-1}\sum\limits_{m=M+1}^{\lfloor \frac{MN+M+(Mn-1)}{N}\rfloor}\delta(\mid l\mid-|Mn-Nm|)-\delta(l)}_\text{D\textsubscript{c}}
\end{split}
\end{equation}
\begin{equation}\label{eq:extend_proto_period}
\begin{split}
  z_{e_p}(l)=\underbrace{\sum\limits_{n=-(N-1)}^{N-1} (N-\mid n\mid +1)\delta(l-Mn)}_\text{A\textsubscript{p}}\\
        +\underbrace{\sum\limits_{m=-(M-1)}^{M-1} (2M-\mid m\mid)\delta(l-Nm)}_\text{B\textsubscript{p}}\\
        +\underbrace{\sum\limits_{n=1}^{N-1}\sum\limits_{m=1}^{M-1} 2 \delta(l-(Mn-Nm))-\delta(l)}_\text{C\textsubscript{p}}\\
        +\underbrace{\sum\limits_{n=1}^{N-1}\sum\limits_{m=M+1}^{\lfloor \frac{MN+(Mn-1)}{N}\rfloor}\delta(\mid l\mid-|Mn-Nm|)-\delta(l)}_\text{D\textsubscript{p}}
\end{split}
\end{equation}
\begin{figure*}[!t]
	\centering
	\subfloat[$M=3$, $N=4$]{\includegraphics[width=0.13\textwidth]{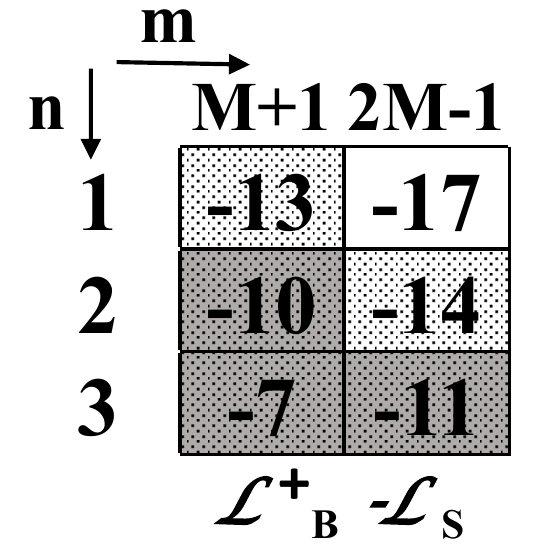}\label{ext_N4M3_LB}}
	\hfil
	\subfloat[$M=3$, $N=5$]{\includegraphics[width=0.13\textwidth]{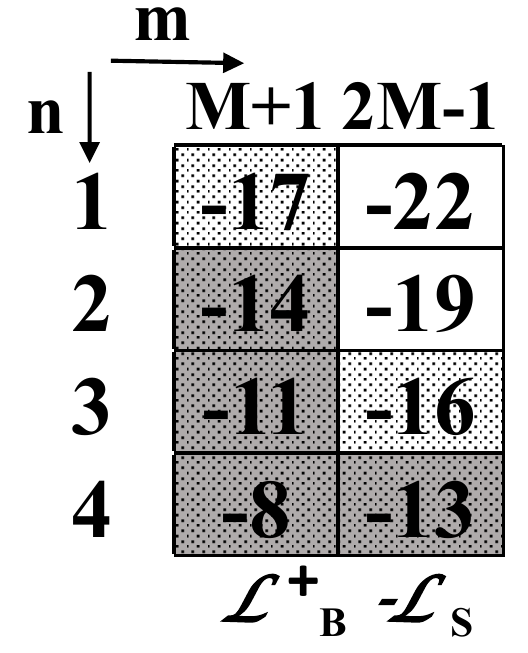}\label{ext_N5M3_LB}}
	\hfil
	\subfloat[$M=3$, $N=7$]{\includegraphics[width=0.1\textwidth]{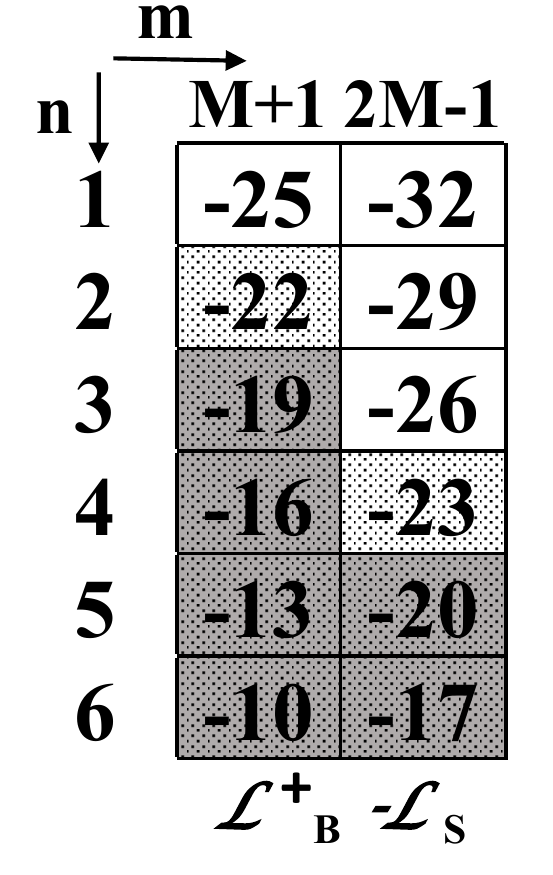}\label{ext_N7M3_LB}}
	\hfil
	\subfloat[$M=3$, $N=8$]{\includegraphics[width=0.09\textwidth]{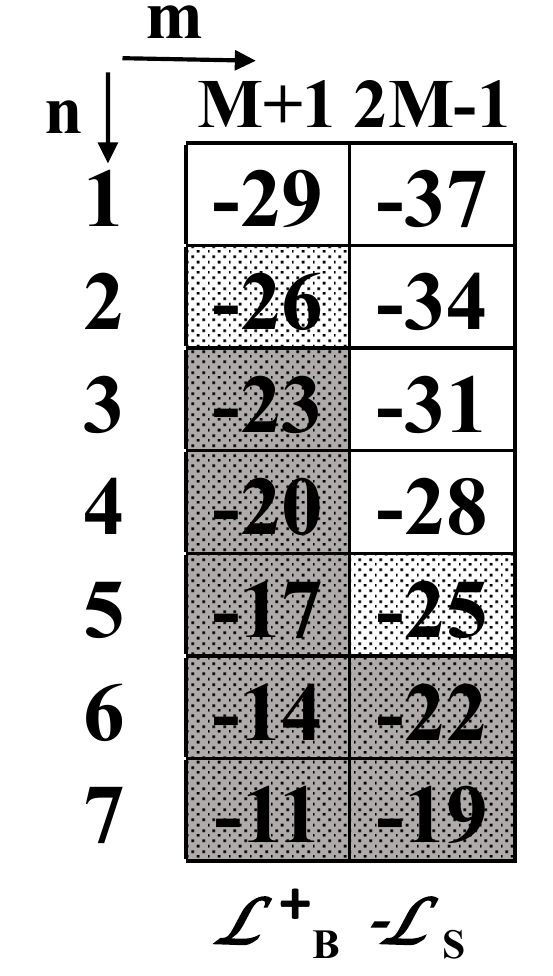}\label{ext_N8M3_LB}}
	\hfil
	\subfloat[$M=4$, $N=5$]{\includegraphics[width=0.15\textwidth]{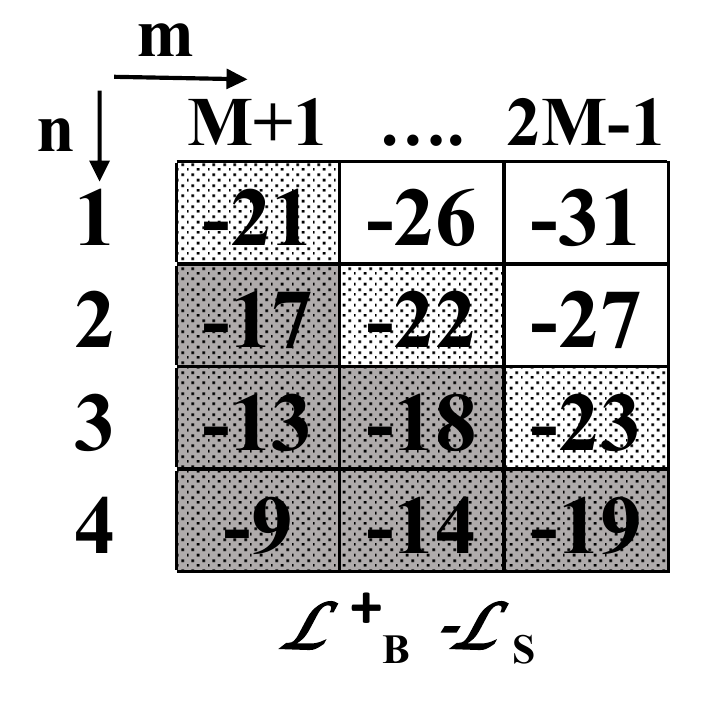}\label{ext_N5M4_LB}}
	\hfil
	\subfloat[$M=4$, $N=7$]{\includegraphics[width=0.13\textwidth]{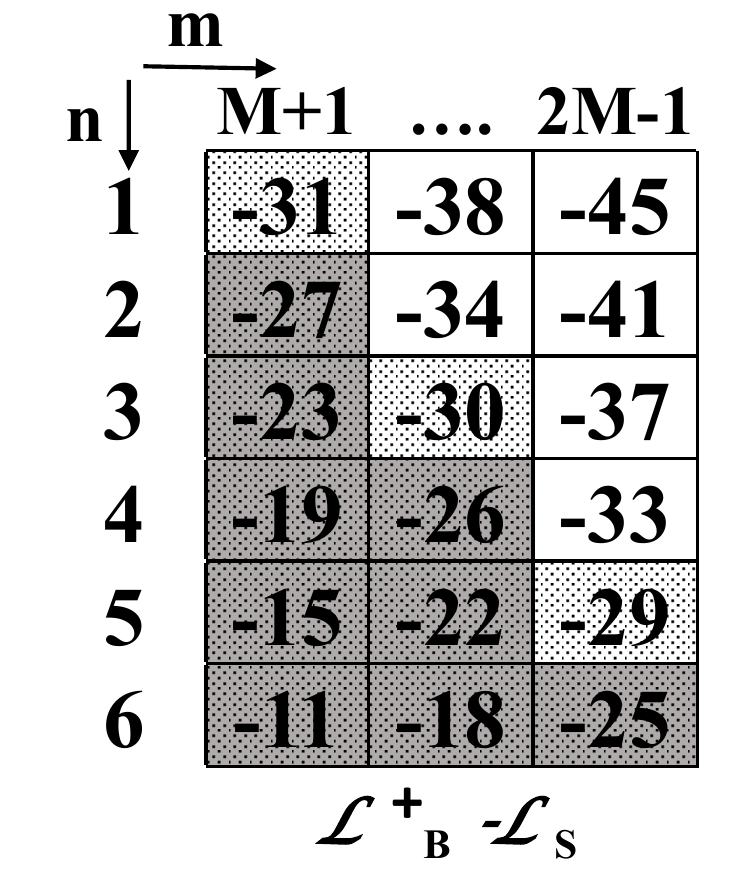}\label{ext_N7M4_LB}}
	\caption{Region for continuous range and co-prime period range of integers in set ${\L}^+_{B}-{\L}_{S}: N>M$.}
	\label{fig:Limits_LB_set_N>M}
\end{figure*}
The expressions for $z_{e_f}$, $z_{e_c}$ and $z_{e_p}$ are directly obtained from Proposition II. $z_{e_f}$ is a function over the entire difference set ($f$ indicates full range). $A_{f}$ is similar to $A$ except for one additional contributor, $B_{f}$ is similar to $B$ with \textit{M} replaced by \textit{2M} and $C_{f}$ is equal to $C$ (where $A$, $B$ and $C$ were described in~\cite{UVD_PHD}). $D_{f}$ represents single contributors in the set ${\L}^+_{B}$ excluding the self differences. The use of $\mid l\mid$ in $\delta(\mid l\mid-|Mn-Nm|)$ is due to the fact that $(Mn-Nm)$ generates values in ${\L}^+_{B}-{\L}_{S}$ which are all negative.

The continuous range of the extended array limits the self differences in the set ${\L}^+_{SN}$ to $Nm$, where $m\in [0, \lfloor \frac{MN+M-1}{N}\rfloor]$, and leads to equation $B_{c}$. $C_{c}$ is same as $C_{f}$ and hence $C$.

The prototype co-prime range is given by $\pm(MN-1)$, hence $A_{p}=A_{c}=A_{f}$ and $C_{p}=C_{c}=C_{f}=C$. Since $m\in[0, M-1]$, we obtain equation $B_{p}$.

\begin{figure*}[!t]
\centering
\subfloat[$M=4$, $N=3$]{\includegraphics[width=0.13\textwidth]{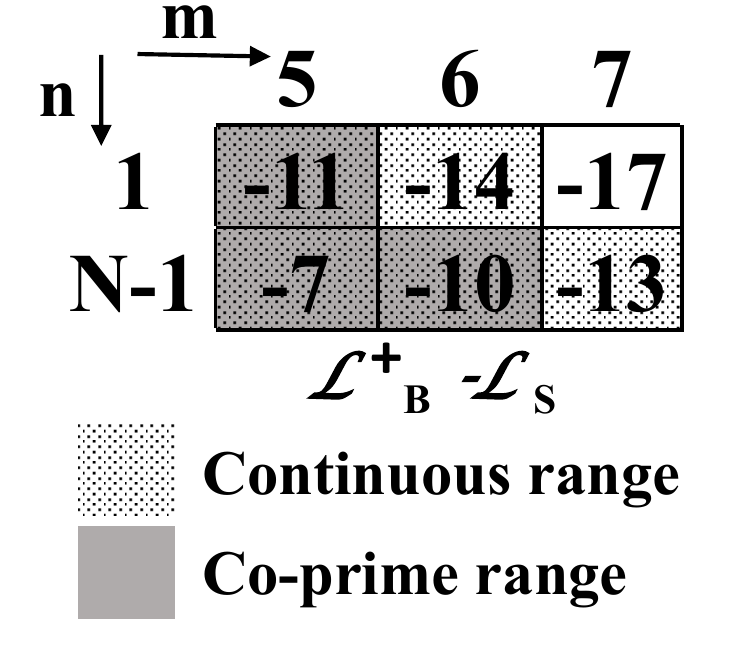}\label{ext_M4N3_LB}}
\hfil
\subfloat[$M=5$, $N=3$]{\includegraphics[width=0.18\textwidth]{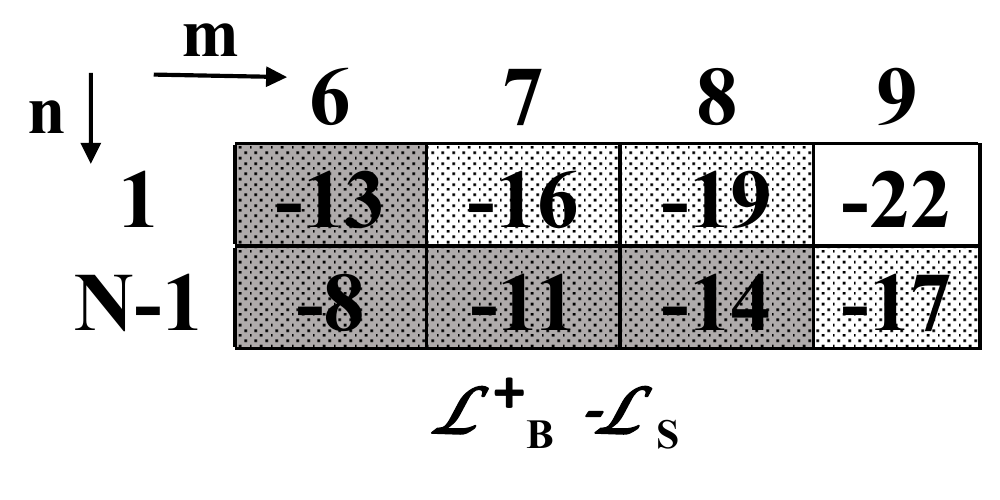}\label{ext_M5N3_LB}}
\hfil
\subfloat[$M=7$, $N=3$]{\includegraphics[width=0.18\textwidth]{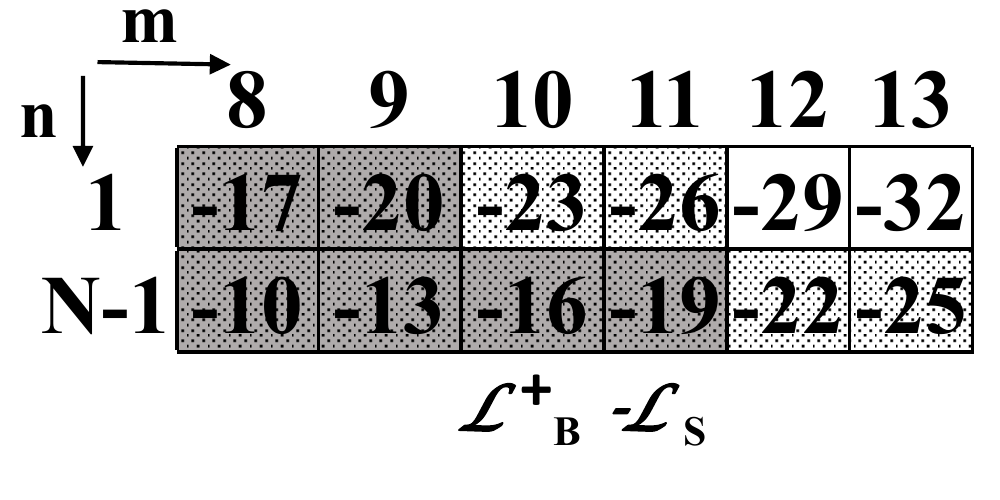}\label{ext_M7N3_LB}}
\hfil
\subfloat[$M=8$, $N=3$]{\includegraphics[width=0.2\textwidth]{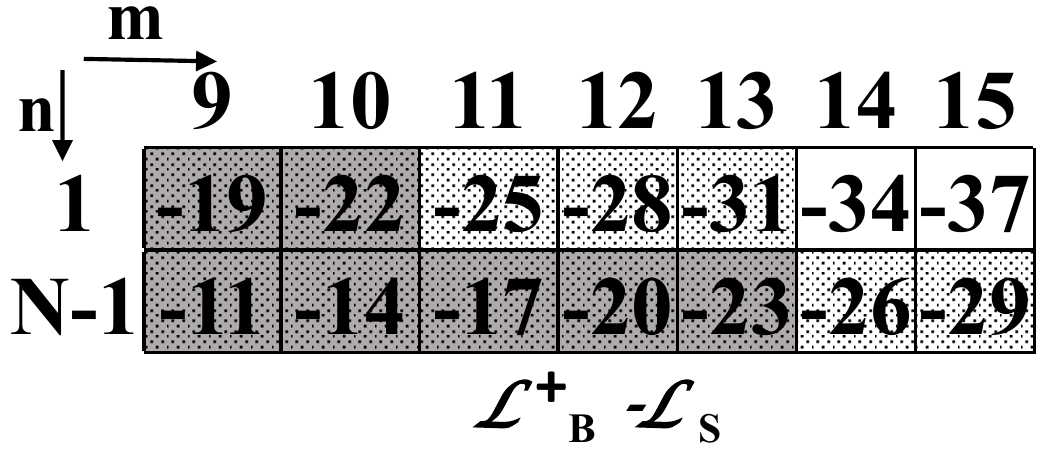}\label{ext_M8N3_LB}}
\hfil
\subfloat[$M=5$, $N=4$]{\includegraphics[width=0.13\textwidth]{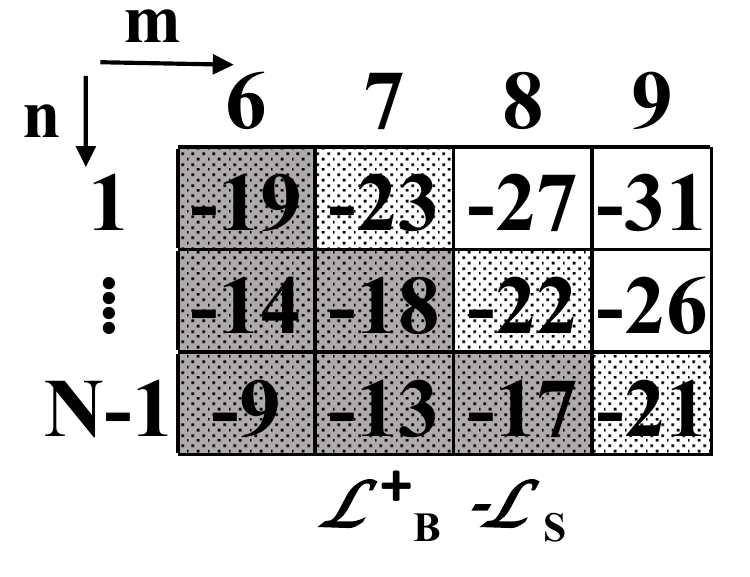}\label{ext_M5N4_LB}}
\hfil
\subfloat[$M=7$, $N=4$]{\includegraphics[width=0.16\textwidth]{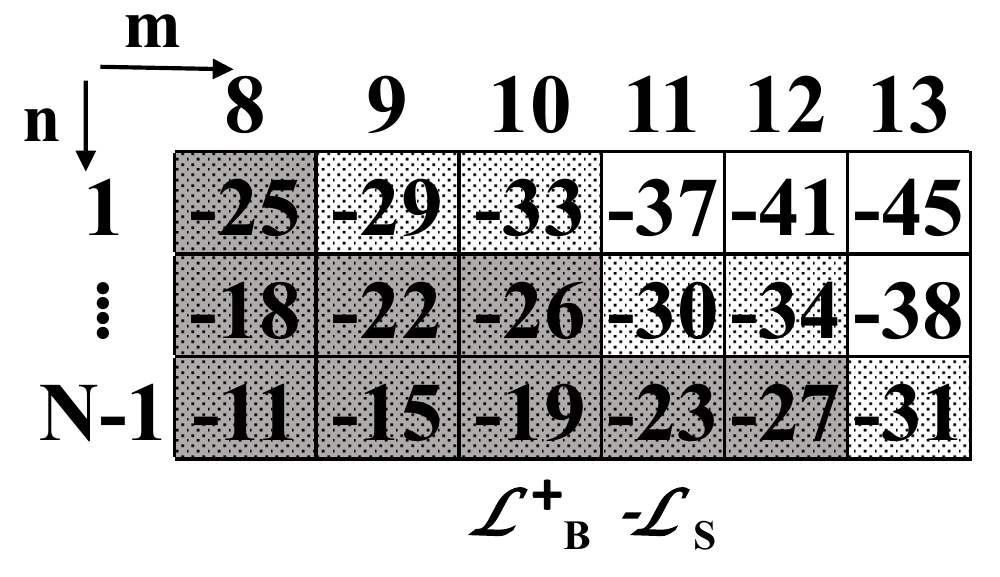}\label{ext_M7N4_LB}}
\hfil
\caption{Region for continuous range and co-prime period range of integers in set ${\L}^+_{B}-{\L}_{S}: M>N$.}
\label{fig:Limits_LB_set_M>N}
\end{figure*}

The range for equations $D_{c}$ and $D_{p}$ cannot be directly inferred and hence, to provide an insight into its formulation consider Fig.~\ref{fig:Limits_LB_set_N>M} and~\ref{fig:Limits_LB_set_M>N}, which displays ${\L}^+_{B}-{\L}_{S}$ for different values of $(M,N)$. Fig.~\ref{fig:Limits_LB_set_M>N} represents the set when $M>N$, while Fig.~\ref{fig:Limits_LB_set_N>M} considers scenarios in which $M<N$. Note that all combinations of $(M, N)$ , i.e, (odd, even), (even, odd) and (odd, odd) are considered. The dotted boxes represent the elements in the continuous range while the shaded boxes represent the prototype range.

\begin{figure}[!t]
\centering
\subfloat[Entire Range]{
\includegraphics[width=0.24\textwidth]{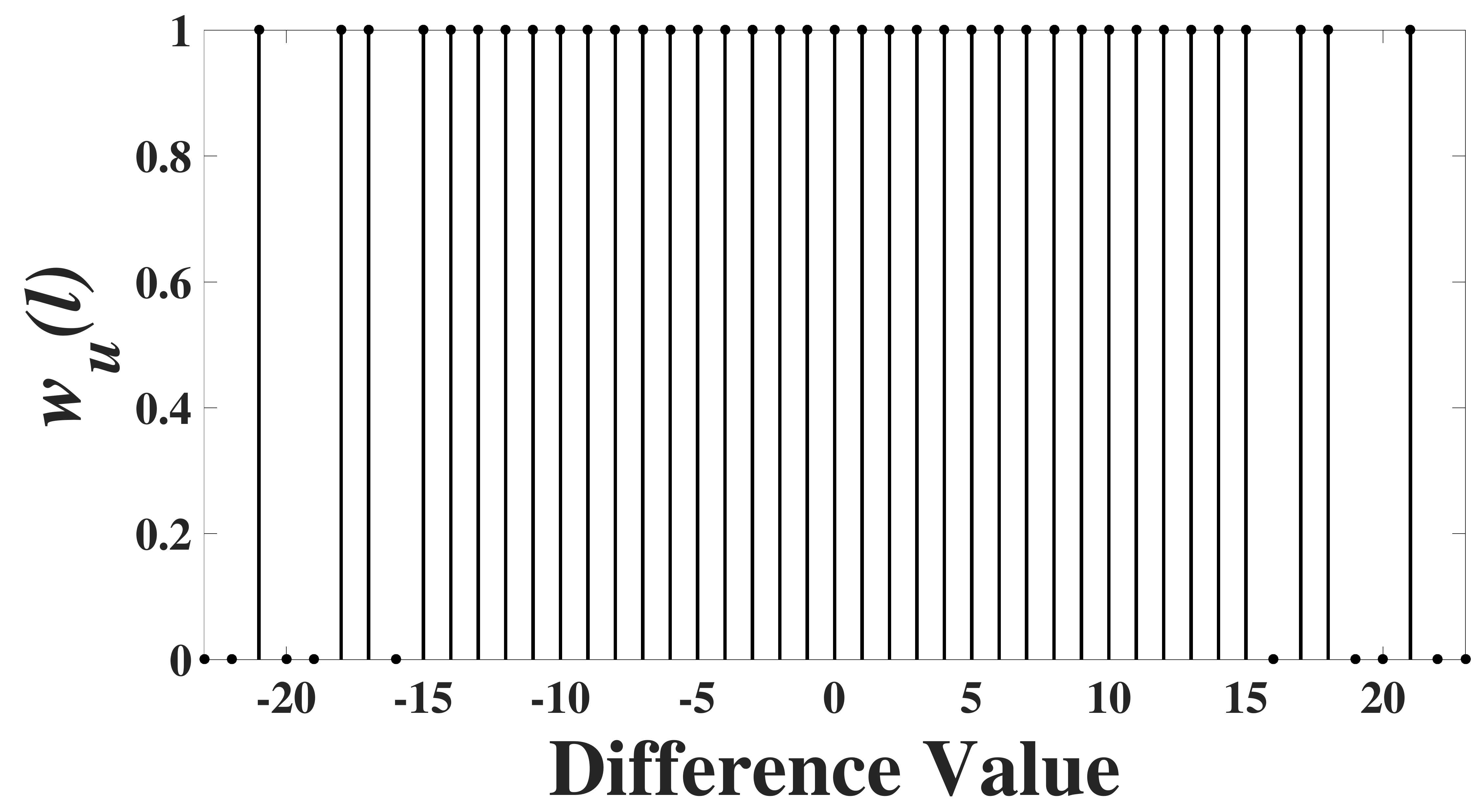}
\includegraphics[width=0.24\textwidth]{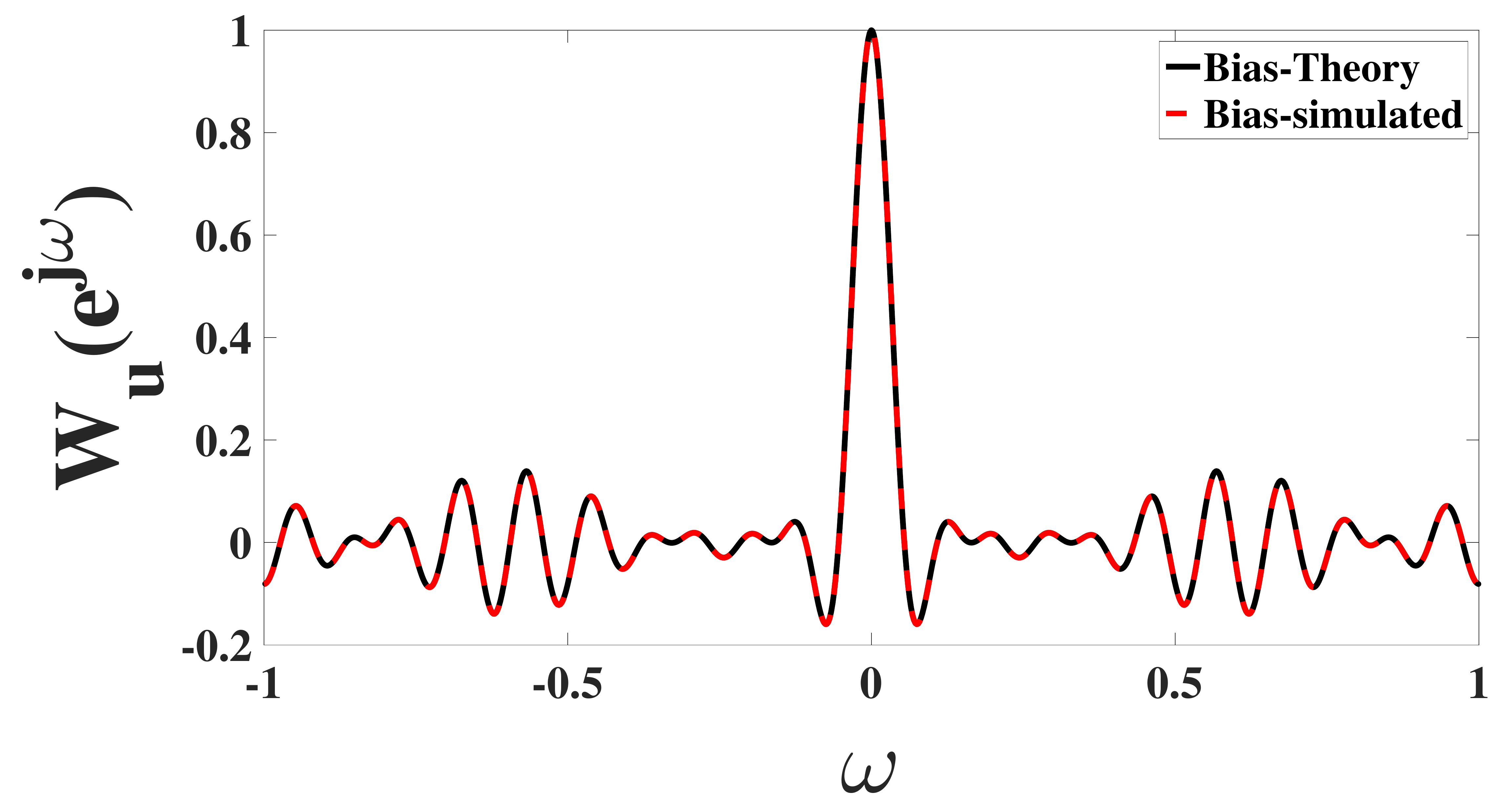}
\label{ext_M4N3_unbiased_Full}}
\hfil
\subfloat[Continuous Range]{
\includegraphics[width=0.24\textwidth]{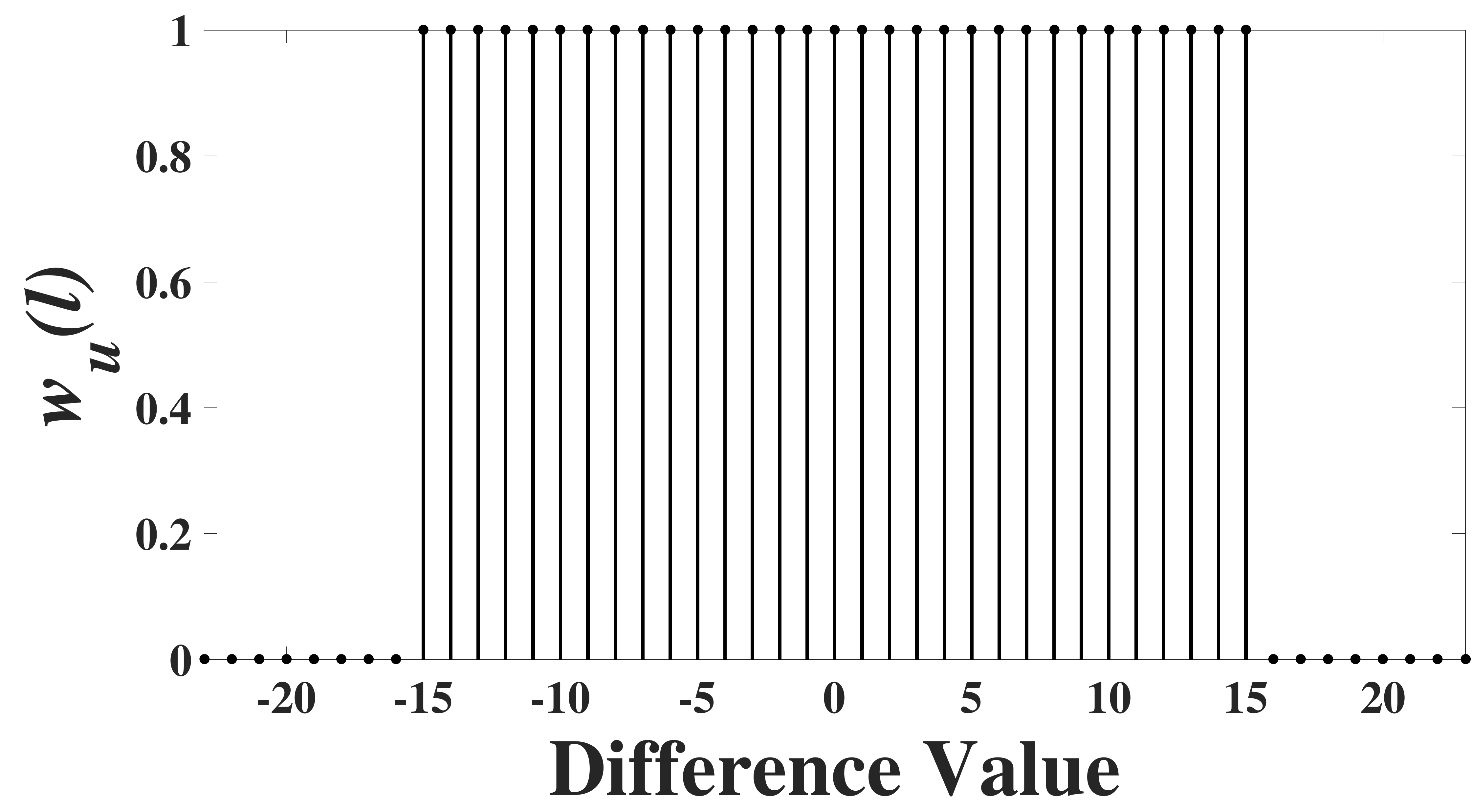}
\includegraphics[width=0.24\textwidth]{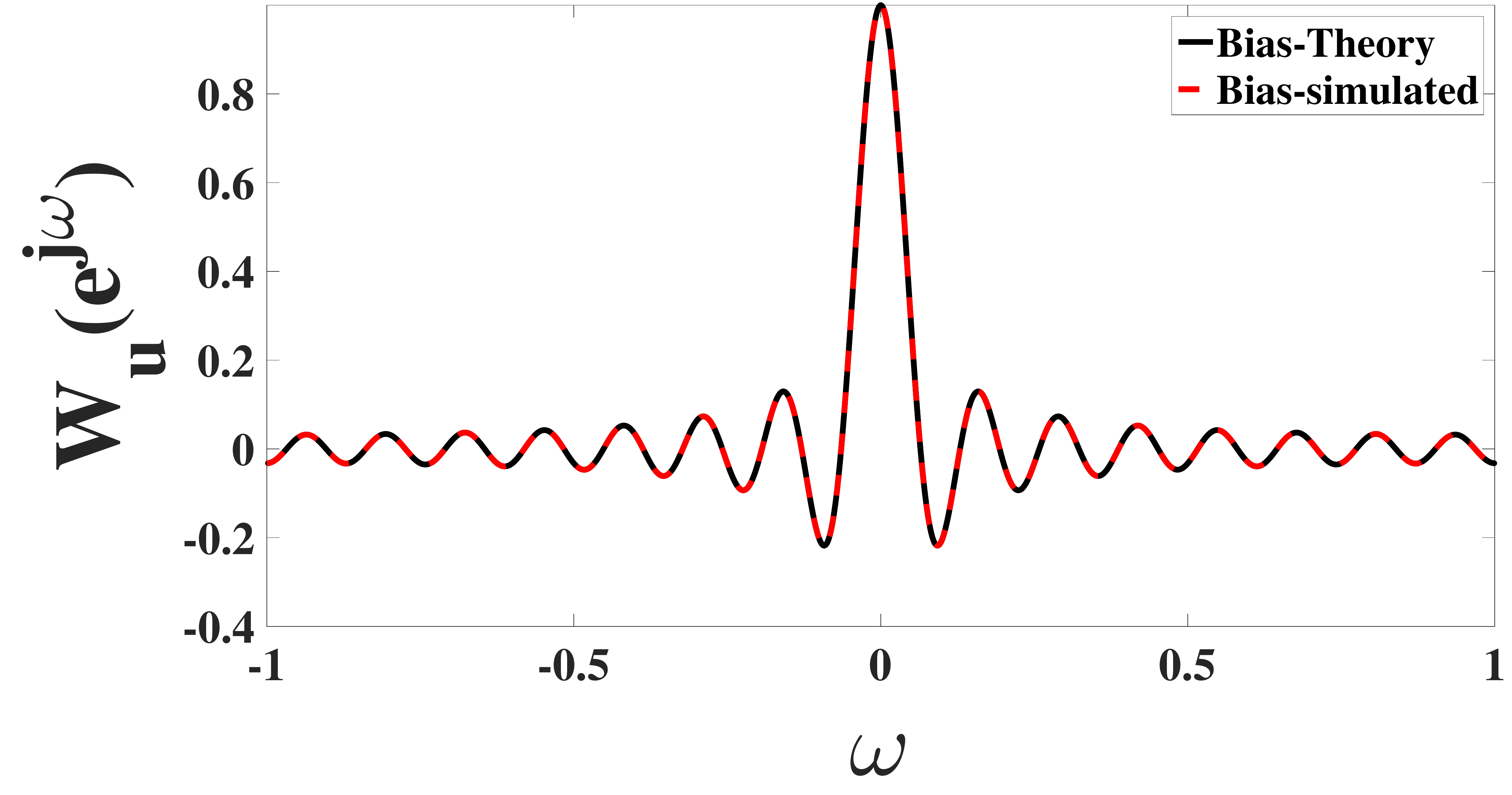}
\label{ext_M3N4_unbiased_Continuous}}
\hfil
\subfloat[Prototype Range]{
\includegraphics[width=0.24\textwidth]{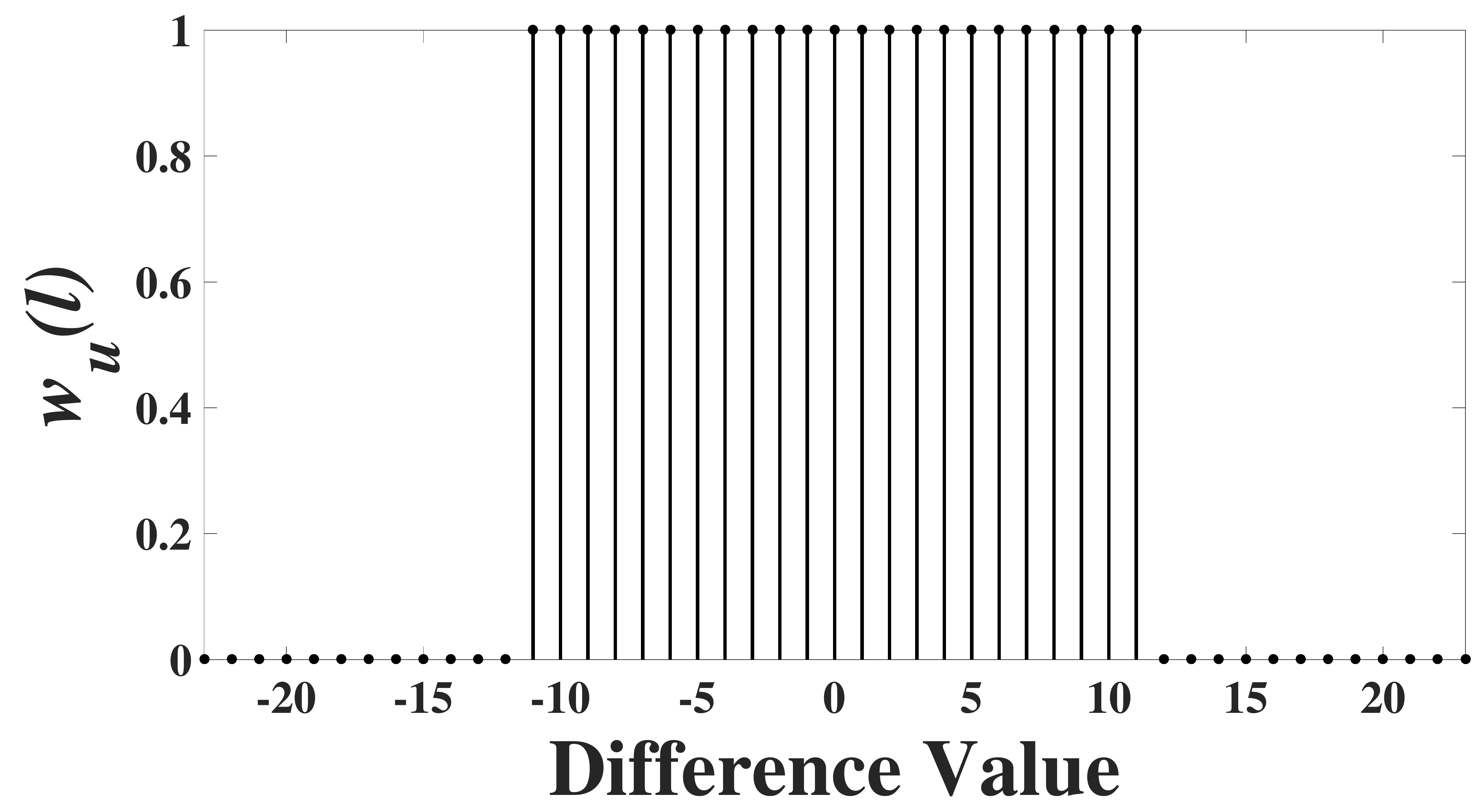}
\includegraphics[width=0.24\textwidth]{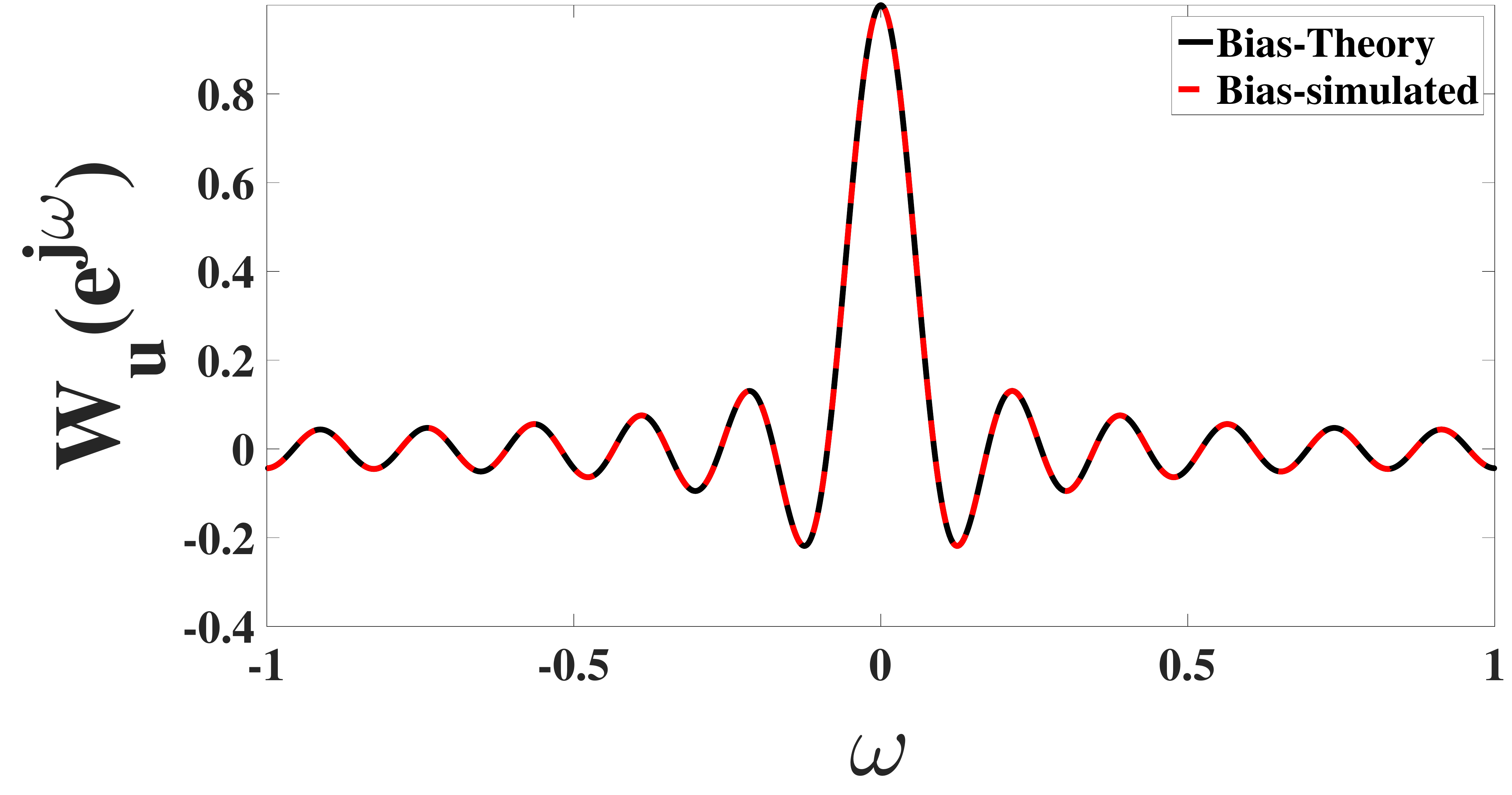}
\label{ext_M4N3_unbiased_Prototype}}
\caption{Window $w_u(l)$ and its corresponding bias $W_u(e^{j\omega})$.}
\label{fig:Extended_window_bias_UNBIASED_AUTO}
\end{figure}

For $M>N$, ${\L}^+_{B}-{\L}_{S}$ has $n\in [1,N-1]$ and the lower limit of $m$ is $M+1$ while the upper limit is given by~\eqref{eq:Dc_rangeM>N} and~\eqref{eq:Dp_rangeM>N} for the continuous and prototype co-prime range respectively.
\begin{eqnarray}\label{eq:Dc_rangeM>N}
 \nonumber Mn-Nm &\geq&-(MN+M-1) \\
  m &\leq&\left\lfloor M+\frac{M}{N}+\frac{1}{N}(Mn-1)\right\rfloor 
\end{eqnarray}
\begin{eqnarray}\label{eq:Dp_rangeM>N}
 \nonumber Mn-Nm &\geq&-(MN-1) \\
  m &\leq&\left\lfloor M+\frac{1}{N}(Mn-1)\right\rfloor 
\end{eqnarray}
For $N>M$, ${\L}^-_{B}-{\L}_{S}$ has $m\in [M+1,2M-1]$ and the upper limit of $n$ is $N-1$ while the lower limit is given by equation \eqref{eq:Dc_rangeN>M} and \eqref{eq:Dp_rangeN>M} for the continuous and co-prime range respectively.
\begin{eqnarray}\label{eq:Dc_rangeN>M}
 \nonumber Mn-Nm &\geq&-(MN+M-1) \\
  n &\geq&\left\lceil -(N+1)+\frac{1}{M}(Nm+1)\right\rceil
\end{eqnarray}
\begin{eqnarray}\label{eq:Dp_rangeN>M}
 \nonumber Mn-Nm &\geq&-(MN-1) \\
  n &\geq&\left\lceil -N+\frac{1}{M}(Nm+1)\right\rceil
\end{eqnarray}
Equations~\eqref{eq:extend_continuous} and~\eqref{eq:extend_proto_period} are presented for the case when $M>N$. Fig.~\ref{fig:Limits_LB_set_N>M} $(N>M)$ has the same elements in the continuous and co-prime range as that in Fig.~\ref{fig:Limits_LB_set_M>N} $(M>N)$, which implies that the equations derived for $M>N$ holds true for $N>M$.
The correlogram bias window is described in~\cite{UVD_PHD} where $w_u(l)$ and $w_b(l)$ represent the window for the case when an unbiased and biased autocorrelation estimator is used to obtain the correlogram spectral estimate respectively. It may be noted that $w_u(l)$ is a rectangular window for the continuous set and the prototype co-prime period. However, it has zeros at locations corresponding to the holes for the entire difference set. The Fourier transform of the windows $w_u(l)$ and $w_b(l)$ represent the bias that perturbs the true spectrum and is denoted by $W_u(e^{j\omega})$ and $W_b(e^{j\omega})$ respectively. The bias  $W_u(e^{j\omega})$ is given by~\eqref{ext_FT_window_u_full} for the entire set with $L=2MN-1$, while~\eqref{ext_FT_window_u_continuous} represents the bias for the continuous and prototype range where $L$ is equal to $MN+M-1$ and $MN-1$ respectively. 
\begin{equation}\label{ext_FT_window_u_full}
\begin{split}
 &W_u(e^{j\omega})=2cos(\frac{\omega MN}{2})\frac{sin(\frac{\omega M(N-1)}{2})}{sin(\frac{\omega M}{2})}\\
&+2cos(\omega MN)\frac{sin(\frac{\omega N(2M-1)}{2})}{sin(\frac{\omega N}{2})}+1\\
&+[1+2cos(\omega MN)]\frac{sin(\frac{\omega M(N-1)}{2})sin(\frac{\omega N(M-1)}{2})}{sin(\frac{\omega M}{2})sin(\frac{\omega N}{2})}
\end{split}
\end{equation}
\begin{equation}\label{ext_FT_window_u_continuous}
  W_u(e^{j\omega})=\frac{sin(\frac{\omega(2L+1)}{2})}{sin(\frac{\omega}{2})}
\end{equation}
Equation~\eqref{ext_FT_window_u_continuous} is obvious since it is the Fourier transform of a rectangular function, while the proof for~\eqref{ext_FT_window_u_full} is given here.\\
\textit{Proof:}
The closed-form expression for the weight function of the unbiased autocorrelation estimator for the entire difference set of an extended co-prime array is given below:
\begin{equation}
\begin{split}
\nonumber w_u(l)=\sum\limits_{m=1}^{2M-1}\delta(\mid l\mid-Nm)
+\sum\limits_{n=1}^{N-1}\sum\limits_{m=1}^{M-1}\delta(l-(Mn-Nm))+\\
\delta(l)+\sum\limits_{n=1}^{N-1}\delta(\mid l\mid-Mn)+\sum\limits_{n=1}^{N-1}\sum\limits_{m=M+1}^{2M-1}\delta(\mid l\mid+(Mn-Nm))
\end{split}
\end{equation}
Note that the first four terms are same as that of the prototype co-prime array~\cite{UVD_PHD}, except that the upper limit of the first term is $2M-1$ in this case. The contribution of the last term to the bias is given by:
\begin{eqnarray}
\nonumber   &&\sum\limits_{l=-(L-1)}^{L-1}\sum\limits_{n=1}^{N-1}\sum\limits_{m=M+1}^{2M-1}\delta(\mid l\mid+(Mn-Nm)) e^{-j\omega l}\\
\nonumber   &=&2cos(\omega MN)\frac{sin(\frac{\omega M(N-1)}{2})sin(\frac{\omega N(M-1)}{2})}{sin(\frac{\omega M}{2})sin(\frac{\omega N}{2})}
\end{eqnarray}
It may be noted that the identity $\sum\limits_{m=M+1}^{2M-1}\alpha ^{m}=\alpha^{M+1}[\frac{1-\alpha^{M-1}}{1-\alpha}]$ is used to arrive at the above equation.
By combining the five terms, we get~\eqref{ext_FT_window_u_full}. 
The weight $w_u(l)$ and the bias $W_u(e^{j\omega})$ for $M=4$ and $N=3$ is given in Fig.~\ref{fig:Extended_window_bias_UNBIASED_AUTO}. It is evident that the derived bias expressions match the simulated FFT of $w_u(l)$.
The bias in this case is not positive and hence cannot guarantee a positive spectral estimate. This was also found to be true for the prototype co-prime and the Nyquist schemes when an unbiased autocorrelation estimator of fixed length is employed.

The bias window that perturbs the true spectrum when a biased autocorrelation estimator is employed in the correlogram method is given by~\eqref{eq:FT_extend_full}-\eqref{eq:FT_extend_proto_period} for the three cases in~\eqref{eq:extend_entire}-\eqref{eq:extend_proto_period} respectively. %
 \newcounter{mytempeqncnt}
 \begin{figure*}[!t]
\small
 \setcounter{mytempeqncnt}{\value{equation}}
 \setcounter{equation}{9}
\begin{equation}\label{eq:FT_extend_full}
\begin{split}
& W_{b_f}(e^{j\omega})=\frac{1}{s_b}\left\{\left |\frac{sin(\frac{\omega MN}{2})}{sin(\frac{\omega M}{2})}\right|^2+\frac{sin(\frac{\omega M(2N-1)}{2})}{sin(\frac{\omega M}{2})}\right.
  \left.+\left|\frac{sin(\omega MN)}{sin(\frac{\omega N}{2})}\right|^2 \right.
  \left.+2(1+cos(\omega MN))\frac{sin(\frac{\omega M(N-1)}{2})sin(\frac{\omega N(M-1)}{2})}{sin(\frac{\omega M}{2})sin(\frac{\omega N}{2})}-2\right\}
\end{split}
\end{equation}
\begin{equation}\label{eq:FT_extend_continuous}
\begin{split}
  &W_{b_c}(e^{j\omega})=\frac{1}{s_b}\left\{\left |\frac{sin(\frac{\omega MN}{2})}{sin(\frac{\omega M}{2})}\right|^2+\frac{sin(\frac{\omega M(2N-1)}{2})}{sin(\frac{\omega M}{2})}
  +\left |\frac{sin(\frac{\omega N(M+\lfloor \frac{M-1}{N}\rfloor+1)}{2})}{sin(\frac{\omega N}{2})}\right|^2\right.
  +(M-\lfloor \frac{M-1}{N}\rfloor-1)\frac{sin(\frac{\omega N(2M+2\lfloor \frac{M-1}{N}\rfloor+1)}{2})}{sin(\frac{\omega N}{2})}\\
&+2\frac{sin(\frac{\omega M(N-1)}{2})sin(\frac{\omega N(M-1)}{2})}{sin(\frac{\omega M}{2})sin(\frac{\omega N}{2})}+\sum\limits_{n=1}^{N-1}2cos(\omega (Mn-\frac{MN}{2}-\frac{N(\lfloor \frac{MN+M+(Mn-1)}{N}\rfloor+1)}{2}))
  \left.\frac{sin(\frac{\omega N(\lfloor \frac{MN+M+(Mn-1)}{N}\rfloor-M)}{2})}{sin(\frac{\omega N}{2})}-2\right\}
\end{split}
\end{equation}
\begin{equation}\label{eq:FT_extend_proto_period}
\begin{split}
  &W_{b_p}(e^{j\omega})=\frac{1}{s_b}\left\{\left |\frac{sin(\frac{\omega MN}{2})}{sin(\frac{\omega M}{2})}\right|^2+\frac{sin(\frac{\omega M(2N-1)}{2})}{sin(\frac{\omega M}{2})}\right.
  +\left |\frac{sin(\frac{\omega MN}{2})}{sin(\frac{\omega N}{2})}\right|^2
  +M\frac{sin(\frac{\omega N(2M-1)}{2})}{sin(\frac{\omega N}{2})}\\
  &+2\frac{sin(\frac{\omega M(N-1)}{2})sin(\frac{\omega N(M-1)}{2})}{sin(\frac{\omega M}{2})sin(\frac{\omega N}{2})}
+\sum\limits_{n=1}^{N-1}2cos(\omega (Mn-\frac{MN}{2}-\frac{N(\lfloor \frac{MN+(Mn-1)}{N}\rfloor+1)}{2}))
  \left.\frac{sin(\frac{\omega N(\lfloor \frac{MN+(Mn-1)}{N}\rfloor-M)}{2})}{sin(\frac{\omega N}{2})}-2\right\}
\end{split}
\end{equation}
\setcounter{equation}{\value{mytempeqncnt}}
\hrulefill
\vspace*{4pt}
\end{figure*}
\addtocounter{equation}{3}
%

\begin{figure*}[!t]
	\centering
	\subfloat[Full: $M=4$, $N=3$]{
		\begin{tabular}{cc}
			\includegraphics[width=0.13\textwidth]{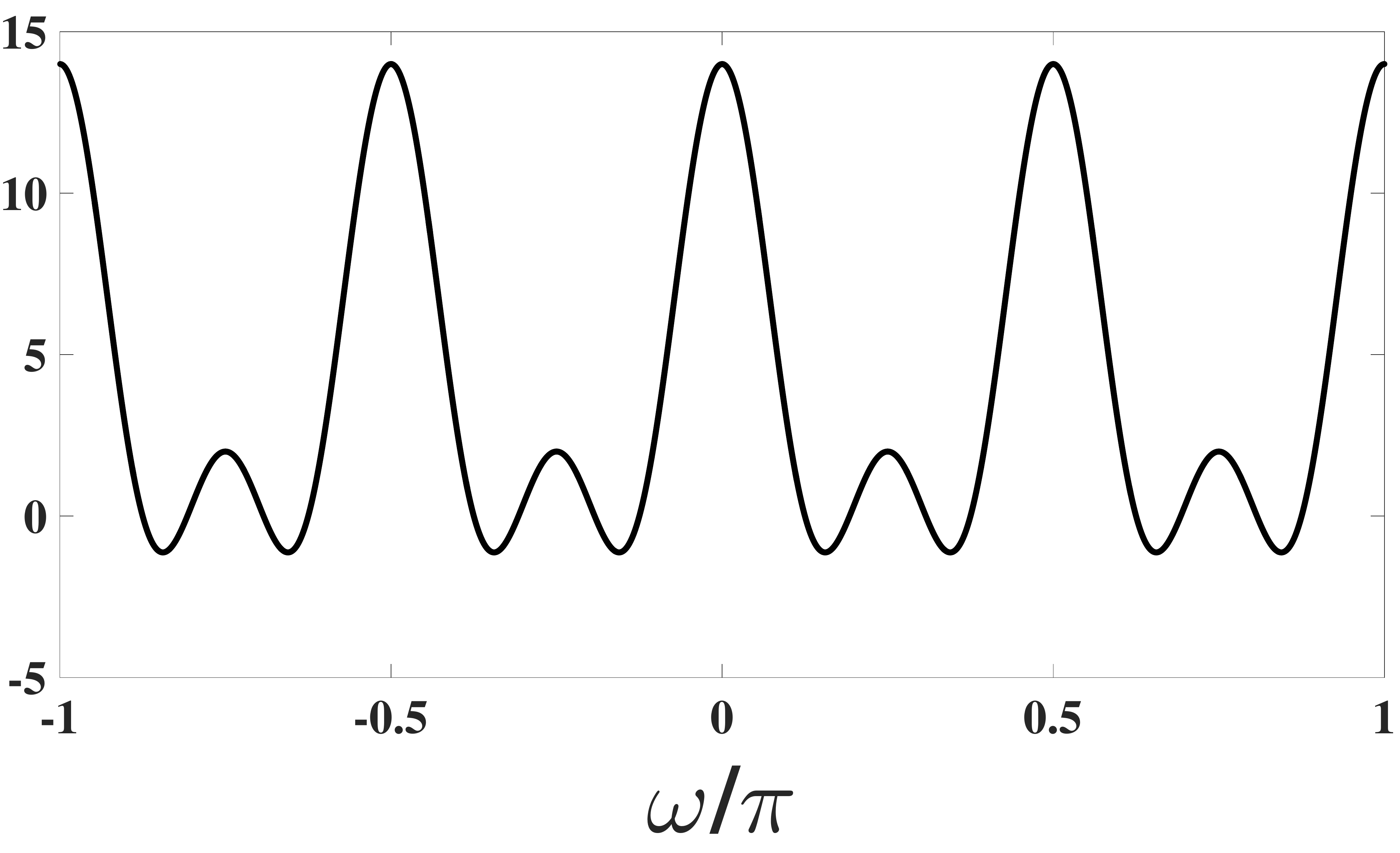}&
			\includegraphics[width=0.13\textwidth]{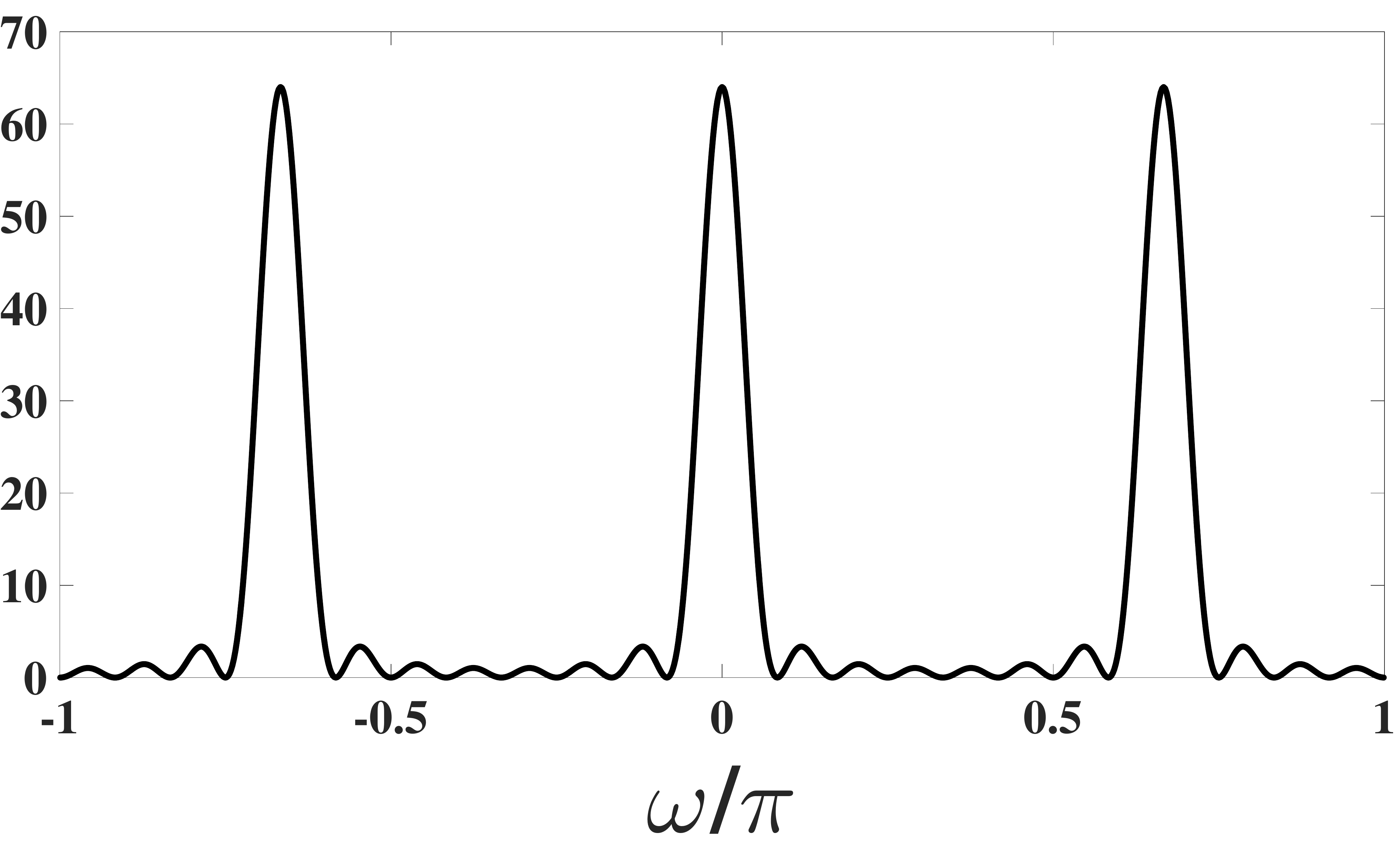}\\
			\includegraphics[width=0.13\textwidth]{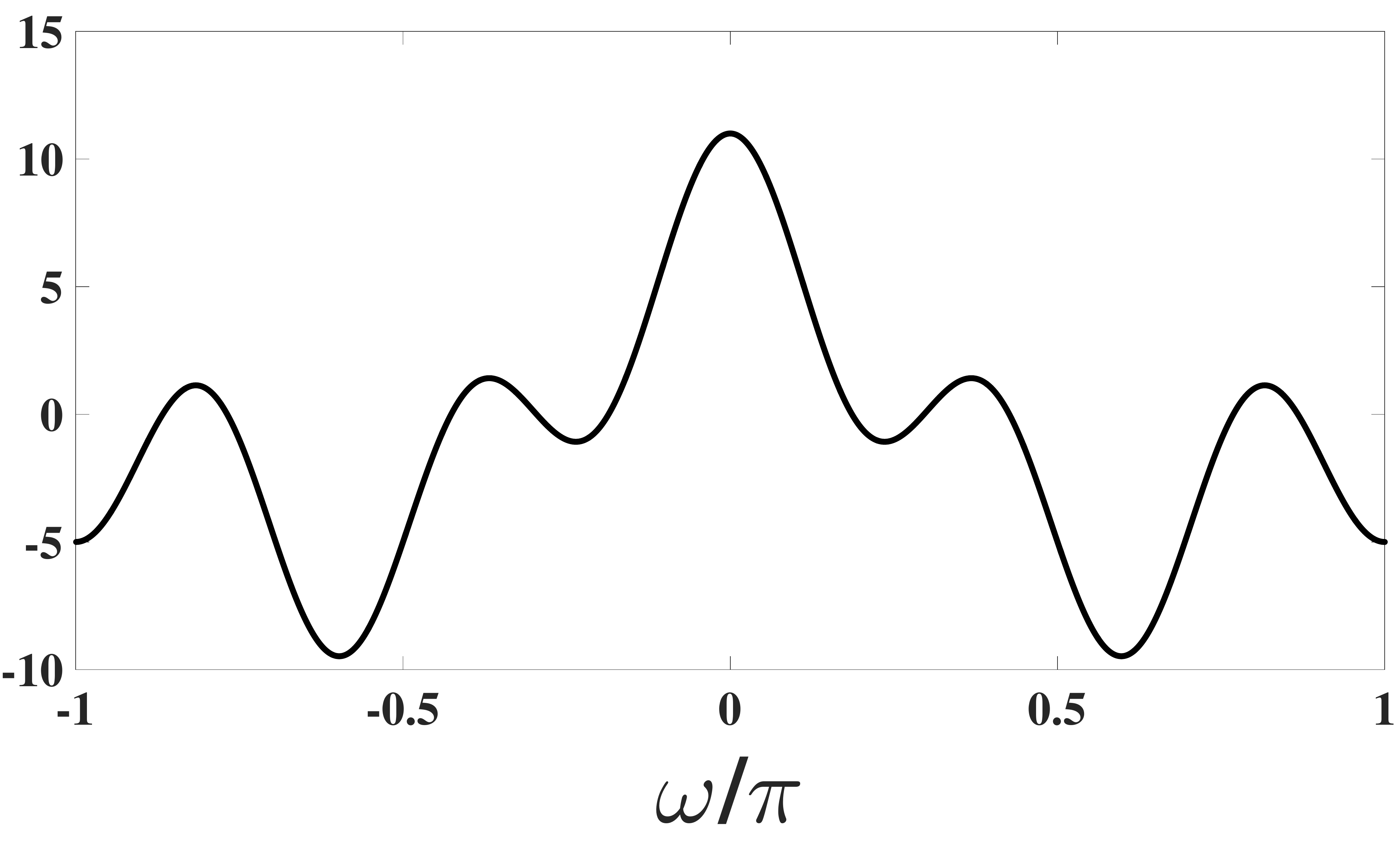}&
			\includegraphics[width=0.13\textwidth]{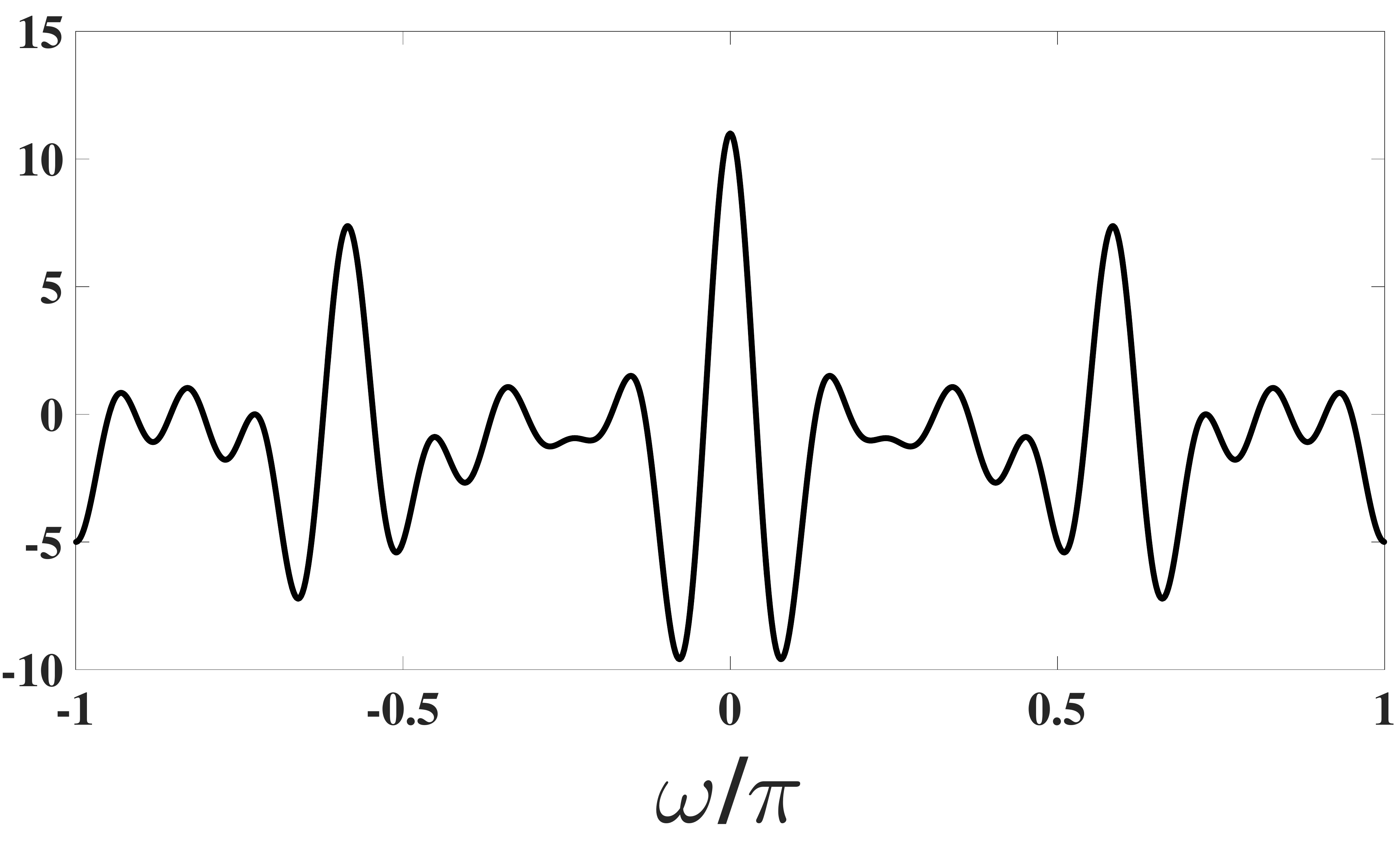} \\
			\multicolumn{2}{c}{
				\includegraphics[width=0.3\textwidth]{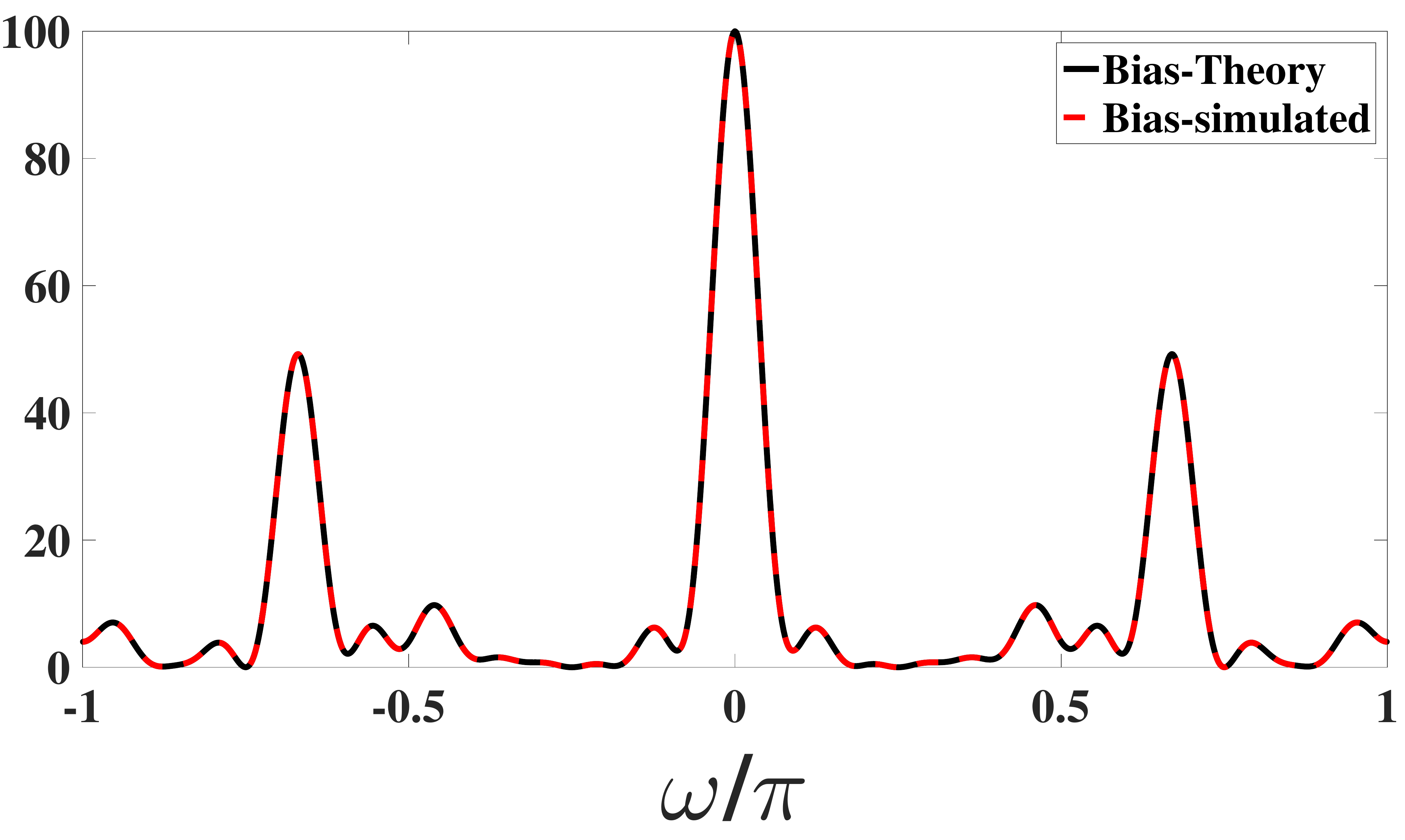}}
		\end{tabular}
		\label{ext_M4N3_fullP}}
	\subfloat[Continuous: $M=4$, $N=3$]{
		\begin{tabular}{cc}
			\includegraphics[width=0.13\textwidth]{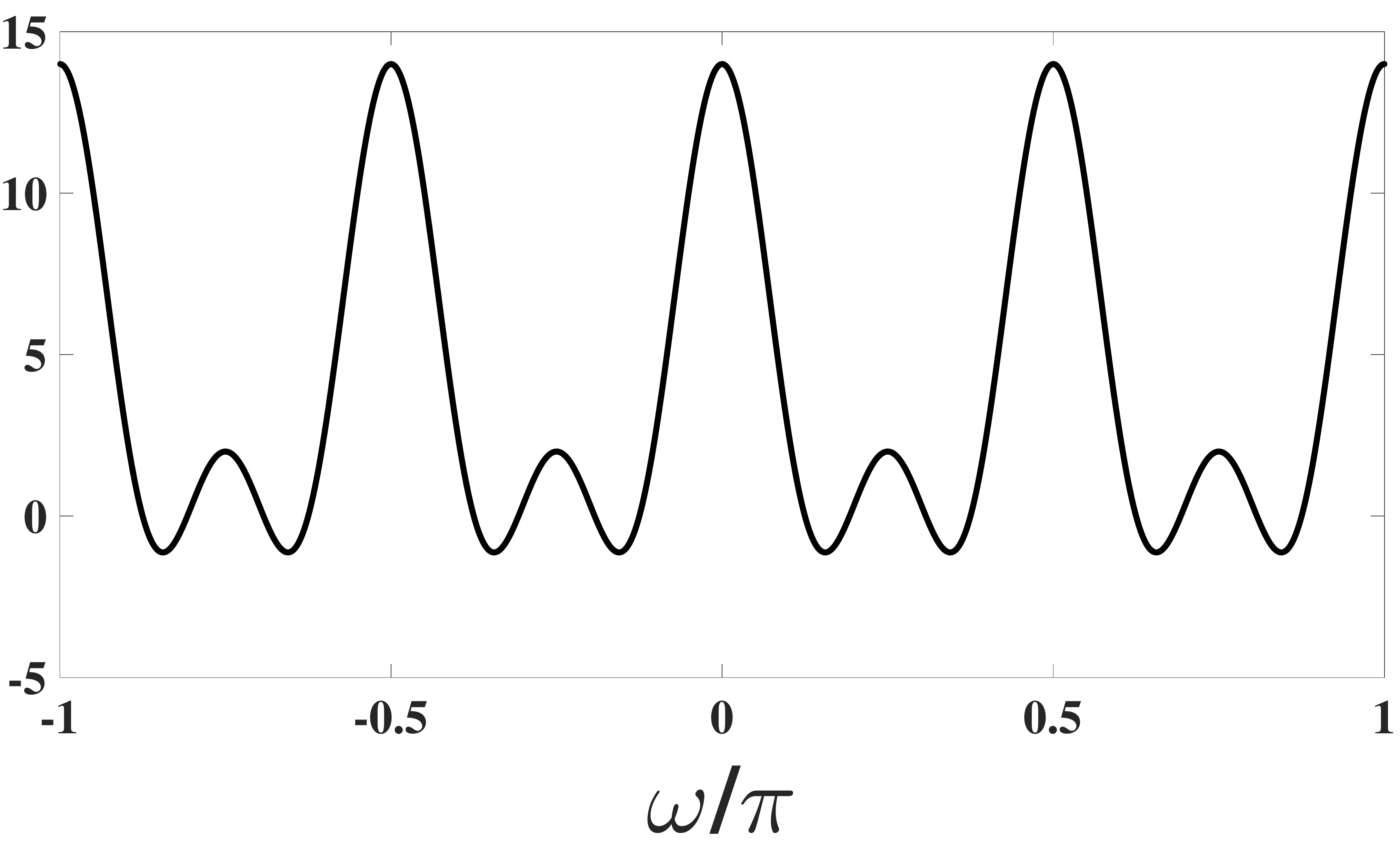}&
			\includegraphics[width=0.13\textwidth]{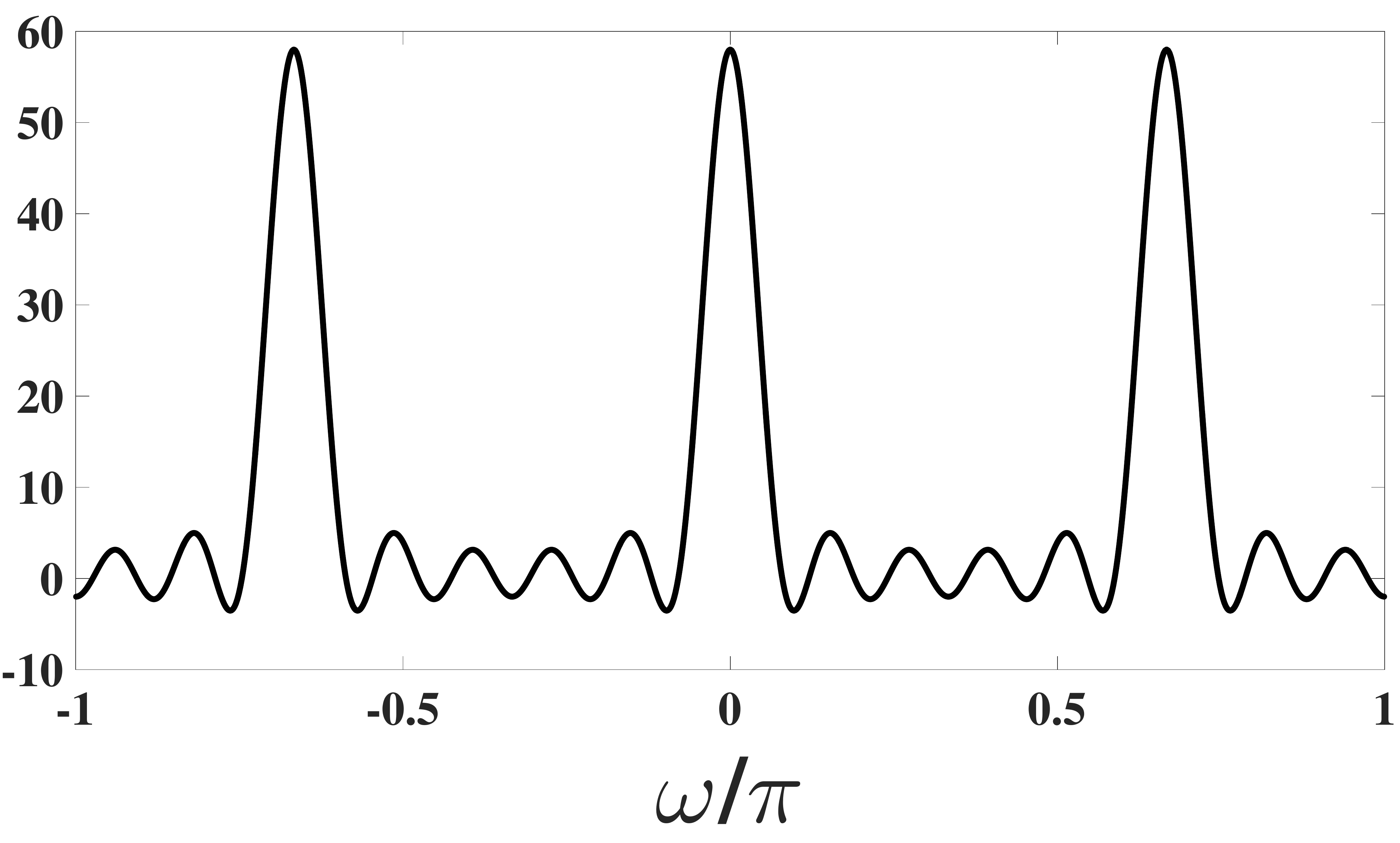}\\
			\includegraphics[width=0.13\textwidth]{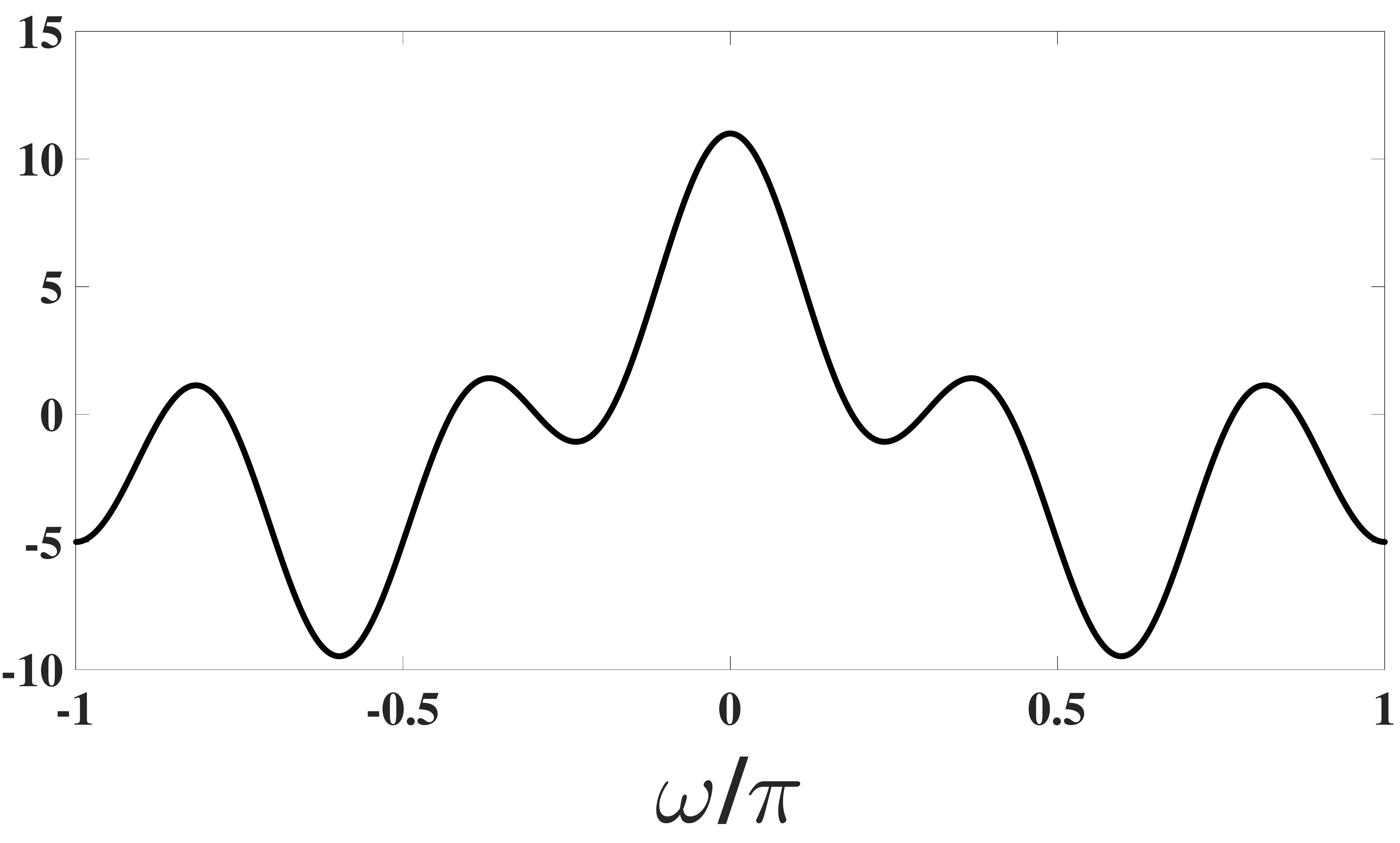}&
			\includegraphics[width=0.13\textwidth]{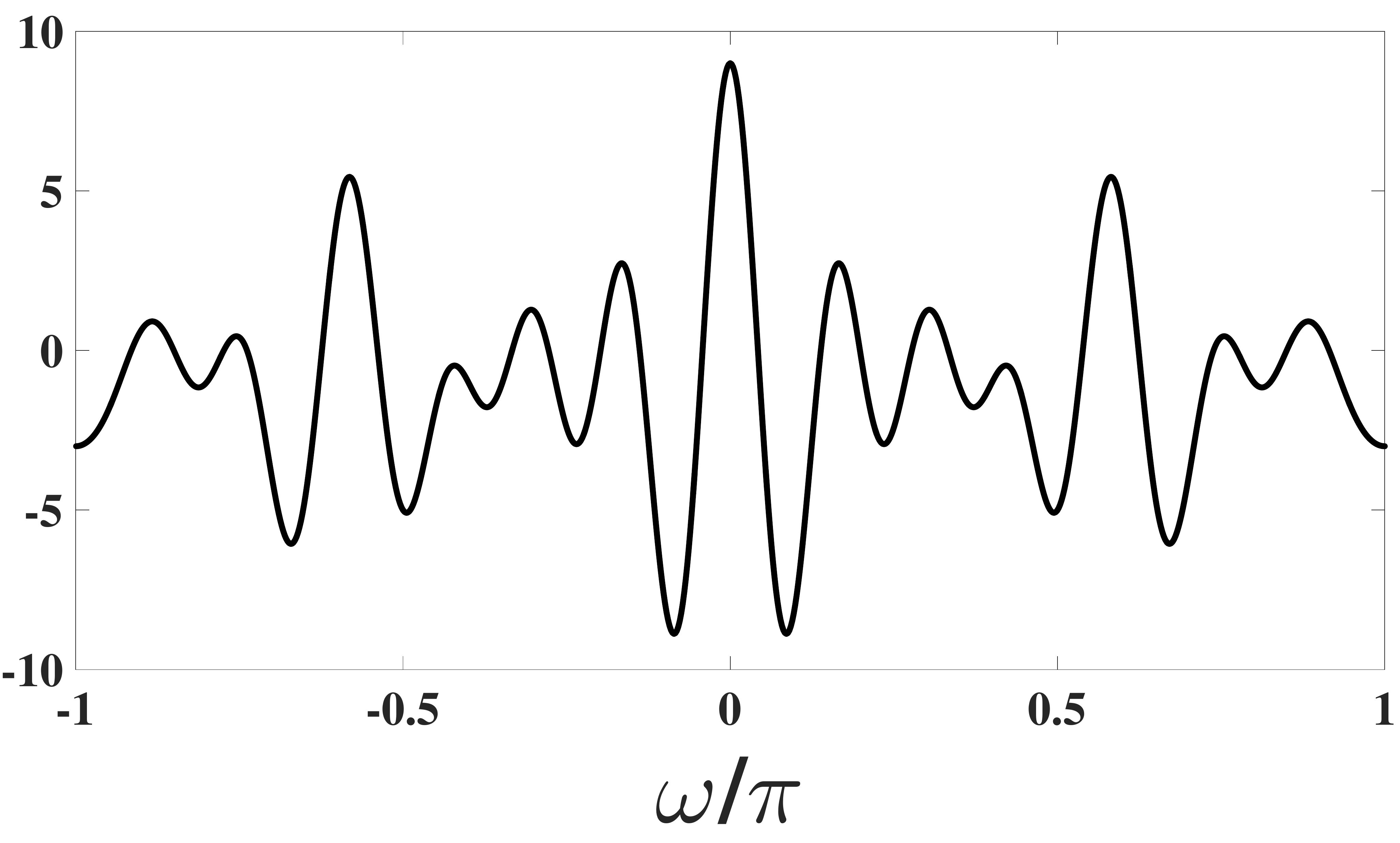} \\
			\multicolumn{2}{c}{
				\includegraphics[width=0.3\textwidth]{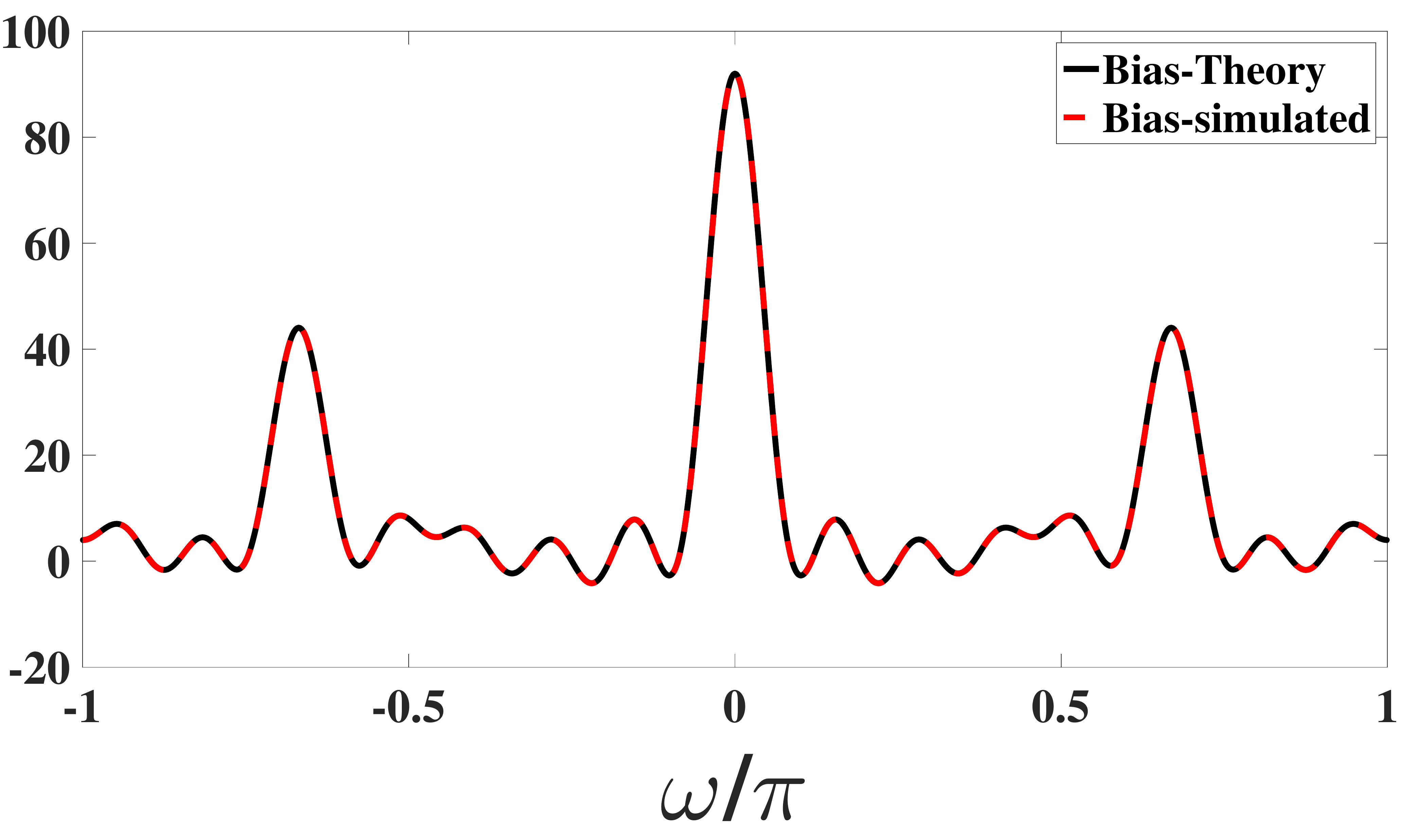}}
		\end{tabular}
		\label{ext_M4N3_continuousP}}
	\subfloat[Prototype: $M=4$, $N=3$]{
		\begin{tabular}{cc}
			\includegraphics[width=0.13\textwidth]{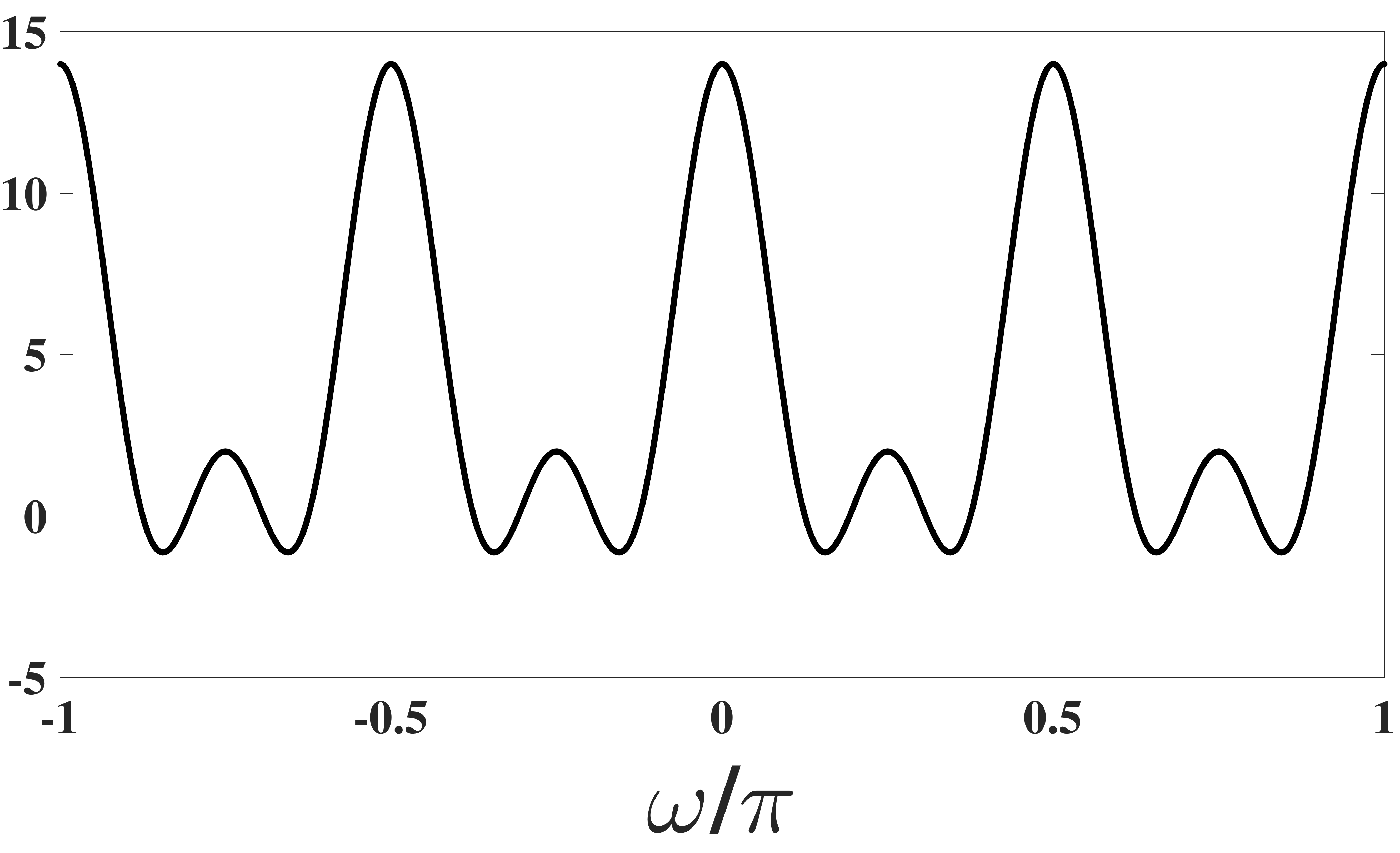}&
			\includegraphics[width=0.13\textwidth]{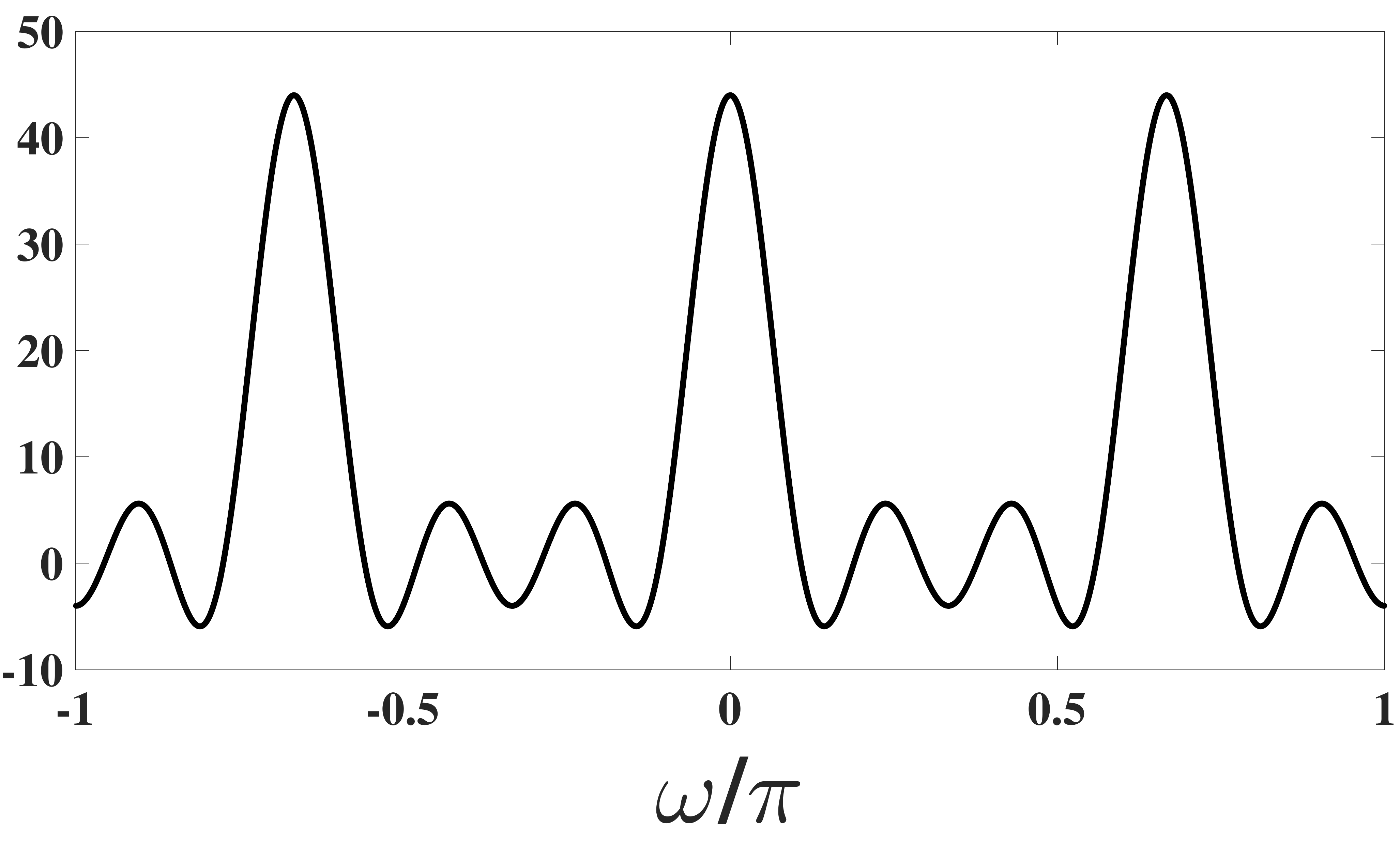}\\
			\includegraphics[width=0.13\textwidth]{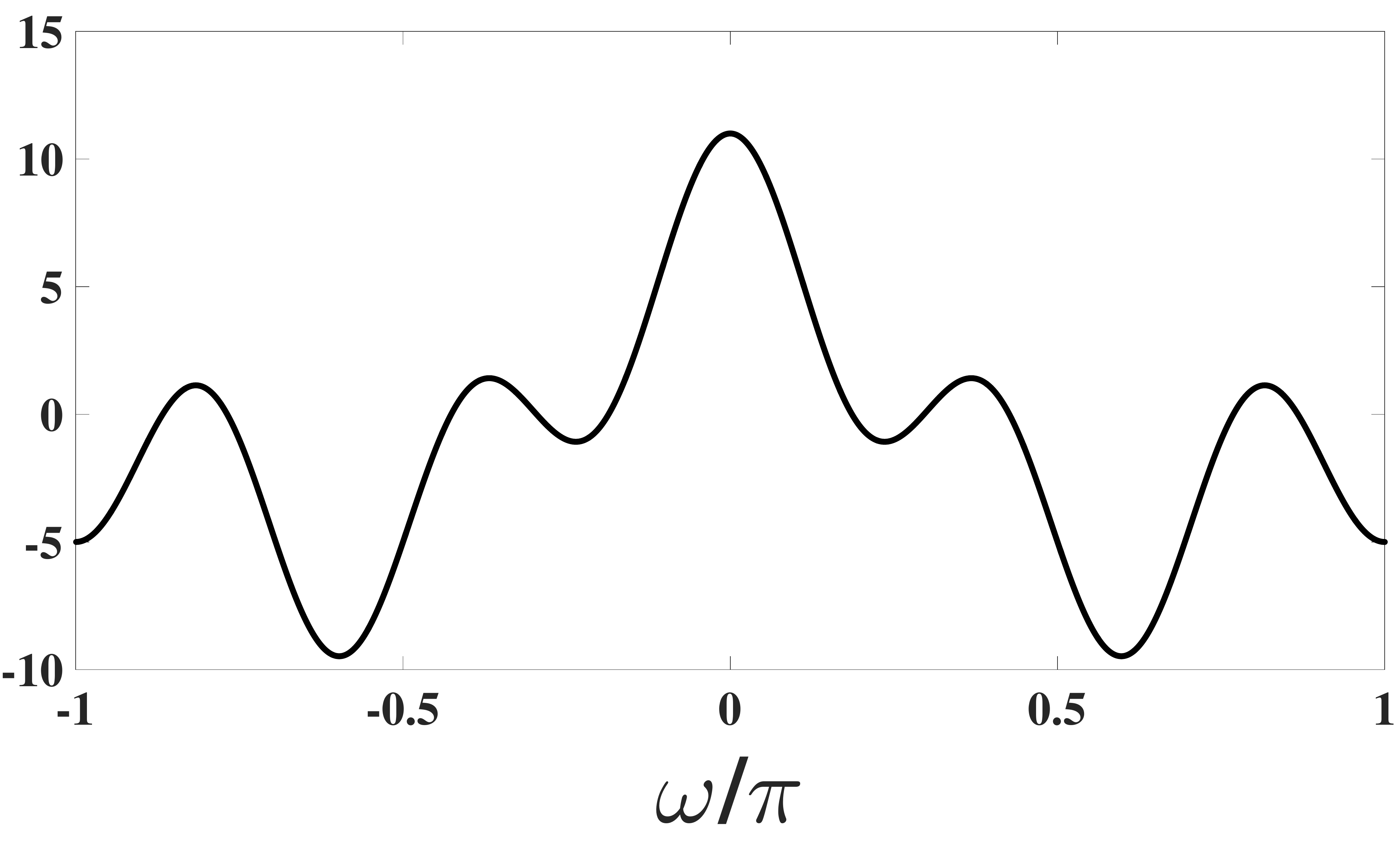}&
			\includegraphics[width=0.13\textwidth]{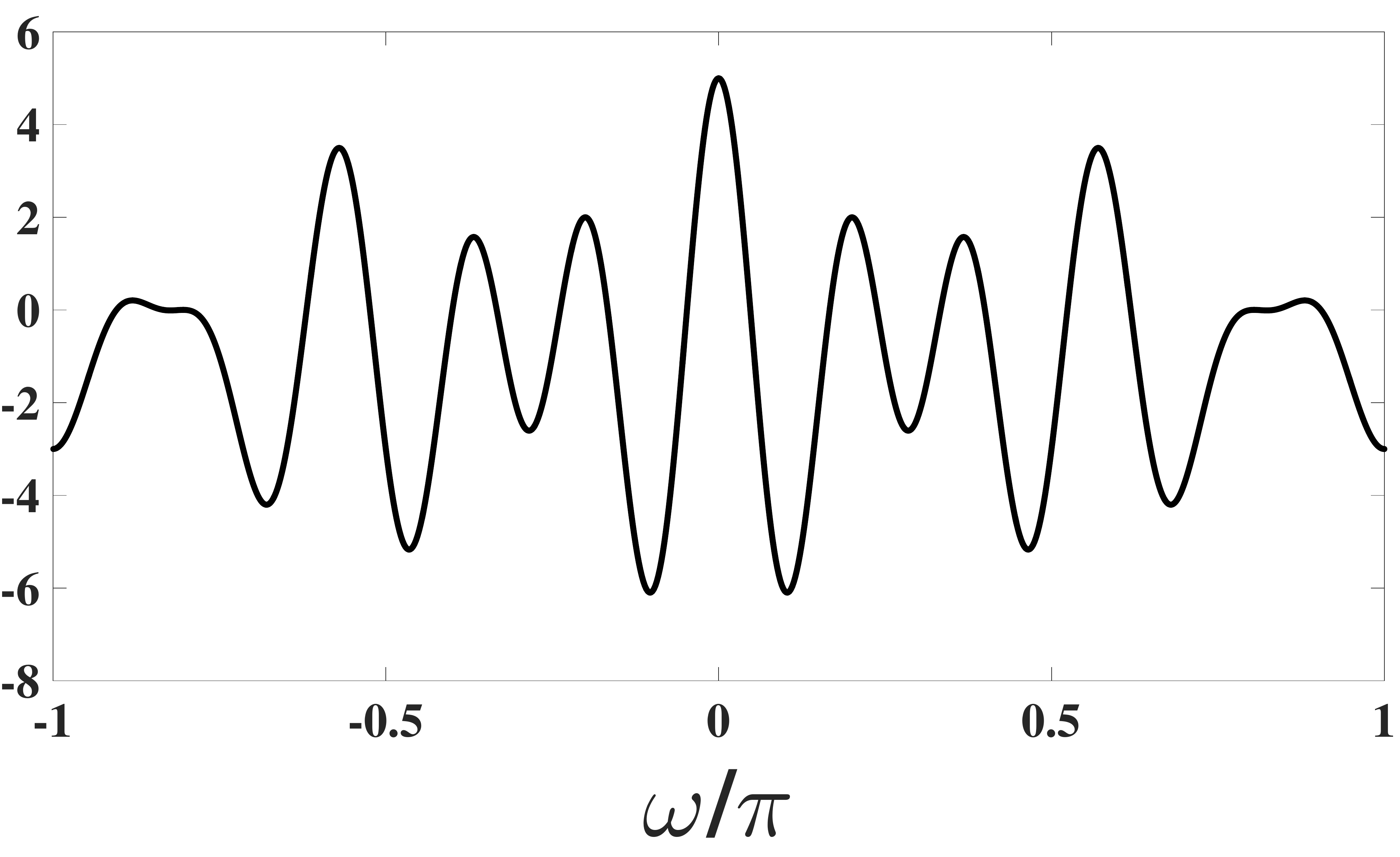} \\
			\multicolumn{2}{c}{
				\includegraphics[width=0.3\textwidth]{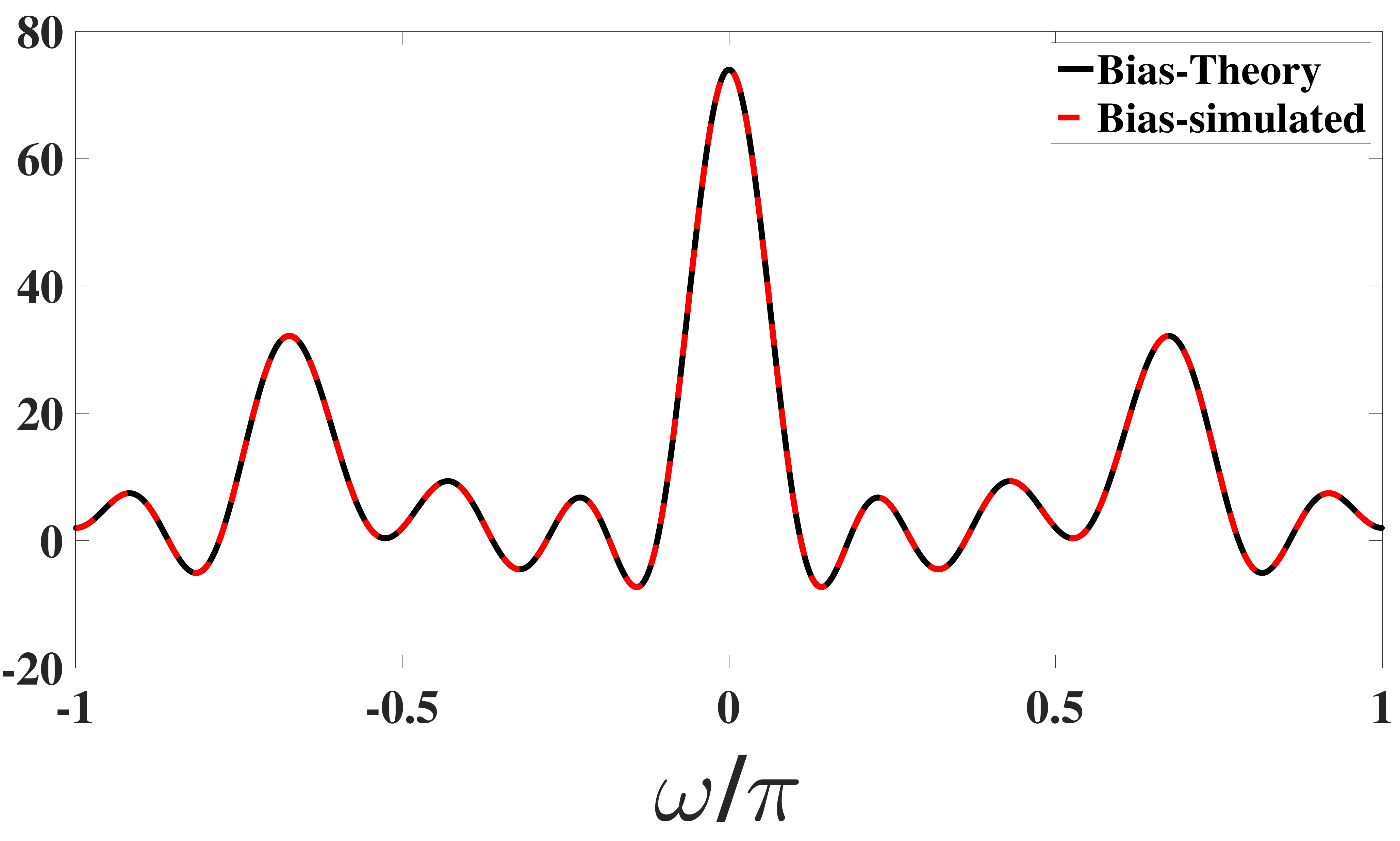}}
		\end{tabular}
		\label{ext_M4N3_prototypeP}}
	\hfil
	\subfloat[Full: $M=3$, $N=4$]{
		\begin{tabular}{cc}
			\includegraphics[width=0.13\textwidth]{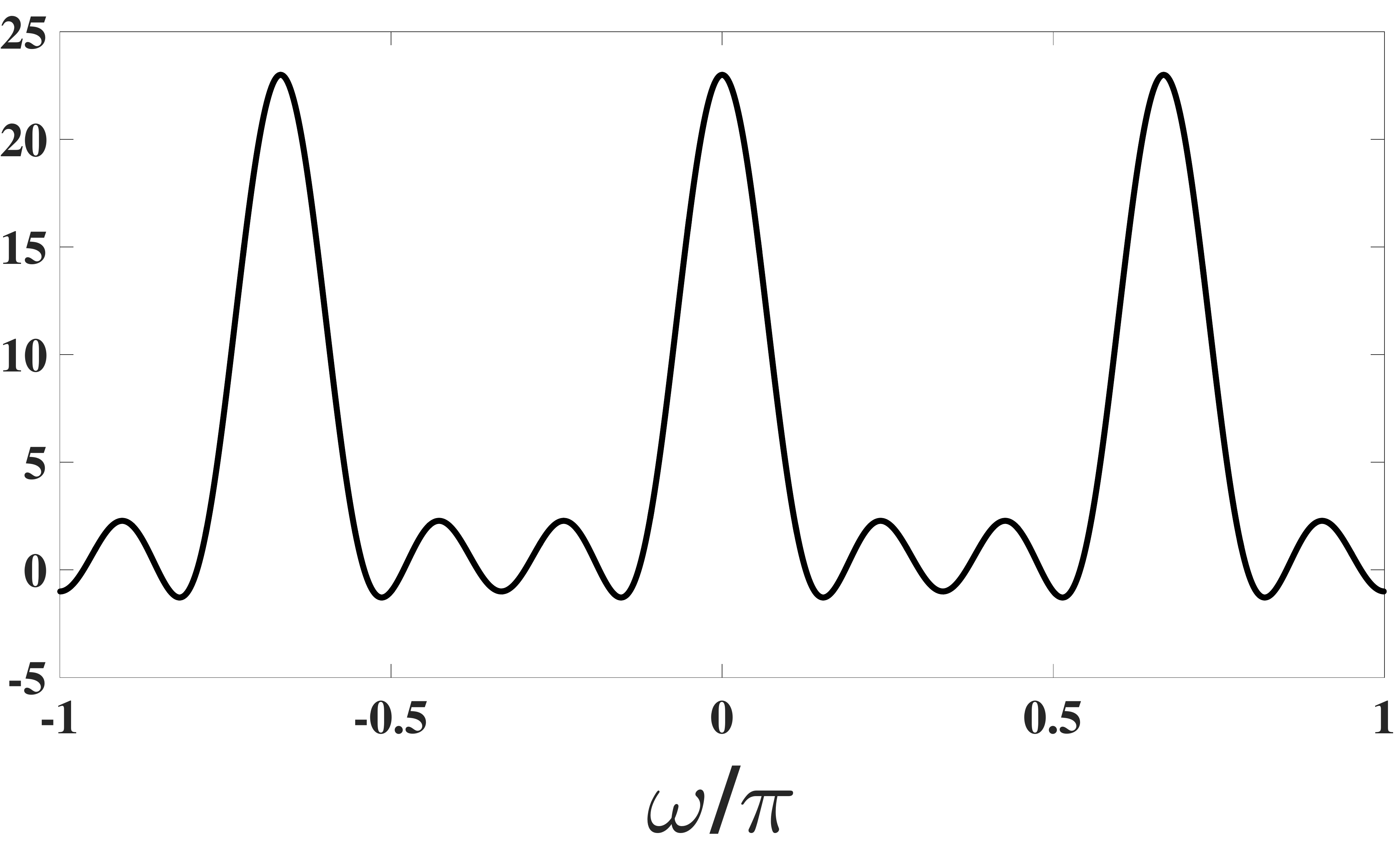}&
			\includegraphics[width=0.13\textwidth]{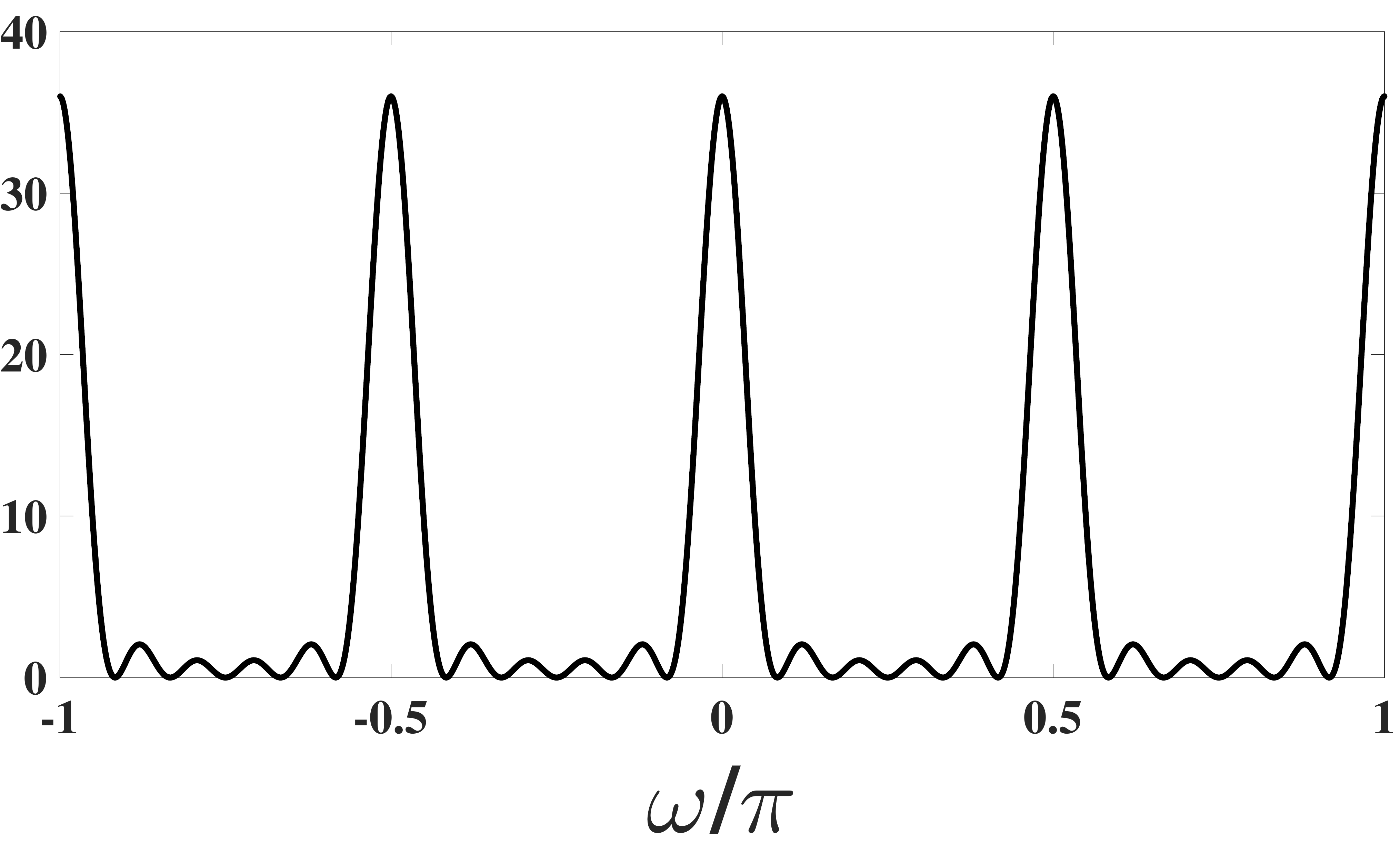}\\
			\includegraphics[width=0.13\textwidth]{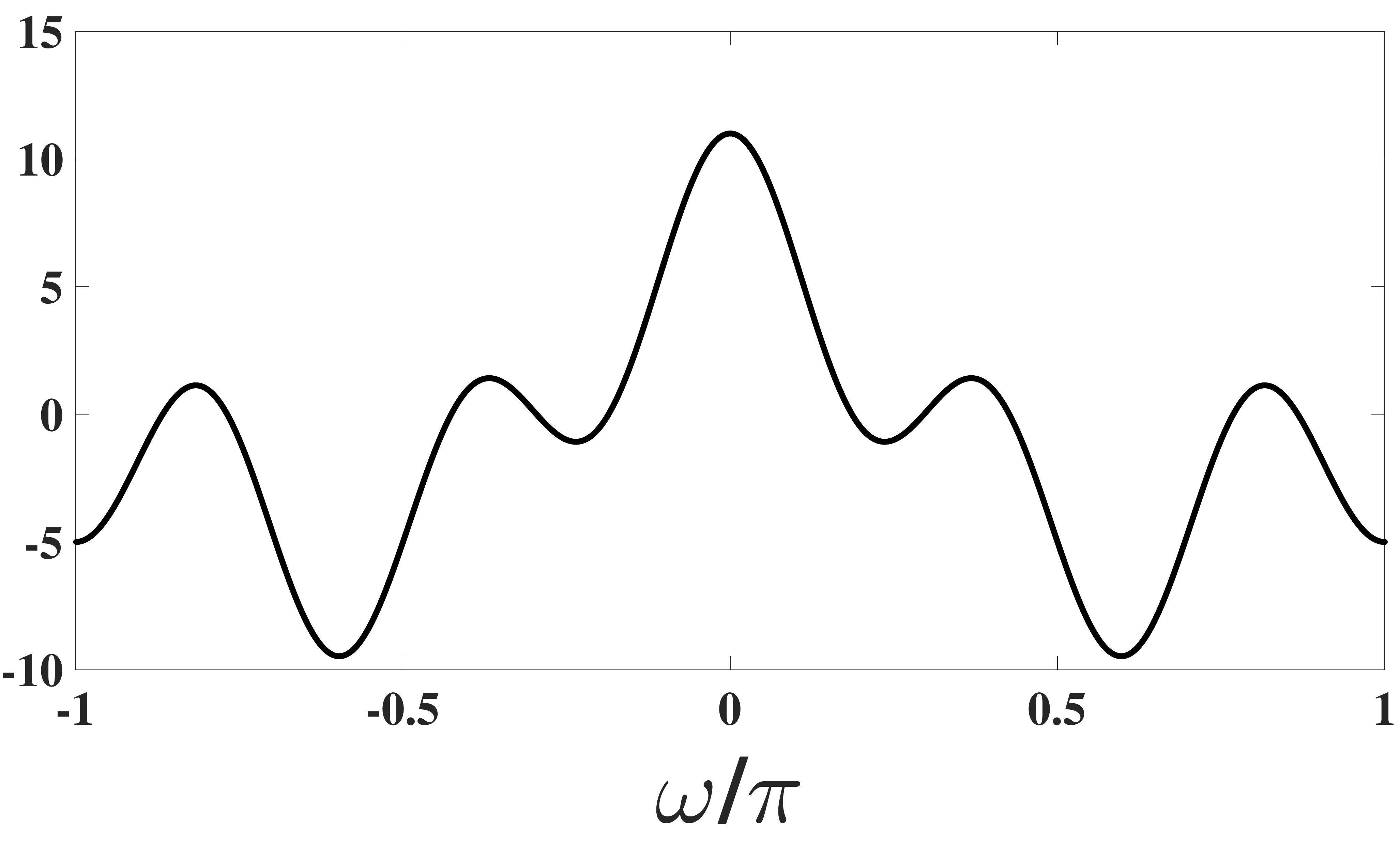}&
			\includegraphics[width=0.13\textwidth]{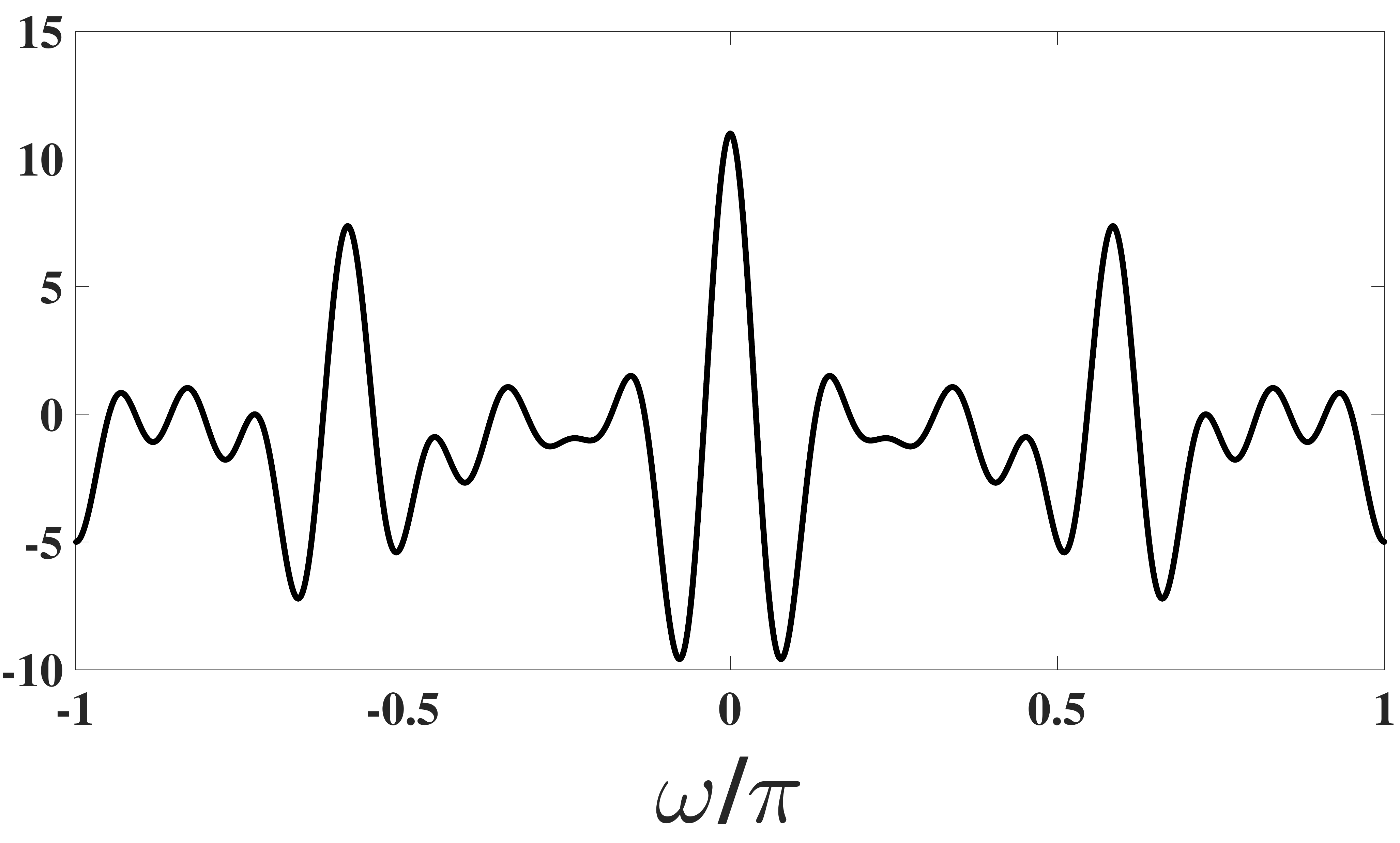} \\
			\multicolumn{2}{c}{
				\includegraphics[width=0.3\textwidth]{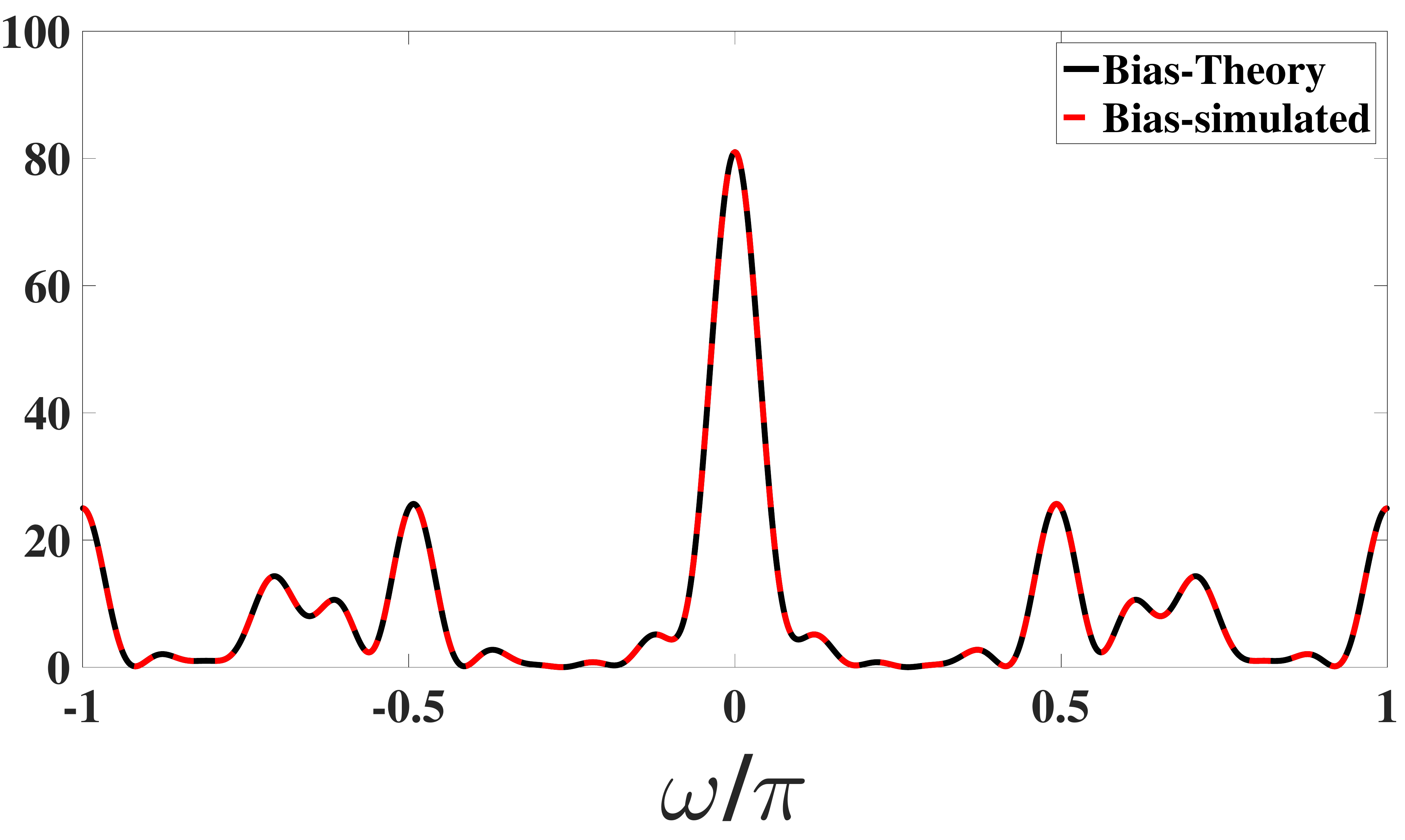}}
		\end{tabular}
		\label{ext_M3N4_fullP}}
	\subfloat[Continuous: $M=3$, $N=4$]{
		\begin{tabular}{cc}
			\includegraphics[width=0.13\textwidth]{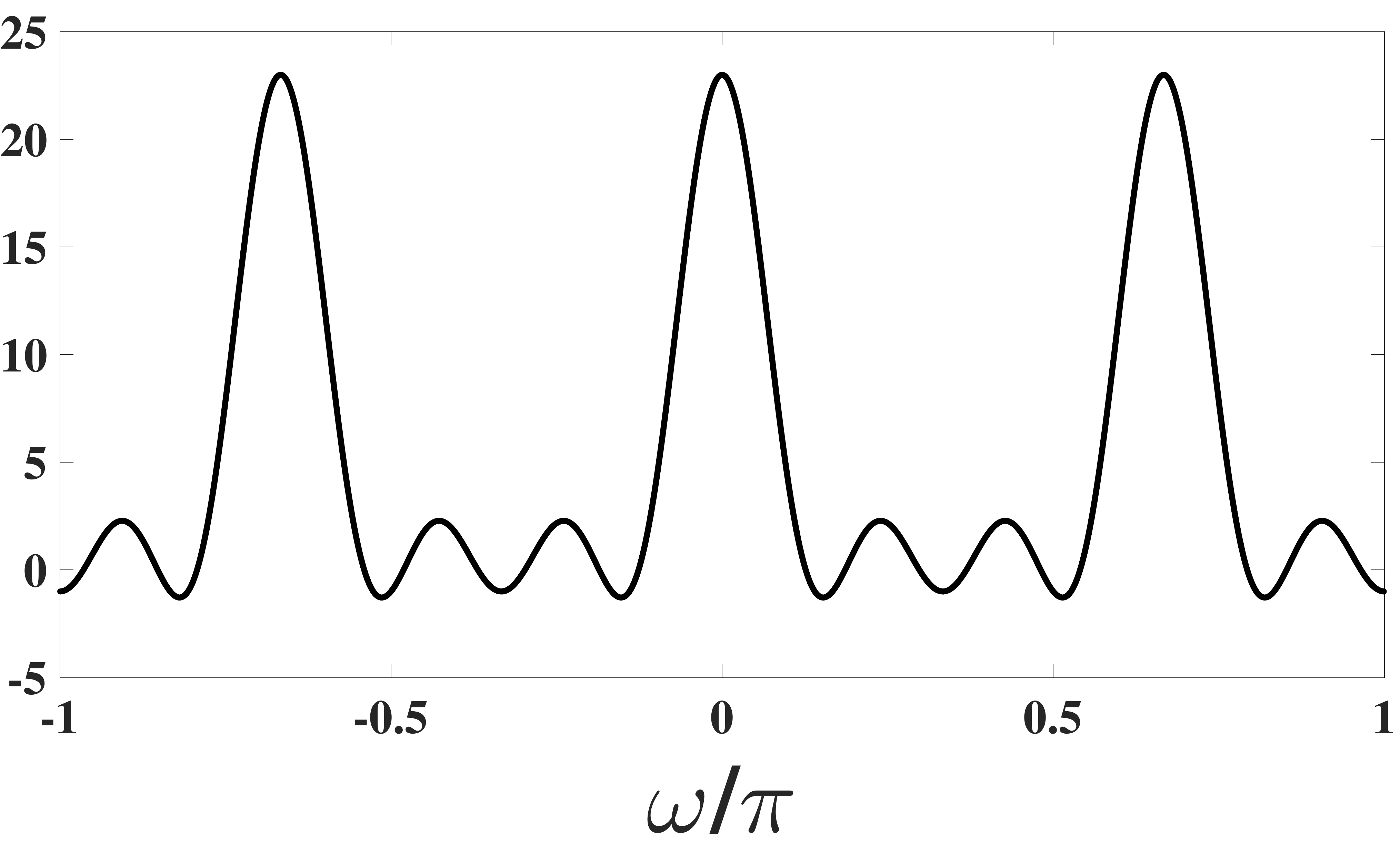}&
			\includegraphics[width=0.13\textwidth]{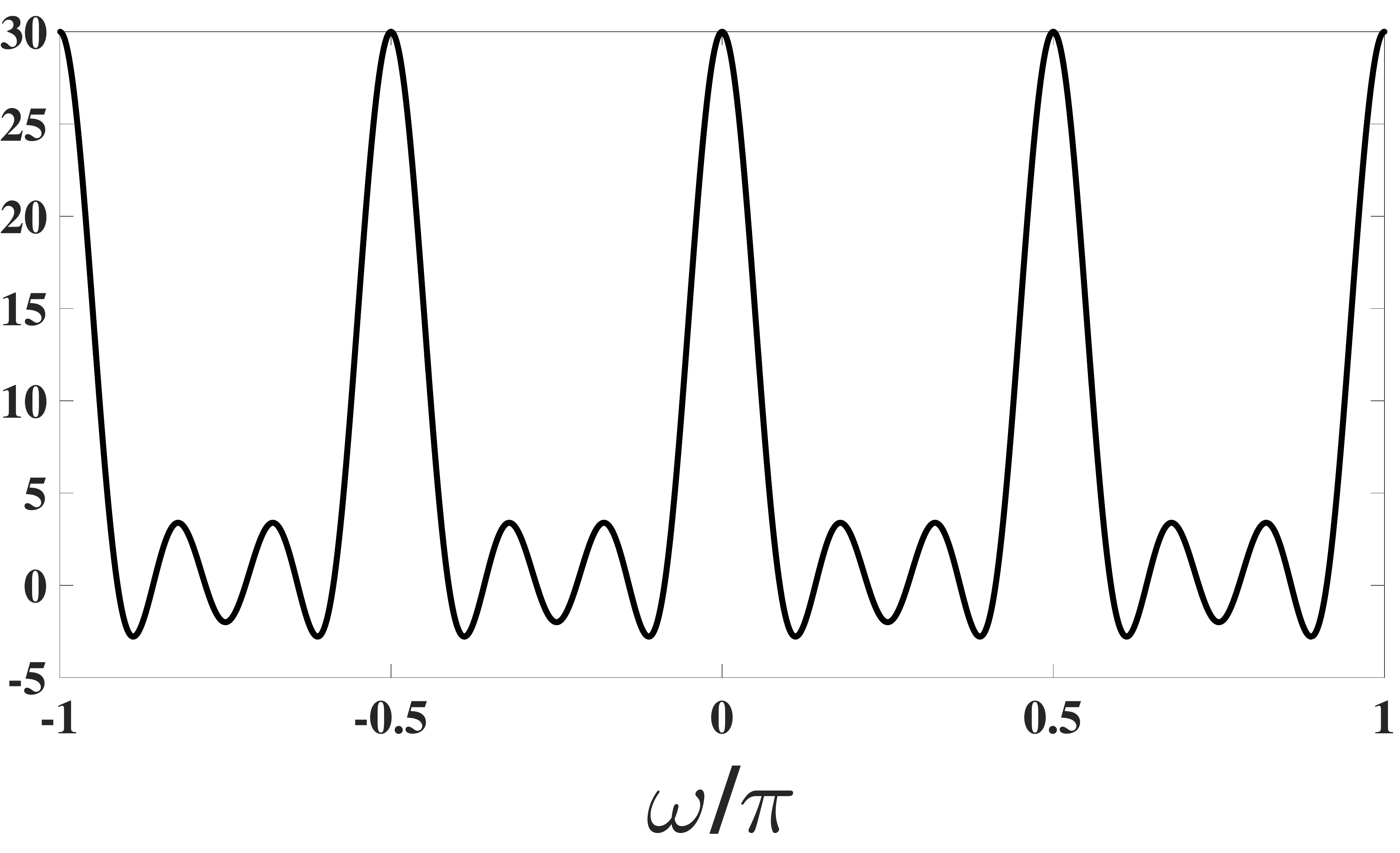}\\
			\includegraphics[width=0.13\textwidth]{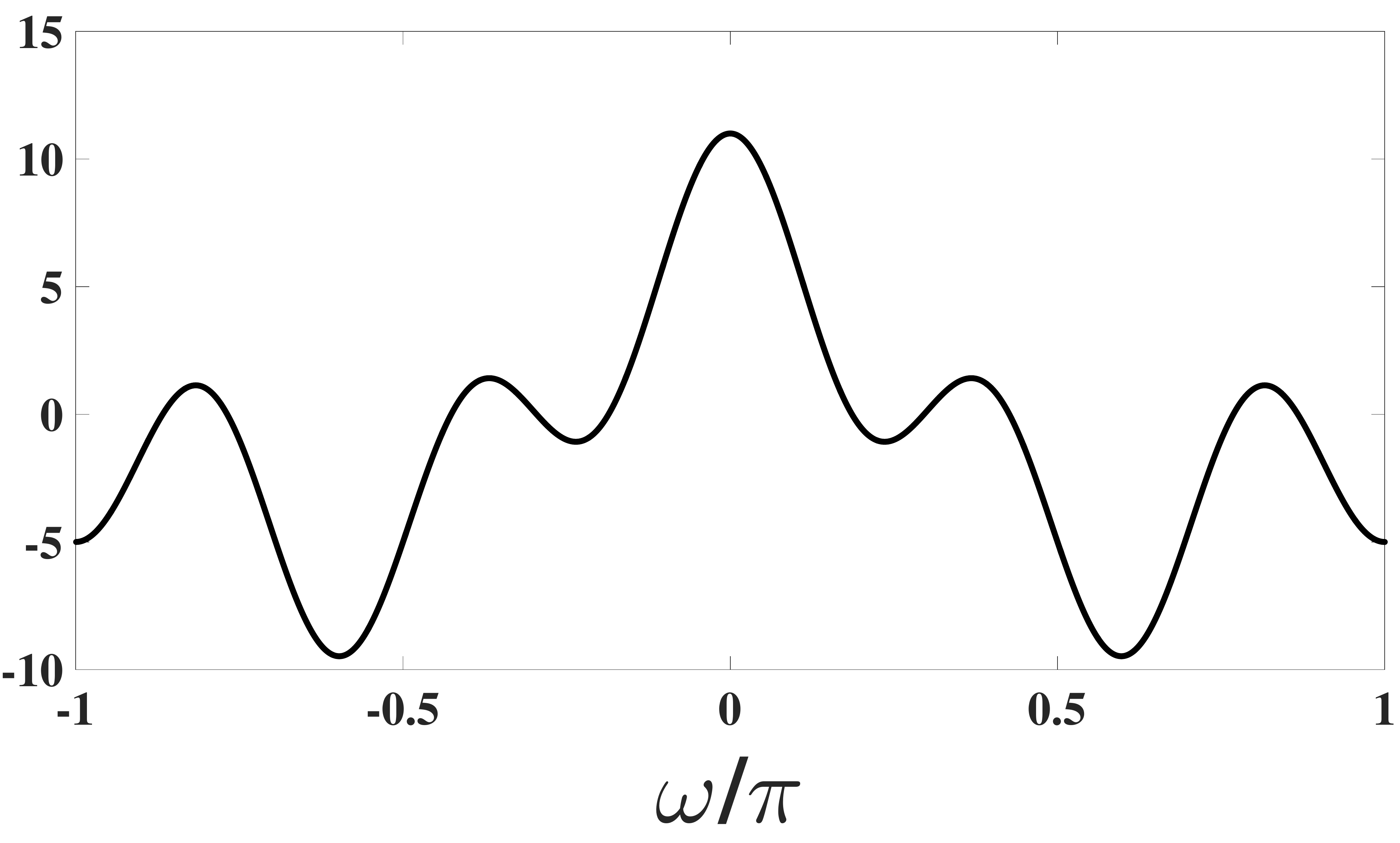}&
			\includegraphics[width=0.13\textwidth]{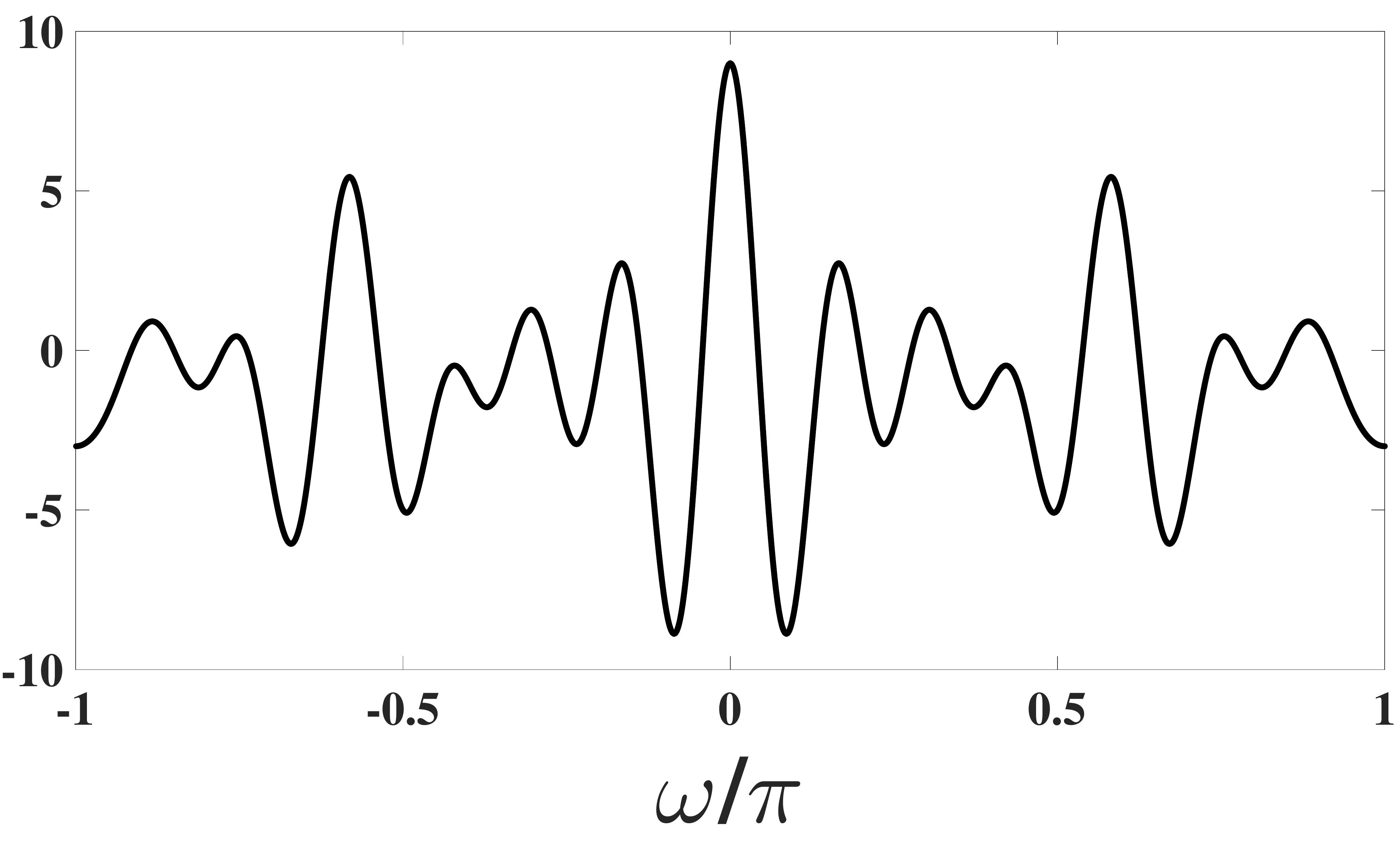} \\
			\multicolumn{2}{c}{
				\includegraphics[width=0.3\textwidth]{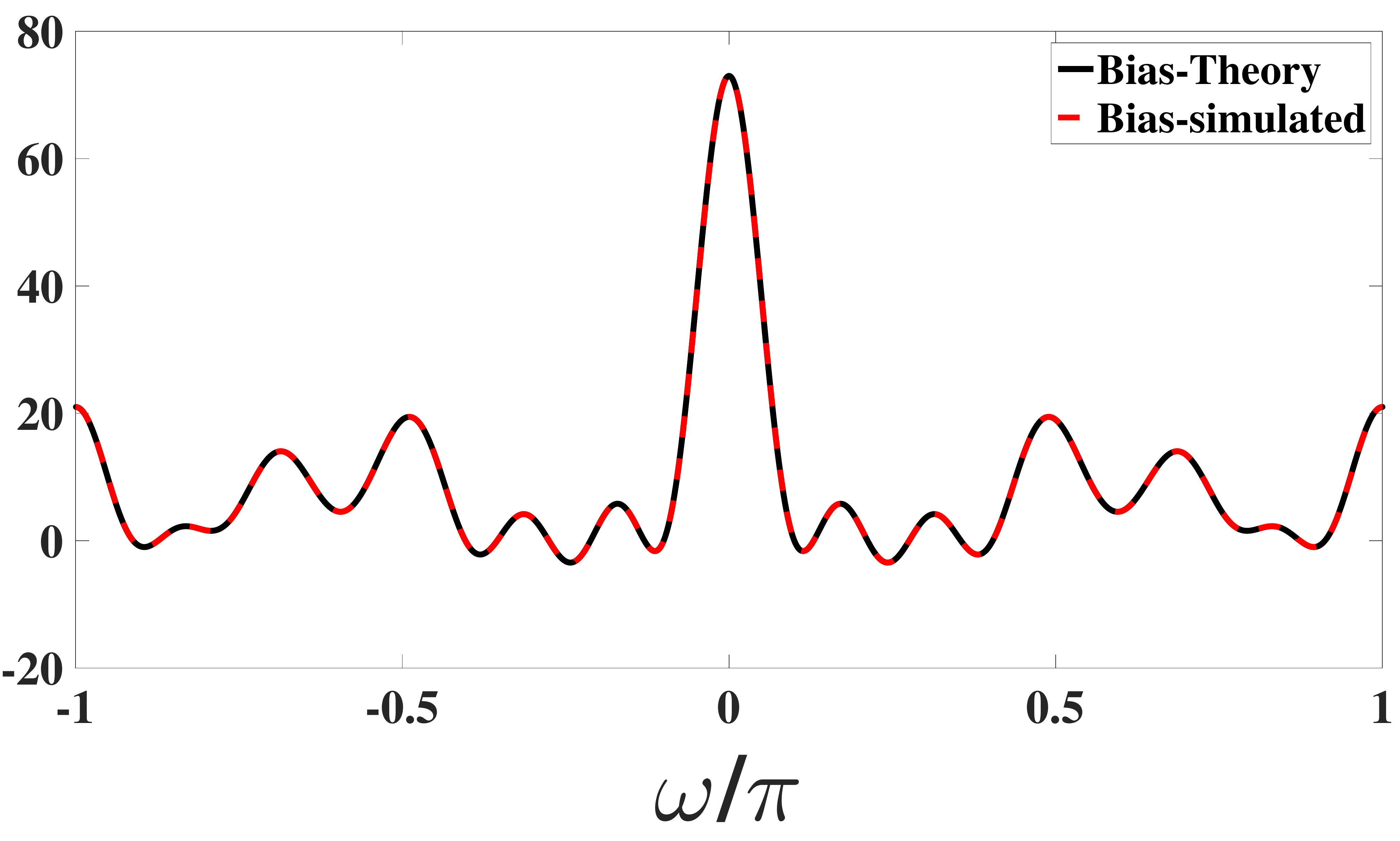}}
		\end{tabular}
		\label{ext_M5N3_continuousP}}
	\subfloat[Prototype: $M=3$, $N=4$]{
		\begin{tabular}{cc}
			\includegraphics[width=0.13\textwidth]{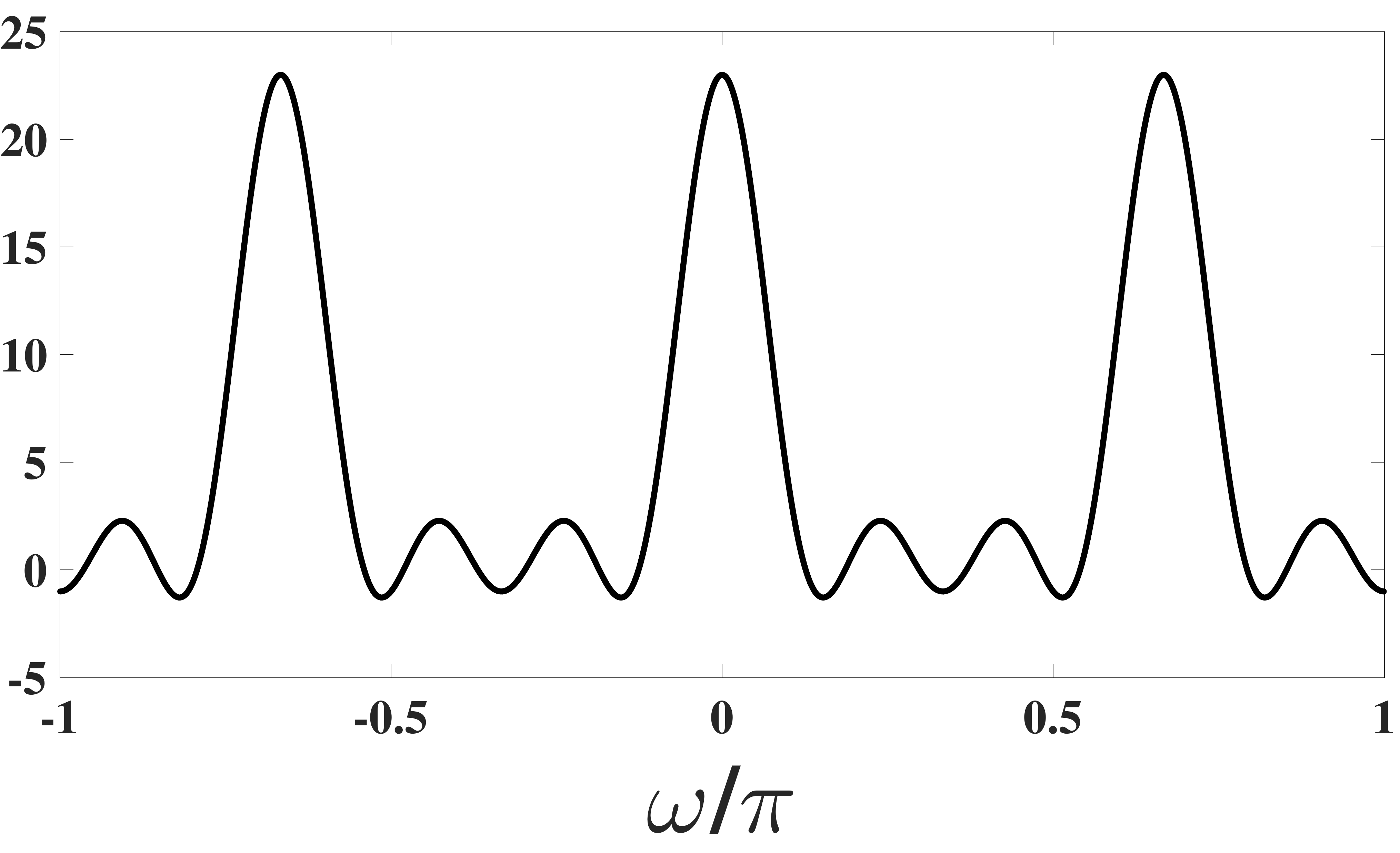}&
			\includegraphics[width=0.13\textwidth]{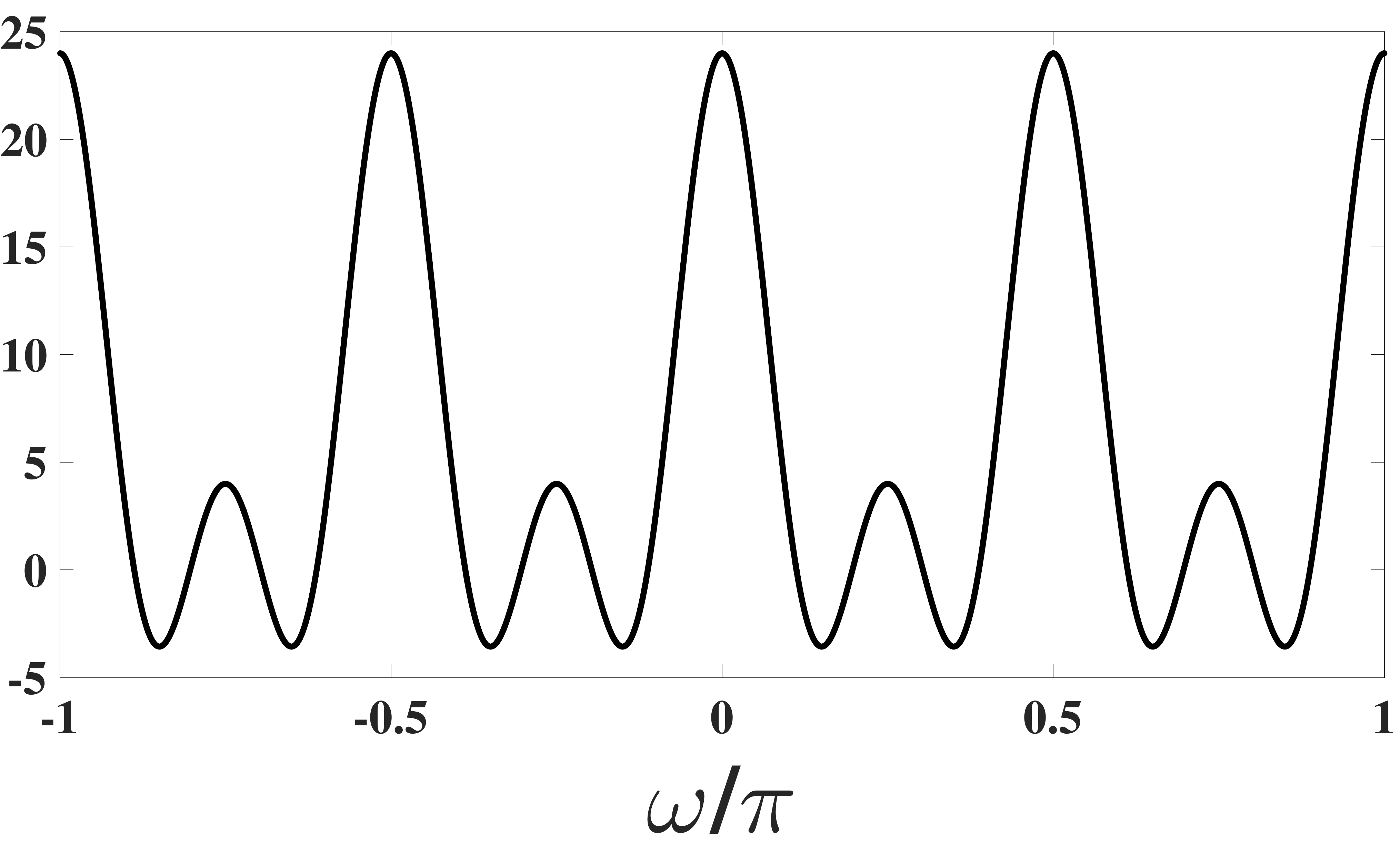}\\
			\includegraphics[width=0.13\textwidth]{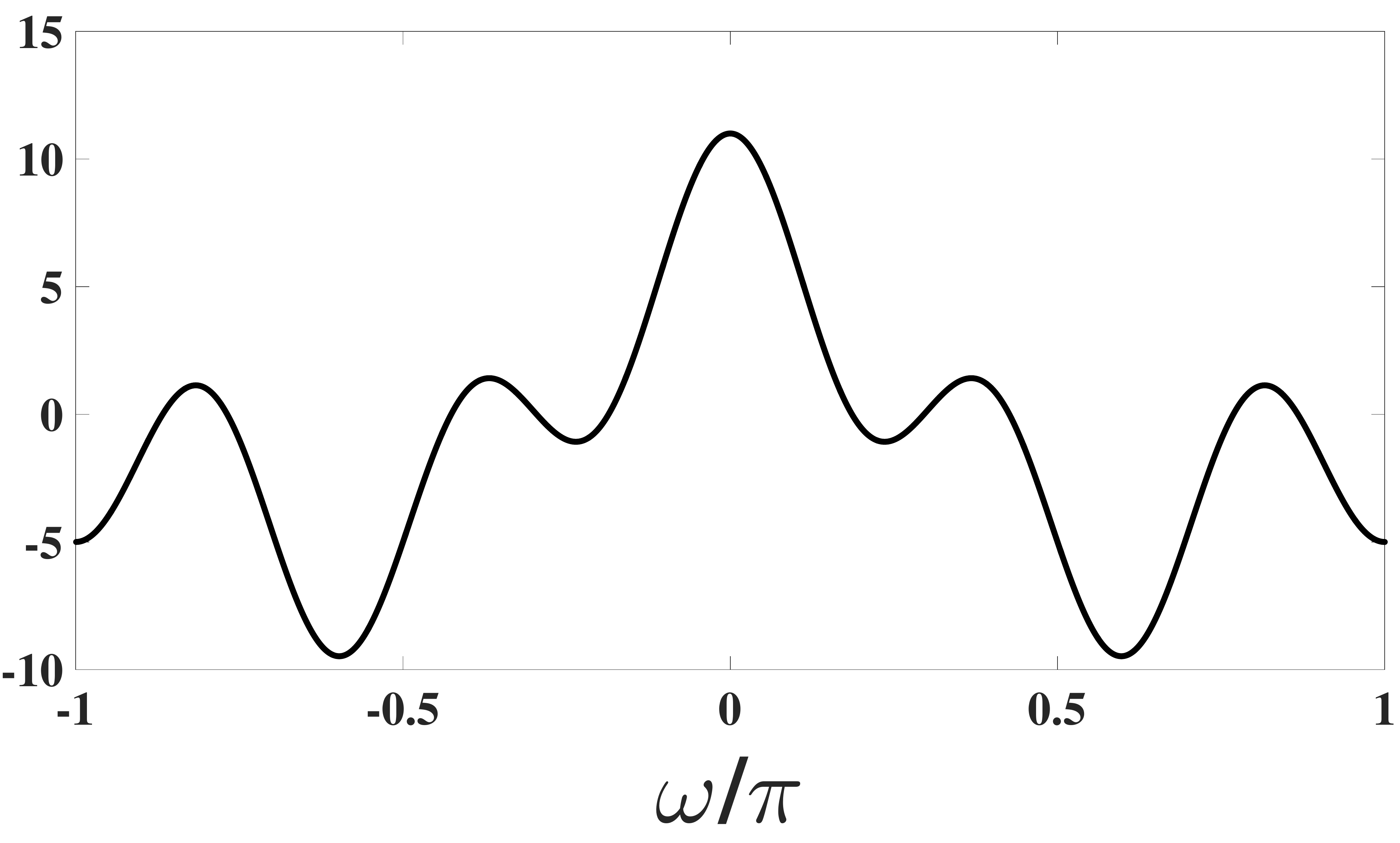}&
			\includegraphics[width=0.13\textwidth]{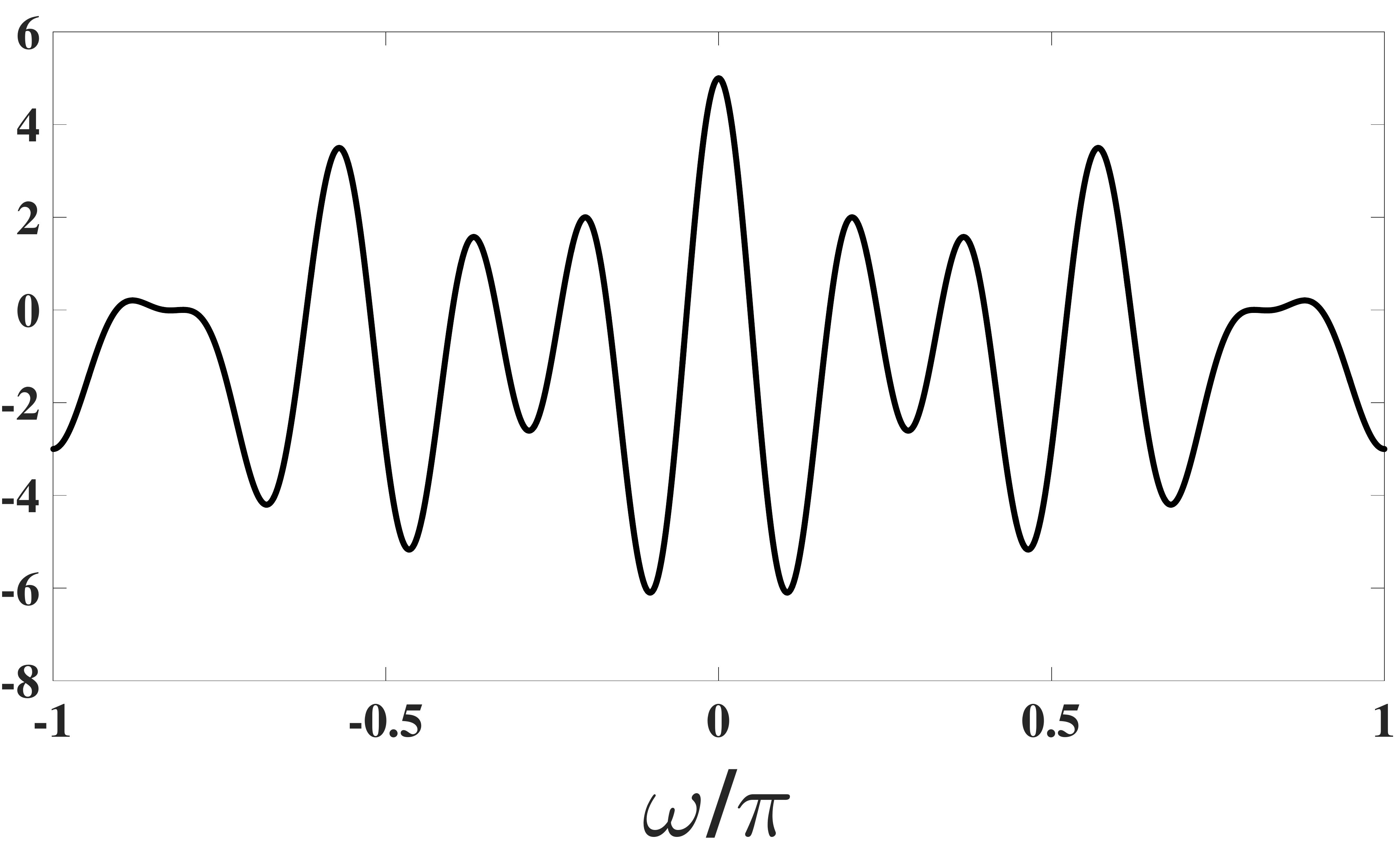} \\
			\multicolumn{2}{c}{
				\includegraphics[width=0.3\textwidth]{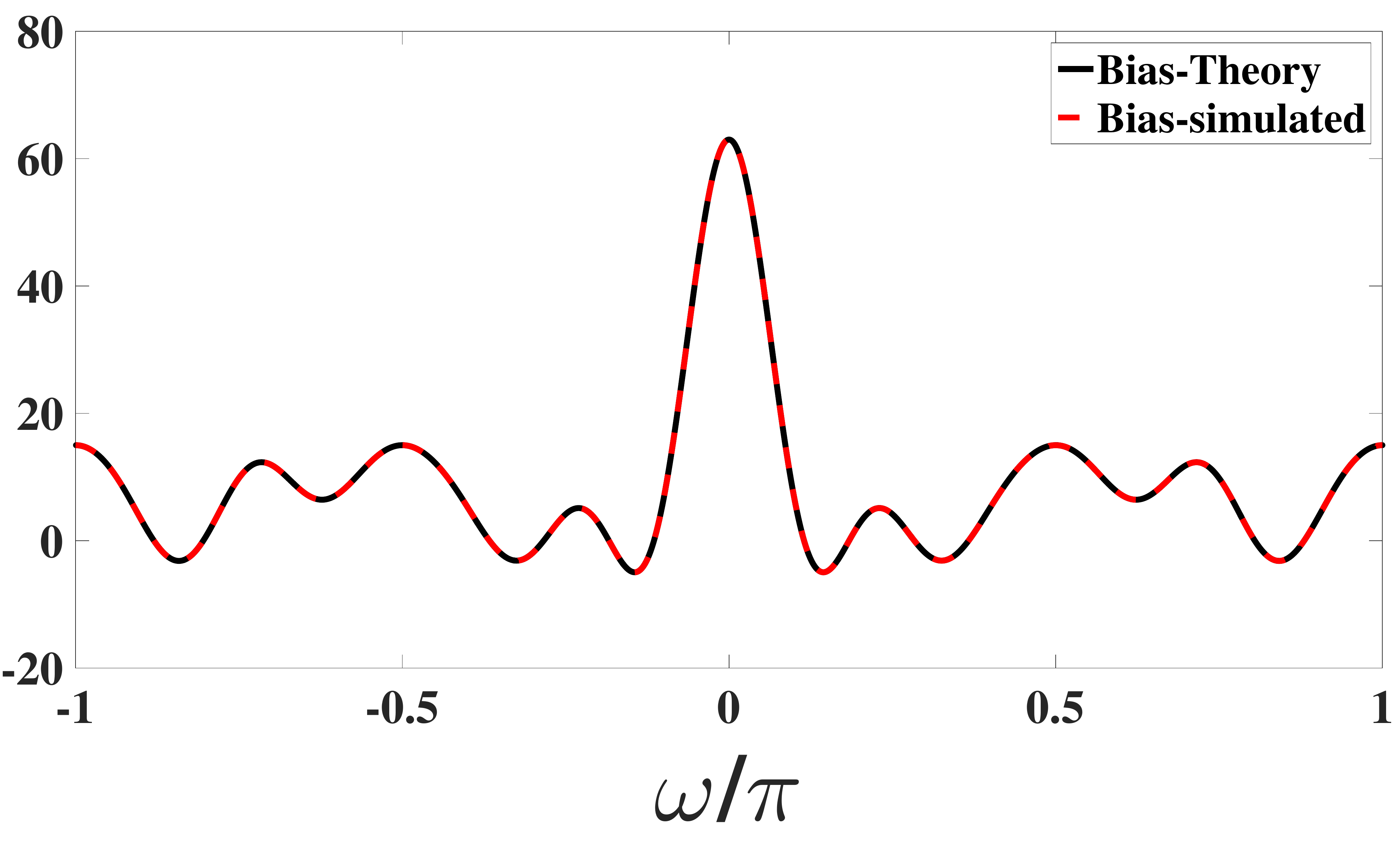}}
		\end{tabular}
		\label{ext_M5N3_prototypeP}}
	\caption{Bias of the correlogram spectral estimate: $M>N$ (top), $N>M$ (bottom)}
	\label{fig:bias_entire_M>N}
\end{figure*}

The proof can be derived by first deriving the Fourier transform of $X_{x}$ (i.e. $\mathfrak{F}\{X_x\}$), where $X$ is one of the elements in $\{A,B,C,D\}$ and $x$ in $\{f,c,p\}$. It follows a similar procedure as in~\cite{UVD_PHD}.
The weight function for the cases in Fig.~\ref{fig:Limits_LB_set_M>N} ($M>N$) and Fig.~\ref{fig:Limits_LB_set_N>M} ($N>M$) is shown in Fig.~\ref{fig:wts_M>N} and~\ref{fig:wts_M<N} respectively. The FFT of these weight functions is the bias of the correlogram estimate for the entire, continuous and the prototype range (i.e. FFT of $z_{e_f}$, $z_{e_c}$ and $z_{e_p}$). These simulated results are compared with the derived bias equations in Fig.~\ref{fig:bias_entire_M>N} for $M>N$ and for $N>M$. Each subplot in this figure contains five images, and represent $\mathfrak{F}\{A_x\}$ (top-left), $\mathfrak{F}\{B_x\}$ (top-right), $\mathfrak{F}\{C_x\}$ (middle-left), $\mathfrak{F}\{D_x\}$ (middle-right) and overall bias $W_{b_x}(e^{j\omega})$ (bottom), where $x$ denotes one of the elements $\{f, c, p\}$. It is evident that the derived equations in all the cases considered match the simulated results. Noted that the derived equations are valid for $M>N$ as well as $N>M$. Though the equations are valid for $M$ greater than $N$ and vice versa, the effect on the bias is not the same. This aspect is analyzed so as to provide an insight into the choice for array extension from a bias perspective. 
%
%
%
%
%
\begin{figure*}[!t]
\centering
\subfloat[$M=4$, $N=3$]{
\includegraphics[width=0.28\textwidth]{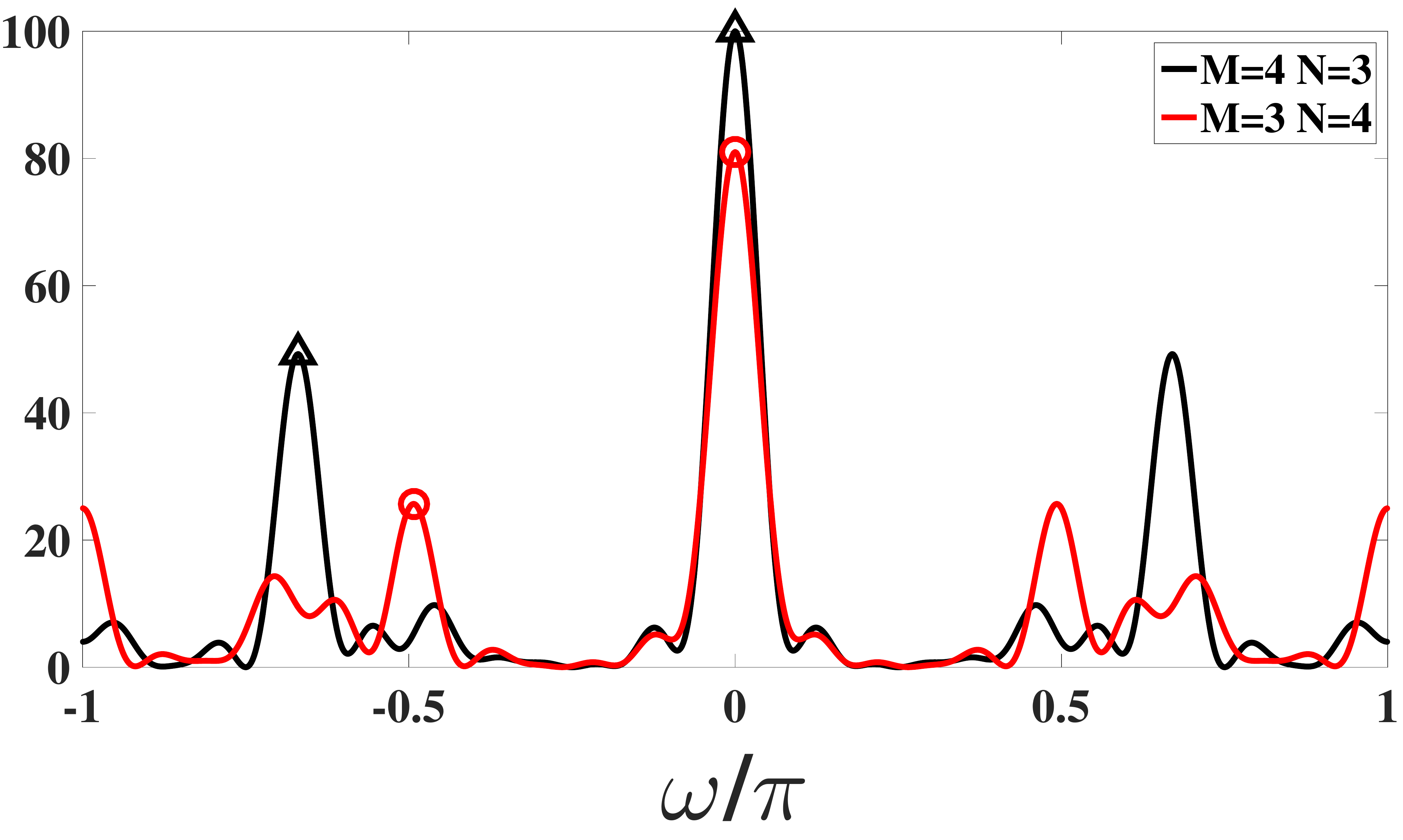}
\includegraphics[width=0.28\textwidth]{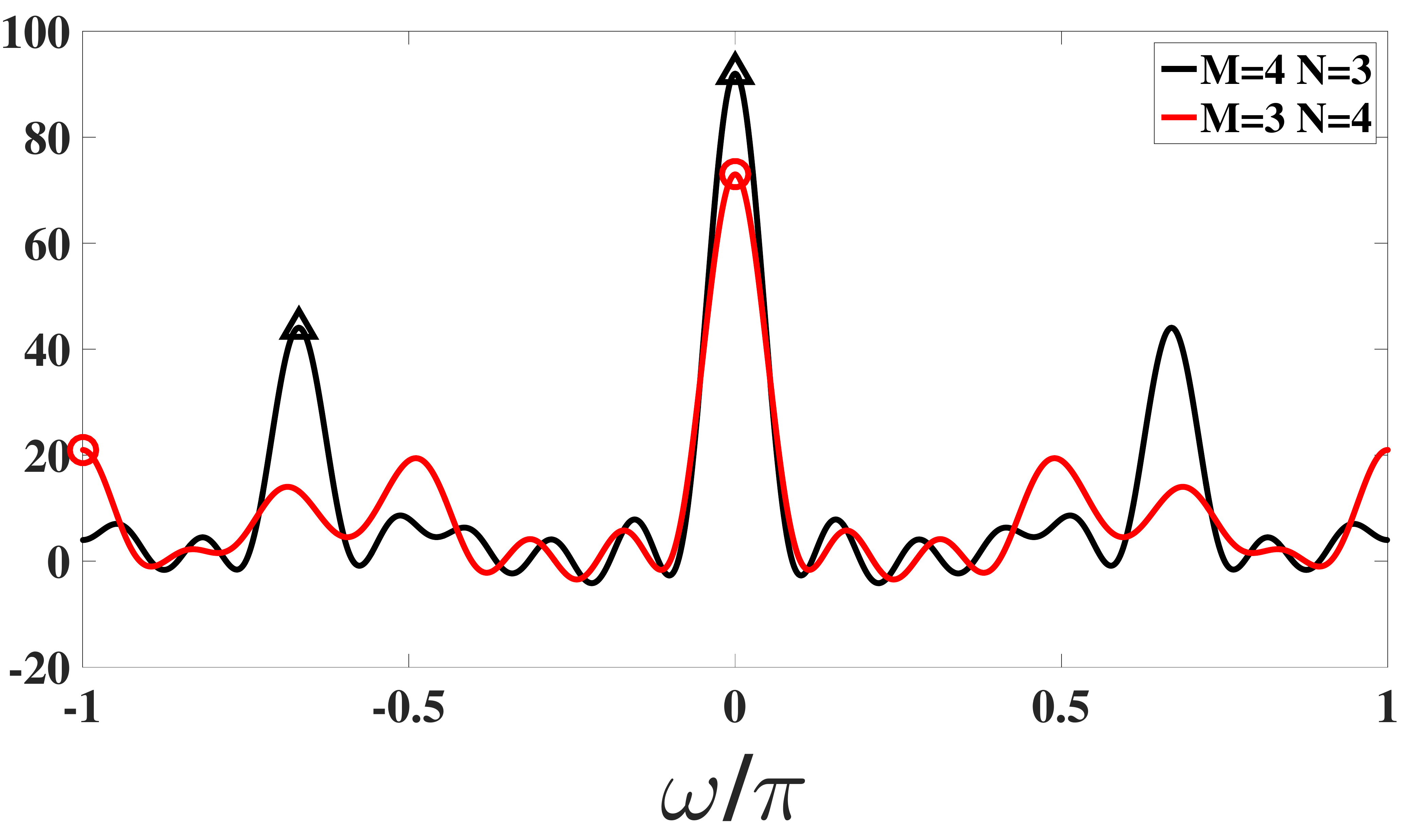}
\includegraphics[width=0.28\textwidth]{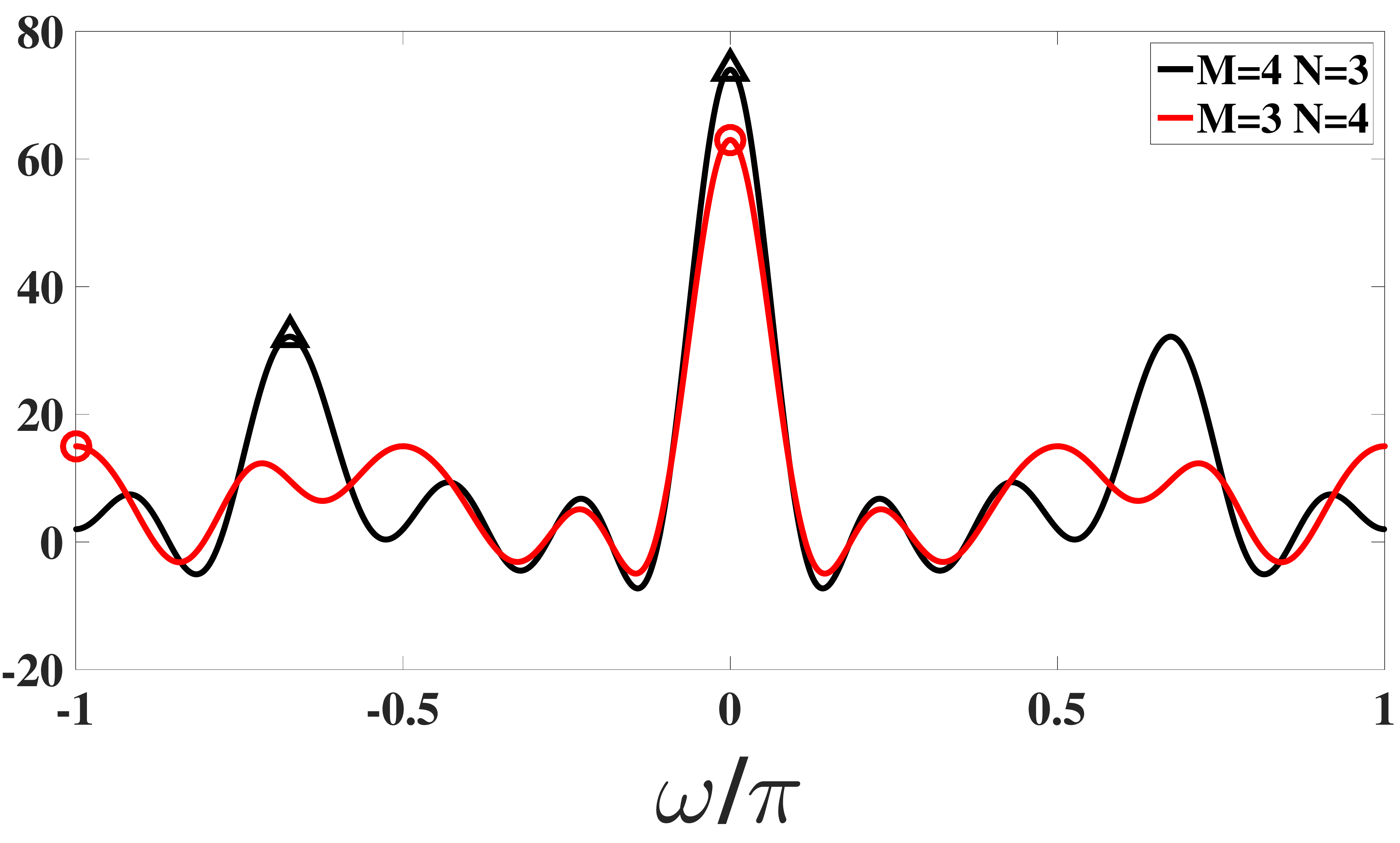}
\label{ext_M4N3_compare}}
\hfil
\subfloat[$M=5$, $N=3$]{
\includegraphics[width=0.28\textwidth]{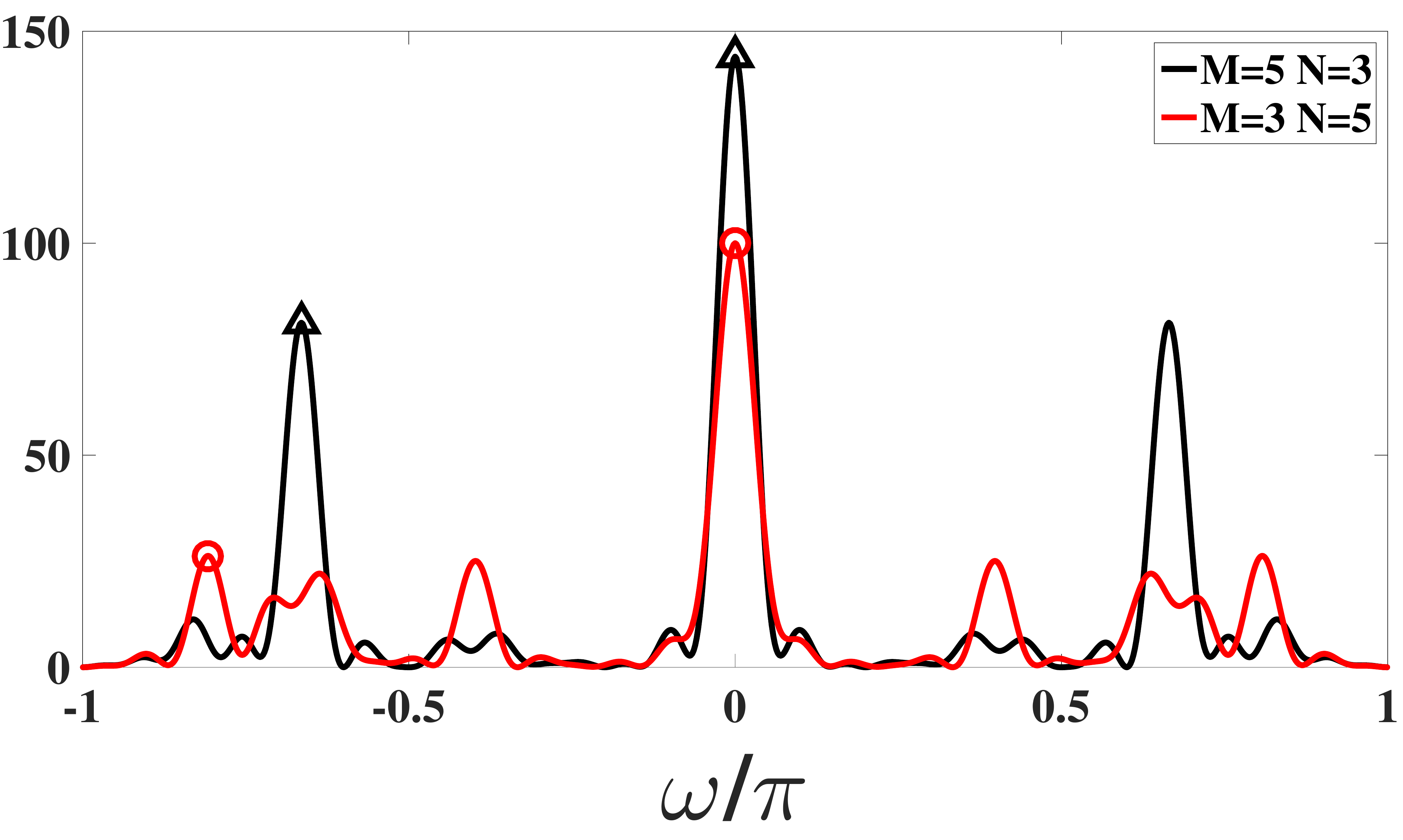}
\includegraphics[width=0.28\textwidth]{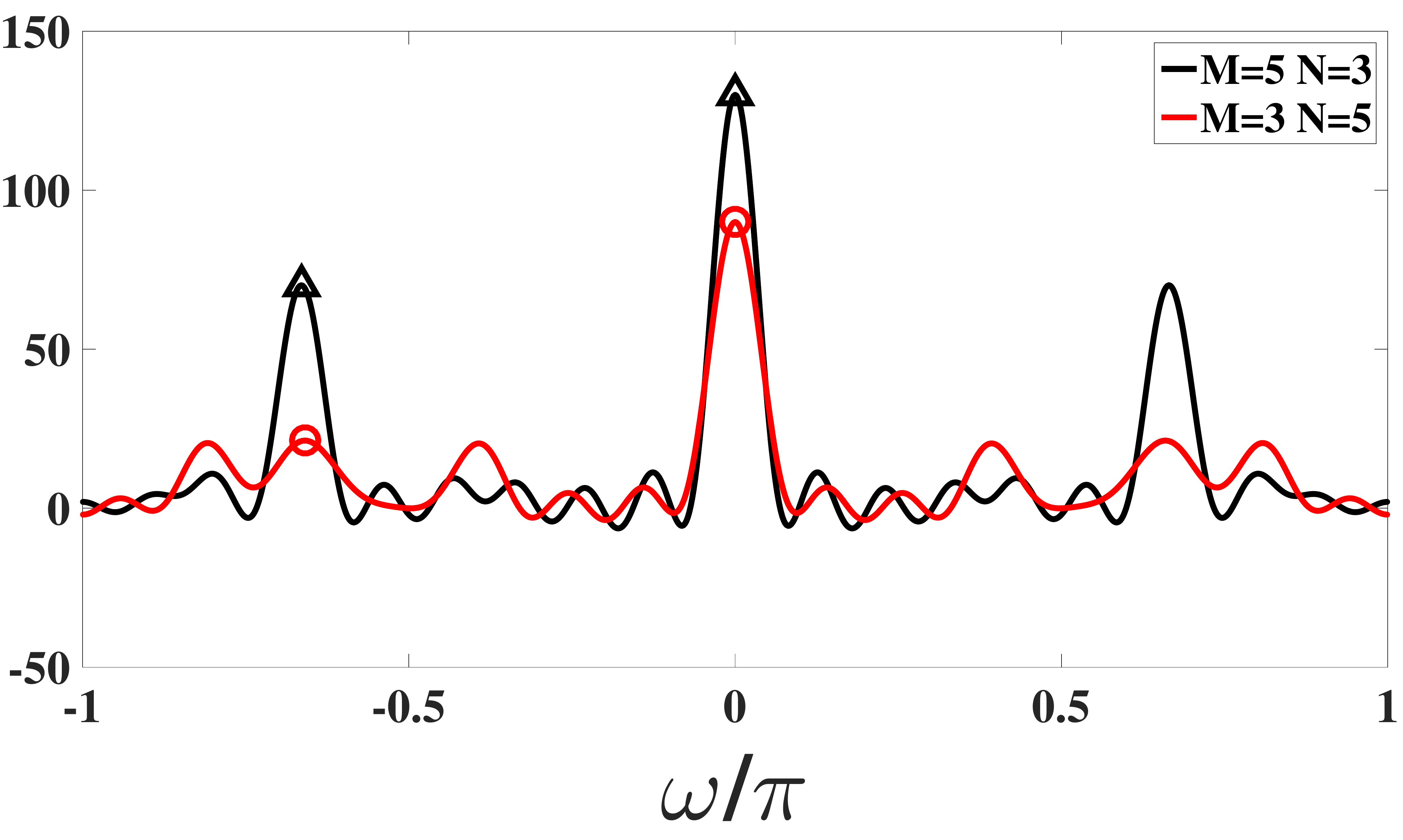}
\includegraphics[width=0.28\textwidth]{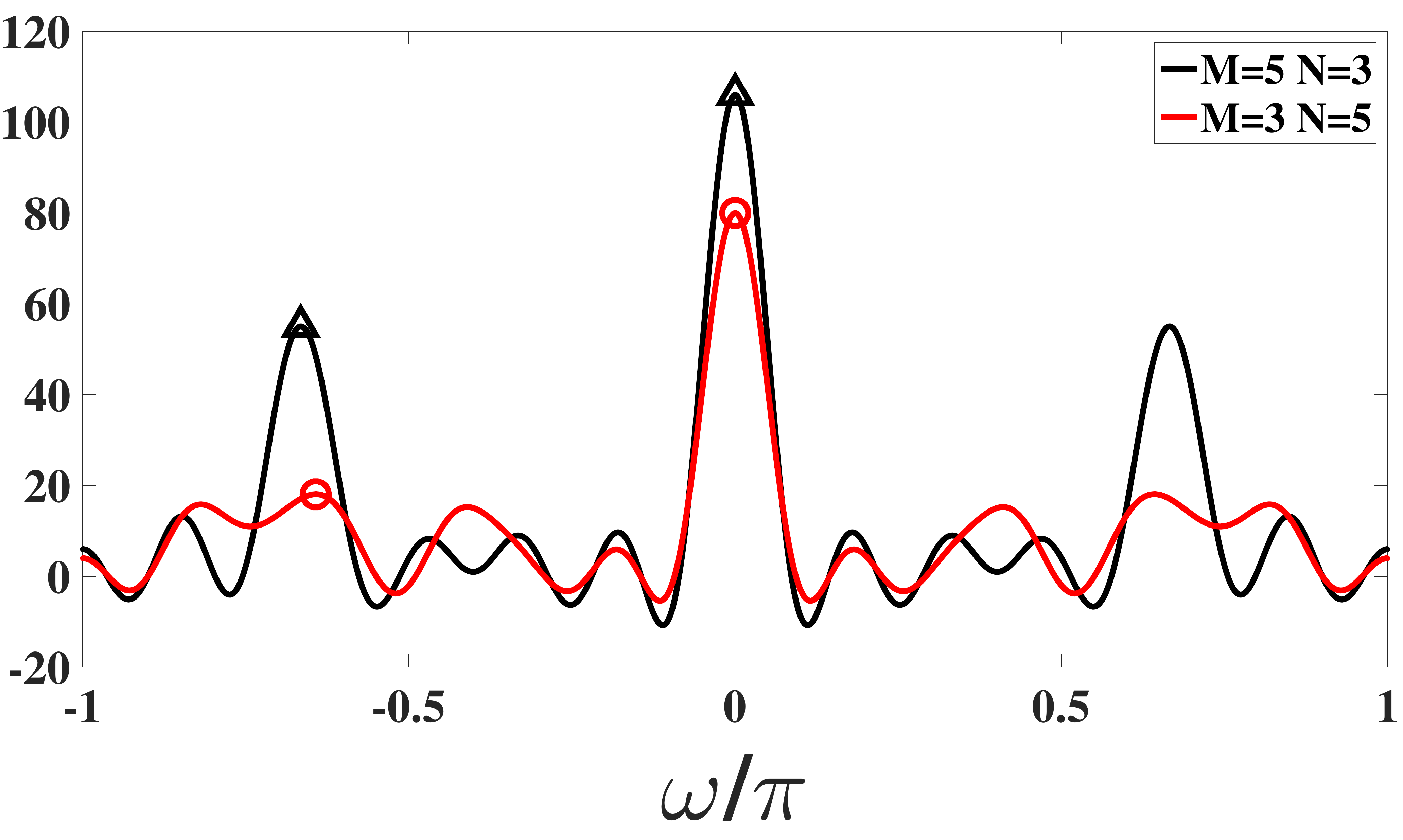}
\label{ext_M5N3_compare}}
\hfil
\subfloat[$M=7$, $N=3$]{
\includegraphics[width=0.28\textwidth]{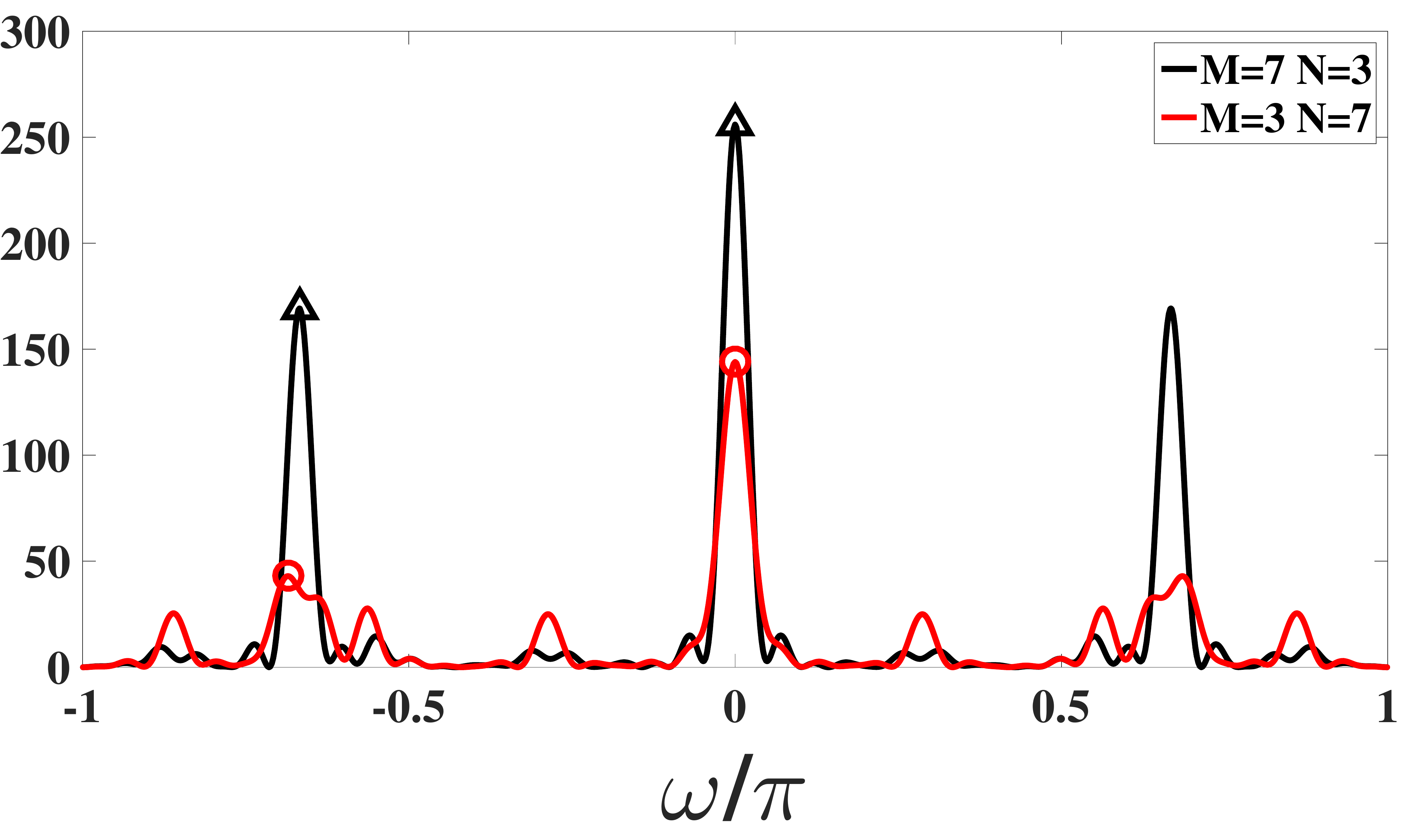}
\includegraphics[width=0.28\textwidth]{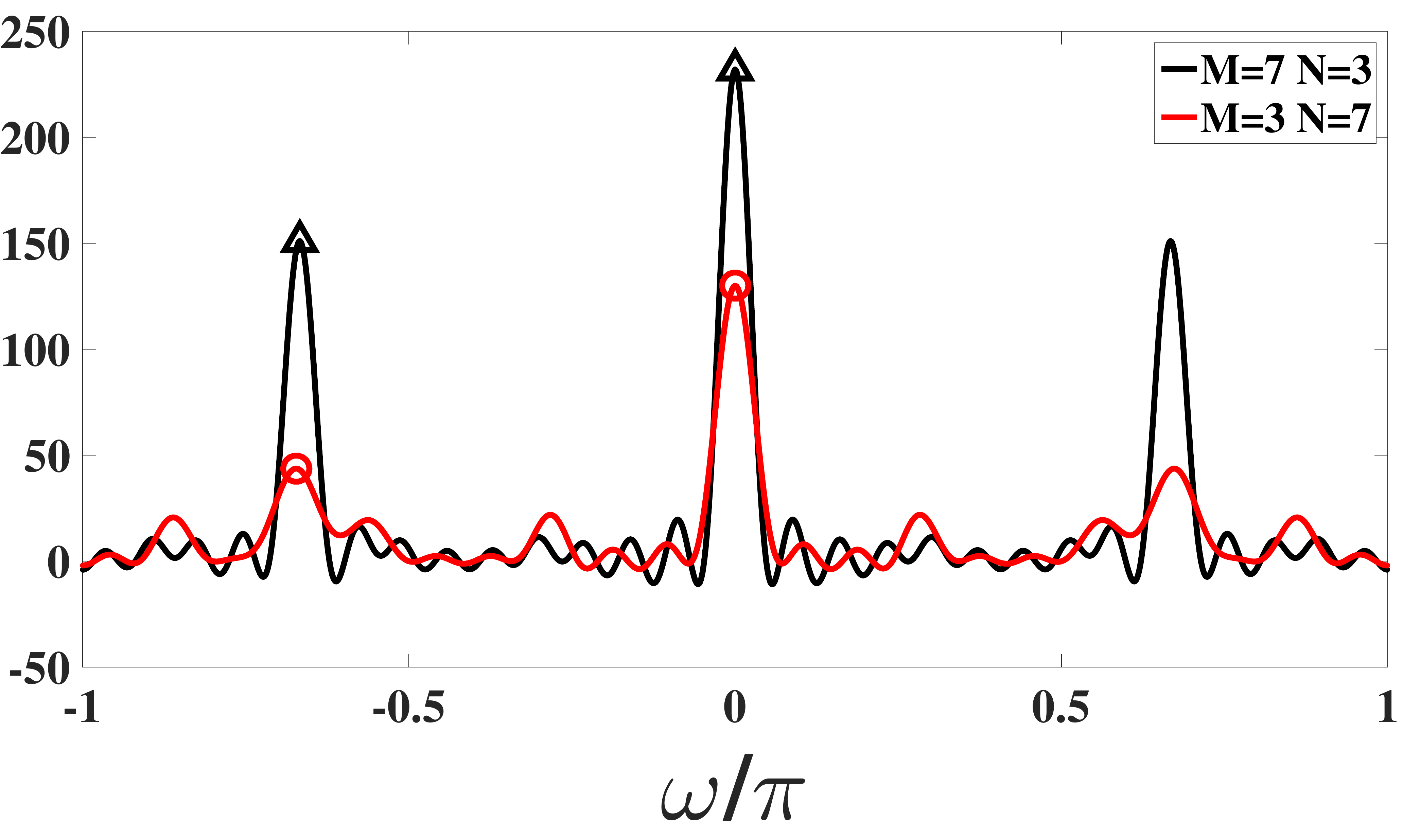}
\includegraphics[width=0.28\textwidth]{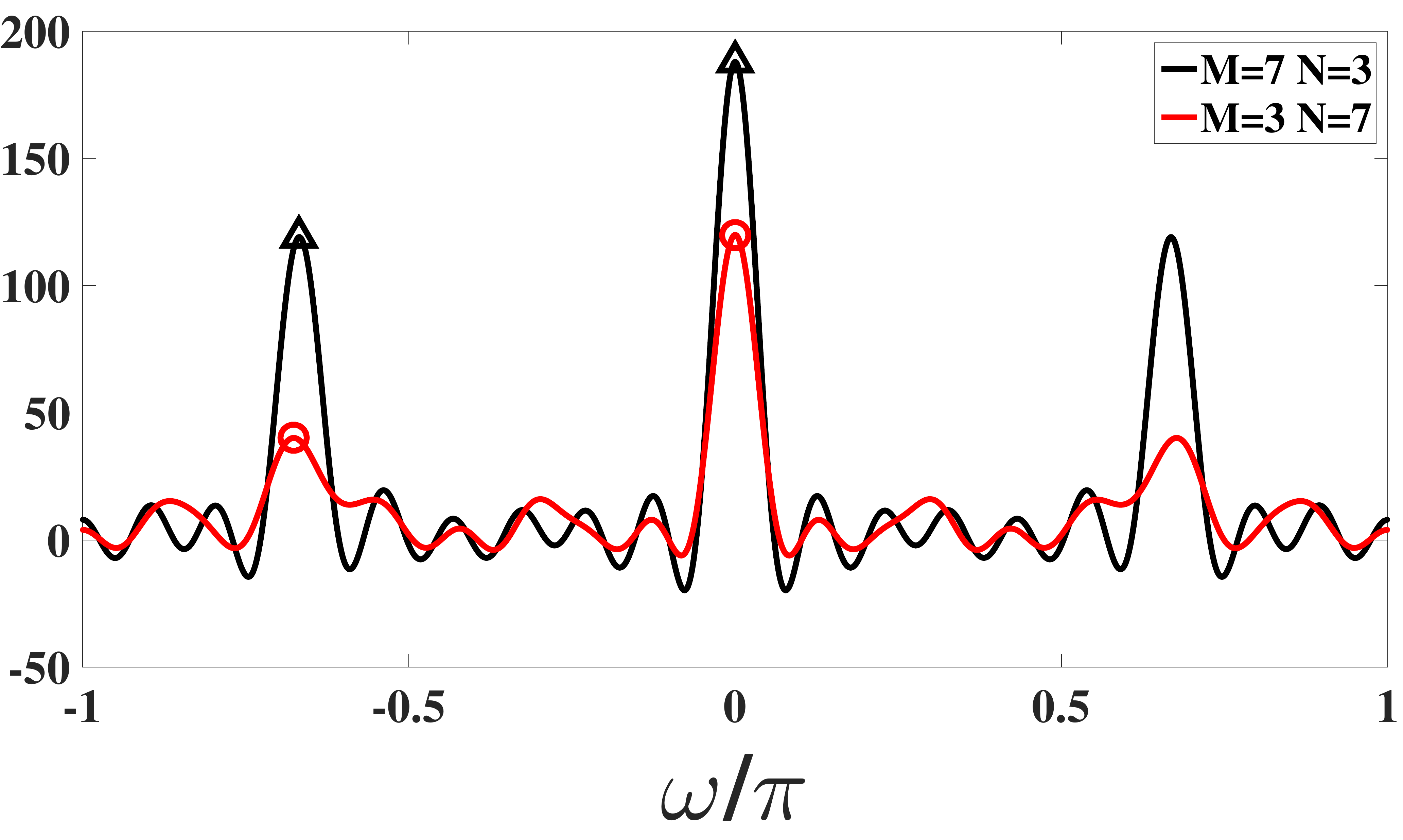}
\label{ext_M7N3_compare}}
\hfil
\subfloat[$M=8$, $N=3$]{
\includegraphics[width=0.28\textwidth]{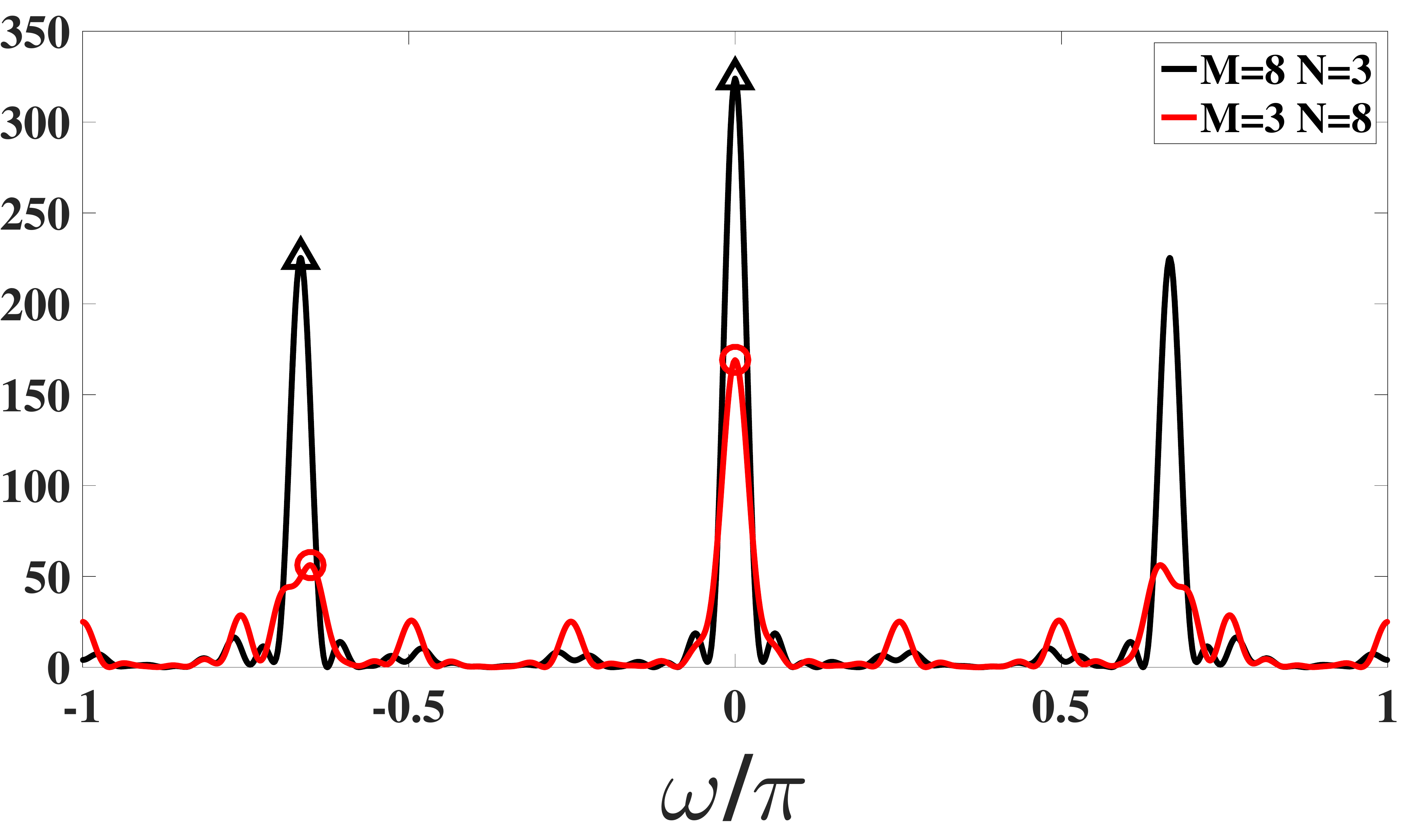}
\includegraphics[width=0.28\textwidth]{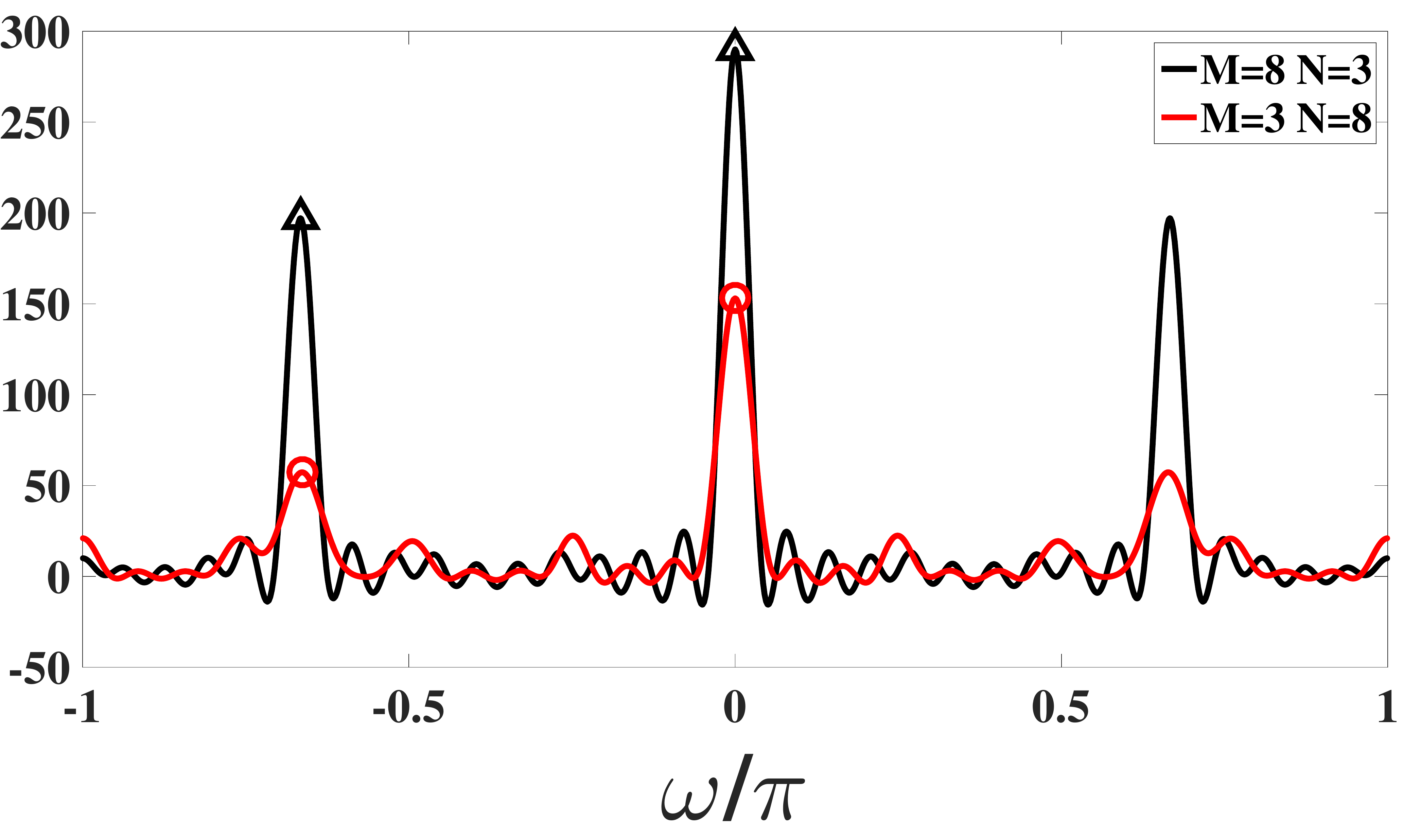}
\includegraphics[width=0.28\textwidth]{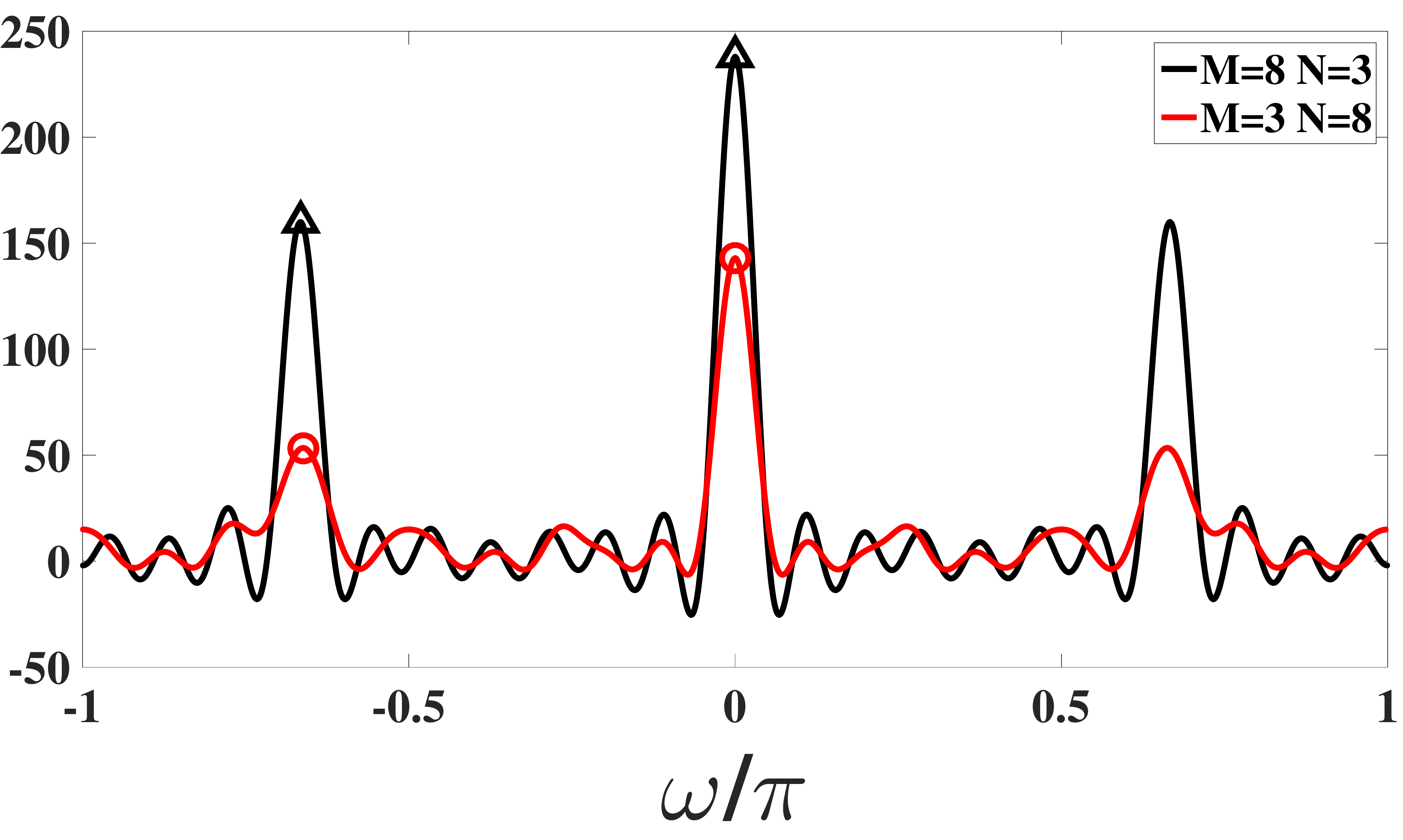}
\label{ext_M8N3_compare}}
\hfil
\subfloat[$M=5$, $N=4$]{
\includegraphics[width=0.28\textwidth]{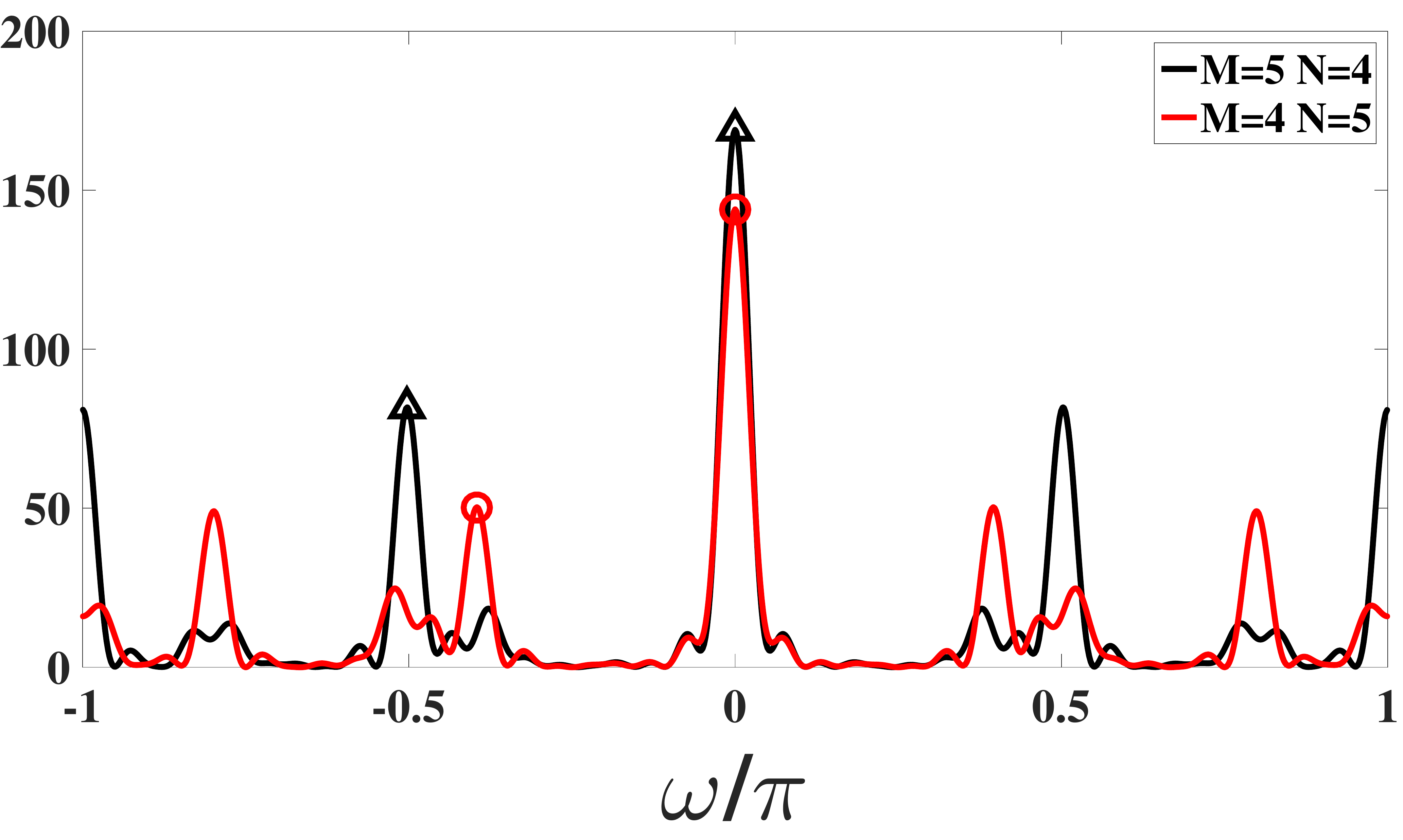}
\includegraphics[width=0.28\textwidth]{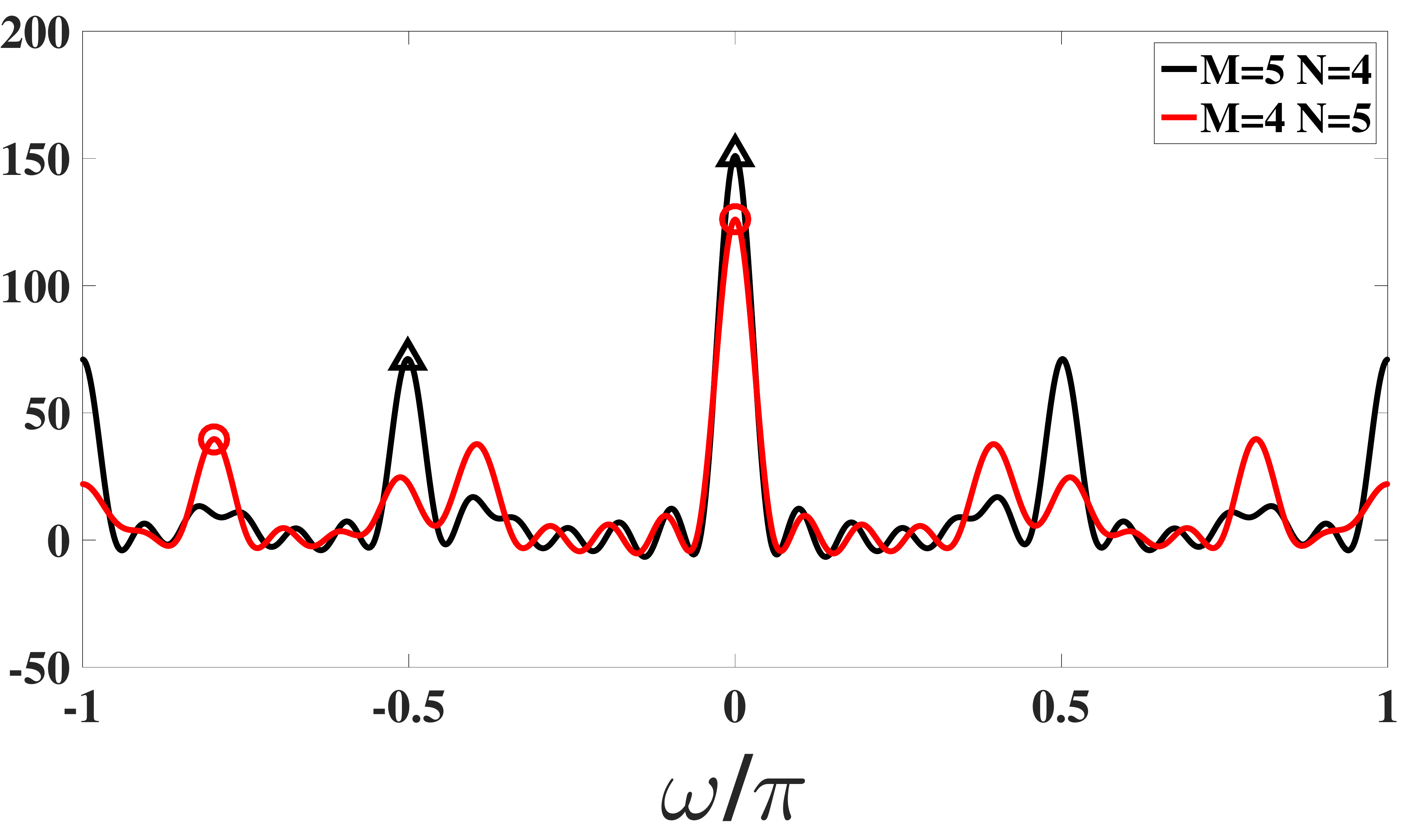}
\includegraphics[width=0.28\textwidth]{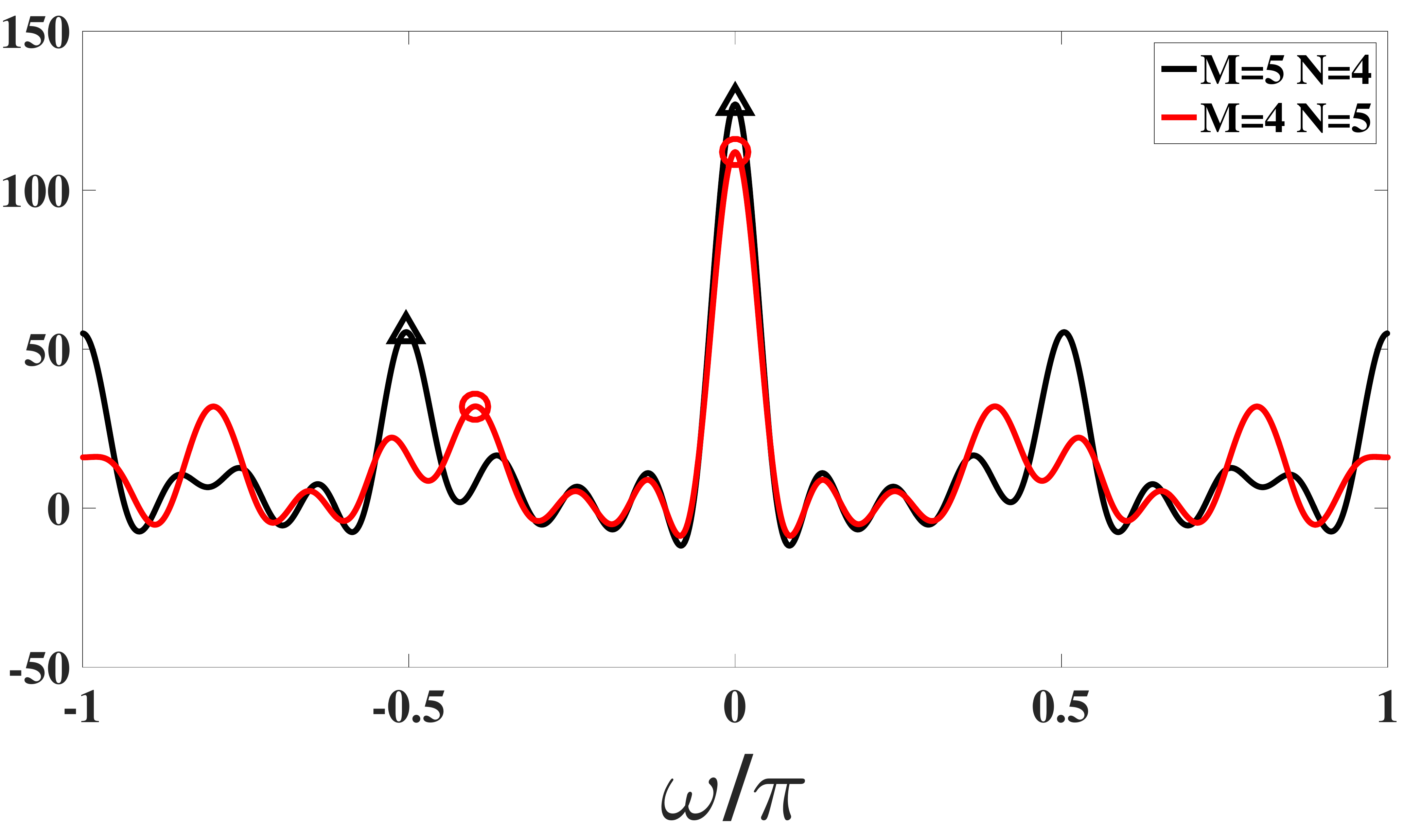}
\label{ext_M5N4_compare}}
\hfil
\subfloat[$M=7$, $N=4$]{
\includegraphics[width=0.28\textwidth]{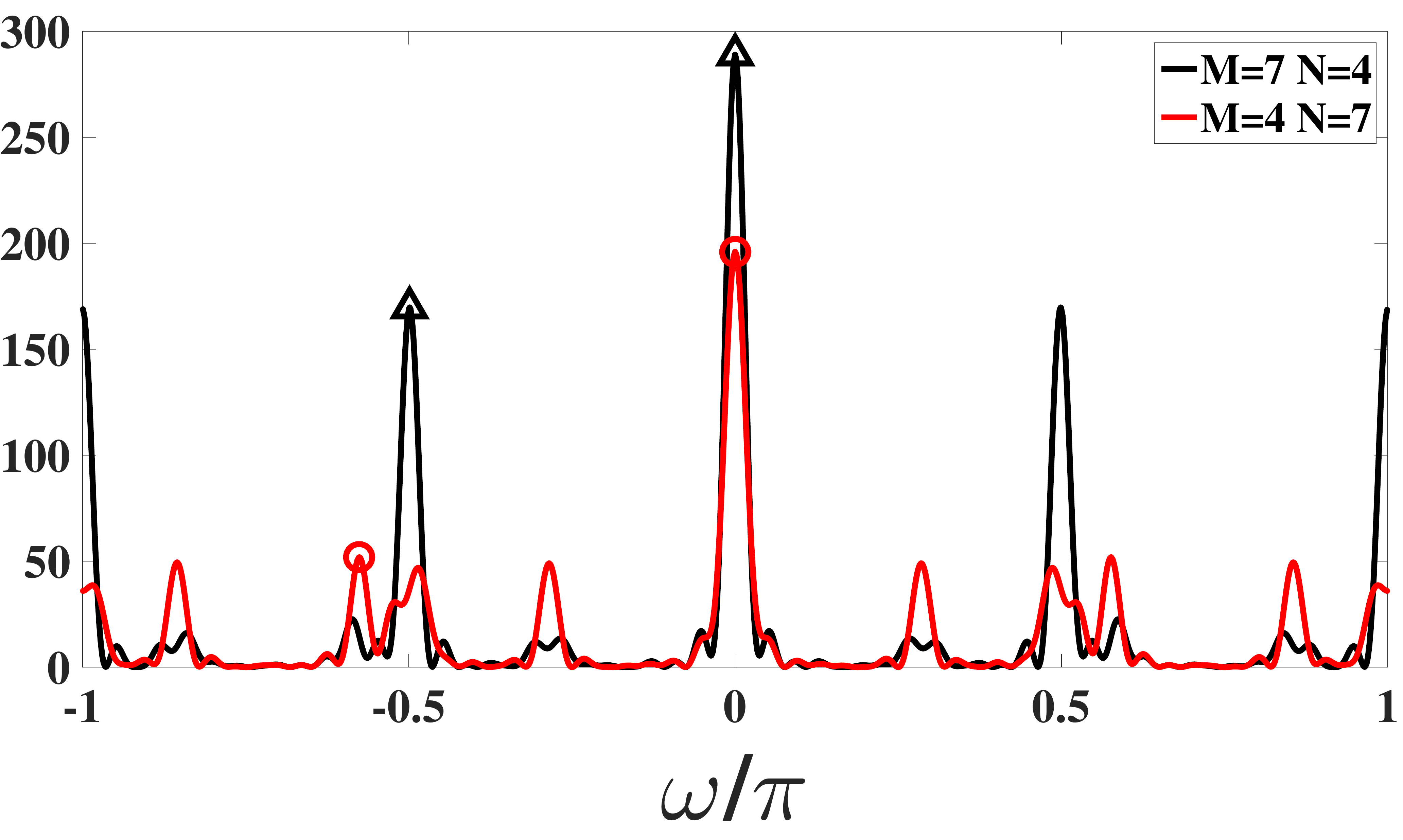}
\includegraphics[width=0.28\textwidth]{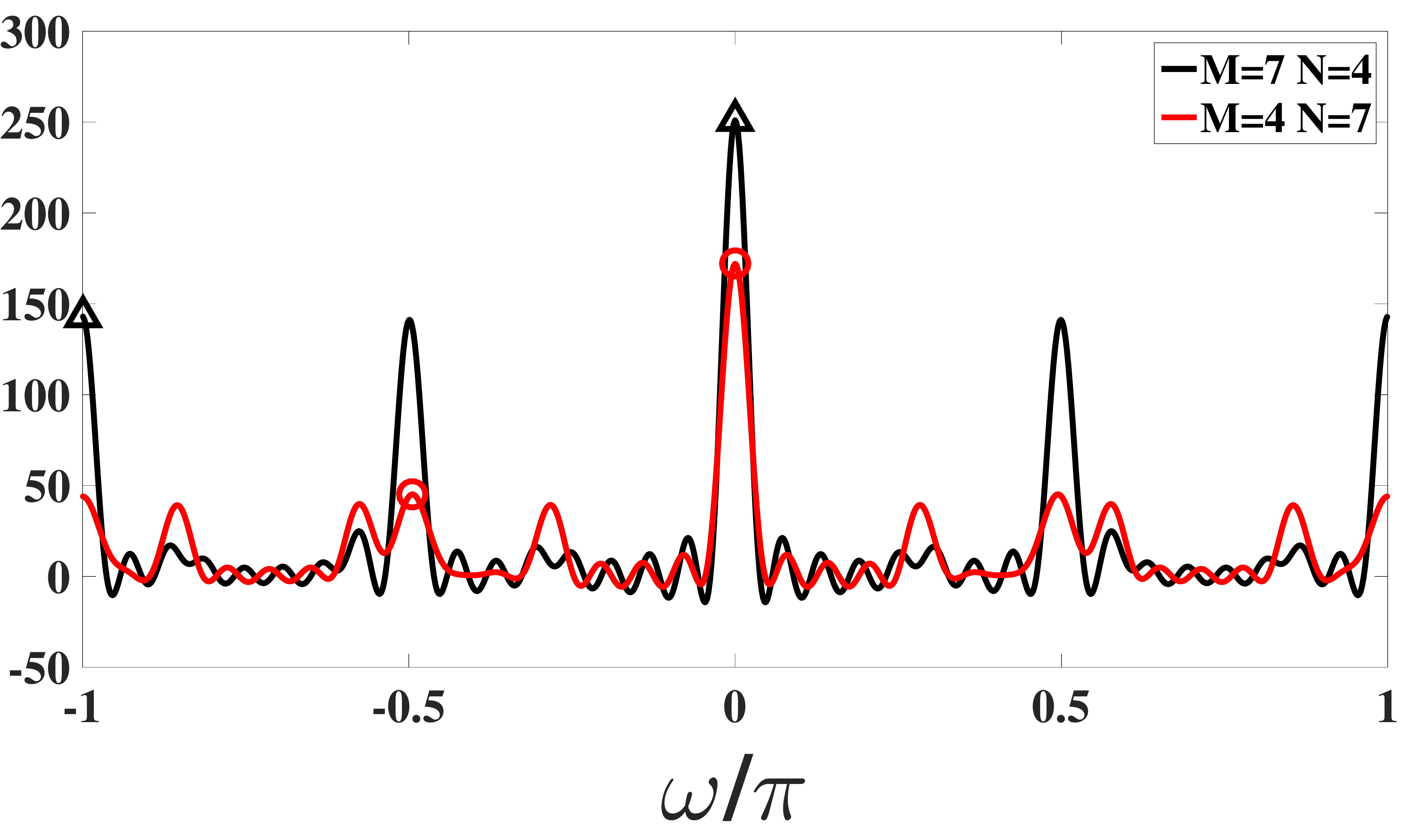}
\includegraphics[width=0.28\textwidth]{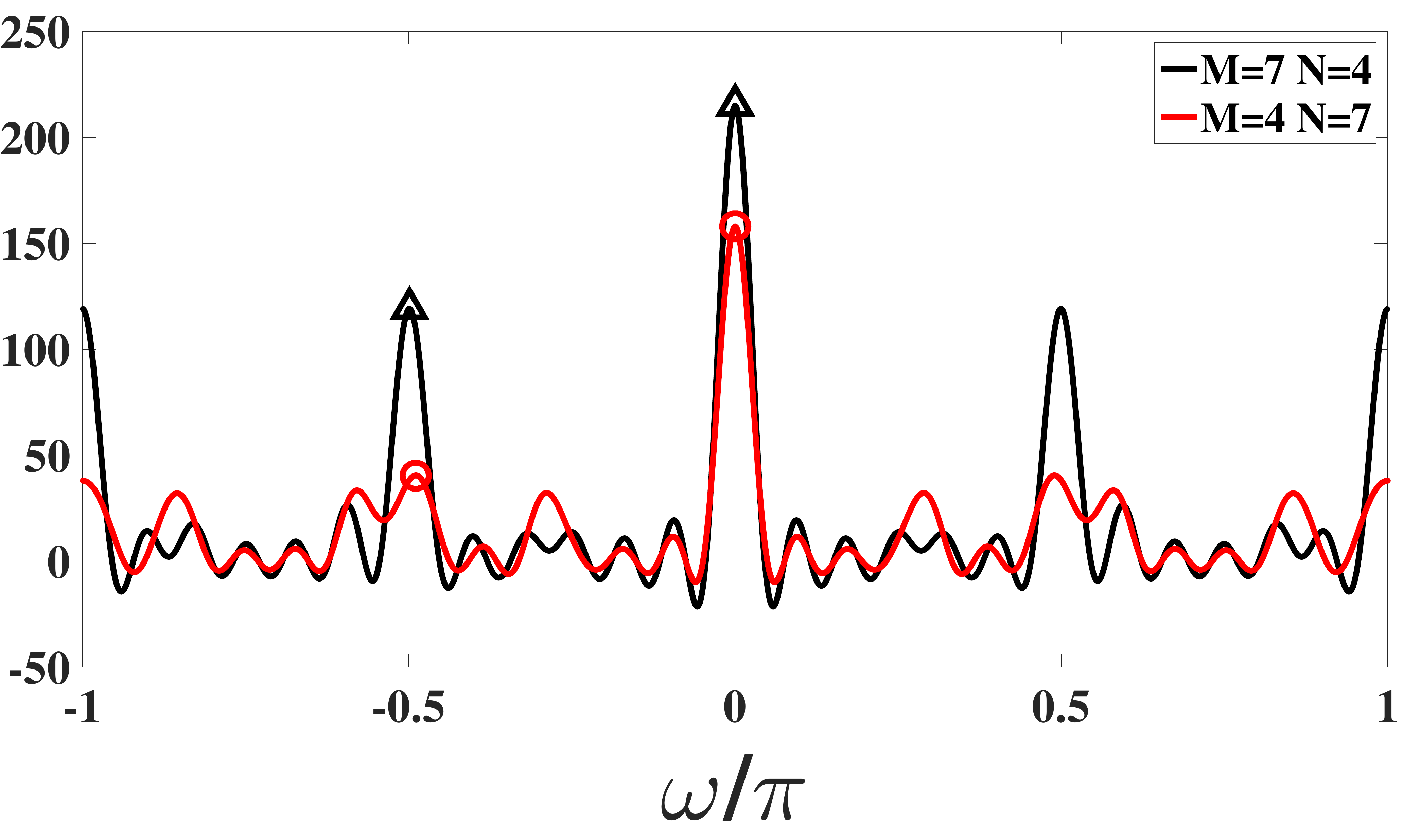}
\label{ext_M7N4_compare}}
\hfil
\caption{Comparison between the bias for $M>N$ and $M<N$: Entire (left), continuous (middle) and prototype (right) range.}
\label{fig:compare_bias_M>Nvice_versa}
\end{figure*}
\begin{figure*}[!t]
	\centering
	\subfloat[$M=14$, $N=13$]{
		\includegraphics[width=0.28\textwidth]{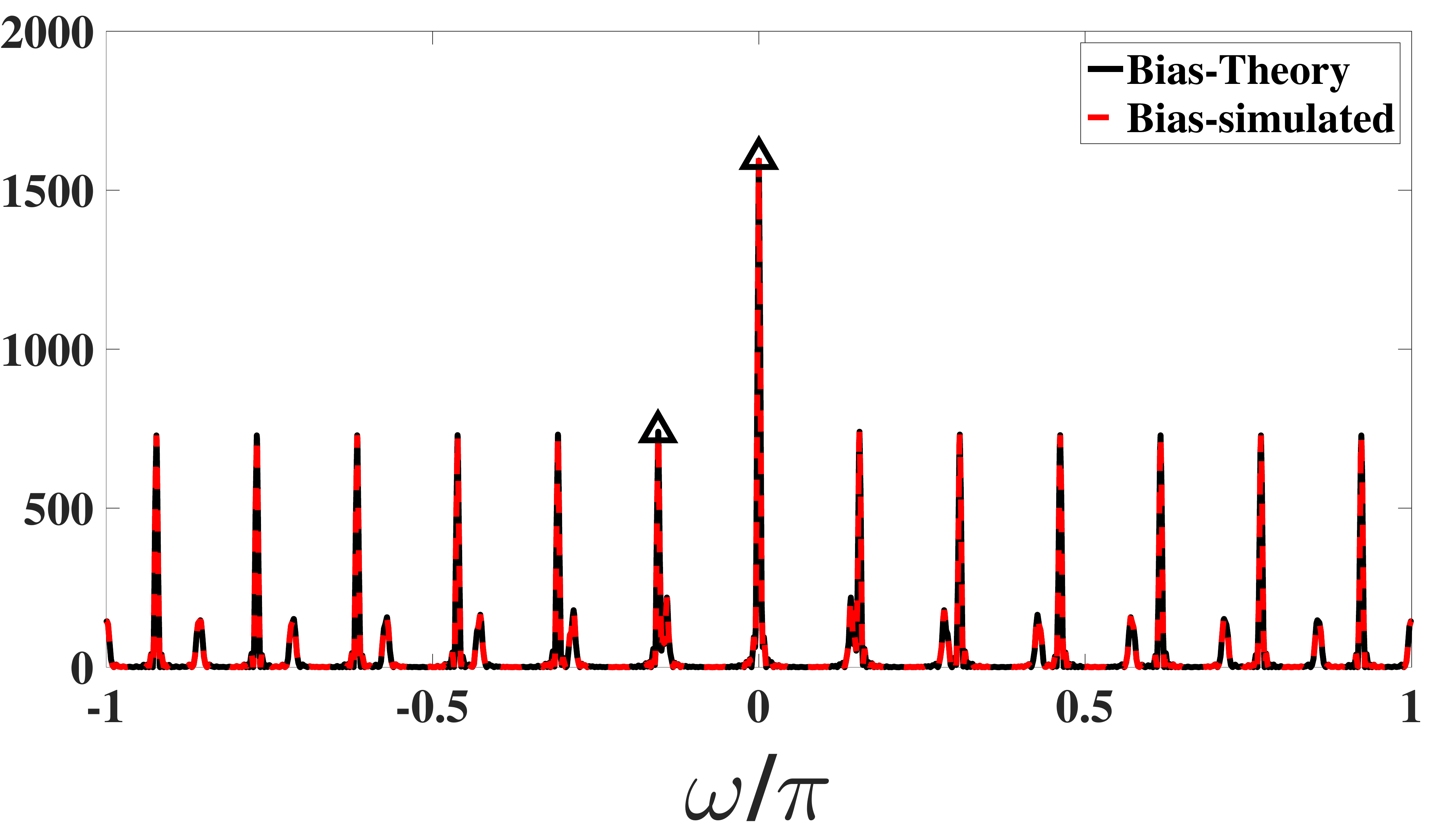}
		\includegraphics[width=0.28\textwidth]{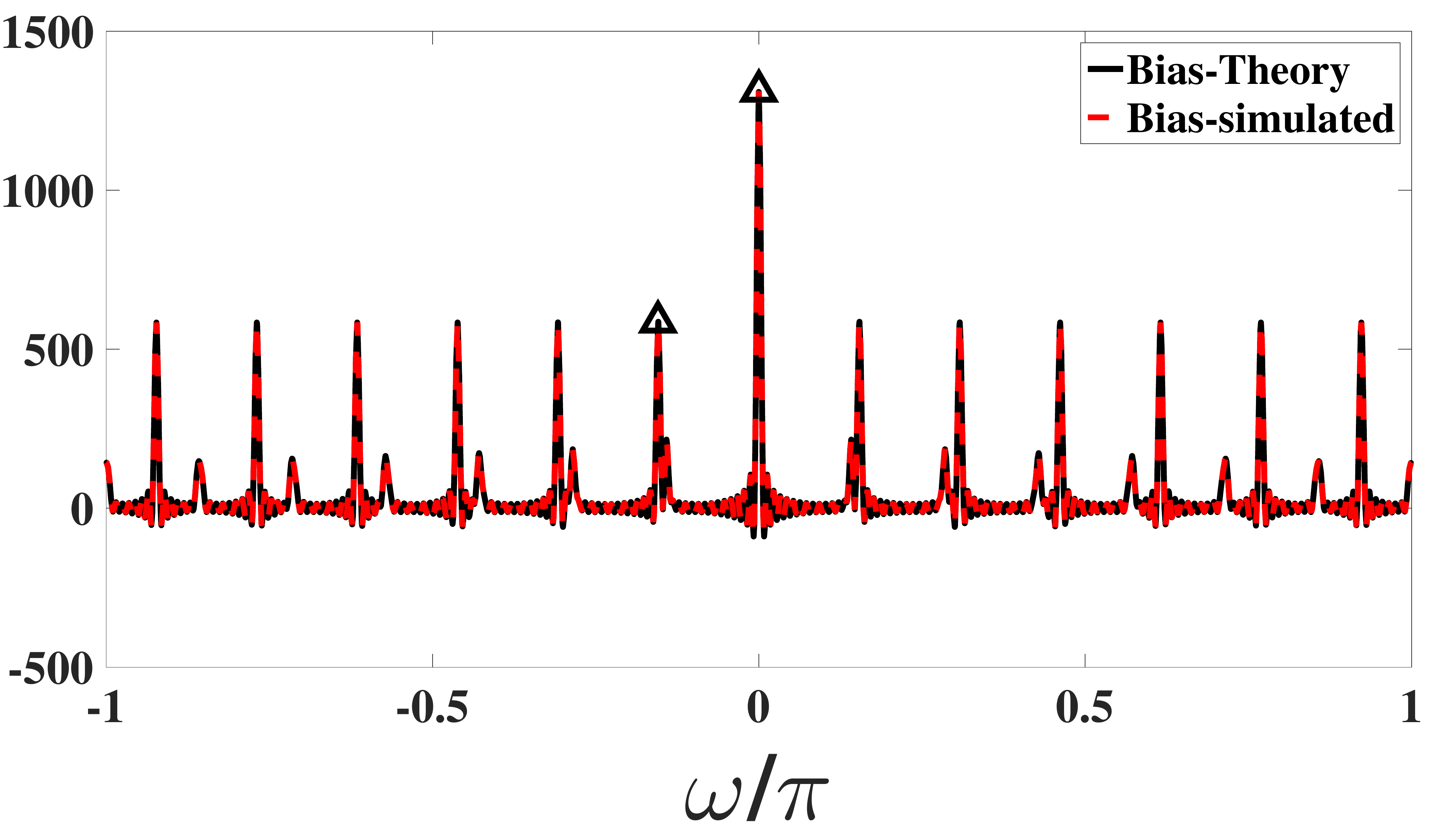}
		\includegraphics[width=0.28\textwidth]{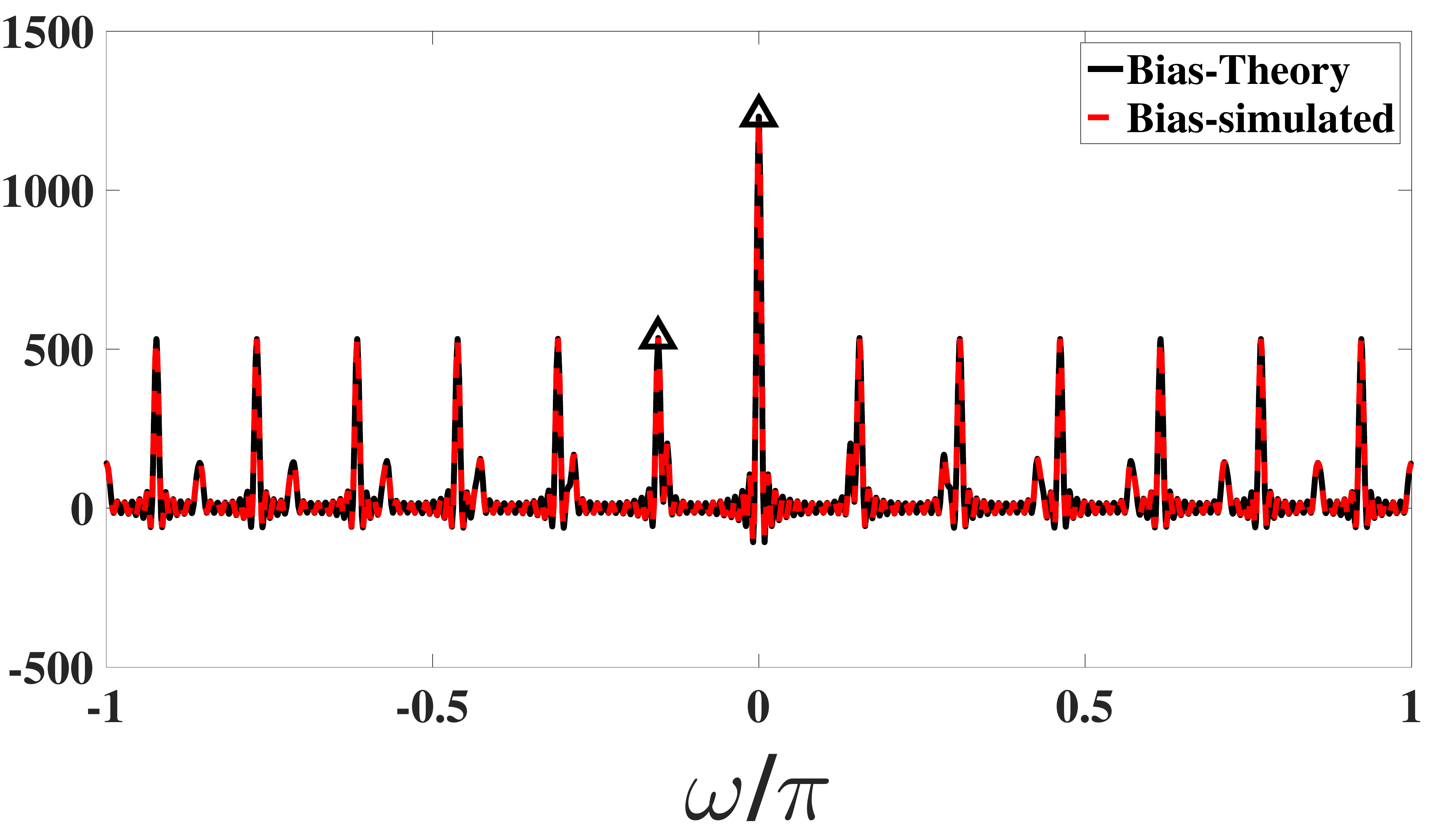}
		\label{ext_M14N13_choice}}
	\hfil
	\subfloat[$M=14$, $N=5$]{
		\includegraphics[width=0.28\textwidth]{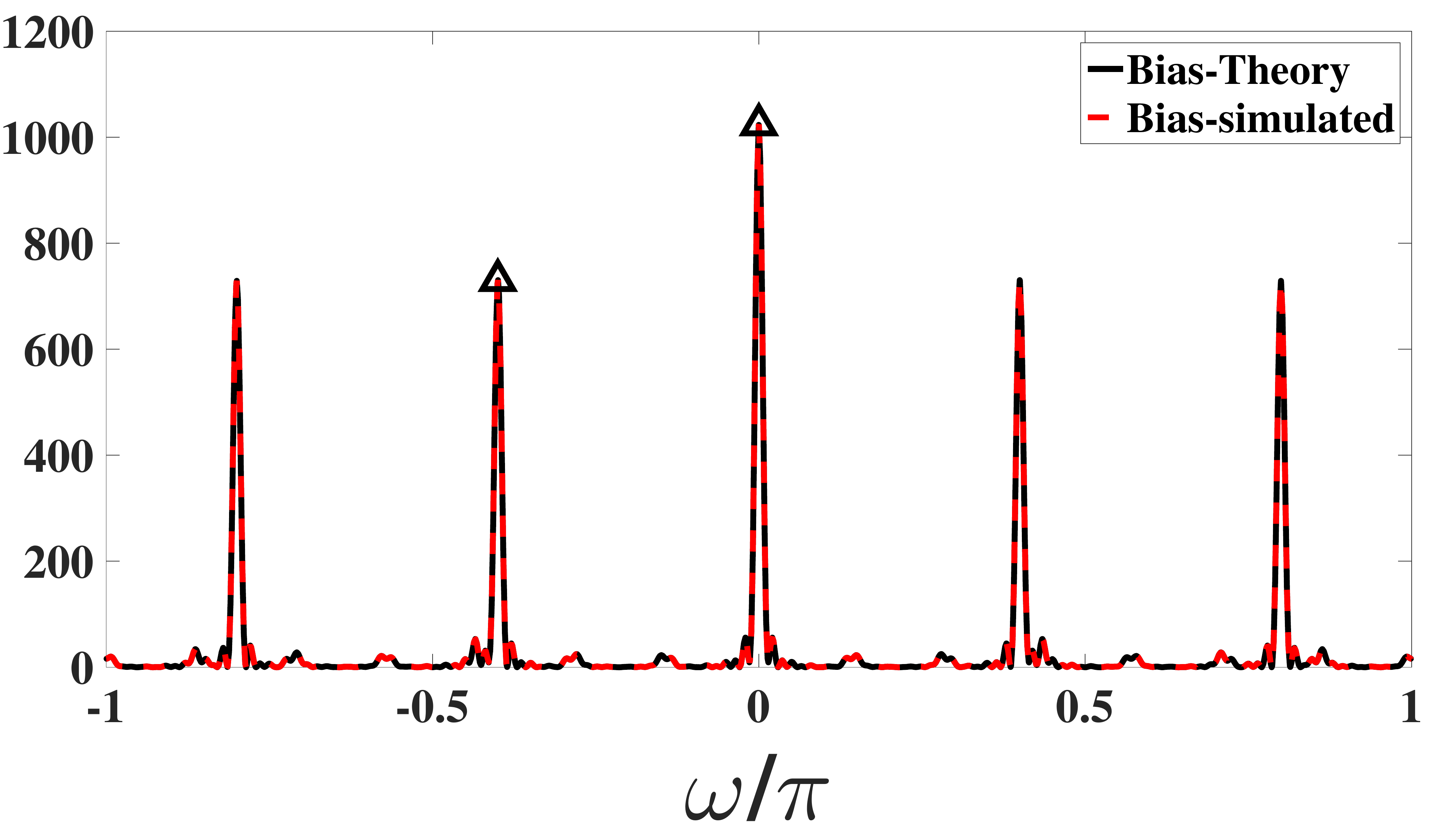}
		\includegraphics[width=0.28\textwidth]{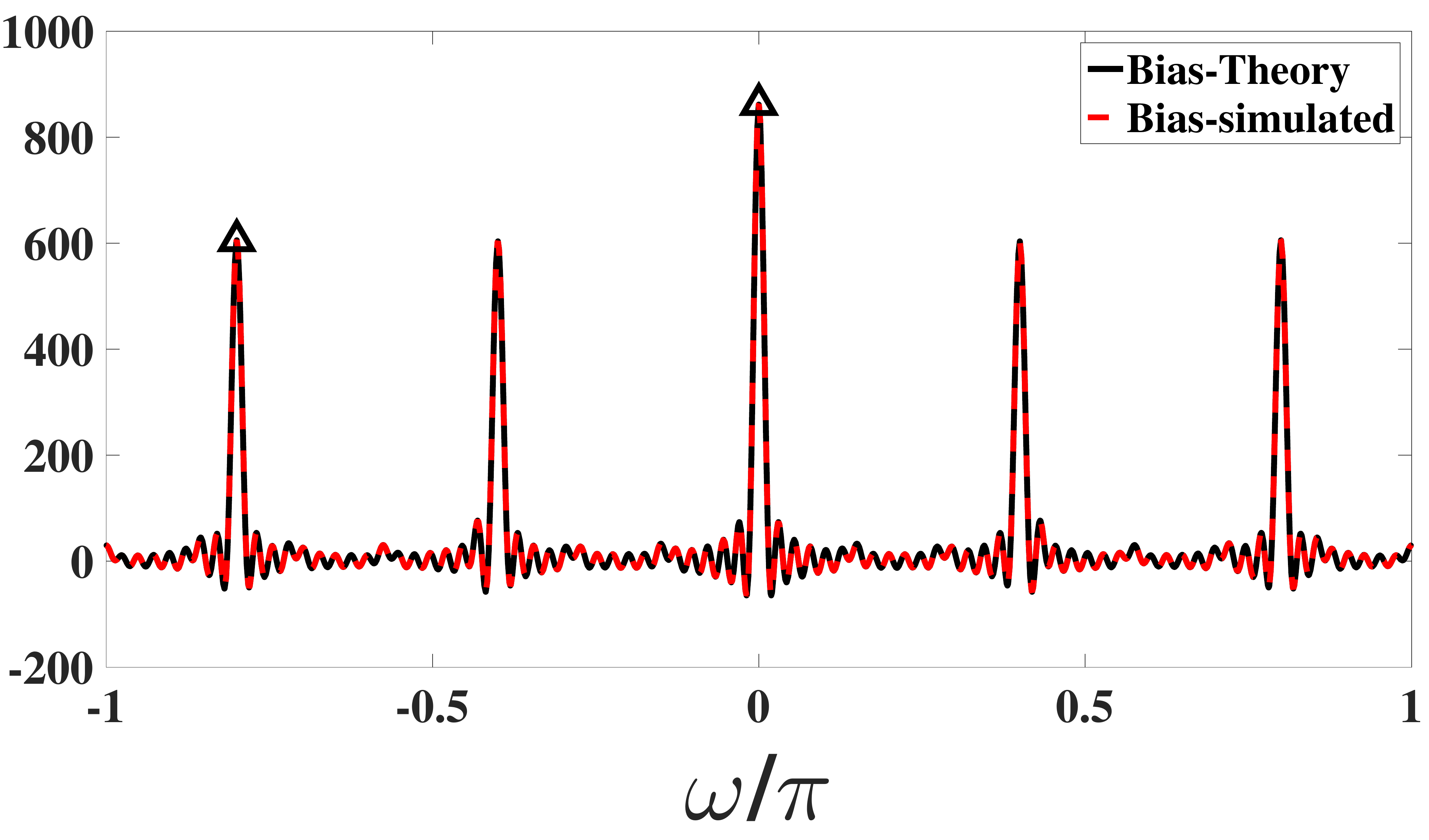}
		\includegraphics[width=0.28\textwidth]{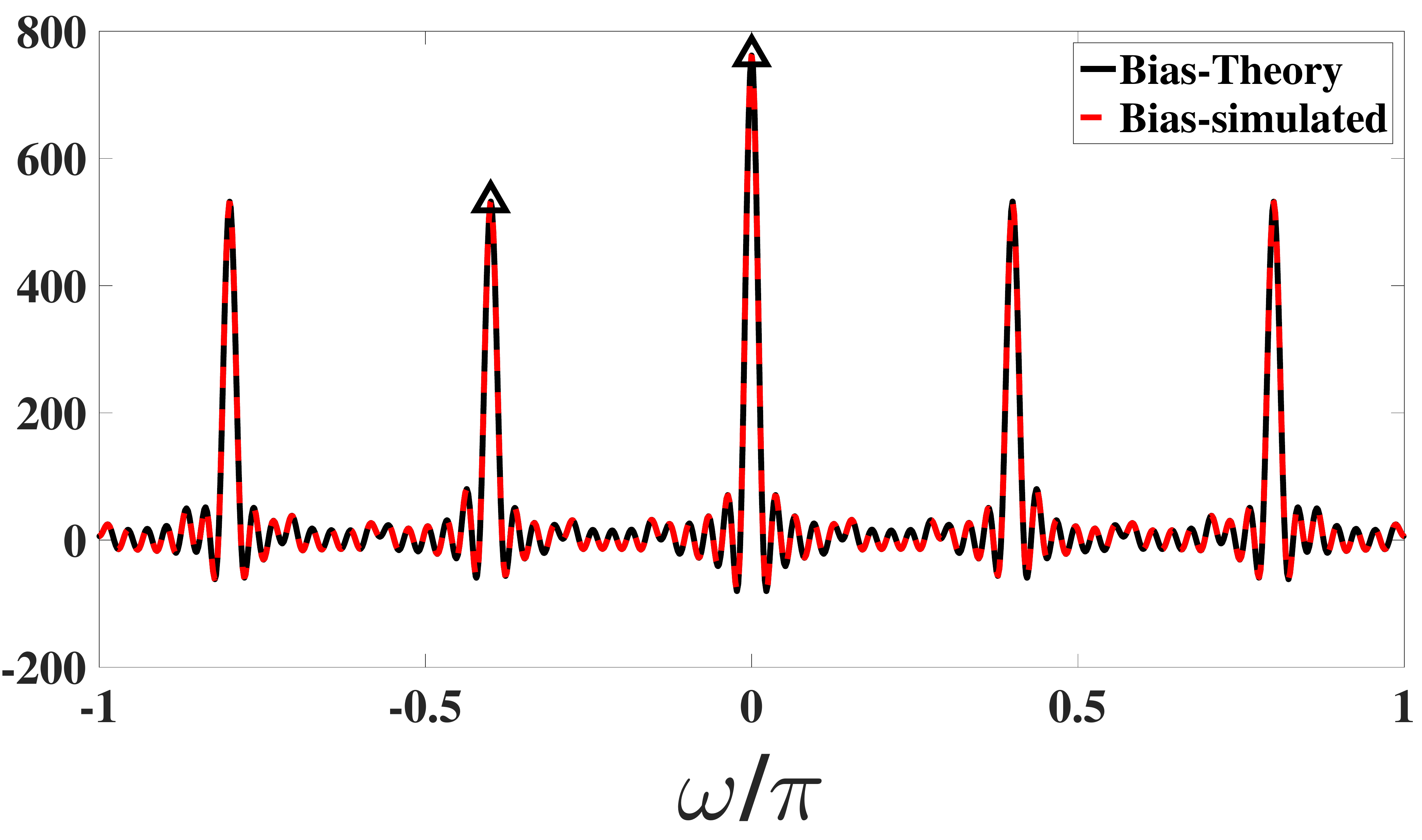}
		\label{ext_M14N5_choice}}
	\hfil
	\subfloat[$M=7$, $N=13$]{
		\includegraphics[width=0.28\textwidth]{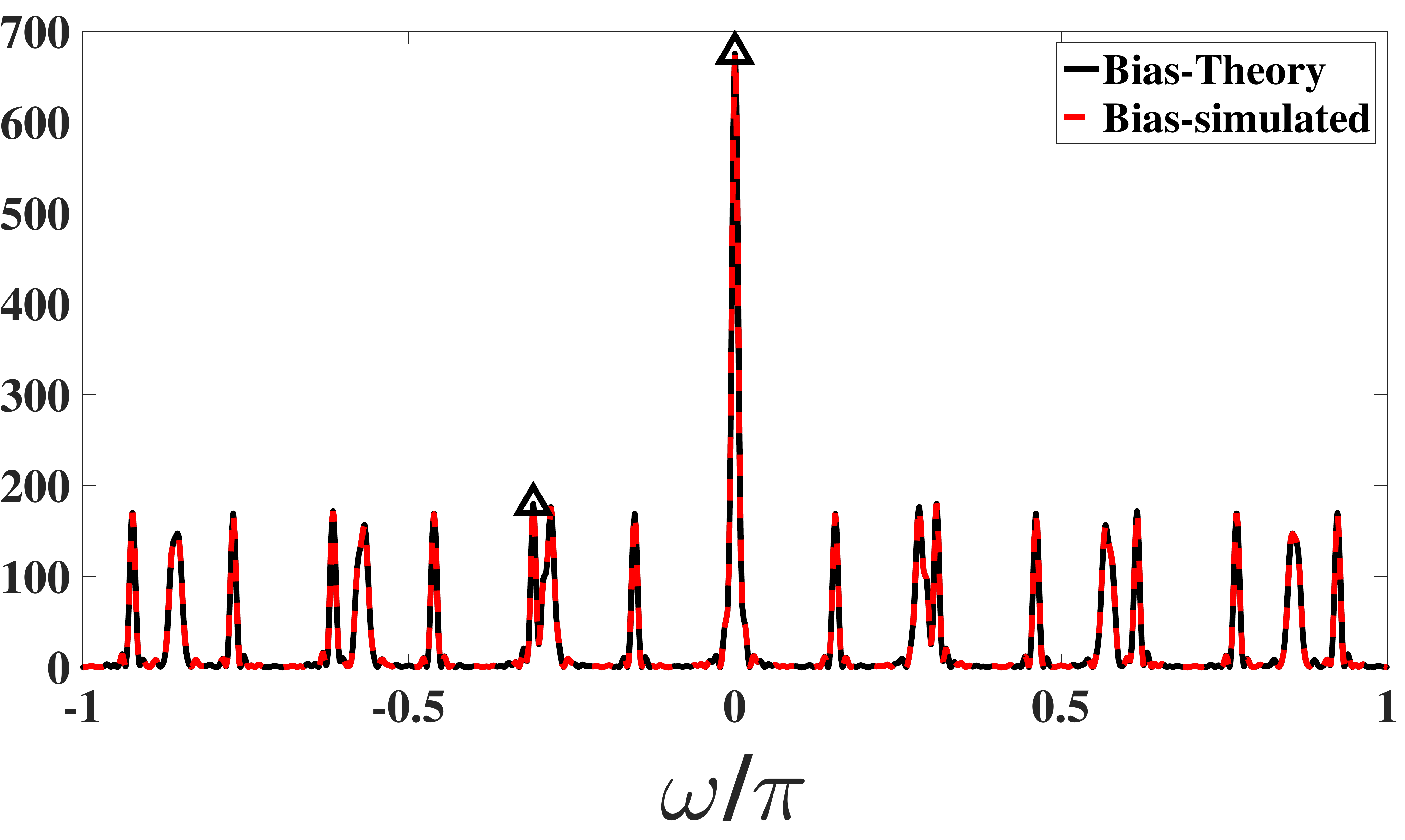}
		\includegraphics[width=0.28\textwidth]{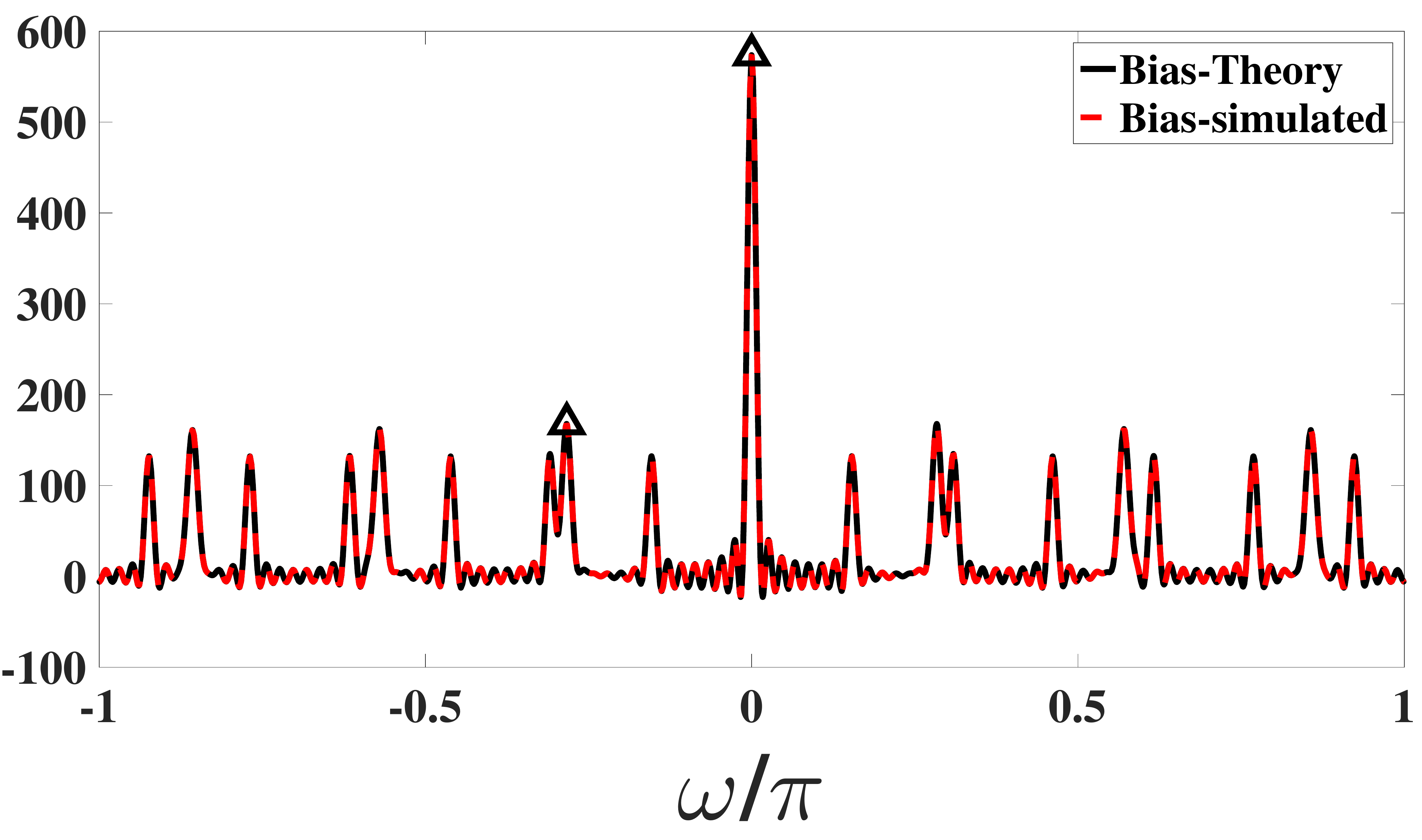}
		\includegraphics[width=0.28\textwidth]{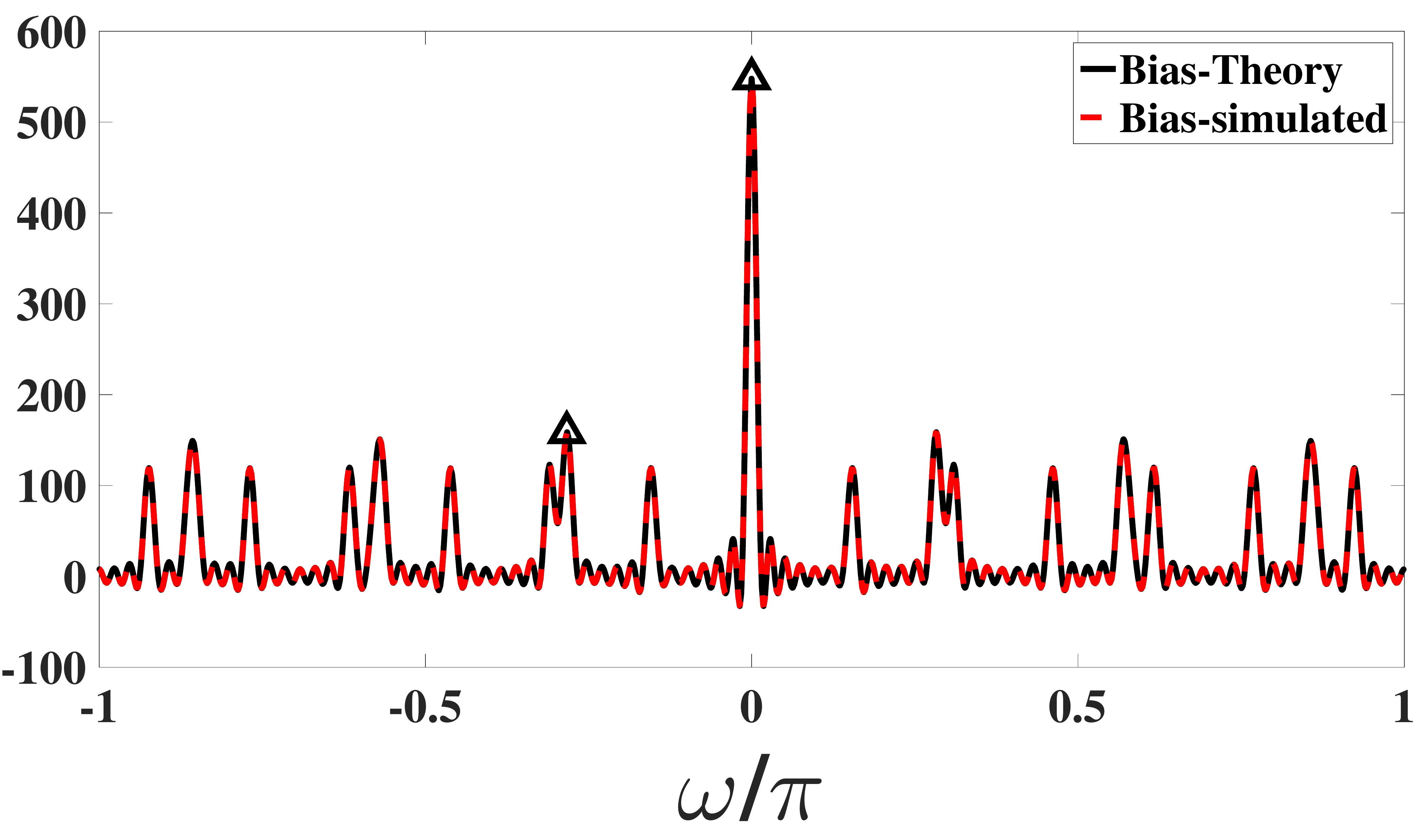}
		\label{ext_M7N13_choice}}
	\hfil
	\subfloat[$M=13$, $N=14$]{
		\includegraphics[width=0.28\textwidth]{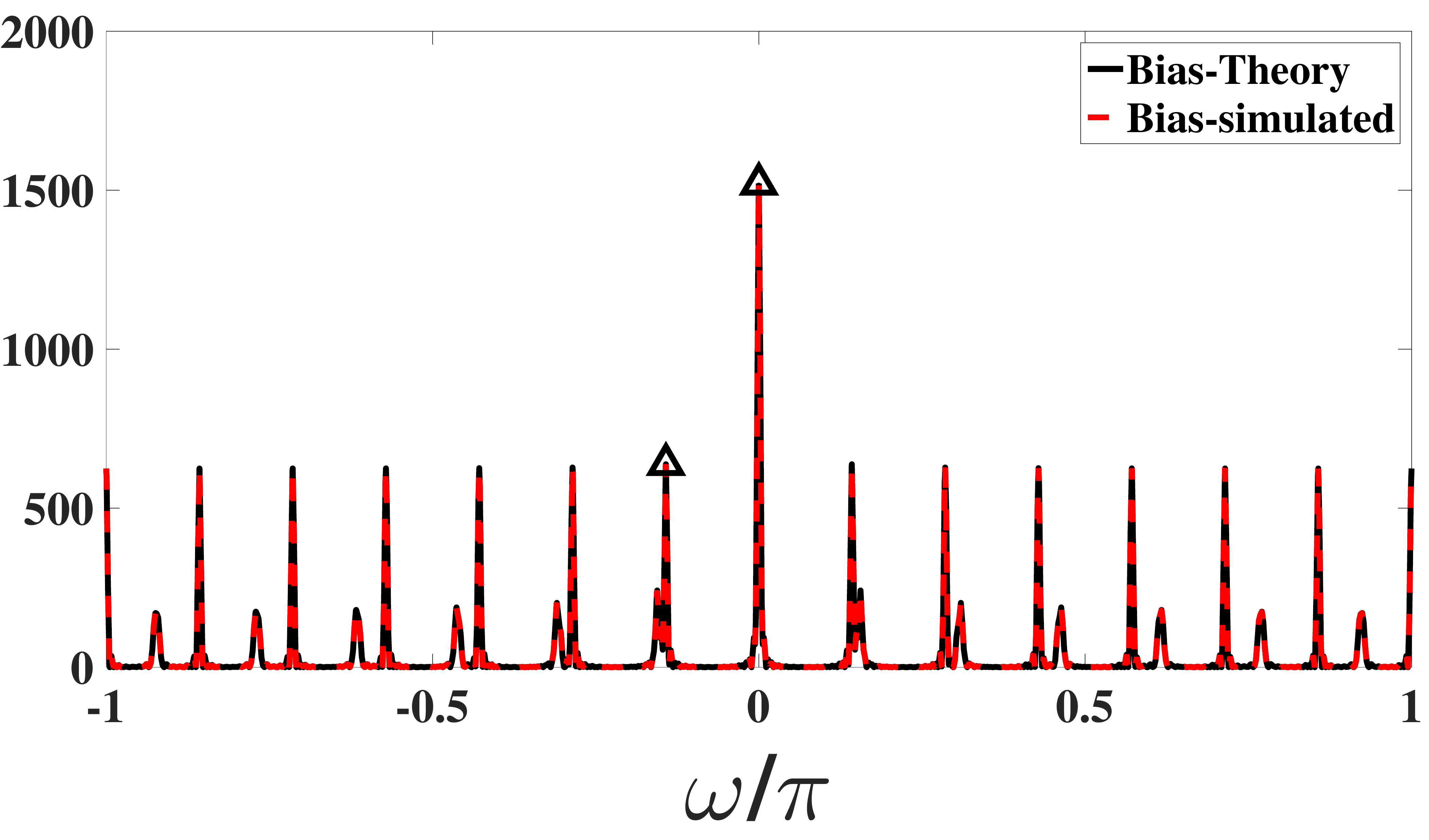}
		\includegraphics[width=0.28\textwidth]{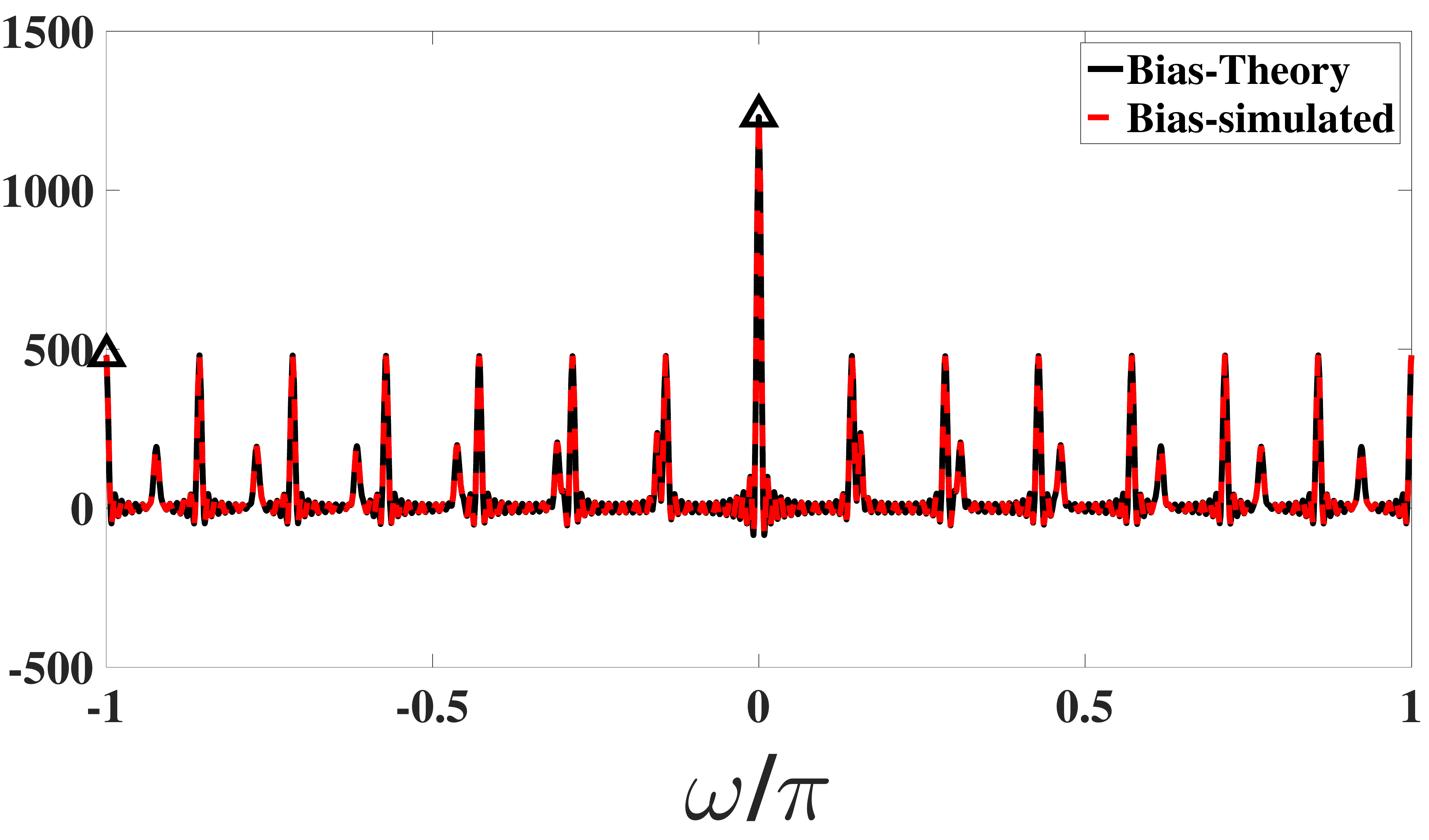}
		\includegraphics[width=0.28\textwidth]{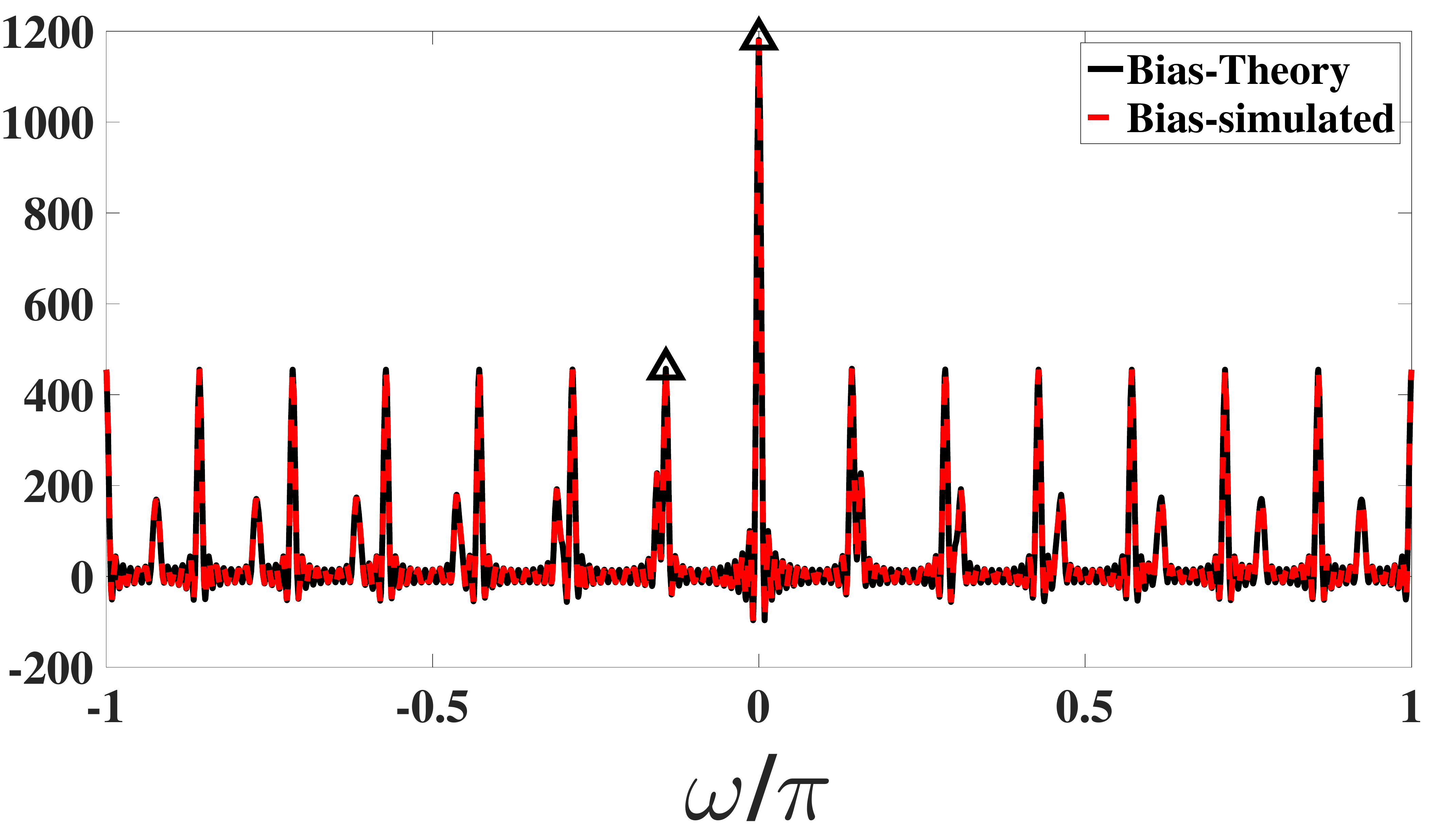}
		\label{ext_M13N14_choice}}
	\hfil
	\subfloat[$M=5$, $N=14$]{
		\includegraphics[width=0.28\textwidth]{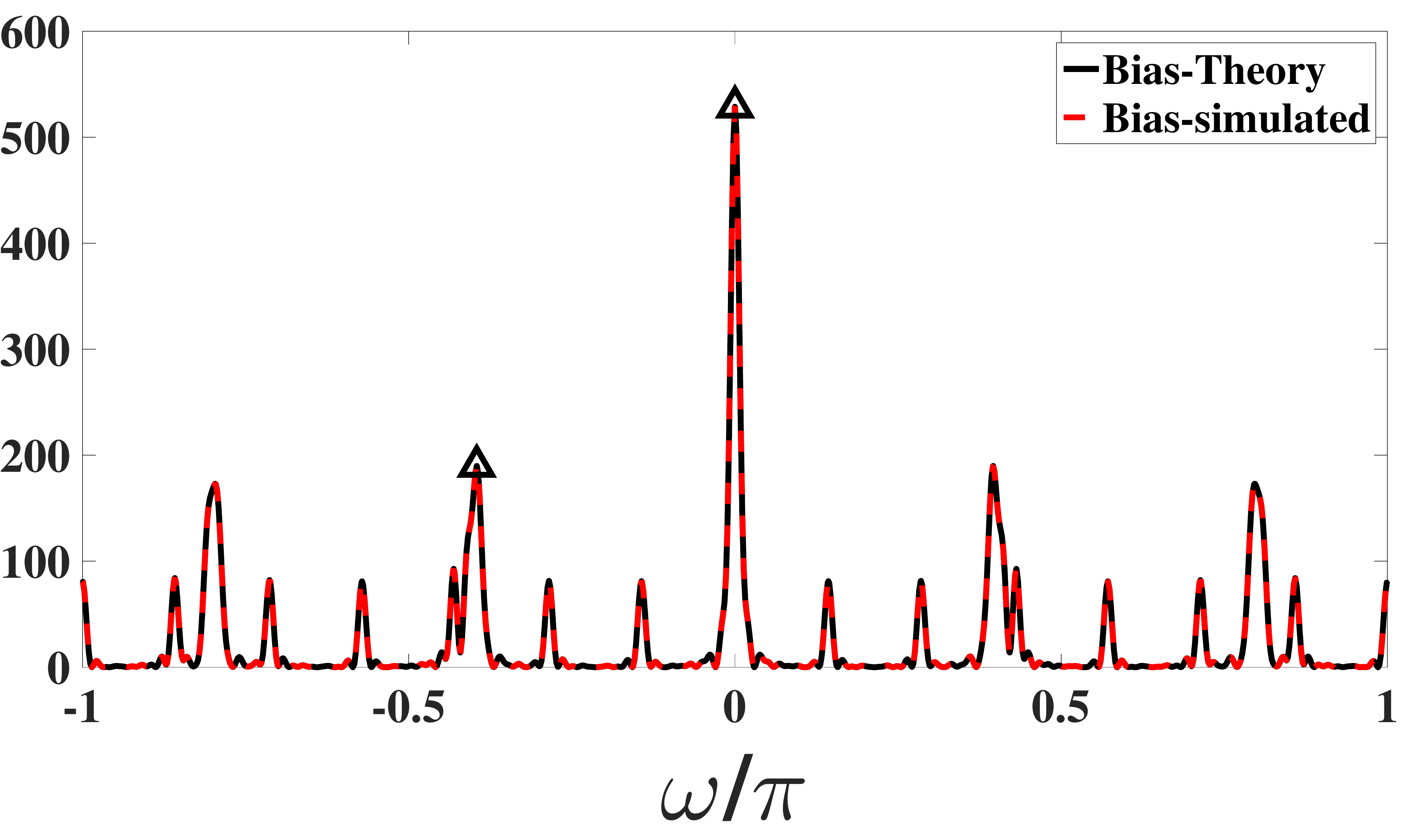}
		\includegraphics[width=0.28\textwidth]{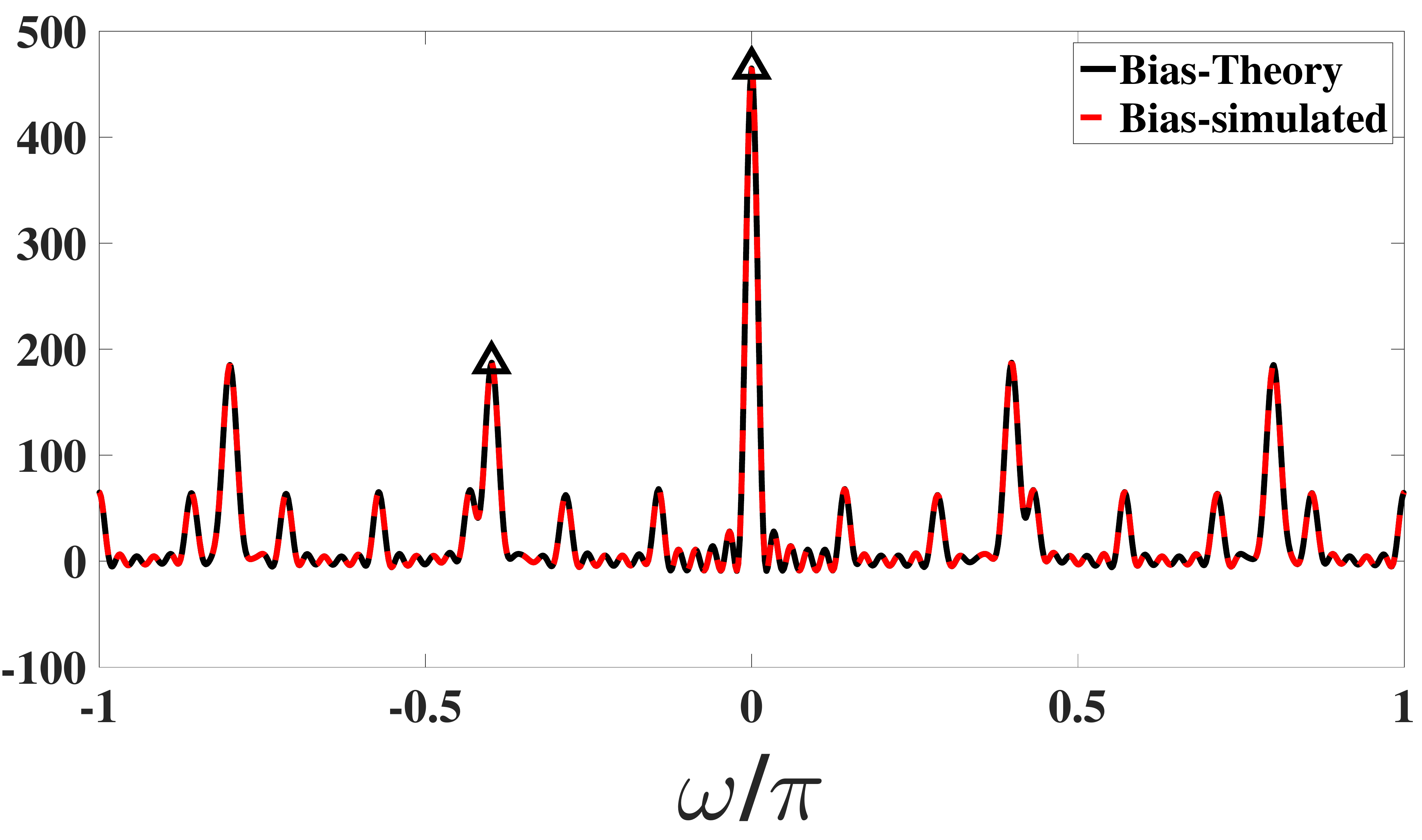}
		\includegraphics[width=0.28\textwidth]{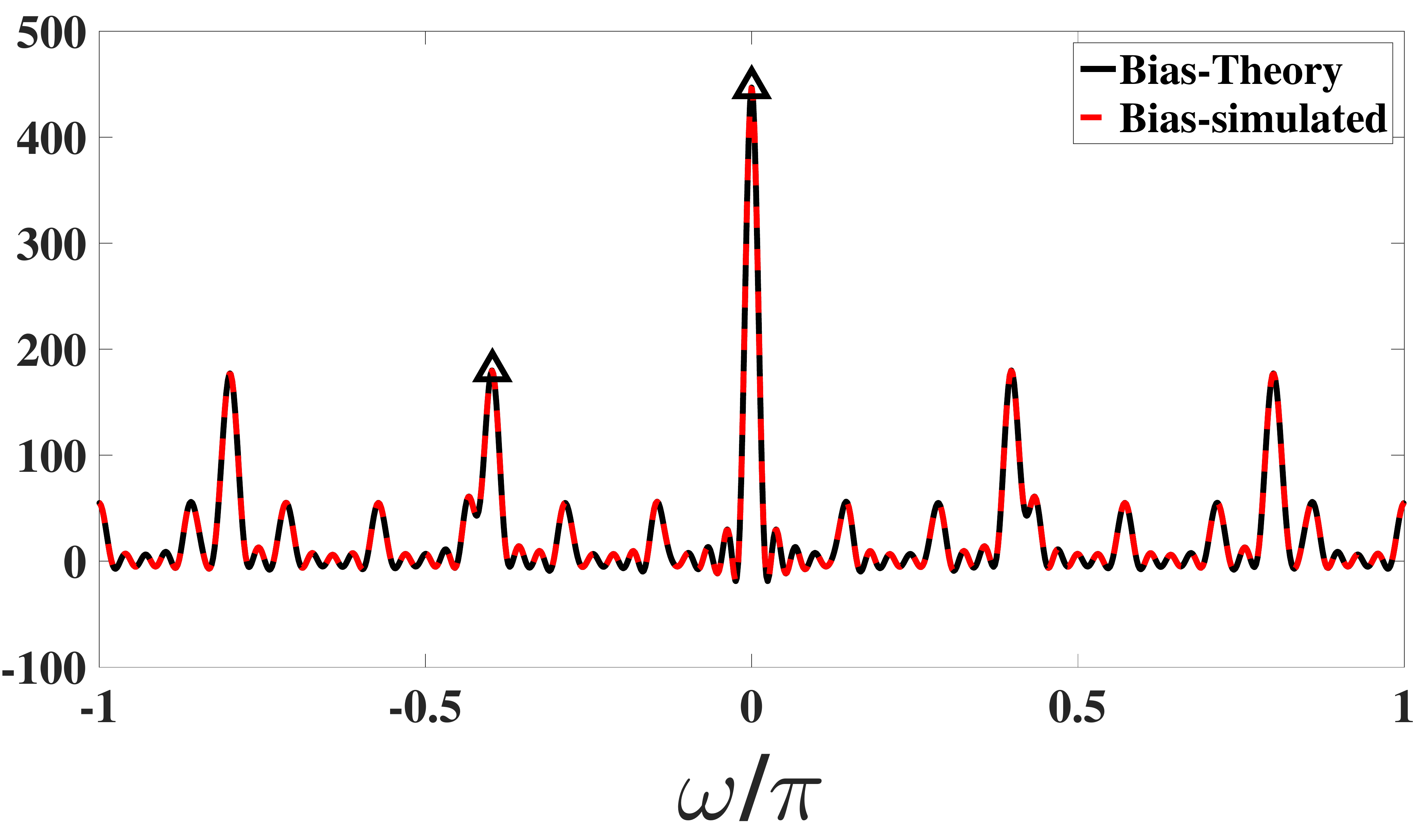}
		\label{ext_M5N14_choice}}
	\hfil
	\subfloat[$M=13$, $N=7$]{
		\includegraphics[width=0.28\textwidth]{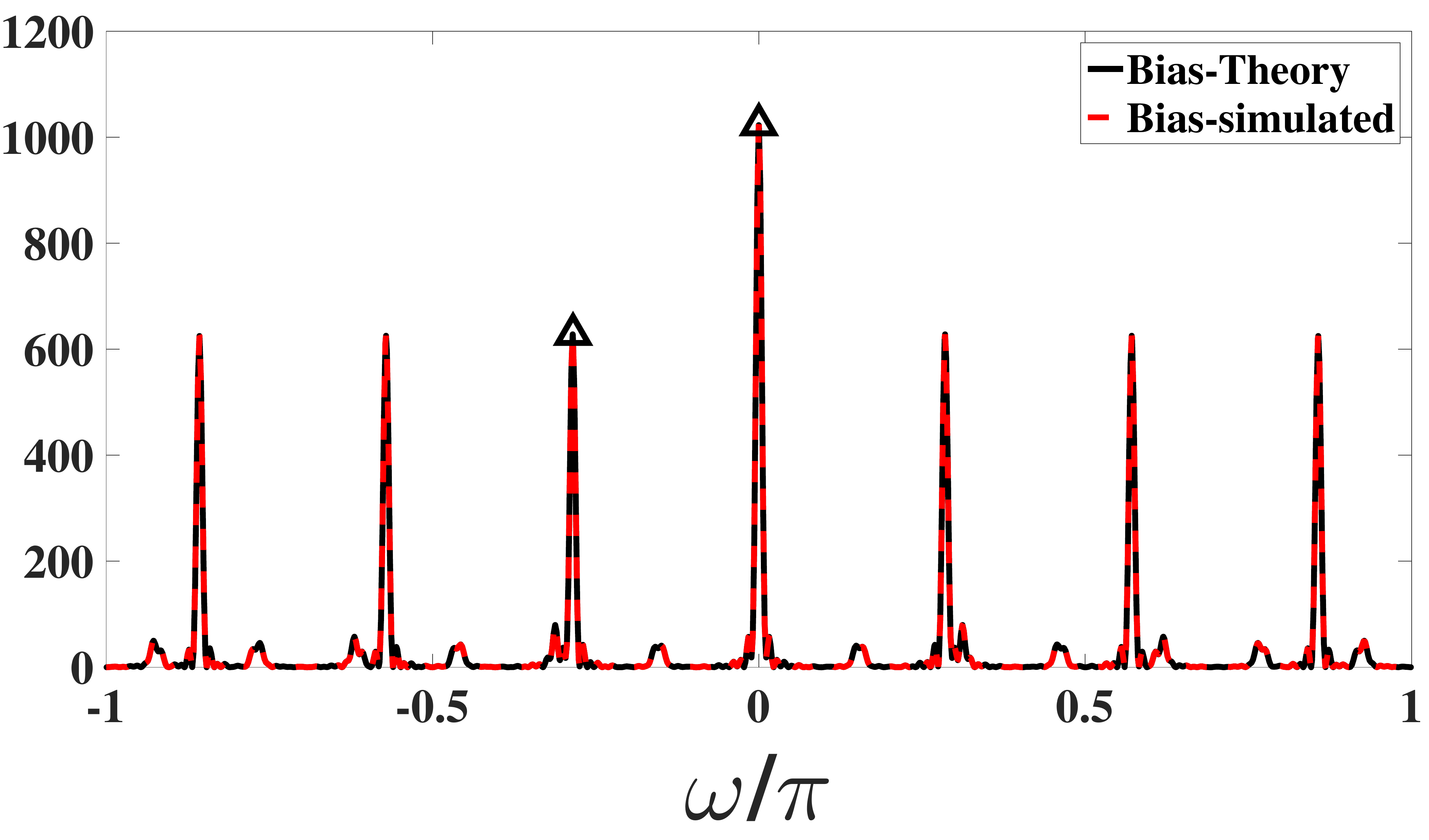}
		\includegraphics[width=0.28\textwidth]{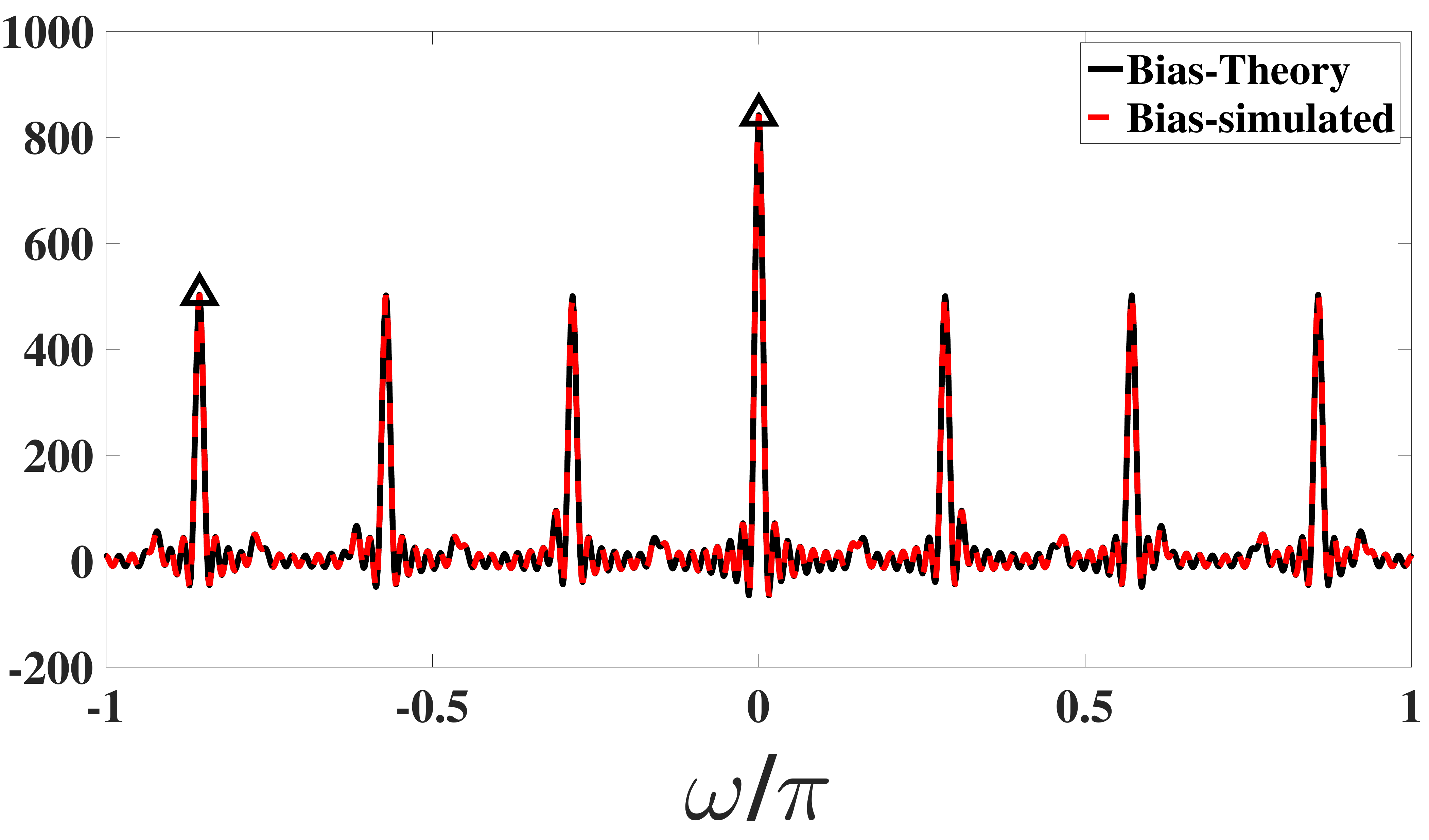}
		\includegraphics[width=0.28\textwidth]{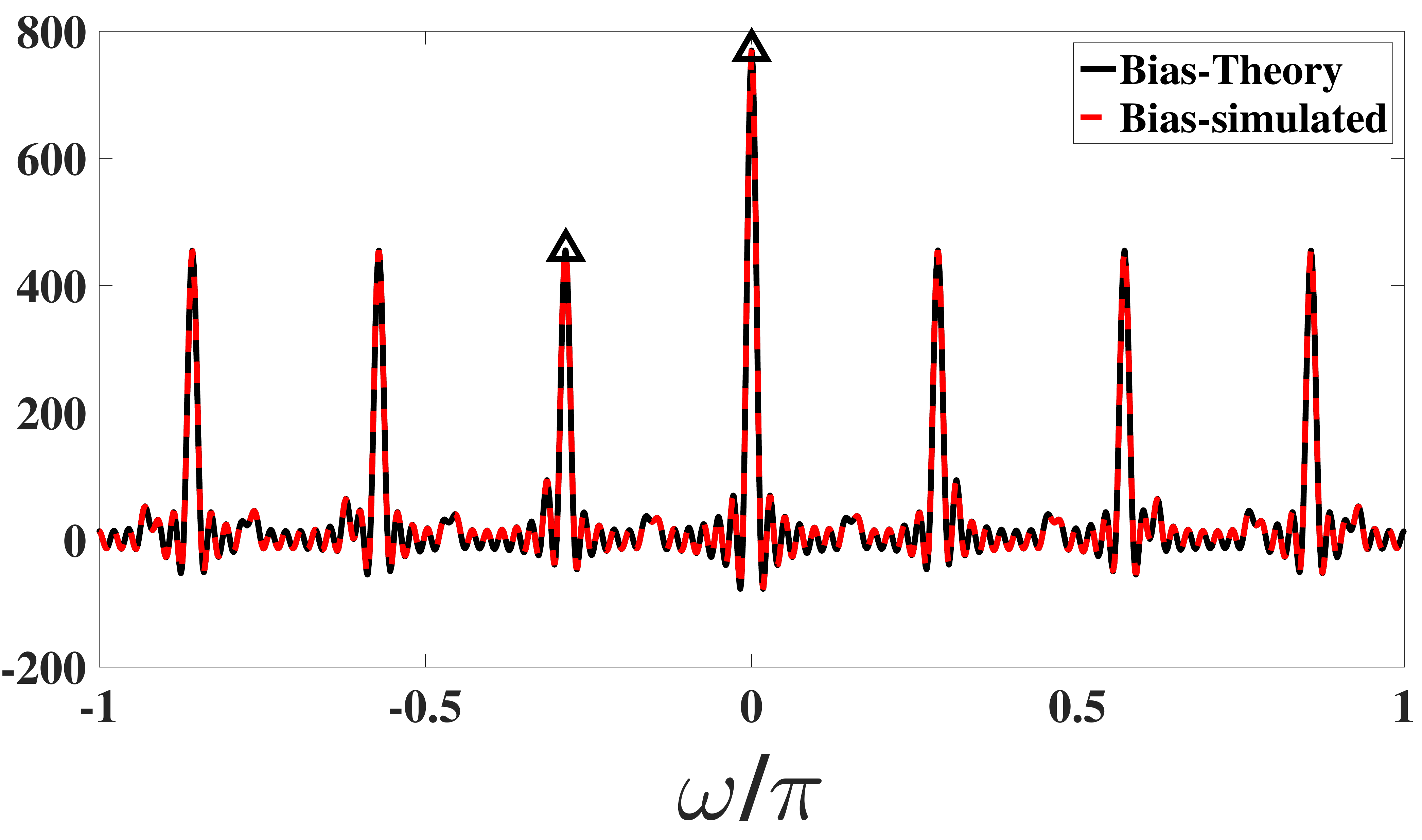}
		\label{ext_M13N7_choice}}
	\hfil
	\caption{Choice of $(M,N)$ for the extended co-prime array: Entire (left), continuous (middle) and prototype (right) range.}
	\label{fig:choiceM_N}
\end{figure*}
%
%
%
To address this issue a comparison between the bias for $M>N$ and $N>M$ is provided in Fig.~\ref{fig:compare_bias_M>Nvice_versa}. The black triangle represents the main-lobe and side-lobe peak when $M>N$ and the red circle represents these peaks when $N>M$. The peak side-lobe is detected using the $\textit{findpeaks}$ command in matlab, and is shown in the range $[-\pi, 0]$. The main-lobe peak is plotted using the formula obtained by substituting $\omega = 0$ in~\eqref{eq:FT_extend_full}-\eqref{eq:FT_extend_proto_period}, and is given by~\eqref{eq:FT_extend_full_peak}-\eqref{eq:FT_extend_proto_period_peak} for the entire, continuous and prototype sets:
\begin{equation}\label{eq:FT_extend_full_peak}
  W_{b_f}(e^{j0})=(2M+N-1)^2
\end{equation}
\begin{equation}\label{eq:FT_extend_continuous_peak}
\begin{split}
  W_{b_c}(e^{j0})=(M+\lfloor \frac{M-1}{N} \rfloor +1)^2 +(M+N)^2+M^2-3M\\
- 3\lfloor \frac{M-1}{N} \rfloor-2\lfloor \frac{M-1}{N} \rfloor^2 -2+\sum\limits_{i=1}^{N-1}2\lfloor \frac{M+Mi-1}{N}\rfloor
\end{split}
\end{equation}
\begin{equation}\label{eq:FT_extend_proto_period_peak}
\begin{split}
  W_{b_p}(e^{j0})=3M^2+N^2-3M+2MN -1+\sum\limits_{i=1}^{N-1}2\lfloor \frac{Mi-1}{N}\rfloor
\end{split}
\end{equation}
The relative amplitude ($R$) between the main lobe peak $P_m$ and side-lobe peak $P_s$ is given by $R=\left(P_m-P_s\right)/P_m$, and is computed for few examples in Table~\ref{table_extened_comparison_MgrtN_viceversa}. It is evident that $R$ is large when $N>M$ for all the three cases, viz. entire, continuous and prototype range. The set-up considered in this paper extends the second array with inter-element spacing $Nd$. In general, it implies that the array with larger inter-element spacing should be extended. It may also be noted from Fig.~\ref{fig:compare_bias_M>Nvice_versa} that the bias is positive only for the case when the entire difference set is employed. Therefore, while using the continuous or the prototype range, a window function would have to be used to guarantee a valid positive power spectrum. Another important observation is that the main-lobe width is the least for the entire set, high for the continuous set and highest for the prototype set. Therefore, the spectral resolution will be high when the entire set is used, and least for the prototype co-prime range.

Despite the detailed analysis, what is the optimum value of $M$ and $N$, is a question that still remains unanswered. For the prototype co-prime array it was observed that small consecutive integers were a good choice to maximize the relative amplitude between the main lobe and side-lobe peaks. For this choice, the side-lobe peaks were approximately of the same height~\cite{UVD_PHD}. Similarly, we wish that the extended co-prime array has side-lobe peaks that are approximately of the same height (contributed by both the sub-arrays). Fig.~\ref{fig:choiceM_N} shows the bias for consecutive values of $M$ and $N$, as well as for the case when $M$ is approximately half the value of $N$ and vice versa. Table~\ref{table_extened_comparison_Choice_MN} compares the relative distance between the main-lobe and side-lobe peaks for the cases considered in Fig~\ref{fig:choiceM_N}. In addition to the fact that $N$ should be greater than $M$, it is now evident that consecutive values of $M$ and $N$ are not an optimum choice for maximizing the distance between the main-lobe and peak side-lobe. $M\approx \frac{N}{2}$ seems to be an optimum choice for the extended co-prime array.
%
%
\begin{table}[!t]
	\caption{Relative distance between main-lobe and side-lobe peak: A comparison between $M>N$ and $N>M$}
	\label{table_extened_comparison_MgrtN_viceversa}
	\centering
	\renewcommand{\arraystretch}{1.2}{
		\begin{tabular}{|c|c|c|c|c|c|c|c|}
			\hline
			&	&\multicolumn{3}{c|}{$M>N$}	&\multicolumn{3}{c|}{$N>M$}\\ 
			&	&\multicolumn{3}{c|}{}		&\multicolumn{3}{c|}{(interchange M and N)}\\ 
			\hline	
			$\textbf{M}$ & $\textbf{N}$ & $\textbf{\textit{f}}$ &$\textbf{\textit{c}}$ & $\textbf{\textit{p}}$ & $\textbf{\textit{f}}$ &$\textbf{\textit{c}}$ & $\textbf{\textit{p}}$ \\
			\hline
			4	&3	&0.508	&0.521	&0.565	&0.683	&0.712	&0.762\\
			\hline
			5	&3	&0.436	&0.461	&0.481	&0.737	&0.764	&0.774\\
			\hline
			7	&3	&0.339	&0.349	&0.367	&0.701	&0.664	&0.665\\
			\hline
			8	&3	&0.305	&0.320	&0.328	&0.667	&0.626	&0.626\\
			\hline
			5	&4	&0.516	&0.529	&0.564	&0.651	&0.685	&0.714\\
			\hline
			7	&4	&0.413	&0.430	&0.446	&0.735	&0.737	&0.744\\
			\hline
	\end{tabular}}
\end{table}
\begin{table}[!t]
	\caption{Relative distance between main-lobe and side-lobe peak: A comparison between $M>N$ and $N>M$}
	\label{table_extened_comparison_Choice_MN}
	\centering
	\renewcommand{\arraystretch}{1}{
		\begin{tabular}{|c|c|c|c|c|c|c|c|}
			\hline
			\multicolumn{2}{|c|}{$\textbf{(M,N)}$} 					&\textbf{(14,13)}			&\textbf{(14,5)}		&\textbf{(7,13)}		&\textbf{(13,14)} 		&\textbf{(5,14)} 		& \textbf{(13,7)}\\
			\hline
			\multirow{3}{*}{\textbf{Relative}}& \textbf{\textit{f}}	&0.537	&0.287	&\textbf{0.734}	&0.580	&0.641	&0.387\\ \cline{2-8}
			\multirow{3}{*}{\textbf{Distance}}& \textbf{\textit{c}}		&0.553	&0.297	&\textbf{0.708}	&0.610	&0.597	&0.403\\ \cline{2-8}
			&\textbf{\textit{p}}		&0.566	&0.302	&\textbf{0.710}	&0.614	&0.597	&0.408\\ \hline
	\end{tabular}}
\end{table}
\section{Variance Analysis}
\label{sec:variance}
The variance of the extended co-prime array is analyzed along similar lines as that of the prototype and multiple period co-prime array~\cite{UVD_PHD}. We assume that $x(n)$ is a circular complex gaussian white process with zero mean and variance $\sigma^2$~\cite{7.1}\cite{7.17}.
The expected value of the extended co-prime array-based correlogram for this process is given by:
\begin{eqnarray}
  \nonumber\mathbb{E}\{\hat{P}(e^{j\omega})\}=\sum\limits_{l}w_b(l)\sigma^2\delta(l)e^{-jwl}=w_b(0)\sigma^2=P(e^{j\omega})
\end{eqnarray}
where $s_b$ is assumed to be $2M+N-1$, and hence gives $w_b(0)=1$. Therefore, the expected value matches the true spectrum.
The covariance of the correlogram estimate for the extended co-prime array is given by the bias equations~\eqref{eq:FT_extend_full},~\eqref{eq:FT_extend_continuous} and~\eqref{eq:FT_extend_proto_period} when $\omega$ is replaced by $\omega_1-\omega_2$ (as noted in~\cite{UVD_PHD}) for the entire, continuous and prototype co-prime range respectively. The scale factor is $\frac{\sigma^4}{{s_b}^2}$ (instead of $\frac{1}{s_b}$).

\begin{figure}[!t]
\centering
%
\includegraphics[width=0.5\textwidth]{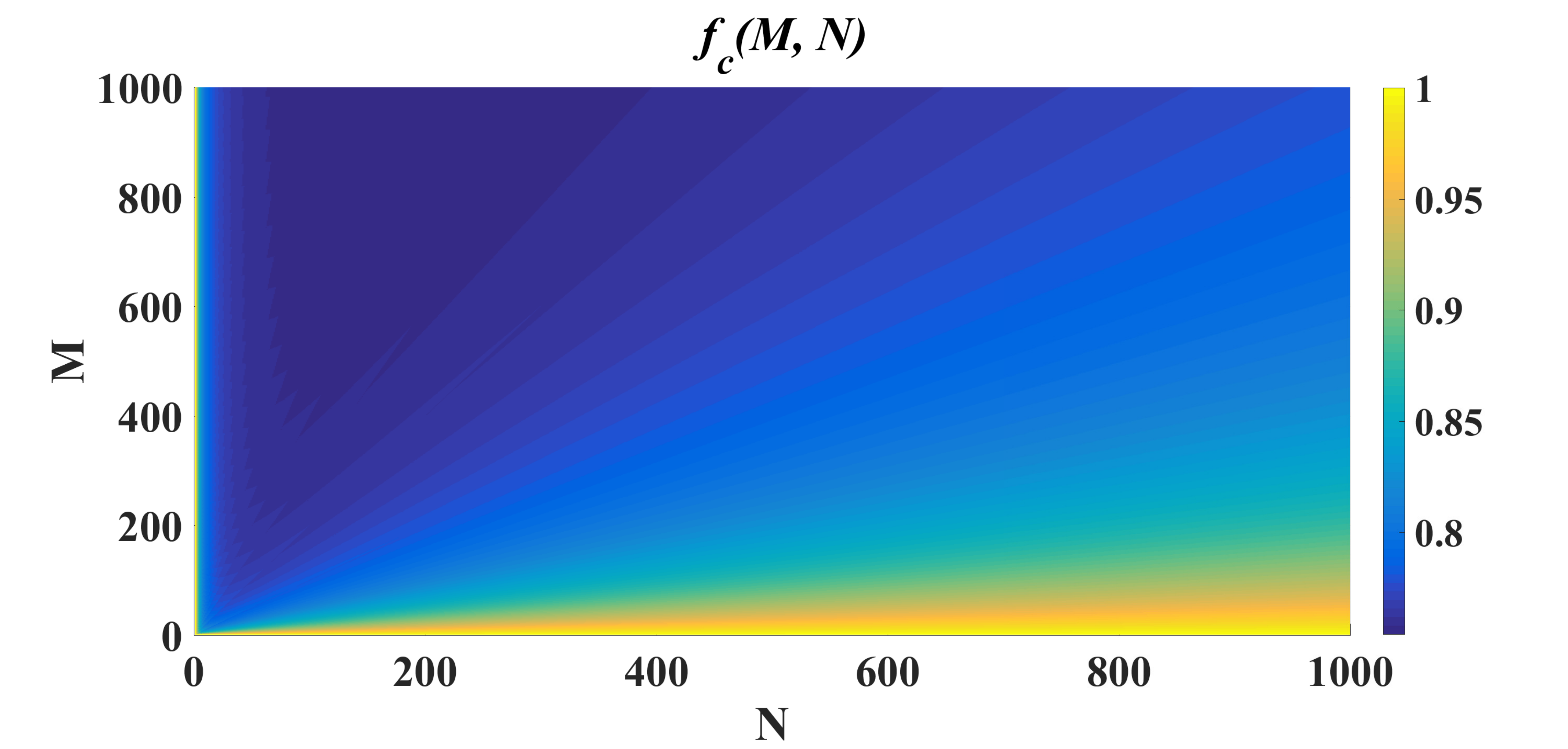}\\
\includegraphics[width=0.5\textwidth]{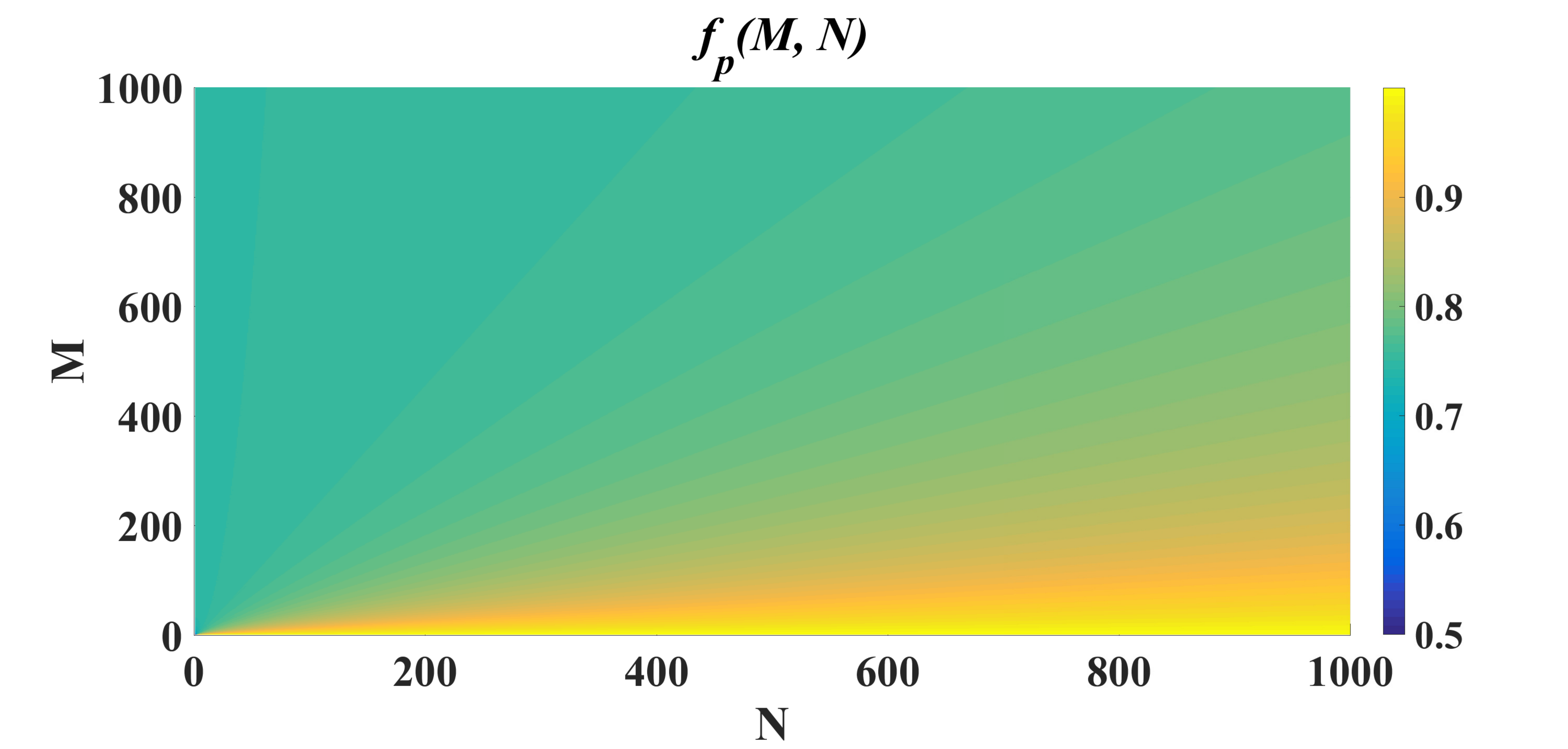}
%
\caption{Variance for the continuous and prototype range: $f_c(M, N)$ and $f_p(M, N)$}
\label{fig:covariance}
\end{figure}

%
For the case when $(\omega_1-\omega_2)=0$, we have, $Cov\{\hat{P_f}(e^{j\omega_1})\hat{P_f}(e^{j\omega_2})\}=\sigma^4$. Therefore, the estimate using the entire difference set is not consistent.
The variance of the correlogram estimate for the continuous and prototype range is:
\begin{eqnarray}
\nonumber Cov\{\hat{P_{c}}(e^{j\omega_1})\hat{P_{c}}(e^{j\omega_2})|(\omega_1-\omega_2)=0\}&=&\sigma^4f_c(M,N)\\
\nonumber Cov\{\hat{P_{p}}(e^{j\omega_1})\hat{P_{p}}(e^{j\omega_2})|(\omega_1-\omega_2)=0\}&=&\sigma^4f_p(M,N)
\end{eqnarray}
where $f_c(M,N)$ and $f_p(M,N)$  are given by~\eqref{eq:FT_extend_continuous_peak} and~\eqref{eq:FT_extend_proto_period_peak} respectively, scaled by $\frac{1}{s_b^2}$. These functions are shown in Fig.~\ref{fig:covariance} for $1\leq M, N \leq 1000$ and includes $(M, N)$ pairs that are not co-prime.

\section{Computational Complexity}
\label{sec:complexity}
The number of multiplications and additions required for the prototype co-prime array-based autocorrelation estimation was derived in~\cite{U_S_2, UVD_PHD} for the entire difference set. The continuous range was not considered. We provide the number of multiplications and additions for the continuous range denoted by $C_M$ and $C_A$ respectively in~\eqref{eq:complexity_pca_continuous} for $M>N$. Since the autocorrelation function is symmetric, the complexity is dervied for $l\in[0, L]$ where $L=M+N-1$. $z_c(l)$ is the weight function of the prototype co-prime array for the continuous range and was derived in~\cite{UVD_PHD}. The total number of multiplications required is the summation of the weight function while the adders is given by the summation of the weights minus one at each difference value.

Next, we consider the computational complexity for the extended co-prime array. Let $C_{M_x}$ and $C_{A_x}$ denote the number of multiplications and additions required for autocorrelation estimation for the extended co-prime array respectively, where $x=\{f, c, p\}$ for the full, continuous and prototype range. The complexity is given by~\eqref{eq:complexity_extend_full}-\eqref{eq:complexity_extend_prototype} for the entire, continuous and prototype range respectively and is derived using the corresponding weights $z_{e_f}$, $z_{e_c}$, and $z_{e_p}$.

\begin{figure*}[!t]
	\small
	\setcounter{mytempeqncnt}{\value{equation}}
	\setcounter{equation}{15}
	%
	\begin{align}\label{eq:complexity_pca_continuous}
	\nonumber   C_{M}&=\sum\limits_{l=0}^{M+N-1}z_{c}(l)=(2M+4N-4)+\left(\left\lfloor \frac{M+N-1}{N}\right\rfloor+1\right)\left(M-\frac{\left\lfloor \frac{M+N-1}{N}\right\rfloor}{2}-2\right)\\
	C_{A}&=\sum\limits_{l\{l|z_{c}(l)>1\}}[z_{c}(l)-1]=\sum\limits_{l=0}^{M+N-1}z_{c}(l)-\sum\limits_{\{l|z_{c}(l)\geq1\}}1=(M+3N-4)+\left(\left\lfloor \frac{M+N-1}{N}\right\rfloor+1\right)\left(M-\frac{\left\lfloor \frac{M+N-1}{N}\right\rfloor}{2}-2\right)
	\end{align}	
	\begin{align}\label{eq:complexity_extend_full}
	C_{M_f}&=\frac{(2M+N)(2M+N-1)}{2}~\text{and}~C_{A_f}=\frac{4M^2+N^2+MN-3M-1}{2}\\
	\label{eq:complexity_extend_continuous}
	\nonumber   C_{M_c}&=\sum\limits_{l=0}^{MN+M-1}z_{e_c}(l)=\frac{N^2+3M^2+2MN+N+M+\lfloor \frac{M-1}{N}\rfloor(2M-1-\lfloor \frac{M-1}{N}\rfloor)}{2}+\sum\limits_{n=1}^{N-1}\lfloor\frac{M+Mn-1}{N}\rfloor-1\\
	C_{A_c}&=\sum\limits_{l=0}^{MN+M-1}z_{e_c}(l)-\sum\limits_{\{l|z_{e_c}(l)\geq1\}}1=\frac{N^2+3M^2+N-M+\lfloor \frac{M-1}{N}\rfloor(2M-1-\lfloor \frac{M-1}{N}\rfloor)}			{2}+\sum\limits_{n=1}^{N-1}\lfloor\frac{M+Mn-1}{N}\rfloor-1\\
	\label{eq:complexity_extend_prototype}
	C_{M_p}&=\frac{N^2+3M^2+N-M+2MN}{2}+\sum\limits_{n=1}^{N-1}\left\lfloor\frac{Mn-1}{N}\right\rfloor-1~\text{and}~C_{A_p}=\frac{N^2+3M^2+N-M}{2}+\sum\limits_{n=1}^{N-1}\left\lfloor\frac{Mn-1}{N}\right\rfloor-1
	\end{align}
	\setcounter{equation}{\value{mytempeqncnt}}
	\hrulefill
	\vspace*{4pt}
\end{figure*}
\addtocounter{equation}{3}

\section{Numerical simulations}\label{sec4}
\begin{figure*}[!t]
	\includegraphics[width=0.33\textwidth]{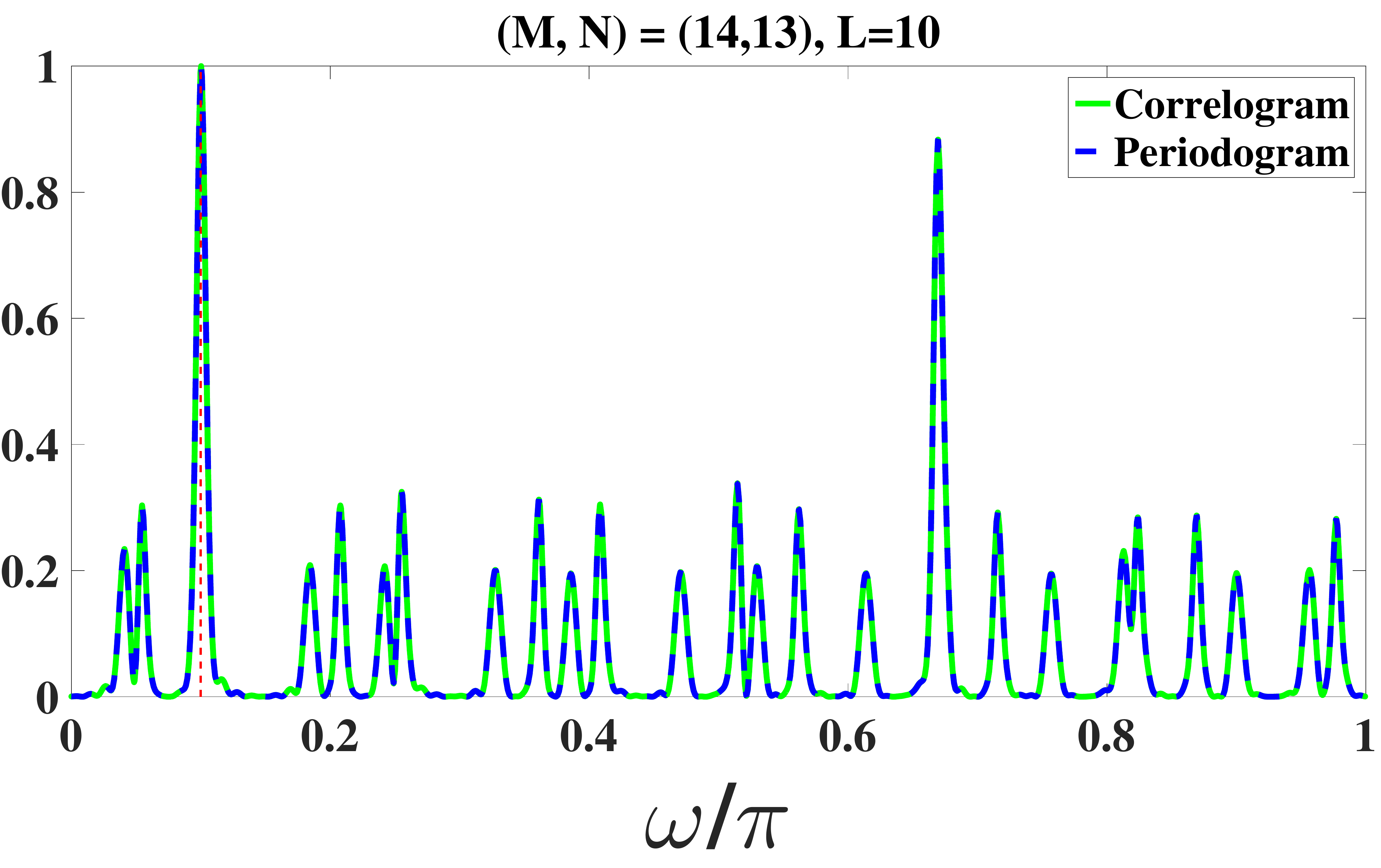}
	\includegraphics[width=0.33\textwidth]{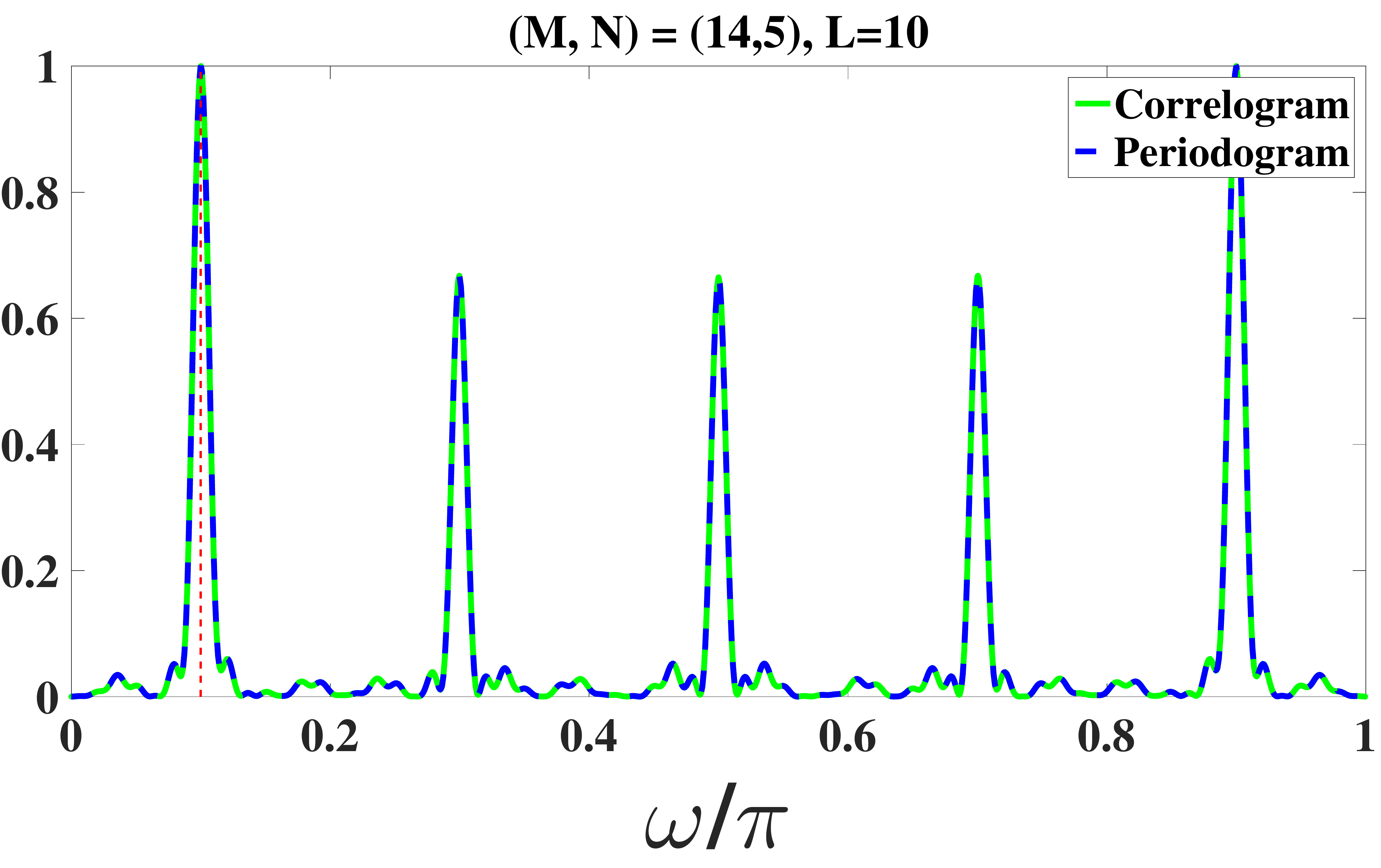}
	\includegraphics[width=0.33\textwidth]{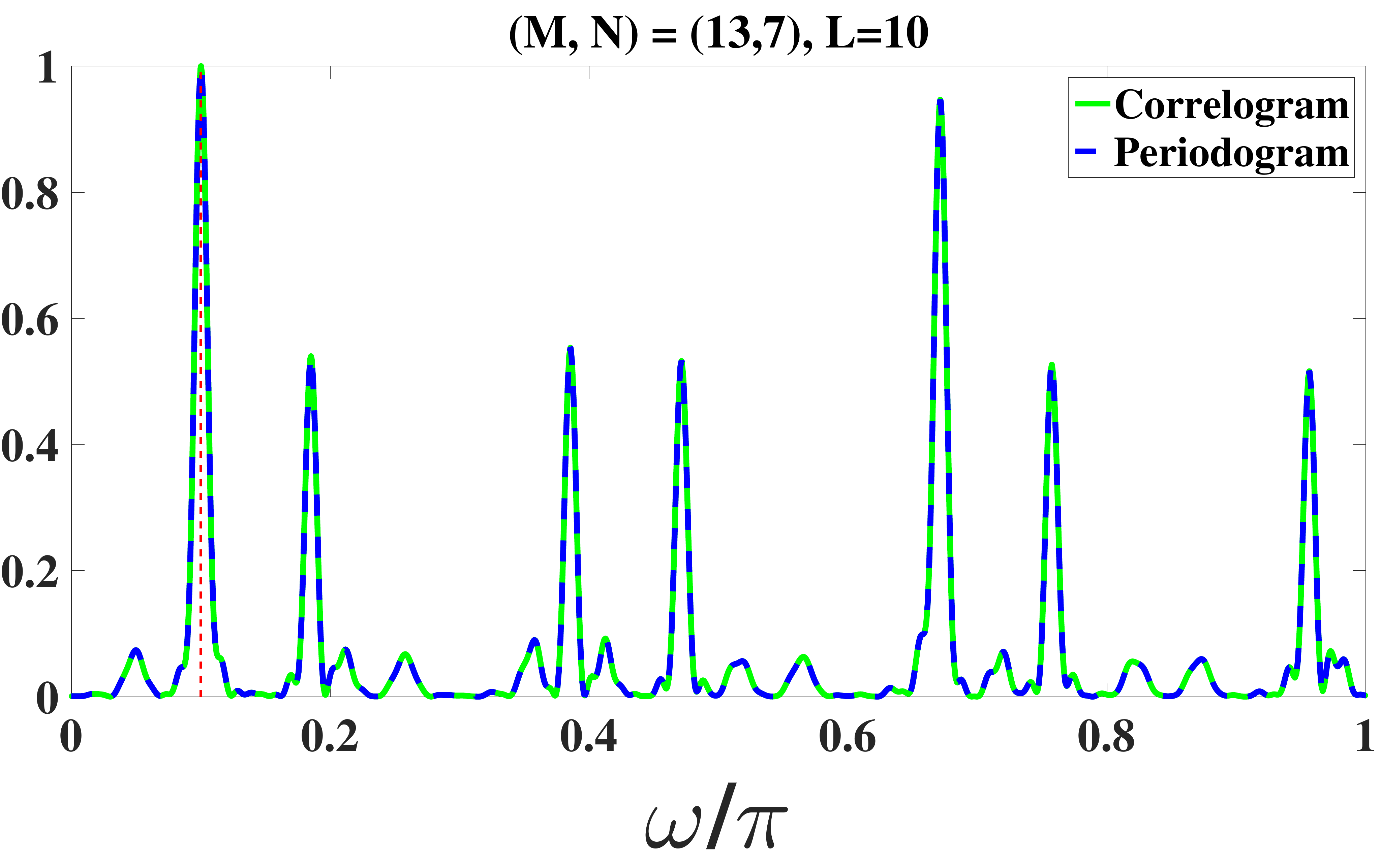}
	\hfil
	\caption{Simulation results for spectral estimation (1 peak) with number of snapshots $L=10$: (M, N) is (14, 13), (14, 5), (13, 7).}
	\label{fig:sim_1peak_1}
\end{figure*}

\begin{figure*}[!t]
	\includegraphics[width=0.33\textwidth]{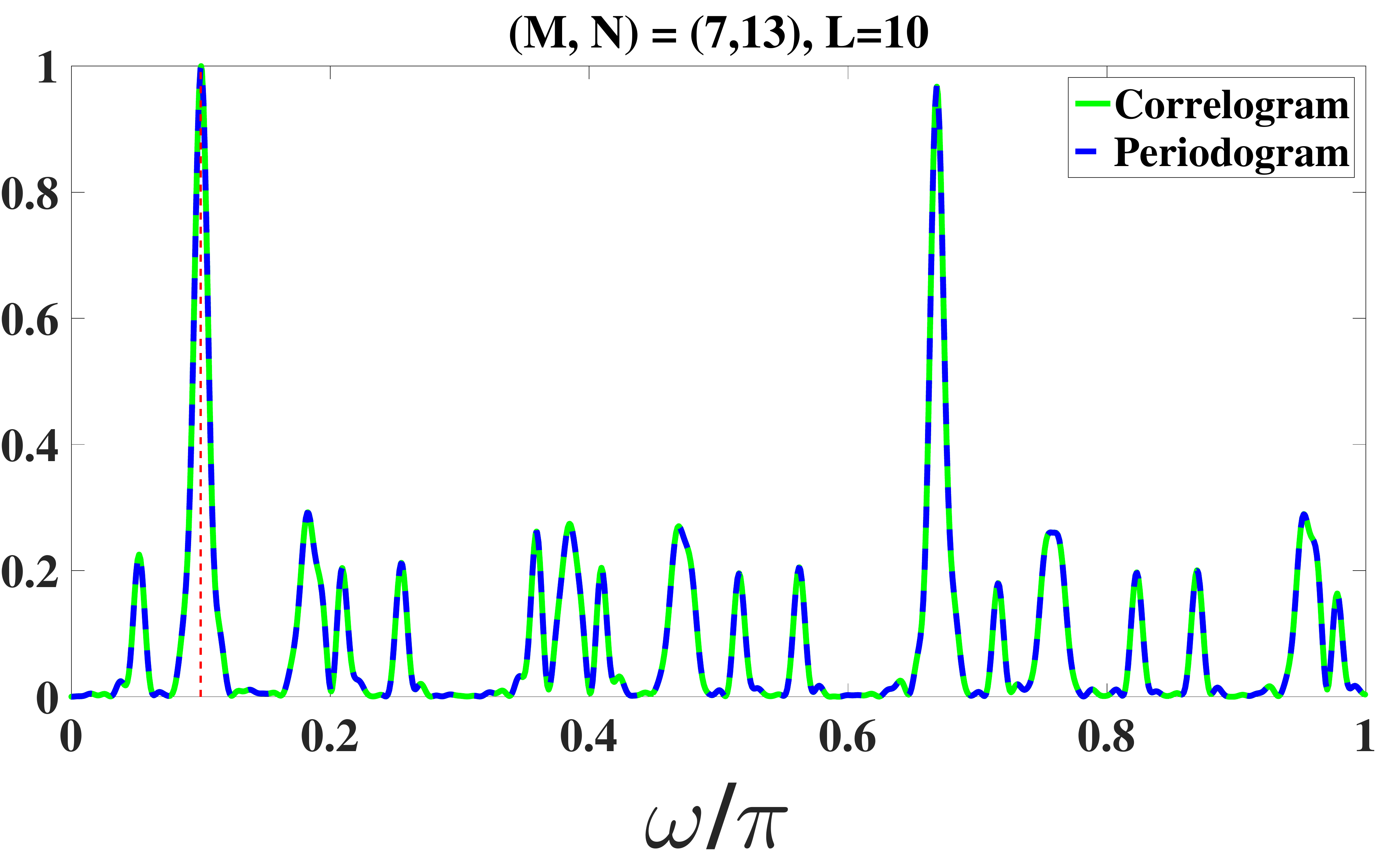}
	\includegraphics[width=0.33\textwidth]{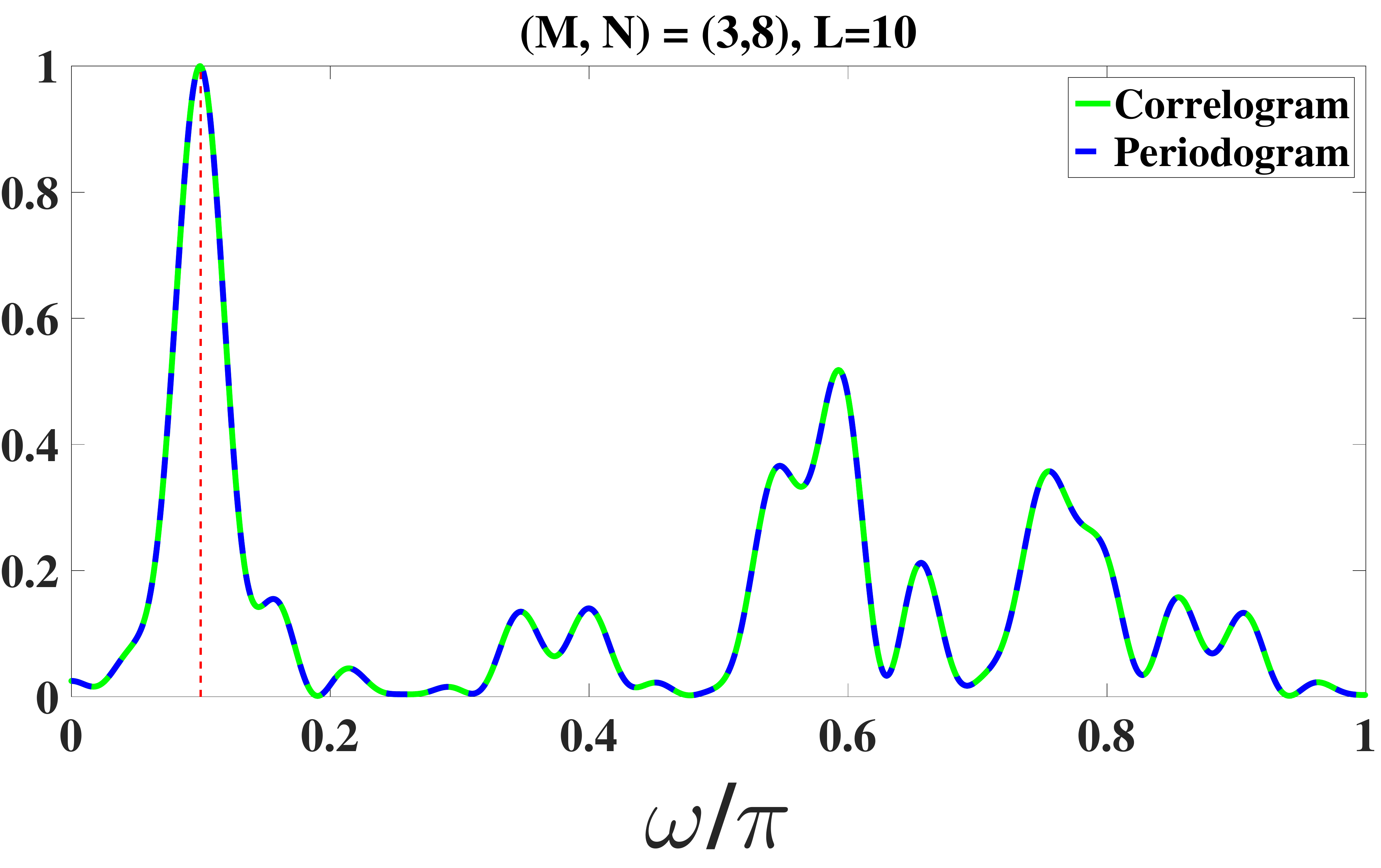}
	\includegraphics[width=0.33\textwidth]{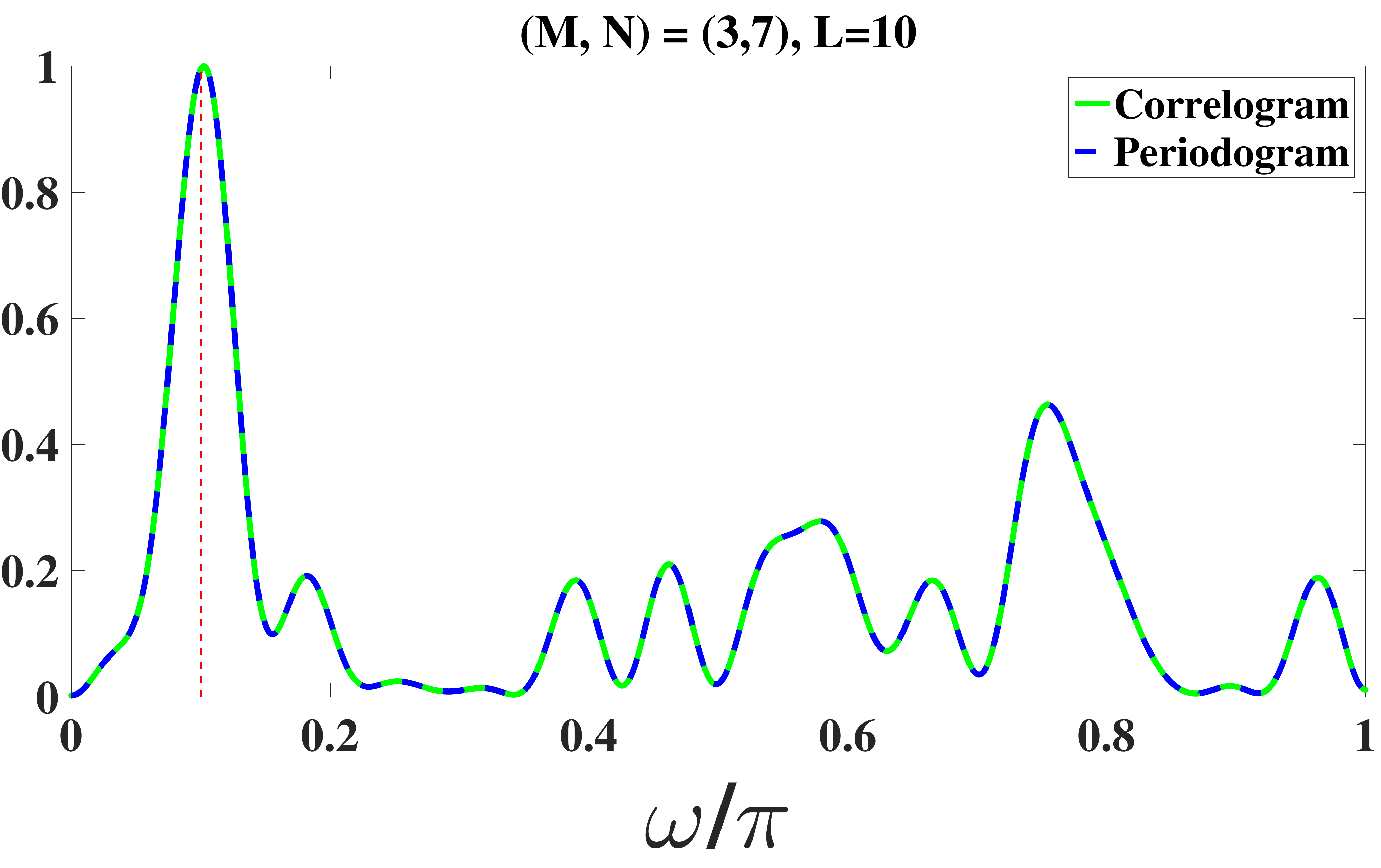}
	\caption{Simulation results for spectral estimation (1 peak) with number of snapshots $L=10$: (M, N) is (7, 13), (3, 8), (3, 7).}
	\label{fig:sim_1peak_2}
\end{figure*}

Simulations are performed on a temporal signal model described in~\cite{UVD_PHD,4.32}. The first step is to estimate the autocorrelation using the combined set. Next take the Fourier transform of the autocorrelation estimate to obtain the correlogram spectral estimate. Repeat this for $L$ snapshots and compute the average correlogram. It may be noted that instead of calculating the correlogram using the autocorrelation, the Fourier transform of the acquired signal with zeros inserted at the holes (missing difference values) can be calculated, i.e. Periodogram. The entire difference range is considered here, since it guarantees a valid power spectrum (Refer~\cite{UVD_PHD}). We demonstrate the results for snapshots, $L=10$. Fig.~\ref{fig:sim_1peak_1} and Fig.~\ref{fig:sim_1peak_2} demonstrates single peak estimation for different values of $(M, N)$. Several spurious peaks are observed. Note that $M\approx \frac{N}{2}$ seems to be a good choice. Furthermore, $(M, N)=(3, 7)$ or $(3, 8)$ works better.

\begin{figure*}[!t]
	\includegraphics[width=0.33\textwidth]{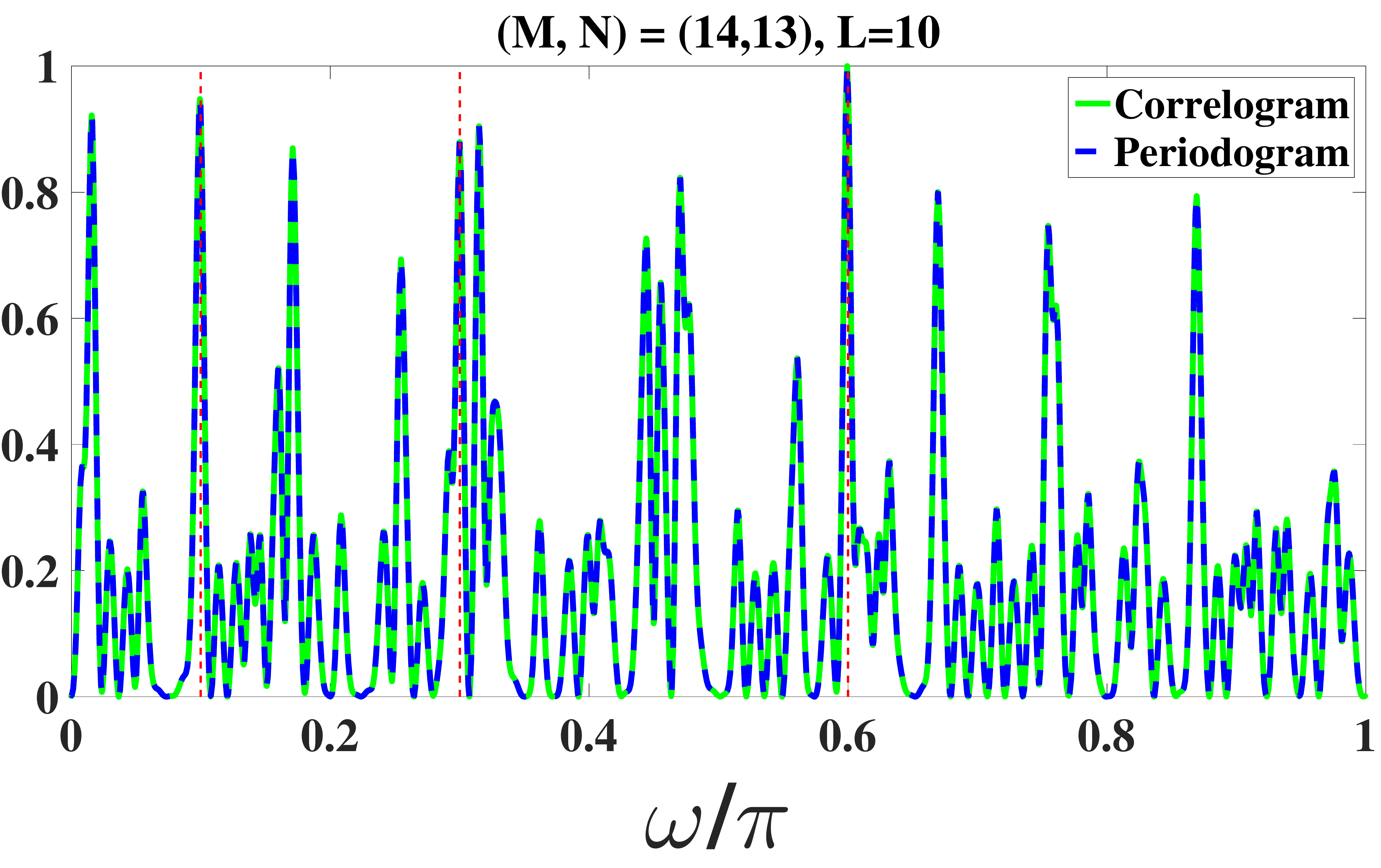}
	\includegraphics[width=0.33\textwidth]{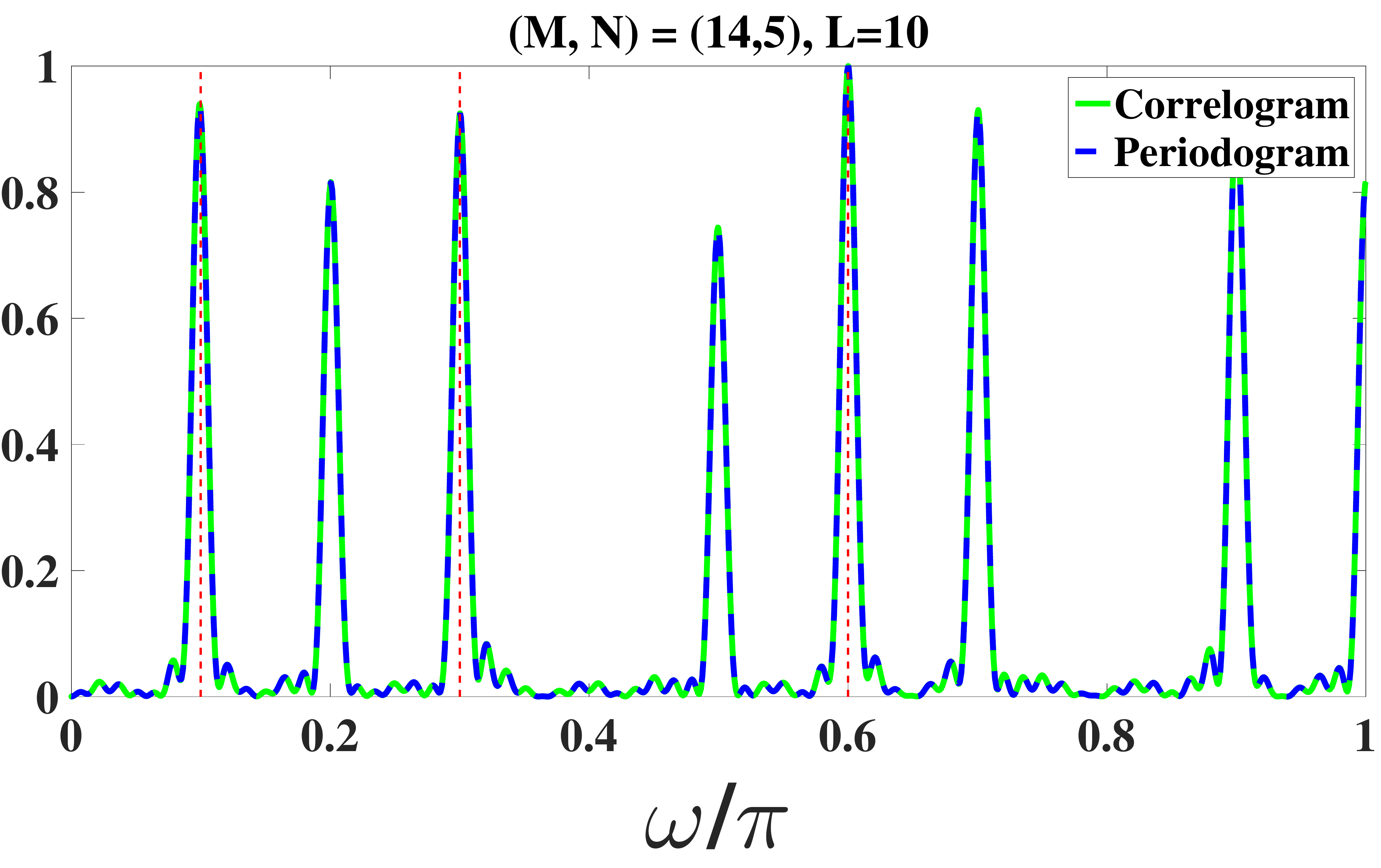}
	\includegraphics[width=0.33\textwidth]{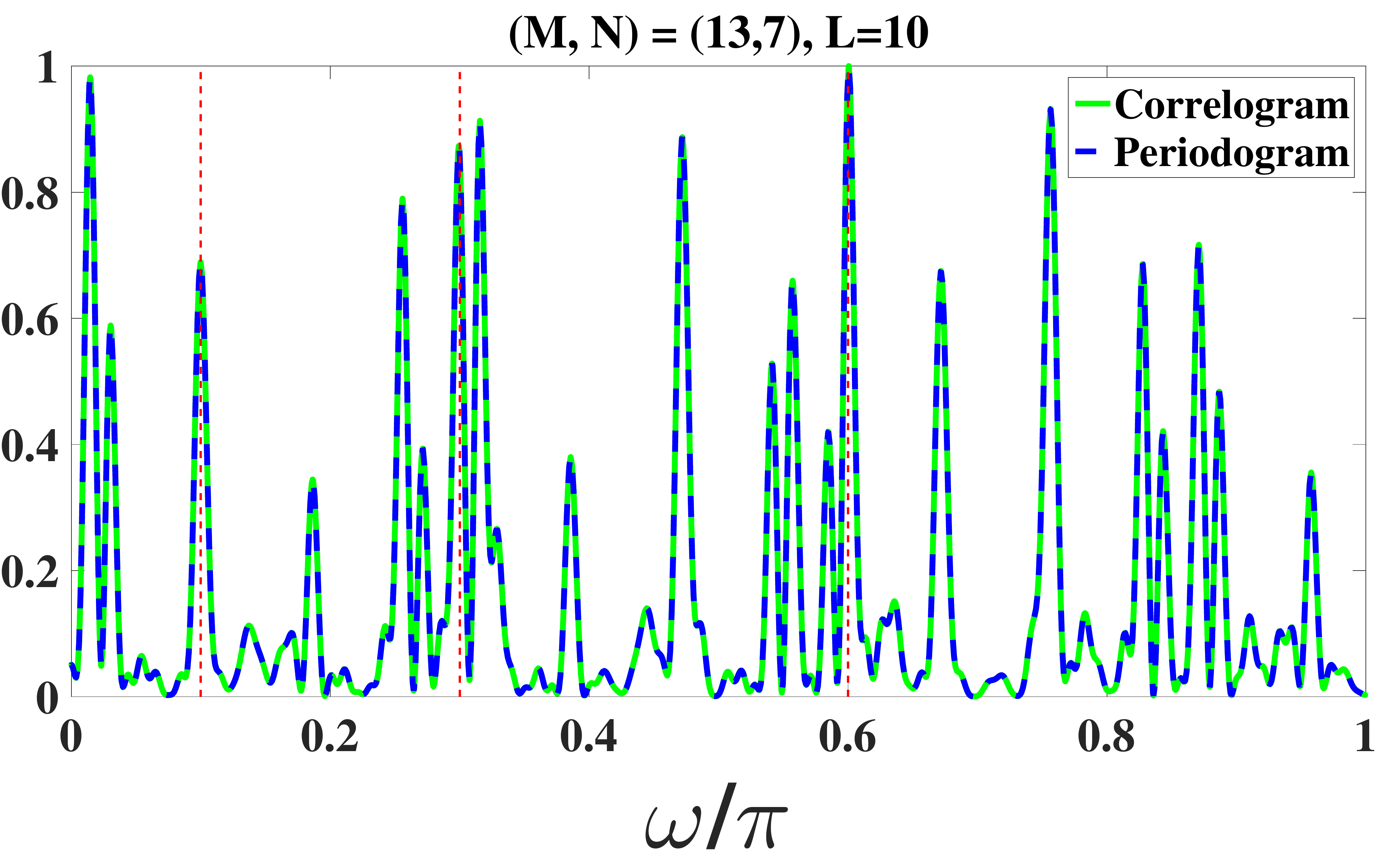}
	\caption{Simulation results for spectral estimation (3 peaks) with number of snapshots $L=10$: (M, N) is (14, 13), (14, 5), (13, 7).}
	\label{fig:sim_3peak_1}
\end{figure*}

\begin{figure*}[!t]
	\includegraphics[width=0.33\textwidth]{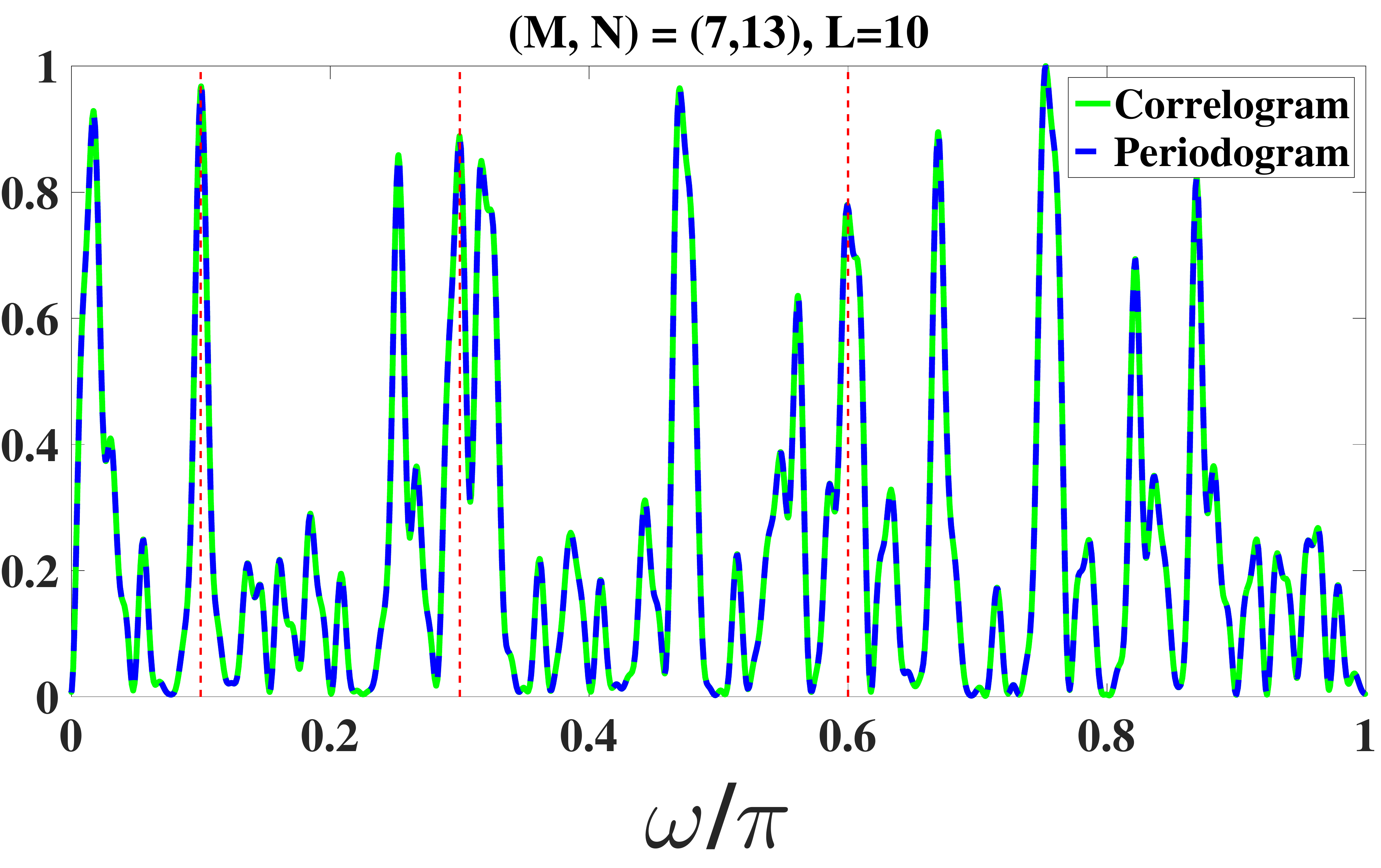}
	\includegraphics[width=0.33\textwidth]{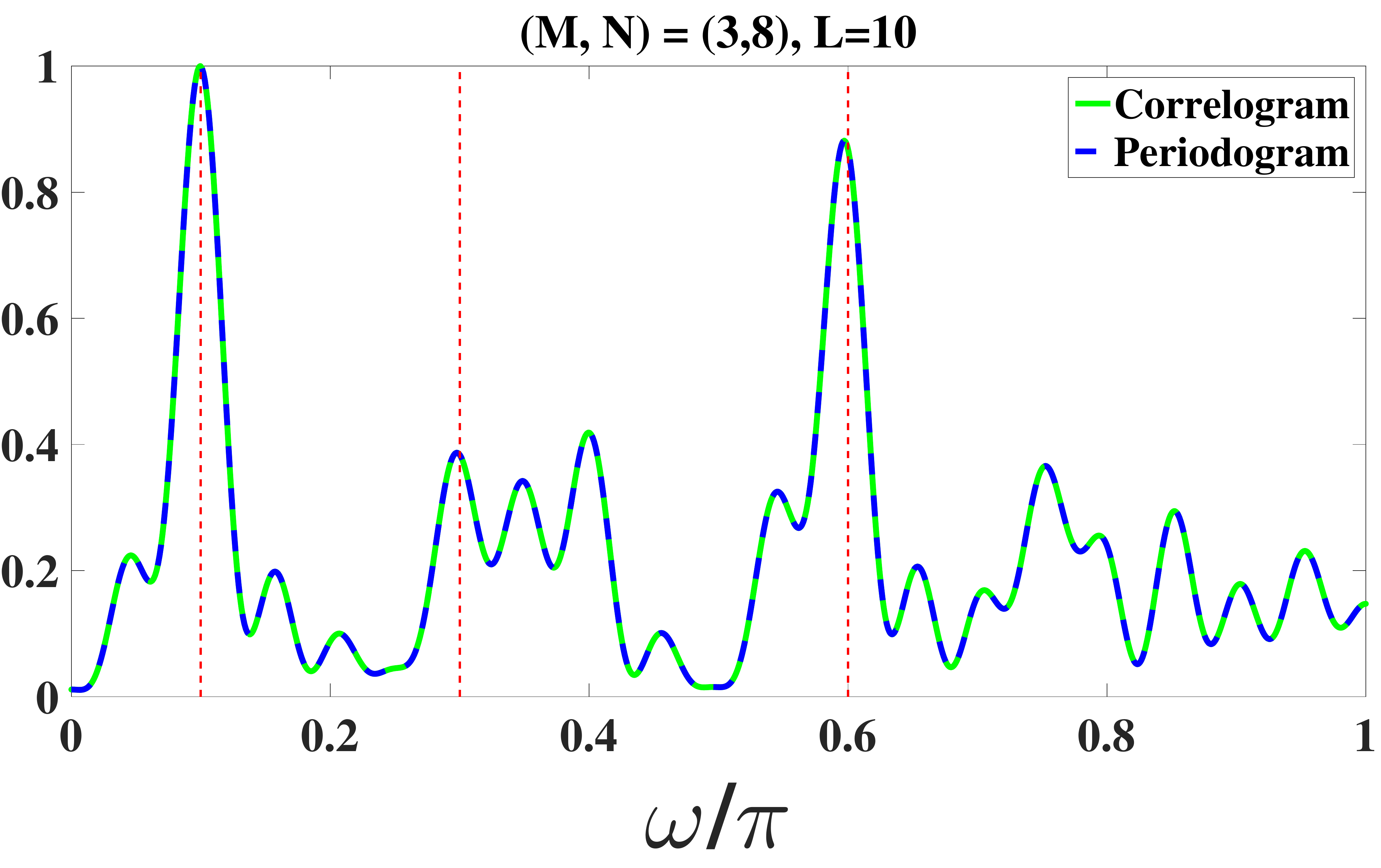}
	\includegraphics[width=0.33\textwidth]{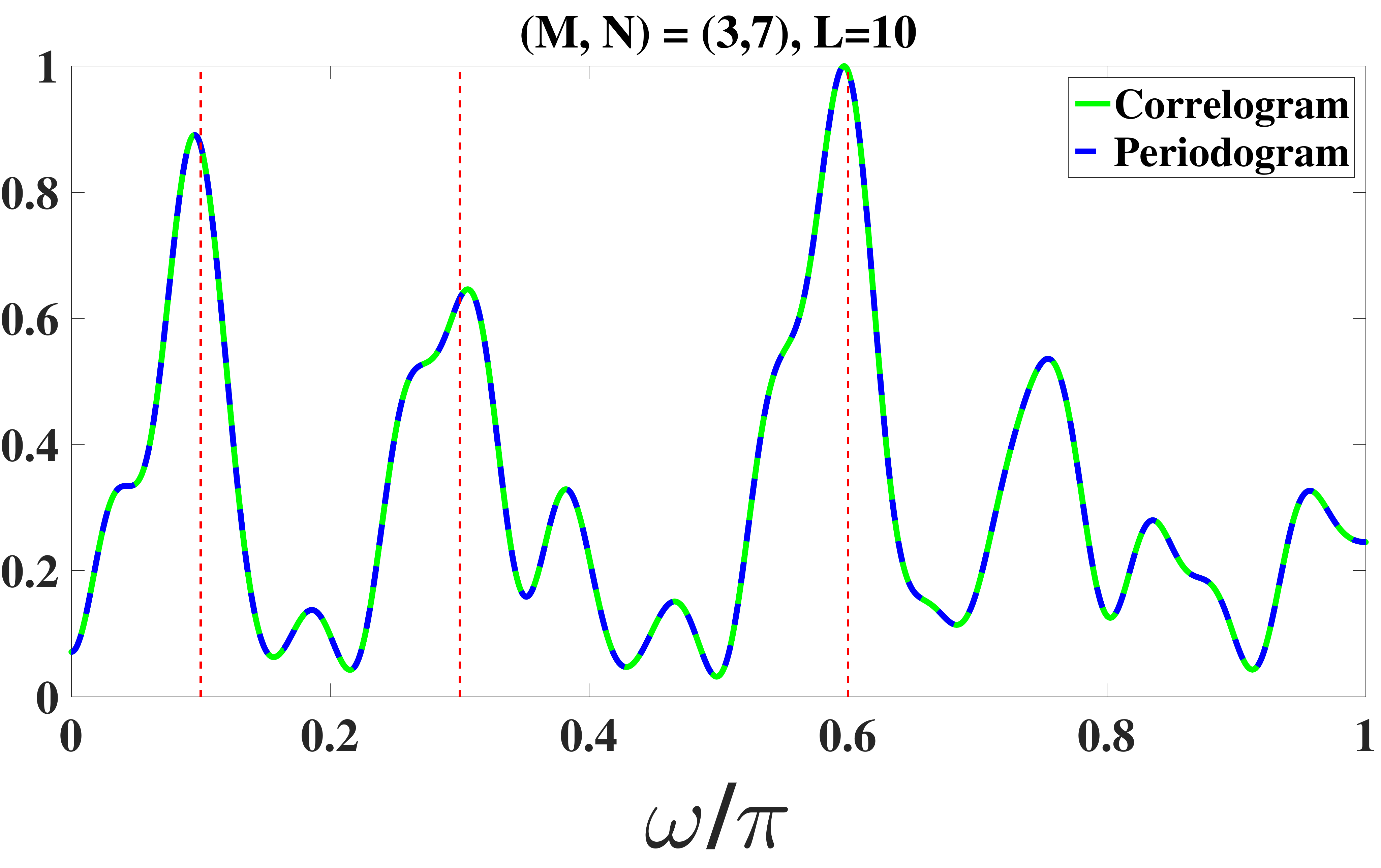}
	\caption{Simulation results for spectral estimation (3 peaks) with number of snapshots $L=10$: (M, N) is (7, 13), (3, 8), (3, 7).}
	\label{fig:sim_3peak_2}
\end{figure*}

Next, consider Fig.~\ref{fig:sim_3peak_1} and~\ref{fig:sim_3peak_2} for the estimation of three spectral peaks. Most of the examples considered here, have spurious peaks. $(M, N)=(3, 7)$ seems to be a good choice. Based on the analysis in this paper, we have the following thoughts:
\begin{enumerate}
	\item On the basis of the bias window expression and the examples considered, we find that $M\approx \frac{N}{2}$ seems to be a good choice to reduce the side-lobes for the extended co-prime arrays.
	\item The simulation results indicate that all values of $M\approx \frac{N}{2}$ will not work. For example, we find lower valued integer, i.e. $(M, N)=(3, 7)$ to be good.
	\item Since the correlogram spectral estimate is the convolution of the bias wiondow with the true spectrum (in a statistical sense), we conclude that the accuracy will also depend on the signal model. For example, $(M, N)=(3, 8)$ works for single peak estimation in Fig.~\ref{fig:sim_1peak_2} but fails for three peak estimation in Fig.~\ref{fig:sim_3peak_2}.
\end{enumerate}

In the next section, concluding remarks and possible questions for future research are considered.
%
%
%
\section{Conclusion}
The fundamentals of the extended co-prime array was developed from a difference set perspective. The closed-form expression for the weight function and the bias of the correlogram spectral estimate is provided. This was derived for the entire difference set, continuous set and the set containing difference values in the prototype co-prime range. Simulations validate the correctness of the derived equations for all the three cases.
The bias equations derived are valid for $M>N$ as well as $N>M$. However, it was found that the choice of $N>M$ provided a large relative amplitude between the main lobe and the side-lobe peaks. Since the structure assumed in this paper extends the array with inter-element spacing $Nd$, it can be concluded that the array with larger inter-element spacing should be extended. Furthermore, $M\approx \frac{N}{2}$ seems to be a good choice. We also analyze the variance of the correlogram estimate, and derive the number of multiplications and additions required for autocorrelation estimation. Low latency spectral estimation is demonstrated using the correlogram method. 

The bias window expression describes the distortion in the estimate. This throws up several challenges which needs further investigation. For example, can the bias distortion be reduced? What could be the ways to achieve this? In addition, other spectral estimation methods can be investigated. Several applications can be explored for low latency estimation along similar lines.
\ifCLASSOPTIONcaptionsoff
  \newpage
\fi

\bibliographystyle{IEEEtran}
\bibliography{refs}

\end{document}